\documentclass[prd,twocolumn,showpacs,showkeys,preprintnumbers,floatfix,nofootinbib,superscriptaddress]{revtex4-1}
\usepackage{amsfonts} 
\usepackage{amssymb} 
\usepackage{amsmath} 
\usepackage{graphicx} 
\usepackage{subfigure} 
\usepackage{array} 
\usepackage{dcolumn} 
\usepackage{lipsum}
\usepackage{bm} 
\usepackage{latexsym} 
\usepackage{longtable} 
\usepackage{bbold}
\usepackage{color}
\usepackage{multirow}
\usepackage{rotating}
\usepackage{comment}

\graphicspath{{./figs/}}

\newcommand{\tsep}{\mathop{\tau}\nolimits}
\newcommand{\tsepi}{\mathop{\tau \to \infty}\nolimits}
\newcommand{\tskip}{\mathop{t_{\rm skip}}\nolimits}

\newcommand{\GeV}{\mathop{\rm GeV}\nolimits}

\newcommand{\lsim}{\raisebox{-0.7ex}{$\stackrel{\textstyle <}{\sim}$ }}

\DeclareMathOperator{\Tr}{Tr}

\providecommand{\matrixe}[3]{\langle#1\lvert#2\rvert#3\rangle}

\definecolor{green}{rgb}{0.1, 0.8, 0.1}

\begin{document}


\title{Isovector Charges of the Nucleon from 2+1+1-flavor Lattice QCD}
\author{Rajan Gupta}
\email{rajan@lanl.gov}
\affiliation{Los Alamos National Laboratory, Theoretical Division T-2, Los Alamos, New Mexico 87545}

\author{Yong-Chull Jang}
\email{ypj@bnl.gov}
\affiliation{Brookhaven National Laboratory, Physics Department, Upton, New York 11973}

\author{Boram Yoon}
\email{boram@lanl.gov}
\affiliation{Los Alamos National Laboratory, Theoretical Division T-2, Los Alamos, New Mexico 87545}

\author{Huey-Wen~Lin}
\email{hwlin@pa.msu.edu}
\affiliation{Department of Physics and Astronomy, Michigan State University, East Lansing, Michigan, 48824, USA}
\affiliation{Department of Computational Mathematics,  Science and Engineering, Michigan State University, East Lansing, Michigan 48824,USA}

\author{Vincenzo Cirigliano}
\email{cirigliano@lanl.gov}
\affiliation{Los Alamos National Laboratory, Theoretical Division T-2, Los Alamos, New Mexico 87545}

\author{Tanmoy Bhattacharya}
\email{tanmoy@lanl.gov}
\affiliation{Los Alamos National Laboratory, Theoretical Division T-2, Los Alamos, New Mexico 87545}

\collaboration{Precision Neutron Decay Matrix Elements (PNDME) Collaboration}
\preprint{LA-UR-18-25335}
\preprint{MSUHEP-18-011}
\date{\today}
\begin{abstract}
We present high statistics results for the isovector
charges $g^{u-d}_A$, $g^{u-d}_S$ and
$g^{u-d}_T$ of the nucleon. Calculations were carried
out on eleven ensembles of gauge configurations generated by the MILC
collaboration using highly improved staggered quarks (HISQ) action
with 2+1+1 dynamical flavors. These ensembles span four lattice
spacings $a \approx$ 0.06, 0.09, 0.12 and 0.15~fm and light-quark
masses corresponding to $M_\pi \approx$ 135, 225 and
315~MeV. Excited-state contamination in the nucleon 3-point
correlation functions is controlled by including up to three-states in
the spectral decomposition. Remaining systematic uncertainties
associated with lattice discretization, lattice volume and light-quark
masses are controlled using a simultaneous fit in these three
variables.  Our final estimates of the isovector charges in the
$\overline{\text{MS}}$ scheme at 2 GeV are $g_A^{u-d} =
1.218(25)(30)$, $g_S^{u-d} = 1.022(80)(60) $ and $g_T^{u-d} =
0.989(32)(10)$. The first error includes statistical and all
systematic uncertainties except that due to the extrapolation ansatz,
which is given by the second error estimate.  We provide a detailed
comparison with the recent result of $g_A^{u-d} = 1.271(13)$ by the
CalLat collaboration and argue that our error estimate is more
realistic.  Combining our estimate for $g_S^{u-d}$ with the difference
of light quarks masses $(m_d-m_u)^{\rm QCD}=2.572(66)$~MeV given by
the MILC/Fermilab/TUMQCD collaboration for 2+1+1-flavor theory, we
obtain $(M_N-M_P)^{\rm QCD} = 2.63(27)$~MeV.  We update the low-energy
constraints on novel scalar and tensor interactions, $\epsilon_{S}$
and $\epsilon_{T}$, at the TeV scale by combining our new estimates
for $g^{u-d}_S$ and $g^{u-d}_T$ with precision low-energy nuclear
experiments, and find them comparable to those from the ATLAS and the CMS
experiments at the LHC.\looseness-1

\end{abstract}
\maketitle
%
%
%
%
\section{Introduction}
\label{sec:into}

The axial, scalar and tensor charges of the nucleon are needed to
interpret the results of many experiments and probe new physics. In
this paper, we extend the calculations presented in
Refs.~\cite{Bhattacharya:2015wna,Bhattacharya:2015esa,Bhattacharya:2016zcn}
by analyzing eleven ensembles of $2+1+1$ flavors of highly improved
staggered quarks (HISQ)~\cite{Follana:2006rc} generated by the MILC
collaboration~\cite{Bazavov:2012xda}.  These now include a second
physical mass ensemble at $a=0.06$~fm, and an ensemble with
$a=0.15$~fm and $M_\pi \approx 310$~MeV. We have also increased the
statistics significantly on six other ensembles using the truncated
solver with bias correction method~\cite{Bali:2009hu,Blum:2012uh}. The
resulting high-statistics data provide better control over various
sources of systematic errors, in particular the two systematics: (i)
excited-state contamination (ESC) in the extraction of the
ground-state matrix elements of the various quark bilinear operators
and (ii) the reliability of the chiral-continuum-finite volume (CCFV)
extrapolation used to obtain the final results that can be compared to
phenomenological and experimental values.  With improved simultaneous
CCFV fits, we obtain $g_A^{u-d} =1.218(25)(30)$, $g_S^{u-d}
=1.022(80)(60)$ and $g_T^{u-d} = 0.989(32)(10)$ for the isovector
charges in the $\overline{MS}$ scheme at 2~GeV. The first error
includes statistical and all systematic uncertainties except that due
to the ansatz used for the final CCFV extrapolation, which is given by
the second error estimate.  We also update our estimates for the
connected contributions to the flavor diagonal charges $g_{A,T}^{u}$
and $g_{A,T}^{d} $, and the isoscalar combination $g_T^{u+d} $.
Throughout the paper, we present results for the charges of the
proton, which by convention are called nucleon charges in the
literature.  From these, results for the neutron, in our isosymmetric
formulation with $m_u = m_d$, are obtained by the $u \leftrightarrow d$
interchange.

The axial charge, $g_A^{u-d}$, is an important parameter that
encapsulates the strength of weak interactions of nucleons. It enters
in many analyses of nucleon structure and of the Standard Model (SM)
and beyond-the-SM (BSM) physics. For example, it impacts the
extraction of the Cabibbo-Kobayashi-Maskawa (CKM) matrix element
$V_{ud}$, tests the unitarity of the CKM matrix, and is needed for the
analysis of neutrinoless double-beta decay.  Also,
the rate of proton-proton fusion, the first step in the thermonuclear
reaction chains that power low-mass hydrogen-burning stars like the
Sun, is sensitive to it. The current best determination of the ratio
of the axial to the vector charge, $g_A/g_V$, comes from measurement
of neutron beta decay using polarized ultracold neutrons (UCN) by the UCNA
collaboration, $1.2772(20)$~\cite{Mendenhall:2012tz,Brown:2017mhw}, and by PERKEO
II, $1.2761{}^{+14}_{-17}$~\cite{Mund:2012fq}. Note that, in the SM,
$g_V=1$ up to second order corrections in isospin
breaking~\cite{Ademollo:1964sr,Donoghue:1990ti} as a result of the
conservation of the vector current.

Given the accuracy with which $g_A^{u-d}$ has been measured in
experiments, our goal is to calculate it directly with $O(1\%)$
accuracy using lattice QCD.  The result presented in this paper,
$g_A^{u-d}=1.218(25)(30)$, is, however, about $1.5\sigma$ ($5\%$)
smaller than the experimental value.  In Sec.~\ref{sec:comparison}, we
compare with the result $g_A^{u-d} = 1.271(13)$ by the CalLat
collaboration. We show that the data on seven HISQ
ensembles analyzed by both collaborations agree within $1\sigma$ and
the final difference is due to the chiral and continuum
extrapolation--the fits are weighted differently by the data points
that are not common. Based on the analysis of the size of the various
systematics in Sec.~\ref{sec:errors}, and on the comparison with
CalLat calculation, we conclude that our analysis of errors is
realistic. Our goal, therefore, is to continue to quantify and control
the various sources of errors to improve precision. 


The Standard Model does not contain fundamental scalar or tensor
interactions. However, loop effects and new interactions at the TeV
scale can generate effective interactions at the hadronic scale that
can be probed in decays of neutrons, and at the TeV scale itself at
the LHC.  Such scalar and tensor interactions contribute to the
helicity-flip parameters $b$ and $b_\nu$ in the neutron decay
distribution~\cite{Bhattacharya:2011qm}. Thus, by combining the
calculation of the scalar and tensor charges with the measurements of
$b$ and $b_\nu$ in low energy experiments, one can put constraints on
novel scalar and tensor interactions at the TeV scale as described in
Ref.~\cite{Bhattacharya:2011qm}.  To optimally bound such scalar and
tensor interactions using measurements of $b$ and $b_\nu$ parameters
in planned experiments targeting $10^{-3}$
precision~\cite{abBA,WilburnUCNB,Pocanic:2008pu}, the level of
precision required in $g_S^{u-d}$ and $g_T^{u-d}$ is at the $10\%$
level as explained in
Refs.~\cite{Bhattacharya:2011qm,abBA,WilburnUCNB,Pocanic:2008pu}.
Future higher-precision measurements of $b$ and $b_\nu$ would require
correspondingly higher-precision calculations of the matrix elements
to place even more stringent bounds on TeV-scale couplings.

In a recent work~\cite{Bhattacharya:2015wna}, we showed that
lattice-QCD calculations have reached a level of control over all
sources of systematic errors needed to yield the tensor charge with
the required precision. The errors in the scalar 3-point functions are
about a factor of 2 larger. In this paper we show that by using the
truncated solver method with bias
correction~\cite{Bali:2009hu,Blum:2012uh}, (for brevity called TSM
henceforth), to obtain high statistics on all ensembles, we are also
able to control the uncertainty in $g_S^{u-d}$ to the required 10\%
level. These higher-statistics results also improve upon our previous
estimates of the axial and the tensor charges.

The matrix elements of the flavor-diagonal tensor operators are needed
to quantify the contributions of the $u,\ d, \ s, \ c$ quark electric
dipole moments (EDM) to the neutron electric dipole moment
(nEDM)~\cite{Bhattacharya:2015wna,Pospelov:2005pr}. The nEDM is a very
sensitive probe of new sources of $T$ and $CP$ violation that arise in
most extensions of the Standard Model designed to explain nature at
the TeV scale. Planned experiments aim to reduce the current bound on
the nEDM of $2.9 \times 10^{-26}\ e$~cm~\cite{Baker:2006ts} to around
$ 10^{-28}\ e$~cm. Improving the bound will put stringent constraints on many BSM
theories provided the matrix elements of novel $CP$-violating
interactions, of which the quark EDM is one, are calculated with the
required precision. In
Refs.~\cite{Bhattacharya:2015wna,Bhattacharya:2016zcn}, we showed that
the disconnected contributions are negligible so we update the
connected contributions to the flavor diagonal tensor charges for the
light $u$ and $d$ quarks that are taken to be degenerate.

The tensor charges are also extracted as the zeroth moment of the
transversity distributions, These are measured in many experiments
including Drell-Yan and semi-inclusive deep inelastic scattering
(SIDIS) and describe the net transverse polarization of quarks in a
transversely polarized nucleon. There exists an active program at
Jefferson Lab (JLab) to measure them~\cite{Dudek:2012vr}. It is,
however, not straightforward to extract the transversity distributions
from the data taken over a limited range of $Q^2$ and Bjorken $x$,
consequently additional phenomenological modeling is required. Lattice QCD 
results for $g_T^{u}$, $g_T^{d}$, $g_T^{s}$ and $g_T^{u-d}$
are the most accurate at present as already discussed in
Ref.~\cite{Bhattacharya:2016zcn}.  Future experiments at JLab and
other experimental facilities worldwide will significantly improve the
extraction of the transversity distributions, and together with
accurate calculations of the tensor charges using lattice QCD
elucidate the structure of the nucleon in terms of quarks and gluons.

The methodology for calculating the isovector charges in an isospin
symmetric theory, that is, measuring the contribution to the matrix
elements of the insertion of the zero-momentum bilinear quark
operators in one of the three valence quarks in the nucleon, is well
developed~\cite{Bhattacharya:2015wna,Bhattacharya:2015esa,Bhattacharya:2016zcn,Lin:2012ev,Syritsyn:2014saa,Constantinou:2014tga}.
Calculation of the flavor-diagonal charges is similar except that it
gets additional contributions from contractions of the operator as a
vacuum quark loop that interacts with the nucleon propagator through
the exchange of gluons.  In Ref.~\cite{Bhattacharya:2015wna}, we
showed that these contributions to $g_T^{u,d,s}$ are small, $O(0.01)$,
and consistent with zero within errors. Thus, within current error
estimates, the connected contributions alone provide reliable
estimates for the flavor diagonal charges $g_{T}^{u,d} $ and the
isoscalar combination $g_T^{u+d} $. A detailed analysis 
of disconnected contributions to the axial, scalar and tensor charges
will be presented in a separate paper.

This paper is organized as follows. In Sec.~\ref{sec:Methodology}, we
describe the parameters of the gauge ensembles analyzed and the
lattice methodology. The fits used to isolate excited-state
contamination are described in Sec.~\ref{sec:excited}.  The
renormalization of the operators is discussed in
Sec.~\ref{sec:renorm}. Our final results for the isovector charges and
the connected parts of the flavor-diagonal charges are presented in
Sec.~\ref{sec:results}.  Our estimation of errors is revisited
in Sec.~\ref{sec:errors}, and a comparison with previous works is given
in Sec.~\ref{sec:comparison}. In Sec.~\ref{sec:est}, we provide
constraints on novel scalar and tensor interactions at the TeV scale
using our new estimates of the charges and precision beta decay experiments and
compare them to those from the LHC.  Our final conclusions are
presented in Sec.~\ref{sec:conclusions}.

\section{Lattice Methodology}
\label{sec:Methodology}

\begin{table*}[tbp]    
\begin{center}
\renewcommand{\arraystretch}{1.2} 
\begin{ruledtabular}
\begin{tabular}{l|ccc|cc|cccc}
Ensemble ID & $a$ (fm) & $M_\pi^{\rm sea}$ (MeV) & $M_\pi^{\rm val}$ (MeV) & $L^3\times T$    & $M_\pi^{\rm val} L$ & $\tau/a$ & $N_\text{conf}$  & $N_{\rm meas}^{\rm HP}$  & $N_{\rm meas}^{\rm LP}$  \\
\hline
$a15m310 $      & 0.1510(20) & 306.9(5) & 320.6(4.3)     & $16^3\times 48$ & 3.93 &  $\{5,6,7,8,9\}$    & 1917 & 7668  & 122,688   \\
\hline
$a12m310 $      & 0.1207(11) & 305.3(4) & 310.2(2.8) & $24^3\times 64$ & 4.55 &  $\{8,10,12\}$      & 1013 & 8104  &  64,832   \\
$a12m220S$      & 0.1202(12) & 218.1(4) & 225.0(2.3) & $24^3\times 64$ & 3.29 & $\{8, 10, 12\}$     & 946  & 3784  &  60,544   \\
$a12m220 $      & 0.1184(10) & 216.9(2) & 227.9(1.9) & $32^3\times 64$ & 4.38 & $\{8, 10, 12\}$     & 744  & 2976  &  47,616   \\
$a12m220L_O$    & 0.1189(09) & 217.0(2) & 227.6(1.7) & $40^3\times 64$ & 5.49 & $\{8,10,12,14\}$    & 1010 & 8080  &  68,680   \\
$a12m220L$      &            &          &            &                 &      & $\{8,10,12,14\}$    & 1000 & 4000  & 128,000   \\
\hline                                                                                                         
$a09m310 $      & 0.0888(08) & 312.7(6) & 313.0(2.8) & $32^3\times 96$ & 4.51 & $\{10,12,14,16\}$   & 2263 & 9052  & 114,832   \\
$a09m220 $      & 0.0872(07) & 220.3(2) & 225.9(1.8) & $48^3\times 96$ & 4.79 & $\{10,12,14,16\}$   & 964  & 7712  & 123,392   \\
$a09m130 $      & 0.0871(06) & 128.2(1) & 138.1(1.0) & $64^3\times 96$ & 3.90 & $\{10,12,14\}$      & 883  & 7064  &  84,768   \\
$a09m130W$      &            &          &            &                 &      & $\{8,10,12,14,16\}$ & 1290 & 5160  & 165,120   \\
\hline                                                                                                         
$a06m310 $      & 0.0582(04) & 319.3(5) & 319.6(2.2) & $48^3\times 144$& 4.52 & $\{16,20,22,24\}$   & 1000 & 8000  &  64,000   \\
$a06m310W$      &            &          &            &                 &      & $\{18,20,22,24\}$   & 500  & 2000  &  64,000   \\
$a06m220 $      & 0.0578(04) & 229.2(4) & 235.2(1.7) & $64^3\times 144$& 4.41 & $\{16,20,22,24\}$   & 650  & 2600  &  41,600   \\
$a06m220W$      &            &          &            &                 &      & $\{18,20,22,24\}$   & 649  & 2596  &  41,546   \\
$a06m135 $      & 0.0570(01) & 135.5(2) & 135.6(1.4) & $96^3\times 192$& 3.7  & $\{16,18,20,22\}$   & 675  & 2700  &  43,200   \\
\end{tabular}
\end{ruledtabular}
\caption{Parameters, including the Goldstone pion mass
  $M_\pi^{\rm sea}$, of the eleven 2+1+1- flavor HISQ ensembles generated
  by the MILC collaboration and analyzed in this study are quoted from
  Ref.~\cite{Bazavov:2012xda}.  All fits are made versus $M_\pi^{\rm
    val}$ and finite-size effects are analyzed in terms of $M_\pi^{\rm
    val} L$.  Estimates of $M_\pi^{\rm val}$, the clover-on-HISQ pion
  mass, are the same as given in Ref.~\cite{Bhattacharya:2015wna} and
  the error is governed mainly by the uncertainty in the lattice
  scale. In the last four columns, we give, for each ensemble, the
  values of the source-sink separation $\tau$ used in the
  calculation of the three-point functions, the number of
  configurations analyzed, and the number of measurements made using
  the high precision (HP) and the low precision (LP) truncation of the
  inversion of the clover operator.  The second set of calculations,
  $a09m130W$, $a06m310W$ and $a06m220W$, have been done with the
  larger smearing size $\sigma$ that is given in
  Table~\protect\ref{tab:cloverparams}. The new $a12m220L$ simulations
  replace $a12m220L_O$ for reasons explained in the text.}
\label{tab:ens}
\end{center}
\end{table*}

\begin{table}[htbp]  
\centering
\begin{ruledtabular}
\begin{tabular}{l|lc|c|c}
\multicolumn1c{ID}  & \multicolumn1c{$m_l$} &  $c_{\text{SW}}$ & Smearing    & RMS smearing \\
                    &                       &                  & Parameters  & radius       \\
\hline
$a15m310 $          & $-0.0893$  & 1.05094 & \{4.2, 36\}   & 4.69  \\
\hline
$a12m310 $          & $-0.0695$  & 1.05094 & \{5.5, 70\}   & 5.96  \\
$a12m220S$          & $-0.075$   & 1.05091 & \{5.5, 70\}   & 5.98  \\ 
$a12m220 $          & $-0.075$   & 1.05091 & \{5.5, 70\}   & 5.96  \\ 
$a12m220L$          & $-0.075$   & 1.05091 & \{5.5, 70\}   & 5.96  \\ 
\hline                                                        
$a09m310 $          & $-0.05138$ & 1.04243 & \{7.0,100\}   & 7.48  \\
$a09m220 $          & $-0.0554$  & 1.04239 & \{7.0,100\}   & 7.48  \\
$a09m130 $          & $-0.058$   & 1.04239 & \{5.5, 70\}   & 6.11  \\
$a09m130W$          & $-0.058$   & 1.04239 & \{7.0,100\}   & 7.50  \\
\hline                                                        
$a06m310 $          & $-0.0398$  & 1.03493 & \{6.5, 70\}   & 7.22  \\
$a06m310W     $     & $-0.0398$  & 1.03493 & \{12, 250\}   & 12.19  \\
$a06m220 $          & $-0.04222$ & 1.03493 & \{5.5, 70\}   & 6.22  \\
$a06m220W     $     & $-0.04222$ & 1.03493 & \{11, 230\}   & 11.24  \\
$a06m135 $          & $-0.044$   & 1.03493 & \{9.0,150\}   & 9.56  \\
\end{tabular}
\end{ruledtabular}
\caption{The parameters used in the
  calculation of the clover propagators.  The hopping parameter for
  the light quarks, $\kappa_l$, in the clover action is given by
  $2\kappa_{l} = 1/(m_{l}+4)$.  $m_l$ is tuned to achieve $M_\pi^{\rm
    val} \approx M_\pi^\text{sea}$. The parameters used to construct
  Gaussian smeared sources, $\{\sigma, N_{\text{KG}}\}$, are given in
  the fourth column where $N_{\text{KG}}$ is the number of
  applications of the Klein-Gordon operator and the width of the
  smearing is controlled by the coefficient $\sigma$, both in Chroma
  convention~\cite{Edwards:2004sx}.  The resulting root-mean-square
  radius of the smearing, defined as $\sqrt{\int r^2 \sqrt{S^\dag S}
    dr /\int \sqrt{S^\dag S} dr} $, is given in the last column.  }
  \label{tab:cloverparams}
\end{table}

The parameters of the eleven ensembles used in the analysis are
summarized in Table~\ref{tab:ens}.  They cover a range of lattice
spacings ($0.06 \, \lsim a \, \lsim 0.15$~fm), pion masses ($135 \,
\lsim M_\pi \, \lsim 320$~MeV) and lattice sizes ($3.3\, \lsim M_\pi
L\, \lsim5.5$) and were generated using 2+1+1-flavors of HISQ
fermions~\cite{Follana:2006rc} by the MILC
collaboration~\cite{Bazavov:2012xda}.  Most of the details of the
methodology, and the strategies for the calculations and the analyses are the
same as described in
Refs.~\cite{Bhattacharya:2015wna,Bhattacharya:2016zcn}.  Here we will
summarize the key points to keep the paper self-contained and
highlight the new features and analysis.

We construct the correlation functions needed to calculate the matrix
elements using Wilson-clover fermions on these HISQ ensembles. Such
mixed-actions, clover-on-HISQ, are a nonunitary formulation and suffer
from the problem of exceptional configurations at small, but 
{\it a priori} unknown, quark masses. We monitor all correlation functions
for such exceptional configurations in our statistical samples.  For
example, evidence of exceptional configurations on three $a15m310$
lattices prevents us from analyzing ensembles with smaller $M_\pi$ at
$a = 0.15$~fm using the clover-on-HISQ approach. The same holds for
the physical mass ensemble $a12m130$.

The parameters used in the construction of
the 2- and 3-point functions with clover fermions are given in
Table~\ref{tab:cloverparams}. The Sheikholeslami-Wohlert
coefficient~\cite{Sheikholeslami:1985ij} used in the clover action is
fixed to its tree-level value with tadpole improvement, $c_\text{sw} =
1/u_0^3$, where $u_0$ is the fourth root of the plaquette expectation
value calculated on the hypercubic (HYP)
smeared~\cite{Hasenfratz:2001hp} HISQ lattices.

The masses of light clover quarks were tuned so that the
clover-on-HISQ pion masses, $M^{\rm val}_\pi$, match the HISQ-on-HISQ
Goldstone ones, $M_\pi^{\rm sea}$. Both estimates are given in
Table~\ref{tab:ens}. All fits in $M_\pi^2$ to study the chiral
behavior are made using the clover-on-HISQ $M^{\rm val}_{\pi}$ since
the correlation functions, and thus the chiral behavior of the
charges, have a greater sensitivity to it. Henceforth, for
brevity, we drop the superscript and denote the clover-on-HISQ pion
mass as $M_\pi$. Performing fits using the HISQ-on-HISQ values,
${M_\pi^{\rm sea}}$, does not change the estimates significantly.

The highlights of the current work, compared to
the results presented in Ref.~\cite{Bhattacharya:2016zcn}, are as follows:
\begin{itemize}
\item
The addition of a second physical pion mass ensemble $a06m135$ and 
the coarse $a15m310$ ensemble. 
\item
The new $a12m220L$ simulations replace the older $a12m220L_O$ data. In
the $a12m220L_O$ calculation, the HP analysis had only been done for
$\tau=10$, while in the new $a12m220L$ data the HP calculation has
been done for all values of source-sink separation $\tau$, and the
bias correction applied. We have also increased the number of LP
measurements on each configurations and both HP and LP source points
are chosen randomly within and between configurations. Even though the
results from the two calculations are consistent, as shown in
Tables~\ref{tab:2ptmulti},~\ref{tab:results3bareu-d}
and~\ref{tab:results3bareu+d}, nevertheless, for the two reasons
stated above, we will, henceforth, only use the $a12m220L$ data in the
analysis of the charges and other quantities in this and future
papers.
\item
All ensembles are analyzed using the TSM method with much higher statistics
as listed in Table~\ref{tab:ens}.  Our implementation of the TSM method is
described in Refs.~\cite{Bhattacharya:2015wna,Yoon:2016dij}.
\item
The new high statistics data for ensembles $a09m310$, $a09m220$ and
$a09m130W$ were generated using the smearing parameter
$\sigma=7$. This corresponds to a r.m.s. radius of $\approx 7.5$ in
lattice units or roughly 0.66~fm.  As discussed in Sec.~\ref{sec:excited} and 
shown in Figs.~\ref{fig:gA2v3a12}--\ref{fig:gT2v3a06}, 
increasing $\sigma$ from $5.5$ to $7.0$ reduces the ESC at a given
source-sink separation $\tau$.\looseness-1
\item
The two-point correlation functions are analyzed keeping up to four
states in the spectral decomposition. Previous work was based on
keeping two states.\looseness-1
\item
The three-point functions are analyzed keeping up to three states in
the spectral decomposition of the spectral functions. Previous work
was based on keeping two states.
\end{itemize}
We find that the new higher precision data significantly improved the
ESC fits and the final combined CCFV fit used to obtain results in the
limits $a \to 0$, the pion mass $M_\pi \to 135$~MeV and the
lattice volume $M_\pi L \to \infty$.

\subsection{Correlation Functions}
\label{sec:CorrelationFunctions}

We use the following interpolating operator $\chi$ to create$/$annihilate the nucleon
state: 
\begin{align}
 \chi(x) = \epsilon^{abc} \left[ {q_1^a}^T(x) C \gamma_5 
            \frac{(1 \pm \gamma_4)}{2} q_2^b(x) \right] q_1^c(x) \,, 
\label{eq:nucl_op}
\end{align}
with $\{a, b, c\}$ labeling the color indices, $C=\gamma_0 \gamma_2$ the charge
conjugation matrix, and $q_1$ and $q_2$ denoting the two different
flavors of light quarks.  The nonrelativistic projection $(1 \pm
\gamma_4)/2$ is inserted to improve the signal, with the plus and
minus signs applied to the forward and backward propagation in
Euclidean time, respectively~\cite{Gockeler:1995wg}.  At zero
momentum, this operator couples only to the spin-$\frac{1}{2}$ state.

The zero momentum 2-point and 3-point nucleon correlation functions 
are defined as 
\begin{align}
{\mathbf C}_{\alpha \beta}^{\text{2pt}}(\tau)
  &= \sum_{\mathbf{x}} 
   \langle 0 \vert \chi_\alpha(\tau, \mathbf{x}) \overline{\chi}_\beta(0, \mathbf{0}) 
   \vert 0 \rangle \,, 
\label{eq:corr_fun2} \\
{\mathbf C}_{\Gamma; \alpha \beta}^{\text{3pt}}(t, \tau)
  &= \sum_{\mathbf{x}, \mathbf{x'}} 
  \langle 0 \vert \chi_\alpha(\tau, \mathbf{x}) \mathcal{O}_\Gamma(t, \mathbf{x'})
  \overline{\chi}_\beta(0, \mathbf{0}) 
   \vert 0 \rangle \,,
\label{eq:corr_fun3}
\end{align}
where $\alpha$ and $\beta$ are spinor indices. The source is
placed at time slice $0$, $\tau$ is the sink time slice, and $t$ is an
intermediate time slice at which the local quark bilinear operator
$\mathcal{O}_\Gamma^q(x) = \bar{q}(x) \Gamma q(x)$ is inserted. The
Dirac matrix $\Gamma$ is $1$, $\gamma_4$, $\gamma_i \gamma_5$ and
$\gamma_i \gamma_j$ for scalar (S), vector (V), axial (A) and tensor
(T) operators, respectively.
In this work, subscripts $i$ and $j$ on gamma matrices run over $\{1,2,3\}$, 
with $i<j$. 

The nucleon charges $g_\Gamma^q$ are obtained from the ground state
matrix element $ \langle N(p, s) \vert \mathcal{O}_\Gamma^q \vert N(p,
s) \rangle$, that, in turn, are extracted using the spectral
decomposition of the 2- and 3-point correlation functions. They are
related as
\begin{align}
 \langle N(p, s) \vert \mathcal{O}_\Gamma^q \vert N(p, s) \rangle
 = g_\Gamma^q \bar{u}_s(p) \Gamma u_s(p)
\end{align}
with spinors satisfying
\begin{equation}
\sum_s u_s(p) \bar{u}_s(p)  = \frac{E_{\mathbf{p}} \gamma_4 - i\vec{\gamma}\cdot \vec{p}  + M_N} {2 E_{\mathbf{p}}}\,.
\end{equation}

To extract the charges, we construct the projected 2- and 3-point correlation functions
\begin{align}
C^{\text{2pt}}(t) & = {\langle \Tr [ \mathcal{P}_\text{2pt} {\mathbf C}^{\text{2pt}}(t) ] \rangle} 
\label{eq:2pt_proj}  \\
C_{\Gamma}^{\text{3pt}}(t, \tau)  & = \langle \Tr [ \mathcal{P}_{\rm 3pt} {\mathbf C}_{\Gamma}^{\text{3pt}}(t, \tau) ]\rangle \, .
 \label{eq:3pt_proj}
\end{align}
The operator $\mathcal{P}_\text{2pt} = (1 \pm \gamma_4)/2$ is used to
project on to the positive parity contribution for the nucleon
propagating in the forward (backward) direction. For the connected
3-point contributions, $\mathcal{P}_{\rm 3pt} =
\mathcal{P}_\text{2pt}(1+i\gamma_5\gamma_3)$ is used.  Note that the
$C_{\Gamma}^{\text{3pt}}(t, \tau)$ defined in Eq.~\eqref{eq:3pt_proj}
becomes zero if $\Gamma$ anticommutes with $\gamma_4$, so only $\Gamma
= 1$, $\gamma_4$, $\gamma_i \gamma_5$ and $\gamma_i \gamma_j$ elements
of the Clifford algebra survive.  The fits used to extract the masses,
amplitudes and matrix elements from the 2- and 3-point functions,
defined in Eqs.~\eqref{eq:2pt_proj} and~\eqref{eq:3pt_proj}, are
discussed in Sec.~\ref{sec:excited}.
%

\subsection{High Statistics Using the Truncated Solver Method}
\label{sec:TSM}

We have carried out high-statistics calculation on all the ensembles
using the truncated solver method with bias
correction~\cite{Bali:2009hu,Blum:2012uh}.  In this method,
correlation functions are constructed using quark propagators inverted
with high precision (HP) and low precision (LP) using the multigrid
algorithm. The bias corrected correlators on each configuration are
then given by
\begin{align}
 C^\text{imp}& 
 = \frac{1}{N_\text{LP}} \sum_{i=1}^{N_\text{LP}} 
    C_\text{LP}(\mathbf{x}_i^\text{LP}) \nonumber \\
  +& \frac{1}{N_\text{HP}} \sum_{i=1}^{N_\text{HP}} \left[
    C_\text{HP}(\mathbf{x}_i^\text{HP})
    - C_\text{LP}(\mathbf{x}_i^\text{HP})
    \right] \,,
  \label{eq:2-3pt_TSM}
\end{align}
where $C_\text{LP}$ and $C_\text{HP}$ are the 2- and 3-point
correlation functions constructed using LP and HP quark propagators,
respectively, and $\mathbf{x}_i^\text{LP}$ and
$\mathbf{x}_i^\text{HP}$ are the source positions for the two kinds of
propagator inversion. The LP stopping criteria, defined as $r_{\rm
  LP} \equiv |{\rm residue}|_{\rm LP}/|{\rm source}|$  varied between $ 10^{-3}$ and $5
\times 10^{-4}$, while that for the HP calculations between $10^{-7}$
and $10^{-8}$.

As discussed in Ref.~\cite{Yoon:2016dij}, to reduce statistical
correlations between measurements, $N_\text{HP}$ maximally separated
time slices were selected randomly on each configuration and on each
of these time slices, $N_\text{LP}/N_\text{HP}$ LP source positions
were again selected randomly.  The number of sources, $N_\text{LP}$
and $N_\text{HP}$, used are given in Table~\ref{tab:ens}. An important
conclusion based on all our calculations with $O(10^5)$ measurements
of nucleon charges and form factors carried out so far (see
Refs.~\cite{Bhattacharya:2015wna,Bhattacharya:2016zcn,Yoon:2016dij,Yoon:2016jzj,Rajan:2017lxk}),
is that the difference between the LP and the bias corrected estimates
(or the HP) is smaller than the statistical errors.

To further reduce the computational cost, we also used the coherent
sequential source method discussed in Ref.~\cite{Yoon:2016dij}.
Typically, we constructed four HP or LP sequential sources on four
sink time slices, and added them to obtain the coherent source. A
single inversion was then performed to construct the coherent
sequential propagator. This was then contracted with the four original
propagators to construct four measurements of each three-point
function. All of these propagators were held in the computer memory to
remove the I/O overhead.

Our final errors are obtained using a single elimination jackknife
analysis over the configurations, that is, we first construct the
average defined in Eq.~\eqref{eq:2-3pt_TSM} on each
configuration. Because of this ``binning'' of the data, we do not need
to correct the jackknife estimate of the error for correlations
between the $N_\text{LP}$ LP measurements per configuration.

\section{Excited-State Contamination}
\label{sec:excited}

To extract the nucleon charges we need to evaluate the matrix
elements of the currents between ground-state nucleons. The
lattice nucleon interpolating operator given in
Eq.~\eqref{eq:nucl_op}, however, couples to the nucleon, all its
excitations and multiparticle states with the same quantum
numbers. Previous lattice calculations have shown that the
ESC can be large. In our earlier 
works~\cite{Bhattacharya:2015wna,Bhattacharya:2016zcn,Yoon:2016jzj,Yoon:2016dij},
we have shown that this can be controlled to within a few percent
using the strategy summarized below. 

The overlap between the nucleon operator and the excited states in the
construction of the two- and three-point functions is reduced by using
tuned smeared sources when calculating the quark propagators on the
HYP smeared HISQ lattices. We construct gauge-invariant Gaussian
smeared sources by applying the three-dimensional Laplacian operator,
$\nabla^2$, $N_{\rm GS}$ number of times, i.e., $(1 +
\sigma^2\nabla^2/(4N_{\rm GS}))^{N_{\rm GS}}$ on a delta function
source.  The input smearing parameters $\{\sigma, N_{\rm GS}\}$ for
each ensemble are given in Table~\ref{tab:cloverparams} along with the
resulting root-mean-square radius defined as $\sqrt{\int r^2 \sqrt{S^\dag S}
dr /\int \sqrt{S^\dag S} dr }$.  We find that, as a function of
distance $r$, the modulus of the sum of the values of the twelve
spin-color components at each site, $\sqrt{S^\dag S}$, is well
described by a Gaussian, and we use this ansatz to fit the data. The
results for the root-mean-square radius given in
Table~\ref{tab:cloverparams} show weak dependence on the lattice
spacing or the pion mass for fixed $\sigma$, and are roughly equal to
the input $\sigma$. Throughout this work, the same smearing is used at
the source and sink points.

The analysis of the two-point functions, $C^\text{2pt}$, was carried
out keeping four states in the spectral decomposition:
\begin{align}
C^\text{2pt}
  &(t,\bm{p}) = \nonumber \\
  &{|{\cal A}_0|}^2 e^{-M_0 t} + {|{\cal A}_1|}^2 e^{-M_1 t}\,+ \nonumber \\
  &{|{\cal A}_2|}^2 e^{-M_2 t} + {|{\cal A}_3|}^2 e^{-M_3 t}\,, 
\label{eq:2pt}
\end{align}
where the amplitudes and the masses of the
four states are denoted by ${\cal A}_i$ and $M_i$, respectively.

In fits including more than two states, the estimates of $M_i$ and the
${\cal A}_i$ for $i \ge 2$ were sensitive to the choice of the
starting time slice $t_{\rm min}$, and the fits were not always
stable. The fits were stabilized using the empirical Bayesian
procedure described in Ref.~\cite{Yoon:2016jzj}.  Examples of the
quality of the fits are shown in Figs.~22--29 in
Ref.~\cite{Rajan:2017lxk}. The new results for masses and amplitudes
obtained from 2-, 3- and 4-state fits are given in
Table~\ref{tab:2ptmulti}.

\begin{figure*}[tb] 
\centering
  \subfigure{
    \includegraphics[width=0.45\linewidth]{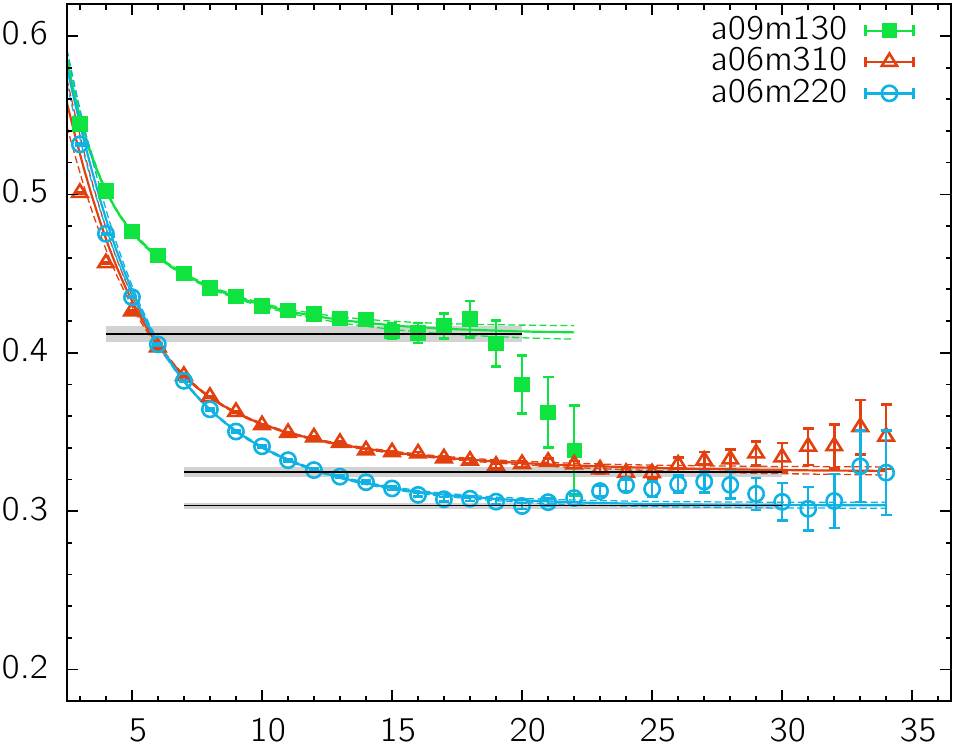} \qquad
    \includegraphics[width=0.45\linewidth]{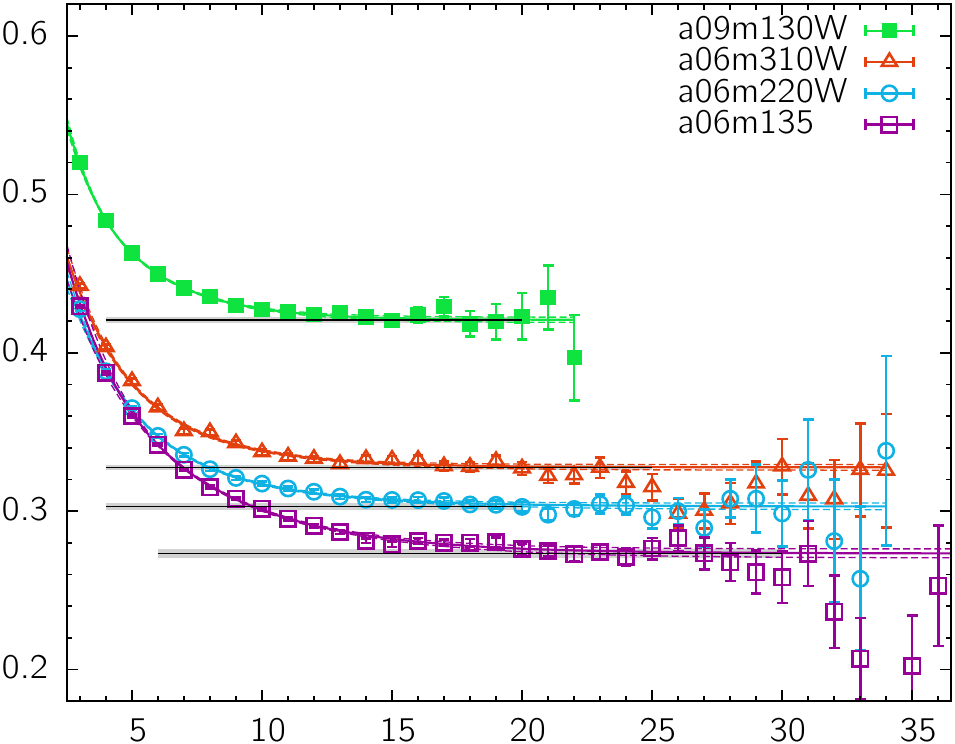}
  }
\caption{Illustration of the
  data for the nucleon $M_{\rm eff}$ versus Euclidean time $t$ and the
  results of the 4-state fit to the 2-point correlation function. We
  compare the data obtained with two different smearing sizes on three
  ensembles.  In the right panel we also show results for the
  $a06m135$ ensemble.  The onset of the plateau in $M_{\rm eff}$
  occurs at earlier $t$ with the larger smearing size but the errors
  at larger $t$ are also larger.
  \label{fig:2pta09m130}}
\end{figure*}

In Fig.~\ref{fig:2pta09m130}, we compare the efficacy of different
smearing sizes in controlling excited states in the 2-point data on
the three ensembles $a09m130$, $a06m310$ and $a06m220$. In each case,
the onset of the plateau with the larger smearing size occurs at
earlier Euclidean time $t$, however, the statistical errors at larger
$t$ are larger. The more critical observation is that, while $M_0$
overlap, the mass gaps $a\Delta M_i$ are significantly different in
two cases.  Thus the excited state parameters are not well determined
even with our high statistics, $O(10^5)$ measurements, data. More
importantly, except for the $a06m310$ case, the mass gap $a \Delta
M_1$ obtained is much larger than $2 a M_\pi$, the value expected if
$N\pi\pi$ is the lowest excitation.  Based on these observations, we
conclude that to resolve the excited state spectrum will require a
coupled channel analysis with much higher statistics data.

The results of different fits for the bare charges extracted from the
three-point data, given in Table~\ref{tab:results3bareu-d}, indicate
that these differences in the mass gaps do not significantly effect
the extraction of the charges. At current level of precision, the
variations in the values of the mass gaps and the corresponding values for the
amplitudes compensate each other in fits to the 2- and
3-point data.\looseness-1

The analysis of the zero-momentum three-point functions, 
$C_\Gamma^{(3\text{pt})} (t;\tau)$  
was carried out retaining three-states in its spectral decomposition:  
\begin{align}
&C^\text{3pt}_{\Gamma}(t_f,t,t_i) = \nonumber\\
  & |{\cal A}_0|^2 \langle 0 | \mathcal{O}_\Gamma | 0 \rangle  e^{-aM_0 (t_f - t_i)} +{}\nonumber\\
  & |{\cal A}_1|^2 \langle 1 | \mathcal{O}_\Gamma | 1 \rangle  e^{-aM_1 (t_f - t_i)} +{}\nonumber\\
  & |{\cal A}_2|^2 \langle 2 | \mathcal{O}_\Gamma | 2 \rangle  e^{-aM_2 (t_f - t_i)} +{}\nonumber\\
  & {\cal A}_1{\cal A}_0^* \langle 1 | \mathcal{O}_\Gamma | 0 \rangle  e^{-aM_1 (t_f-t)} e^{-aM_0 (t-t_i)} +{}\nonumber\\
  & {\cal A}_0{\cal A}_1^* \langle 0 | \mathcal{O}_\Gamma | 1 \rangle  e^{-aM_0 (t_f-t)} e^{-aM_1 (t-t_i)} +{}\nonumber\\
  & {\cal A}_2{\cal A}_0^* \langle 2 | \mathcal{O}_\Gamma | 0 \rangle  e^{-aM_2 (t_f-t)} e^{-aM_0 (t-t_i)} +{}\nonumber\\
  & {\cal A}_0{\cal A}_2^* \langle 0 | \mathcal{O}_\Gamma | 2 \rangle  e^{-aM_0 (t_f-t)} e^{-aM_2 (t-t_i)} +{}\nonumber\\
  & {\cal A}_1{\cal A}_2^* \langle 1 | \mathcal{O}_\Gamma | 2 \rangle  e^{-aM_1 (t_f-t)} e^{-aM_2 (t-t_i)} +{}\nonumber\\
  & {\cal A}_2{\cal A}_1^* \langle 2 | \mathcal{O}_\Gamma | 1 \rangle  e^{-aM_2 (t_f-t)} e^{-aM_1 (t-t_i)} + \ldots \,,
\label{eq:3pt}
\end{align}
where the source point is at $t_i$, the operator is inserted at time
$t$, and the nucleon state is annihilated at the sink time slice
$t_f$. The source-sink separation is $\tau \equiv t_f-t_i$.  The
state $|0\rangle$ represents the ground state and $|n\rangle$, with $n
> 0$, the higher states. The ${\cal A}_i$ are the amplitudes for the
creation of state $i$ with zero momentum by the nucleon interpolating
operator $\chi$.  To extract the matrix elements, the amplitudes
${\cal A}_i$ and the masses $M_i$ are obtained from the 4-state fits to
the two-point functions.  Note that the insertion of the nucleon at
the sink time slice $t_f$ and the insertion of the current at time $t$
are both at zero momentum. Thus, by momentum conservation, only the
zero momentum projections of the states created at the source
time slice contribute to the three-point function.

We calculate the three-point correlation functions for a number of
values of the source-sink separation $\tau$ that are listed in
Table~\ref{tab:ens}.  To extract the desired matrix element $\langle 0
| \mathcal{O}_\Gamma | 0 \rangle$, we fit the data at all $\tau$ and
$t$ simultaneously using the ansatz given in Eq.~\eqref{eq:3pt}. In
this work, we examine three kinds of fits, $2^\ast$-, 2- and
$3^\ast$-state fits.  The $2^\ast$-state fit corresponds to keeping
terms of the type $\matrixe{0}{\mathcal{O}_\Gamma}{0}$ and
$\matrixe{0}{\mathcal{O}_\Gamma}{1}$. The 2-state fits also include
$\matrixe{1}{\mathcal{O}_\Gamma}{1}$, and the $3^\ast$-state fits
further add the $\matrixe{0}{\mathcal{O}_\Gamma}{2}$ and
$\matrixe{1}{\mathcal{O}_\Gamma}{2}$ type terms.\looseness-1

In the simultaneous fit to the data versus $t$ and multiple $\tau$ to
obtain $\matrixe{0}{\mathcal{O}_\Gamma}{0}$, we skip $\tskip$ points
adjacent to the source and the sink to remove points with the largest
ESC. The same $\tskip$ is used for each $\tau$. The $\tskip$ selected
is a compromise between wanting to include as many points as possible
to extract the various terms given in Eq.~\eqref{eq:3pt} with
confidence, and the errors in and stability of the full covariance
matrix used in the fit.  In particular, the choice of $\tskip$ on the
$a=0.06$~fm ensembles is the smallest value for which the covariance
matrix was invertable and reasonable. These values of $\tskip$, tuned
for each ensemble, are given in Table~\ref{tab:results3bareu-d}.

To visualize the ESC, we 
plot the data for the following ratio of correlation functions 
\begin{equation}
R_\Gamma(t,\tau) = \frac{C_{\Gamma}^{\text{3pt}}(t, \tau) }{C^{\text{2pt}}(\tau)} \to g_\Gamma \,, 
 \label{eq:ratio}
\end{equation}
in Figs.~\ref{fig:gA2v3a12}--\ref{fig:gT2v3a06} and show the various
fits corresponding to the results in Table~\ref{tab:results3bareu-d}.
In the limit $t \to \infty$ and $\tau-t \to \infty$, this ratio
converges to the charge $g_\Gamma $.  At short times, the ESC is
manifest in all cases. For sufficiently large $\tau$, the data should
exhibit a flat region about $\tau/2$, and the value should become
independent of $\tau$. The current data for $g_A$, $g_S$ and $g_T$,
with $\tau$ up to about 1.4~fm, do not provide convincing evidence of
this desired asymptotic behavior. To obtain
$\matrixe{0}{\mathcal{O}_\Gamma}{0}$, we use the three-state ansatz given
in Eq.~\eqref{eq:3pt}.

On the three ensembles, $a09m130$, $a06m310$ and $a06m220$, we can
compare the data with two different smearing sizes given in
Table~\ref{tab:ens}. We find a significant reduction in the ESC in the
axial and scalar charges on increasing the smearing
size. Nevertheless, the 2- and $3^\ast$-state fits and the two
calculations give consistent estimates for the ground state matrix
elements. The agreement between these four estimates has increased our
confidence in the control over ESC.  The results for $g_S^{u-d}$,
obtained using $2$-state fits, have larger uncertainty as discussed in
Sec.~\ref{sec:poor}, but are again consistent except those from the
$a06m220$ ensemble.

This higher statistics study of the ESC confirms many features discussed in Ref.~\cite{Bhattacharya:2016zcn}:
\begin{itemize}
\item
The ESC is large in both $g_A^{u-d}$ and $g_S^{u-d}$, and the
convergence to the $\tsepi$ value is monotonic and from below.
\item
The ESC is $g_T^{u-d}$ is $\lesssim 10\%$ for $\tau > 1$~fm, and the
convergence to the $\tsepi$ value is also monotonic but from above.
\item
The ESC in $g_A^{u-d}$ and $g_S^{u-d}$ is reduced on increasing the 
size of the smearing, but $g_T^{u-d}$ is fairly insensitive to the smearing 
size. 
\item
For a given number of measurements at the same $\tau$ and $t$, the
statistical precision of $g_T^{u-d}$ is slightly better than that of
$g_A^{u-d}$.  The data for $g_S^{u-d}$ is noisy, especially at the
larger values of $\tau$.  On many ensembles, it does not exhibit a
monotonic increase with $\tau$. To get $g_S^{u-d}$ with the same precision as 
in $g_A^{u-d}$ currently will require $\approx 5$ times the statistics.
\item
The data for each charge and for each source-sink separation $\tau$
becomes symmetric about $\tau/2$ with increasing statistical
precision.  This is consistent with the $\cosh(t-\tau/2)$ behavior
predicted by Eq.~\eqref{eq:3pt} for each transition matrix element. 
\item
The variations in the results with the fit ranges selected for fits to
the two-point functions and the number, $\tskip$, of points skipped in
the fits to the three-point data decrease with the increased
statistical precision.
\item
Estimates from the $2$- and the $3^\ast$-state fits overlap for all 
fourteen measurements of $g_A^{u-d}$ and $g_T^{u-d}$. 
\item
The $3^\ast$-state fits for $g_S^{u-d}$ are not stable in all cases and many
of the parameters are poorly determined. To extract our best estimates, we use
2-state fits.
\item
The largest excited-state contribution comes from the $\langle 0 |
O_\Gamma | 1 \rangle$ transition matrix elements. We, therefore
discuss a poor person's recipe to get estimates based on the $2^\ast$
fits in Sec.~\ref{sec:poor} that are useful when data at only one 
value of $\tau$ are available.
\end{itemize}

Our conclusion on ESC is that with $O(10^5)$ measurements, $3^\ast$
fits, the choice of smearing parameters used and the values of $\tau$
simulated, the excited-state contamination in $g_A^{u-d}$ and
$g_T^{u-d}$ has been controlled to within a couple of percent, i.e.,
the size of the quoted errors.  The errors in $g_S^{u-d}$ are at the
5\%--10\% level, and we take results from the 2-state fit as our best
estimates. In general, for calculations by other groups when data
with reasonable precision are available only at a single value of
$\tau$, we show that the $2^\ast$ fit gives a much better estimate
than the plateau value.

\subsection{A poor person's recipe and $g_S^{u-d}$}
\label{sec:poor}

Our high statistics calculations allow us to develop the following
poor person's recipe for estimating the ground state matrix element
when data are available only at a single value of $\tau$.  To
illustrate this, we picked two values with $\tau \approx 1$~fm ($\tau
=\{6,7\}, \{8,10\}, \{10,12\}, \{16,18,20\}$ in lattice units for the $a\approx
0.15, 0.12, 0.09, 0.06$ ensembles) for which we have reasonably
precise data at all values of $t$ and for all three isovector
charges.  We then compared the estimates of the charges from the
$2^\ast$ fit to data at these values of $\tau$ with our best estimate
from the $3^\ast$ fit (2-state for $g_S^{u-d}$) to the data at
multiple $\tau$ and $t$.  Fits for all ensembles  are shown in
Figs.~\ref{fig:gA2v3a12}--\ref{fig:gT2v3a06} and the results collected
in Table~\ref{tab:results3bareu-d}.  

In the case of $g_A^{u-d}$ and $g_T^{u-d}$ we get overlapping results
results converging to the $3^\ast$ value. This suggests that, within
our statistical precision, all the excited-state terms that behave as
$\cosh \Delta M(t-\tau/2)$ in the spectral decomposition are
well-approximated by the single term proportional to $\langle 0|{
  \cal{O}} | 1 \rangle$ in the $2^\ast$ fit. Isolating this ESC is,
therefore, essential.  Also, the remainder, the sum of all the terms
independent of $t$ is small. This explains why the values of the
excited state matrix elements $\langle 1| {\cal{O} } | 1 \rangle$ and
$\langle 0| {\cal{O} } | 2 \rangle$, given in Table~\ref{tab:bareEME},
are poorly determined.

We further observe that in our implementation of the lattice
calculations---HYP smoothing of the lattices plus the Gaussian
smearing of the quark sources---the product $(M_1-M_0) \times \tau$ is
$ \gtrsim 1$ for $\tau \approx 1$~fm, i.e., $(M_1-M_0) \gtrsim
200$~MeV. Since this condition holds for the physical nucleon
spectrum, it is therefore reasonable to expect that the charges
extracted from a $2^\ast$ fit to data with $\tau \gtrsim 1$~fm are a
good approximation to the $\tsepi$ value, whereas the value at the
midpoint $t=\tau/2$ (called the plateau value) is not. This is
supported by the data for $g_A^{u-d}$ and $g_T^{u-d}$ shown in
Table~\ref{tab:results3bareu-d}; there is much better consistency
between the $3^\ast$ results and $2^\ast$ fits to data with a
single values of $\tau \gtrsim 1$~fm versus the plateau value.

In this work, the reason for considering such a recipe is that
estimates of $g_S^{u-d}$ have much larger statistical errors, because
of which the data at the larger values of $\tau$ do not, in all cases, 
exhibit the expected monotonic convergence in $\tau$ and have large
errors. As a result, on increasing $n$ in an $n$-state fit to data
with multiple values of $\tau$ does not always give a
better or converged value. We, therefore, argue that to obtain the
best estimates of $g_S^{u-d}$ one can make judicious use of this
recipe, i.e., use $2^\ast$ fits to the data with the largest value of
$\tau$ that conforms with the expectation of monotonic convergence
from below.  In our case, based on such analyses we conclude that the
2-state fits are more reliable than $3^\ast$ fits for
$g_S^{u-d}$. These fourteen values of $g_S^{u-d}$ used in the final
analysis are marked with the superscript ${}^\dag$ in
Table~\ref{tab:results3bareu-d}.  The same strategy is followed for
obtaining the connected contribution to the isoscalar charges,
$g_{S}^{u+d}$, that are given in Table~\ref{tab:results3bareu+d}.

\begin{table*}                
\centering
\begin{ruledtabular}
\begin{tabular}{c|ccc|ccc|ccc}
ID           & $g_A^{u}$  & $g_A^{d}$    & $g_A^{u-d}$ & $g_S^{u}$ & $g_S^{d}$ & $g_S^{u-d}$ & $g_T^{u}$ & $g_T^{d}$    & $g_T^{u-d}$ \\ 
\hline
$a15m310 $   &  0.937(06) & $-$0.313(04) & 1.250(07)   & 3.10(08)  & 2.23(06)  & 0.87(03)    & 0.901(06) & $-$0.219(04) & 1.121(06)  \\   
\hline                                                                                       
$a12m310 $   &  0.946(15) & $-$0.328(09) & 1.274(15)   & 3.65(13)  & 2.69(09)  & 0.96(05)    & 0.859(12) & $-$0.206(07) & 1.065(13)  \\   
$a12m220S$   &  0.934(43) & $-$0.332(27) & 1.266(44)   & 5.23(49)  & 4.23(40)  & 1.00(26)    & 0.816(44) & $-$0.249(33) & 1.065(39)  \\   
$a12m220 $   &  0.947(22) & $-$0.318(13) & 1.265(21)   & 4.83(35)  & 3.72(29)  & 1.11( 9)    & 0.847(17) & $-$0.201(11) & 1.048(18)  \\ 
$a12m220L$   &  0.942(09) & $-$0.347(08) & 1.289(13)   & 4.21(29)  & 3.34(26)  & 0.87(04)    & 0.846(11) & $-$0.203(05) & 1.069(11)  \\  
\hline                                                                                                                               
$a09m310 $   &  0.930(07) & $-$0.308(04) & 1.238(08)   & 3.60(12)  & 2.58(10)  & 1.02(03)    & 0.824(07) & $-$0.203(03) & 1.027(07)  \\   
$a09m220 $   &  0.945(12) & $-$0.334(06) & 1.279(13)   & 4.46(19)  & 3.41(16)  & 1.05(04)    & 0.799(10) & $-$0.203(05) & 1.002(10)  \\   
$a09m130 $   &  0.919(20) & $-$0.350(16) & 1.269(28)   & 5.87(49)  & 4.71(41)  & 1.16(13)    & 0.765(20) & $-$0.196(10) & 0.961(22)  \\
$a09m130W$   &  0.935(14) & $-$0.336(08) & 1.271(15)   & 5.28(17)  & 4.23(14)  & 1.05(06)    & 0.797(12) & $-$0.203(06) & 1.000(12)  \\
\hline                                                                                                                               
$a06m310 $   &  0.923(25) & $-$0.320(15) & 1.243(27)   & 4.48(33)  & 3.24(24)  & 1.24(11)    & 0.785(20) & $-$0.197(11) & 0.982(20)  \\   
$a06m310W$   &  0.906(22) & $-$0.310(16) & 1.216(21)   & 4.06(16)  & 2.94(11)  & 1.12(07)    & 0.784(15) & $-$0.192(08) & 0.975(16)  \\   
$a06m220 $   &  0.912(13) & $-$0.323(13) & 1.235(18)   & 4.40(13)  & 3.29(09)  & 1.11(07)    & 0.779(10) & $-$0.197(10) & 0.975(12)  \\  
$a06m220W$   &  0.917(24) & $-$0.341(15) & 1.257(24)   & 4.32(21)  & 3.55(18)  & 0.77(09)    & 0.764(21) & $-$0.198(11) & 0.962(22)  \\  
$a06m135 $   &  0.917(22) & $-$0.323(13) & 1.240(26)   & 5.26(22)  & 4.26(15)  & 1.00(13)    & 0.768(17) & $-$0.183(10) & 0.952(19)  \\  
\end{tabular}
\end{ruledtabular}
\caption{Results for the
  bare connected contributions to the various charges.}
\label{tab:resultsbare}
\end{table*}

\subsection{Transition and excited state matrix elements}
\label{sec:excitedME}

The only transition matrix element that has been estimated with some
degree of confidence is $\langle 0 | \mathcal{O}_\Gamma | 1 \rangle$
as can be inferred from the results given in
Table~\ref{tab:bareEME}. Also including information from
Figs.~\ref{fig:gA2v3a12}--\ref{fig:gT2v3a06}, our qualitative
conclusions on it are as follows:
\begin{itemize}
\item
Estimates of $\langle 0 | \mathcal{O}_A | 1 \rangle$ vary between
$-0.1$ and $-0.3$ and account for the negative 
curvature evident in the figures. All 
ground-state estimates of $g_A^{u-d}$ converge from below.
\item
Estimates of $\langle 0 | \mathcal{O}_S | 1 \rangle$ vary between
$-0.2$ and $-0.5$ and account for the larger 
negative curvature observed in the figures. All  
ground-state estimates of $g_S^{u-d}$ also converge from below.  
\item
Estimates of $\langle 0 | \mathcal{O}_T | 1 \rangle$ vary between 0.1
and 0.3 and account for the positive curvature evident in the
figures. The ground-state estimates of $g_T^{u-d}$ converge from
above in all cases.
\end{itemize}
Our long term goal is to improve the precision of these calculations
to understand and extract an infinite volume continuum limit value for the
transition matrix elements.

\subsection{A caveat in the analysis of the isoscalar charges $g_{A,S,T}^{u+d}$ keeping only the connected contribution}
\label{sec:PQ}

In this paper, we have analyzed only the connected contributions to
the isoscalar charges $g_{A,S,T}^{u+d}$. The disconnected
contributions are not included as they are not available for all the
ensembles, and are analyzed for different, typically smaller, values
of source-sink separation $\tau$ because of the lower 
quality of the statistical signal.  Since the proper way to extract
the isoscalar charges is to first add the connected and disconnected
contributions and then perform the fits using the lattice QCD spectral
decomposition to remove excited state contamination, analyzing only
the connected contribution introduces an approximation. Isoscalar
charges without a disconnected contribution can be defined in a
partially quenched theory with an additional quark with flavor
$u^\prime$.  However, in this theory the Pauli exclusion principle does
not apply between the $u$ and $u^\prime$ quarks.  The upshot of this is
that the spectrum of states in the partially quenched theory is
larger, for example, an intermediate $u^\prime u d$ state would be the
analogue of a $\Lambda$ baryon\footnote{We thank Stephen Sharpe for providing
  a diagrammatic illustration of such additional states.}. Thus, the
spectral decomposition for this partially quenched theory and QCD is
different. The problem arises because our n-state fits assume the QCD
spectrum since we take the amplitudes and masses of states from the
QCD 2-point function when fitting the 3-point function using
Eq.~\eqref{eq:3pt}. One could make fits to 3-point functions leaving
all the parameters in Eq.~\eqref{eq:3pt} free, but then even 2-state
fits become poorly constrained with current data.

We assume that, in practice, the effect due to using the QCD rather
than the partially quenched QCD spectra to fit the connected
contribution versus $t$ and $\tau$ to remove ESC is smaller than the
quoted errors. First, the difference between the plateau value in our
largest $\tau$ data and the $\tau \to \infty$ value is a few percent
effect, so that any additional systematic is well within the quoted
uncertainty.  Furthermore, for the tensor charges the disconnected
contribution is tiny and consistent with zero, so for the tensor
charges one can ignore this caveat.  For the axial and scalar charges,
the disconnected contribution is between 10\%--20\% of the connected, so
we are neglecting possible systematic effects due to extrapolating the
connected and disconnected contributions separately.

\begin{table*}
\centering
\begin{ruledtabular}
\begin{tabular}{c|ccc|cc|ccc}
         &  \multicolumn{3}{c|} {Axial}  &   \multicolumn{2}{c|} {Scalar}  &   \multicolumn{3}{c} {Tensor}     \\
ID       & $\langle 0 | \mathcal{O}_A | 1 \rangle$ & $\langle 1 | \mathcal{O}_A | 1 \rangle$  & $\langle 0 | \mathcal{O}_A | 2 \rangle$
         & $\langle 0 | \mathcal{O}_S | 1 \rangle$ & $\langle 1 | \mathcal{O}_S | 1 \rangle$ 
         & $\langle 0 | \mathcal{O}_T | 1 \rangle$ & $\langle 1 | \mathcal{O}_T | 1 \rangle$  & $\langle 0 | \mathcal{O}_T | 2 \rangle$ \\
\hline
$a15m310 $     &  $-$0.044( 37) &  $-$2.06(1.3) &   $-$0.08( 5)  &  $-$0.37( 3)  &  $ $ 3.6(4.6)  &  0.31( 4) &    $-$2.72(1.2) &   $-$0.18( 7)  \\
\hline
$a12m310 $     &  $-$0.208( 94) &  $ $1.40(2.4) &   $ $0.07( 4)  &  $-$0.72( 9)  &  $ $ 8.5(10.)  &  0.32( 8) &    $-$0.82(2.2) &   $ $0.08( 4)  \\
$a12m220S$     &  $-$0.119( 77) &  $ $1.46(60)  &   $ $0.03(10)  &  $-$0.42(13)  &  $ $ 3.8(5.7)  &  0.19( 8) &    $ $0.13(62)  &   $ $0.10(11)  \\
$a12m220 $     &  $-$0.047( 52) &  $ $0.33(76)  &   $-$0.08( 5)  &  $-$0.38(11)  &  $-$ 2.8(3.6)  &  0.21( 5) &    $ $0.07(59)  &   $ $0.12( 4)  \\
$a12m220L$     &  $-$0.084( 25) &  $-$0.21(73)  &   $-$0.05( 3)  &  $-$0.38(12)  &  $ $ 4.6(2.7)  &  0.19( 2) &    $-$0.04(43)  &   $ $0.09( 4)  \\
\hline                                                                                                                                            
$a09m310 $     &  $-$0.095( 20) &  $-$1.45(1.9) &   $ $0.11( 6)  &  $-$0.39( 4)  &  $ $ 0.7(1.5)  &  0.20( 2) &    $ $0.17(1.1) &   $ $0.04( 6)  \\
$a09m220 $     &  $-$0.153( 34) &  $-$0.44(98)  &   $ $0.07( 4)  &  $-$0.47( 5)  &  $ $ 1.4(1.0)  &  0.16( 3) &    $ $0.44(60)  &   $ $0.13( 3)  \\
$a09m130 $     &  $-$0.092( 26) &  $ $0.65(19)  &   $ $0.03( 4)  &  $-$0.42( 7)  &  $ $ 2.0(1.2)  &  0.17( 3) &    $ $0.78(14)  &   $ $0.08( 4)  \\
$a09m130W$     &  $-$0.098( 26) &  $-$0.46(94)  &   $ $0.06( 6)  &  $-$0.28( 4)  &  $ $ 2.2(2.2)  &  0.18( 3) &    $ $0.37(71)  &   $ $0.11( 6)  \\
\hline                                                                                                                                            
$a06m310 $     &  $-$0.075( 41) &  $ $0.18(51)  &   $-$0.00( 1)  &  $-$0.41( 6)  &  $ $ 1.2(1.4)  &  0.14( 5) &    $-$0.20(60)  &   $-$0.08( 9)  \\
$a06m310W$     &  $-$0.093(124) &  $-$0.56(4.5) &   $-$0.02(35)  &  $-$0.44( 9)  &  $ $10.6(15.)  &  0.22(12) &    $ $0.41(3.9) &   $ $0.04(36)  \\
$a06m220 $     &  $-$0.184( 40) &  $ $0.43(38)  &   $ $0.28(13)  &  $-$0.32( 4)  &  $-$ 0.3(1.1)  &  0.09( 4) &    $ $0.33(32)  &   $ $0.05(12)  \\
$a06m220W$     &  $-$0.249(127) &  $ $1.2(2.2)  &   $ $0.32(25)  &  $-$0.33(14)  &  $ $23.4(20.)  &  0.29(13) &    $-$1.86(3.0) &   $-$0.17(25)  \\
$a06m135 $     &  $-$0.137( 47) &  $ $0.81(41)  &   $ $0.20(13)  &  $-$0.32( 6)  &  $ $ 2.4(3.1)  &  0.12( 5) &    $ $0.82(39)  &   $ $0.07(12)  \\
\end{tabular}
\end{ruledtabular}
\caption{Estimates of the leading ratios $\langle 0 |
  \mathcal{O}_\Gamma | 1 \rangle /\langle 0 | \mathcal{O}_\Gamma | 0
  \rangle$, $\langle 1 | \mathcal{O}_\Gamma | 1 \rangle /\langle 0 |
  \mathcal{O}_\Gamma | 0 \rangle$, and $\langle 0 | \mathcal{O}_\Gamma
  | 2 \rangle /\langle 0 | \mathcal{O}_\Gamma | 0 \rangle$ for the
  transition and excited state matrix elements in the case of the
  isovector charges. For the scalar charge, $\langle 0 |
  \mathcal{O}_\Gamma | 2 \rangle /\langle 0 | \mathcal{O}_\Gamma | 0
  \rangle$ is not given since our final results are from the $2$-state
  fit that are marked with ${}^\dag$ in
  Table~\protect\ref{tab:results3bareu-d}.  }
\label{tab:bareEME}
\end{table*}

\section{Renormalization of Operators}
\label{sec:renorm}

The renormalization constants $Z_A$, $Z_V$, $Z_S$ and $Z_T$ of the
isovector quark bilinear operators are calculated in the
regularization-independent symmetric momentum-subtraction (RI-sMOM)
scheme~\cite{Martinelli:1994ty,Sturm:2009kb}.  We followed the
methodology given in
Refs.~\cite{Bhattacharya:2015wna,Bhattacharya:2016zcn} and refer the
reader to it for details.  Results based on the six ensembles, 
{\it a12m310, a12m220, a09m310, a09m220, a06m310} and {\it a06m220},
obtained in Refs.~\cite{Bhattacharya:2015wna,Bhattacharya:2016zcn} 
are summarized in Table~\ref{tab:Zfinal} along with the new results
on the $a15m310$ ensemble.  We briefly summarize
the method below for completeness.\looseness-1

The calculation was done as follows: starting with the lattice results
obtained in the RI-sMOM scheme at a given Euclidean four-momentum
squared $Q^2$, we first convert them to the $\overline{\text{MS}}$
scheme at the same scale (horizontal matching) using two-loop
perturbative relations expressed in terms of the coupling constant
$\alpha_{\overline{\text{MS}}}(Q^2)$~\cite{Gracey:2011fb}. This
estimate at $\mu^2=Q^2$, is then run in the continuum in the
$\overline{\text{MS}}$ scheme to $2\GeV$ using the 3-loop anomalous
dimension relations for the scalar and tensor
bilinears~\cite{Gracey:2000am,Agashe:2014kda}.  These data are labeled
by the $Q^2$ in the original RI-sMOM scheme and suffer from artifacts
due to nonperturbative effects and the breaking of the Euclidean
$O(4)$ rotational symmetry down to the hypercubic group. To get the
final estimate, we fit these data versus $Q^2$ using an ansatz
motivated by the form of possible artifacts as discussed in
Refs.~\cite{Bhattacharya:2015wna,Bhattacharya:2016zcn}.

We find that the final renormalization factors on ensembles with constant $a$ 
show no significant dependence versus $M_\pi$. We, therefore,
average the results at different $M_\pi$ to get the mass-independent
values at each $a$.

In Table~\ref{tab:Zfinal}, we also give the results for the ratios
$Z_A/Z_V$, $Z_S/Z_V$, and $Z_T/Z_V$ that show much smaller $O(4)$
breaking, presumably because some of the systematics cancel. From the
individual data and the two ratios, $Z_\Gamma /Z_V$ and
$g_\Gamma/g_V^{u-d}$, we calculate the renormalized charges in two
ways: $Z_\Gamma \times g_\Gamma$ and $(Z_\Gamma /Z_V) \times
(g_\Gamma/g_V^{u-d})$ with $Z_V g_V^{u-d} = 1$ since the
conservation of the vector current. These two sets of renormalized
charges are given in Table~\ref{tab:resultsrenormIV}.

\begin{table*}  
\centering
\begin{ruledtabular}
\begin{tabular}{c|cccc|ccc}
ID   &   $Z_A^{u-d}$& $Z_S^{u-d}$& $Z_T^{u-d}$& $Z_V^{u-d}$& $Z_A^{u-d}/Z_V^{u-d}$   & $Z_S^{u-d}/Z_V^{u-d}$   & $Z_T^{u-d}/Z_V^{u-d}$   \\
\hline                                                      
$a=0.15$~fm  &  $0.96(2)$   & $0.94(4)$ &  $0.95(3)$ & $0.92(2)$ & $1.05(2)$   & $1.02(5)$   & $1.02(3)$   \\
$a=0.12$~fm  &  $0.95(3)$   & $0.90(4)$ &  $0.94(4)$ & $0.91(2)$ & $1.045(09)$ & $0.986(09)$ & $1.034(34)$ \\
$a=0.09$~fm  &  $0.95(4)$   & $0.88(2)$ &  $0.98(4)$ & $0.92(2)$ & $1.034(11)$ & $0.955(49)$ & $1.063(29)$ \\
$a=0.06$~fm  &  $0.97(3)$   & $0.86(3)$ &  $1.04(3)$ & $0.95(1)$ & $1.025(09)$ & $0.908(40)$ & $1.100(25)$ \\
\end{tabular}
\end{ruledtabular}
\caption{The final mass-independent isovector
  renormalization constants $Z_A^{u-d}$, $Z_S^{u-d}$, $Z_T^{u-d}$,
  $Z_V^{u-d}$ and the ratios $Z_A^{u-d}/Z_V^{u-d}$,
  $Z_S^{u-d}/Z_V^{u-d}$ and $Z_T^{u-d}/Z_V^{u-d}$ in the
  $\overline{\text{MS}}$ scheme at 2~GeV at the four values of the
  lattice spacing used in our analysis. Results for the $a=0.12$, $a=0.09$ and $a=0.06$~fm ensembles are reproduced from
  Ref.~\cite{Bhattacharya:2016zcn}.}
\label{tab:Zfinal}
\end{table*}

\begin{table*}   
\centering
\begin{ruledtabular}
\begin{tabular}{c|ccc|ccc|cc}
                    &  \multicolumn{3}{c|} {$g_\Gamma^{u-d,{\rm bare}}/g_V^{u-d,{\rm bare}}\times Z_\Gamma^{u-d}/Z_V^{u-d}$} &   \multicolumn{3}{c|} {$g_\Gamma^{u-d,{\rm bare}} \times Z_\Gamma^{u-d}$}  & \multicolumn{2}{c} {}     \\
ID                  & $g_A^{u-d}$ & $g_S^{u-d}$ & $g_T^{u-d}$  & $g_A^{u-d}$ & $g_S^{u-d}$ & $g_T^{u-d}$  &  $g_V^{u-d,{\rm bare}}$ & $Z_V g_V^{u-d,{\rm bare}}$  \\ 
\hline             
$a15m310 $          & 1.228(25)   & 0.828(049)   & 1.069(32)  & 1.200(26)  & 0.816(044)  & 1.065(34)  & 1.069(04) & 0.983(22) \\   
\hline                                                                                 
$a12m310 $          & 1.251(19)   & 0.891(045)   & 1.035(37)  & 1.210(41)  & 0.865(058)  & 1.001(44)  & 1.064(05) & 0.968(22) \\   
$a12m220S$          & 1.224(44)   & 0.916(233)   & 1.019(53)  & 1.203(56)  & 0.903(237)  & 1.001(56)  & 1.081(18) & 0.983(27) \\   
$a12m220 $          & 1.234(25)   & 1.024(086)   & 1.011(38)  & 1.202(43)  & 1.001(096)  & 0.985(45)  & 1.071(09) & 0.975(23) \\ 
$a12m220L$          & 1.262(17)   & 0.807(039)   & 1.035(36)  & 1.225(41)  & 0.786(052)  & 1.005(44)  & 1.067(04) & 0.971(21) \\
\hline                                                                                                                          
$a09m310 $          & 1.235(15)   & 0.936(054)   & 1.054(30)  & 1.176(50)  & 0.893(031)  & 1.007(42)  & 1.045(03) & 0.962(20) \\   
$a09m220 $          & 1.260(19)   & 0.958(063)   & 1.015(30)  & 1.215(53)  & 0.926(044)  & 0.982(41)  & 1.053(03) & 0.969(21) \\   
$a09m130 $          & 1.245(32)   & 1.050(128)   & 0.969(35)  & 1.206(57)  & 1.019(116)  & 0.942(44)  & 1.052(08) & 0.969(22) \\
$a09m130W$          & 1.249(21)   & 0.952(074)   & 1.011(30)  & 1.207(53)  & 0.923(058)  & 0.980(44)  & 1.052(06) & 0.968(22) \\
\hline                                                                                                                          
$a06m310 $          & 1.233(30)   & 1.090(104)   & 1.046(33)  & 1.205(46)  & 1.065(100)  & 1.021(36)  & 1.043(06) & 0.991(12) \\   
$a06m310W$          & 1.205(24)   & 0.984(074)   & 1.037(30)  & 1.180(42)  & 0.964(071)  & 1.014(34)  & 1.035(11) & 0.983(15) \\   
$a06m220 $          & 1.206(21)   & 0.959(071)   & 1.022(27)  & 1.198(41)  & 0.953(066)  & 1.014(32)  & 1.050(07) & 0.997(12) \\  
$a06m220W$          & 1.241(26)   & 0.672(082)   & 1.018(34)  & 1.220(45)  & 0.661(080)  & 1.000(37)  & 1.039(09) & 0.987(13) \\  
$a06m135 $          & 1.220(27)   & 0.876(120)   & 1.005(30)  & 1.203(45)  & 0.864(118)  & 0.990(35)  & 1.042(10) & 0.990(14) \\  
\hline 
11-point fit        & 1.218(25)   &  1.022(80)   & 0.989(32)  & 1.197(42)  & 1.010(74)   & 0.966(37)  &           &           \\  
$\chi^2/$d.o.f.     & 0.21        &  1.43        & 0.10       & 0.05       & 1.12        & 0.20       &           &           \\  
10-point fit        & 1.215(31)   &  0.914(108)  & 1.000(41)  & 1.200(56)  & 0.933(108)  & 0.994(48)  &           &           \\  
$\chi^2/$d.o.f.     & 0.24        &  1.30        & 0.09       & 0.06       & 1.15        & 0.09       &           &           \\  
$10^\ast$-point fit & 1.218(25)   &  1.021(80)   & 0.989(32)  & 1.197(43)  & 1.009(74)   & 0.966(37)  &           &           \\  
$\chi^2/$d.o.f.     & 0.23        &  1.67        & 0.11       & 0.06       & 1.31        & 0.17       &           &           \\  
$8$-point fit       & 1.245(42)   &  1.214(130)  & 0.977(67)  & 1.172(94)  & 1.123(105)  & 0.899(86)  &           &           \\  
$\chi^2/$d.o.f.     & 0.20        &  1.14        & 0.13       & 0.06       & 0.87        & 0.13       &           &           \\  
\end{tabular}
\end{ruledtabular}
\caption{Results for the renormalized isovector charges calculated in
  two ways, $g_\Gamma^{u-d,{\rm bare}}/g_V^{u-d,{\rm bare}} \times
  Z_\Gamma^{u-d}/Z_V^{u-d}$ and $g_\Gamma^{u-d,{\rm bare}} \times
  Z_\Gamma^{u-d}$. The errors are obtained by adding in quadrature the
  errors in the bare matrix elements and in the renormalization constants
  given in Table~\protect\ref{tab:Zfinal}. The unrenormalized charges
  are given in Table~\protect\ref{tab:resultsbare}.  In the last two
  columns, we also give the results for the bare, $g_V^{u-d,{\rm
      bare}}$ and the renormalized, $Z_V g_V^{u-d,{\rm bare}}$, vector
  charge. The latter should be unity as it is conserved. The
  deviations are found to be up to 4\%. Results of the four CCFV
  fits (11-point, 10-point, $10^\ast$-point, and the $8$-point 
  defined in the text) are given in the bottom eight rows.  }
\label{tab:resultsrenormIV}
\end{table*}

We are also interested in extracting flavor diagonal charges which can
be written as a sum over isovector ($u-d$) and isoscalar ($u+d$)
combinations. These combinations renormalize with the corresponding
isovector, $Z^{\rm isovector}$, and isoscalar, $Z^{\rm isoscalar}$,
factors that are, in general,
different~\cite{Bhattacharya:2005rb}.~\footnote{In general, one
  considers the singlet and non-singlet combinations in a $N_f$-flavor
  theory. In this paper, we are only analyzing the insertions on $u$
  and $d$ quarks that are taken to be degenerate, so it is convenient
  to use the 2-flavor labels, isosinglet ($u+d$) and isovector
  ($u-d$).} Only the isovector renormalization constants are given in
Table~\ref{tab:Zfinal}.

In perturbation theory, the difference between $Z^{\rm isovector}$ and
$Z^{\rm isoscalar}$ appears at two loops, and is therefore expected to
be small.  Explicit calculations in
Refs.~\cite{Alexandrou:2017qyt,Alexandrou:2017oeh,Green:2017keo} show
that $Z^{\rm isosinglet} \approx Z^{\rm isovector}$ for the axial and
tensor charges.  Since the two agree to within a percent, we will
assume $Z_{A,T}^{\rm isoscalar} = Z_{A,T}^{\rm isovector}$ in this
work, and renormalize both isovector ($u-d$) and isoscalar ($u+d$)
combinations of charges using $ Z^{\rm isovector}$. In the case of
the tensor charges, this approximation is even less significant since
the contribution of the disconnected diagrams to the charges is
consistent with zero within errors~\cite{Bhattacharya:2015wna}.

In the case of the scalar charge, the difference between $Z^{\rm
  isosinglet}$ and $Z^{\rm isovector}$ can be large due to the
explicit breaking of the chiral symmetry in the Wilson-clover action
which induces mixing between flavors.  This has not been fully
analyzed for our clover-on-HISQ formulation, so only the bare results
for $g_S^{u-d}$ and $g_S^{u+d}$, and the renormalized results for
$g_S^{u-d}$ are presented in this work.

\begin{table*}   
\centering
\begin{ruledtabular}
\begin{tabular}{c|cc|ccc}
ID                  & $g_A^{u}$  & $g_A^{d}$      & $g_T^{u}$ & $g_T^{d}$     &  $g_T^{u+d}$  \\ 
\hline
$a15m310 $          &  0.920(19) & $-$0.307(07)   & 0.860(26) & $-$0.209(07)  & 0.649(21)  \\   
\hline
$a12m310 $          &  0.929(17) & $-$0.322(09)   & 0.835(30) & $-$0.200(10)  & 0.635(26)  \\   
$a12m220S$          &  0.904(42) & $-$0.321(27)   & 0.781(51) & $-$0.238(33)  & 0.543(68)  \\   
$a12m220 $          &  0.924(24) & $-$0.311(14)   & 0.818(32) & $-$0.194(12)  & 0.624(30)  \\ 
$a12m220L$          &  0.922(12) & $-$0.340(09)   & 0.819(29) & $-$0.216(08)  & 0.600(26)  \\  
\hline                                                                                   
$a09m310 $          &  0.928(12) & $-$0.308(05)   & 0.845(24) & $-$0.208(07)  & 0.637(19)  \\   
$a09m220 $          &  0.931(15) & $-$0.329(08)   & 0.810(24) & $-$0.205(08)  & 0.604(20)  \\   
$a09m130 $          &  0.901(23) & $-$0.344(17)   & 0.772(29) & $-$0.198(12)  & 0.574(28)  \\
$a09m130W$          &  0.919(17) & $-$0.330(09)   & 0.806(25) & $-$0.205(09)  & 0.601(23)  \\
\hline                                                                                   
$a06m310 $          &  0.916(27) & $-$0.317(16)   & 0.836(29) & $-$0.210(13)  & 0.626(31)  \\   
$a06m310W$          &  0.897(24) & $-$0.307(17)   & 0.833(26) & $-$0.204(10)  & 0.629(25)  \\   
$a06m220 $          &  0.890(16) & $-$0.316(13)   & 0.816(22) & $-$0.206(11)  & 0.609(21)  \\  
$a06m220W$          &  0.905(25) & $-$0.336(16)   & 0.809(30) & $-$0.209(12)  & 0.600(30)  \\  
$a06m135 $          &  0.902(23) & $-$0.318(13)   & 0.811(26) & $-$0.193(11)  & 0.618(26)  \\  
\hline 
11-point fit        &  0.895(21) &  $-$0.320(12)  & 0.790(27) & $-$0.198(10)  & 0.590(25)  \\  
$\chi^2/$d.o.f.     &  0.29      &  0.52          & 0.20      & 0.67          & 0.38       \\  
10-point fit        &  0.890(27) &  $-$0.324(17)  & 0.810(36) & $-$0.201(16)  & 0.608(37)  \\  
$\chi^2/$d.o.f.     &  0.33      &  0.59          & 0.12      & 0.77          & 0.37       \\  
$10^\ast$-point fit &  0.895(21) &  $-$0.319(12)  & 0.790(27) & $-$0.197(10)  & 0.592(25)  \\  
$\chi^2/$d.o.f.     &  0.34      &  0.57          & 0.09      & 0.57          & 0.16       \\  
\end{tabular}
\end{ruledtabular}
\caption{Results for the renormalized connected part of the flavor
  diagonal charges, $g_\Gamma^{\rm bare}/g_V^{{u-d},{\rm bare}} \times
  Z_\Gamma^{u-d}/Z_V^{u-d}$.  The final errors are obtained by adding
  in quadrature the errors in estimates of the ratios $g_\Gamma^{\rm
    bare}/g_V^{{u-d},{\rm bare}}$ to the errors in the ratios of the
  renormalization constants, $Z_\Gamma^{u-d}/Z_V^{u-d}$ given in
  Table~\protect\ref{tab:Zfinal}. Results for $g_T^{u+d}$ are
  presented assuming that the disconnected contributions, shown to be
  tiny in Ref.~\protect\cite{Bhattacharya:2015wna}, can be
  neglected. Results of three CCFV fits (the 11-point, the 10-point, and 
  the $10^\ast$-point defined in the text) are given
  in the bottom six rows. }
\label{tab:resultsrenormFD}
\end{table*}

\section{Continuum, chiral and finite volume fit for the charges $g_A$, $g_S$, $g_T$}
\label{sec:results}

To obtain estimates of the renormalized charges given in
Tables~\ref{tab:resultsrenormIV} and~\ref{tab:resultsrenormFD} in the
continuum limit ($a\rightarrow 0$), at the physical pion mass ($M_{\pi^0}
= 135$~MeV) and in the infinite volume limit ($L \rightarrow \infty$), we
need an appropriate physics motivated fit ansatz. To parametrize the
dependence on $M_\pi$ and the finite volume parameter $M_\pi L$, we
resort to results from finite volume chiral perturbation theory
($\chi$FT)~\cite{Bernard:1992qa,Bernard:1995dp,Bernard:2006gx,Bernard:2006te,Khan:2006de,Colangelo:2010ba,deVries:2010ah}.
For the lattice discretization effects, the corrections start with the
term linear in $a$ since the action and the operators in our
clover-on-HISQ formalism are not fully $O(a)$ improved.  Keeping just
the leading correction term in each, plus possibly the chiral
logarithm term discussed below, our approach is to make a simultaneous
fit in the three variables to the data from the eleven ensembles. We call
these the CCFV fits. For the isovector charges and the flavor diagonal
axial and tensor charges, the ansatz is
\begin{align}
  g_{A,S,T}^{u-d} (a,M_\pi,L) &= c_1 + c_2a + c_3 M_\pi^2 + c_3^{\rm log} M_\pi^2 \ln \left(\frac{M_\pi}{M_\rho}\right)^2  \nonumber \\
&+ c_4 M_\pi^2 \frac{e^{-M_\pi L}}{X(M_\pi L)} \,, 
\label{eq:extrapgAST} 
\end{align}
%
where $M_\rho$ in the chiral logarithm is the renormalization scale.  

\begin{figure*}[tb]   
\subfigure{
    \includegraphics[width=0.32\linewidth]{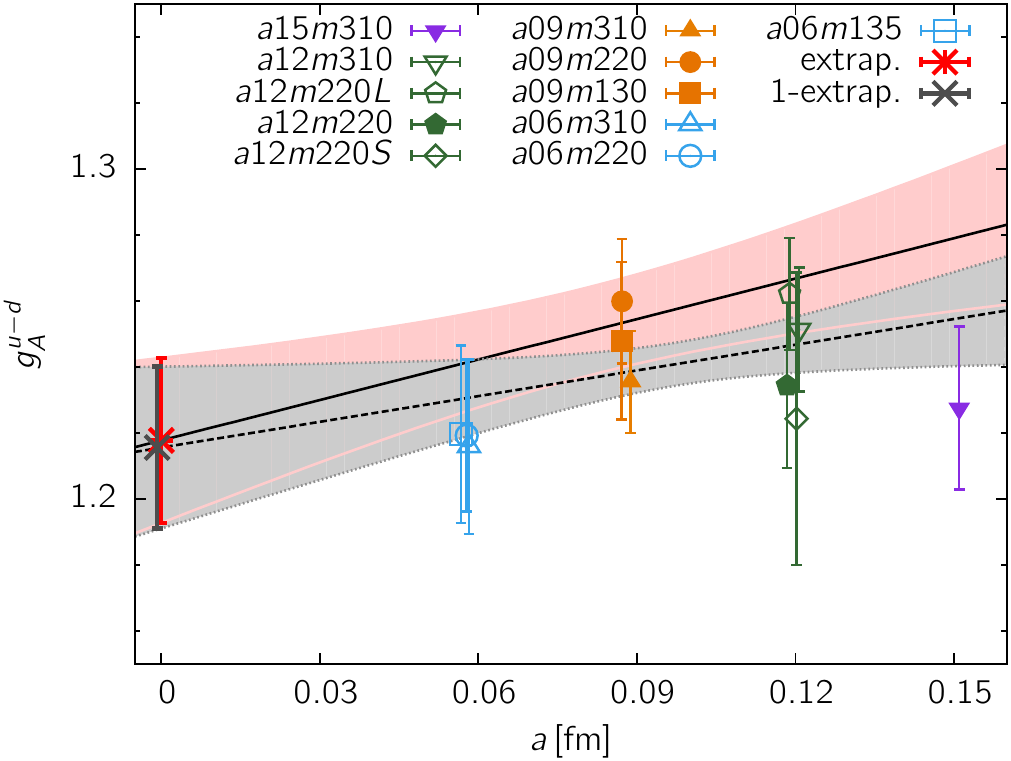}
    \includegraphics[width=0.32\linewidth]{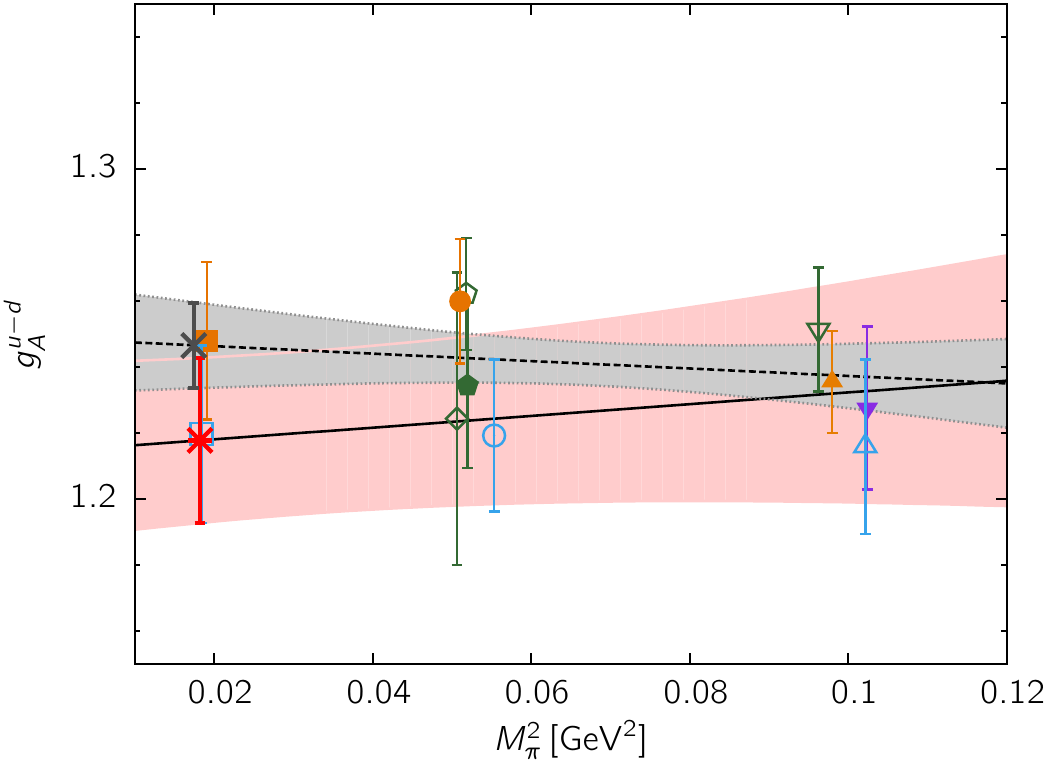}
    \includegraphics[width=0.32\linewidth]{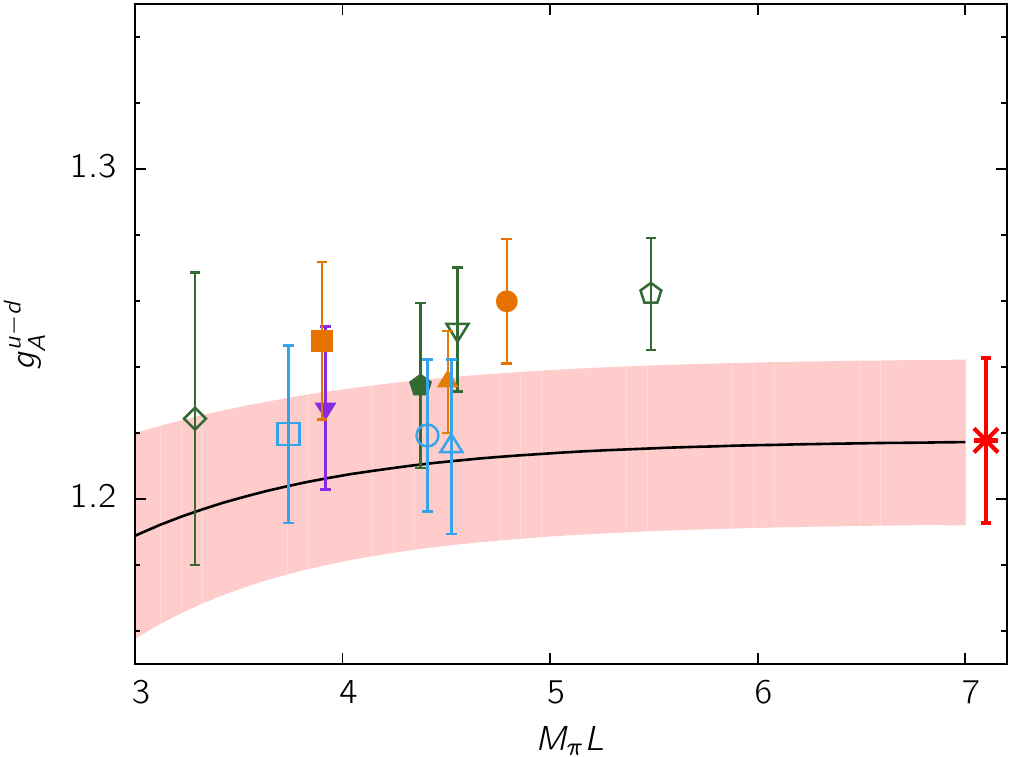}
}
\subfigure{
    \includegraphics[width=0.32\linewidth]{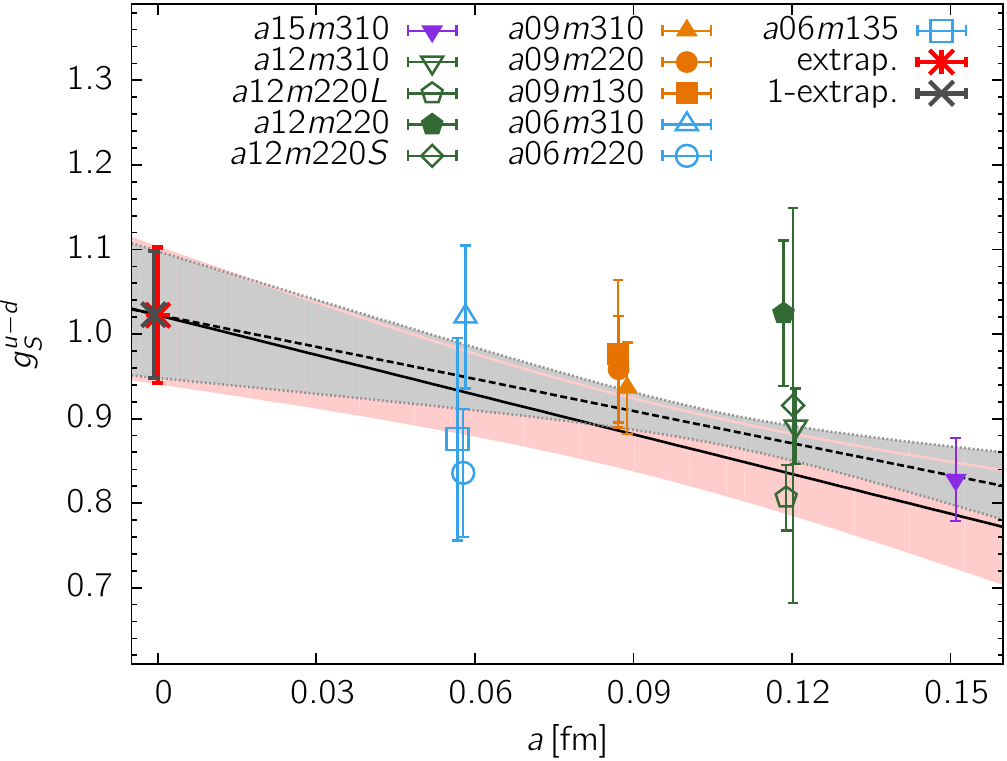}
    \includegraphics[width=0.32\linewidth]{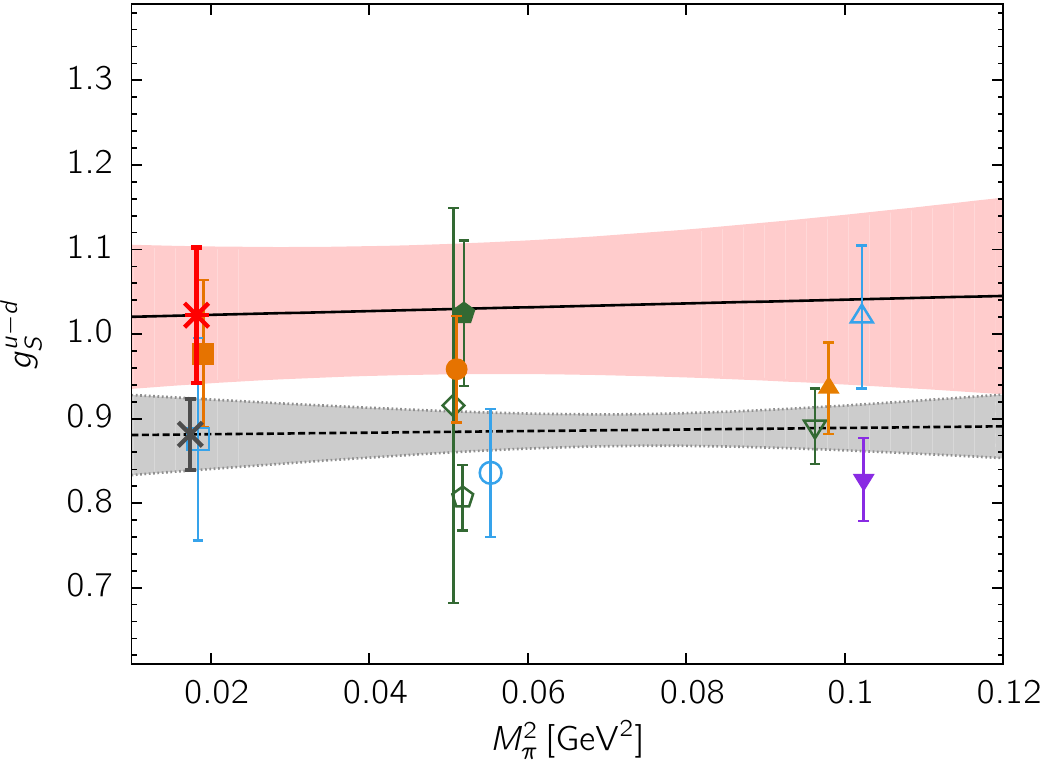}
    \includegraphics[width=0.32\linewidth]{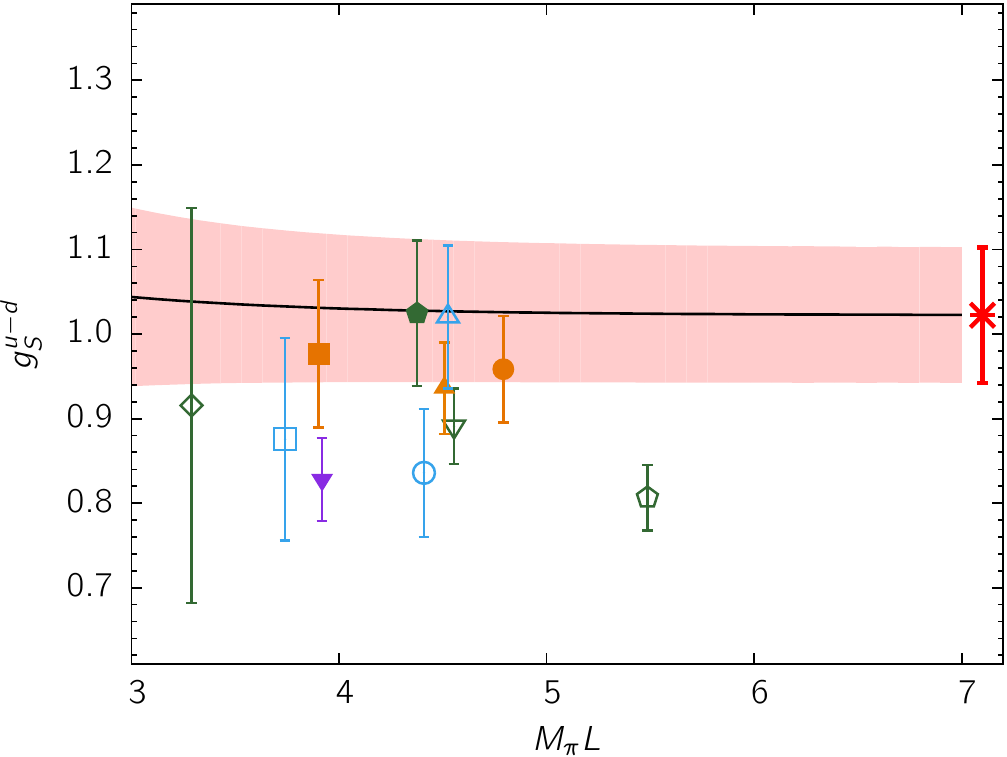}
}
\subfigure{
    \includegraphics[width=0.32\linewidth]{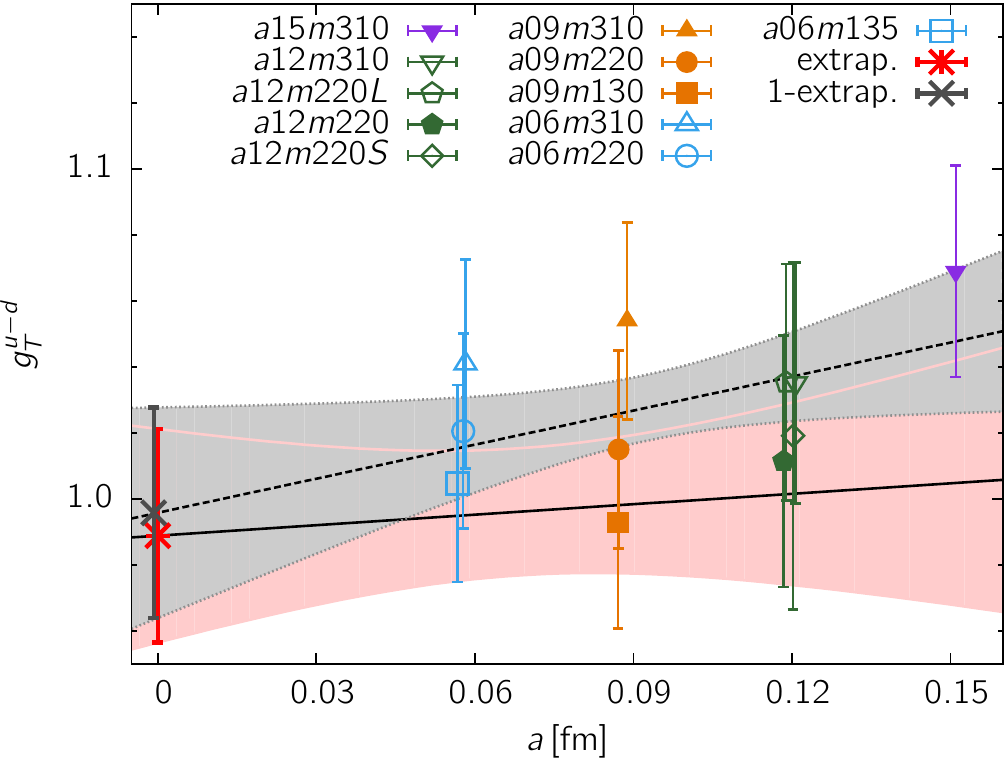}
    \includegraphics[width=0.32\linewidth]{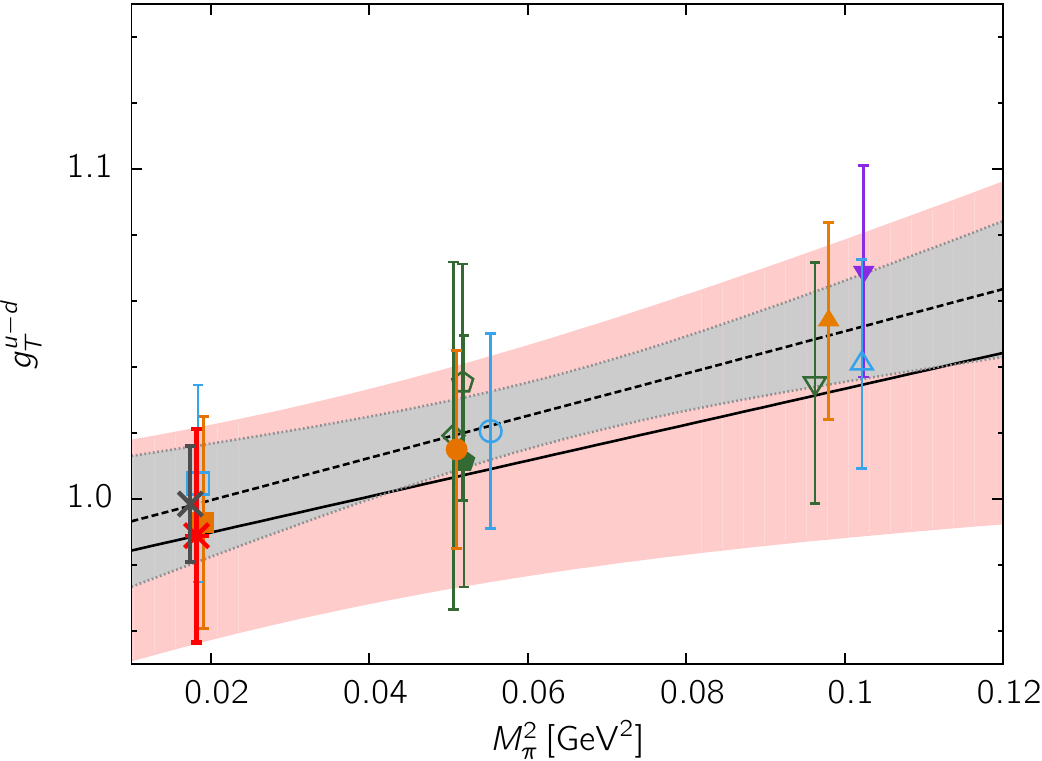}
    \includegraphics[width=0.32\linewidth]{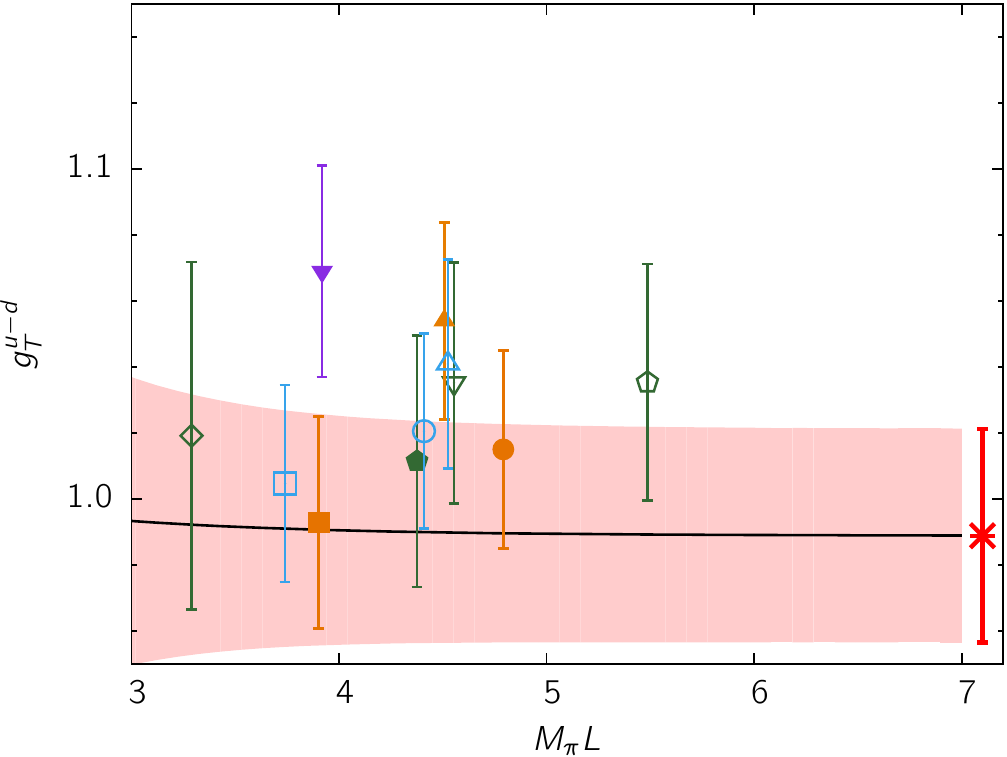}
}
\caption{The 11-point CCFV
  fit using Eq.~\protect\eqref{eq:extrapgAST} to the data for the
  renormalized isovector charges $g_A^{u-d}$, $g_S^{u-d}$, and
  $g_T^{u-d}$ in the $\overline{{\rm MS}}$ scheme at 2~GeV. The
  result of the simultaneous extrapolation to the physical point
  defined by $a\rightarrow 0$, $M_\pi \rightarrow M_{\pi^0}^{{\rm
      phys}}=135$~MeV and $M_\pi L \rightarrow \infty$ are marked by a red
  star.  The pink error band in each panel is the result of the
  simultaneous fit but shown as a function of a single variable. The
  overlay in the left (middle) panels with the dashed line within the
  grey band is the fit to the data versus $a$ ($M_\pi^2$), i.e.,
  neglecting dependence on the other two variables. The symbols used to plot the data are 
  defined in the left panels.}
  \label{fig:conUmD-extrap11}
\end{figure*}

The coefficients, $c_3^{\rm log}$, are known in $\chi$PT, and with
lattice QCD data at multiple values of $M_\pi$ and at fixed $a$ and
$M_\pi L$ one can compare them against values obtained from the
fits. As shown in Fig.~\ref{fig:conUmD-extrap11}, the $M_\pi$
dependence of all three isovector charges is mild and adequately fit
by the lowest order term. Since the $c_3^{\rm log}$ predicted by
$\chi$PT are large, including it requires also including still higher
order terms in $M_\pi$ to fit the mild dependence. In our case, with
data at just three values of $M_\pi$ and the observed mild dependence
between 320 and 135~MeV, including more than one free parameter is not
justified based on the Akaike Information Criterion (AIC) that requires the
reduction of $\chi^2$ by two units for each extra parameter. In short,
we cannot test the predictions of $\chi$PT. For example, in a fit
including the chiral log term and a $M_\pi^3$ term, the two additional
terms essentially negate each other over the range of the data, i.e.,
between 320--135~MeV. If the large $\chi$PT value for the coefficient $c_3^{\rm log}$ 
of the chiral log is used as an input, then the fit pushes the
coefficient of the $M_\pi^3$ term to also be large to keep the net
variation within the interval of the data small. Furthermore, as can be seen from
Table~\ref{tab:chiralfit}, even the coefficients of the leading order
terms are poorly determined for all three charges. This is because the
variations between points and the number of points are both small. For
these reasons, including the chiral logarithm term to analyze the
current data does not add predictive capability, nor does it provide a
credible estimate of the uncertainty due to the fit ansatz, nor tests
the $\chi$PT value of the coefficient $c_3^{\rm log}$. Consequently,
the purpose of our chiral fit reduces to getting the value at
$M_\pi=135$~MeV.  We emphasize that this is obtained reliably with
just the leading chiral correction since the fits are anchored by the
data from the two physical pion mass ensembles.

The finite-volume correction, in general, consists of a number of
terms, each with different powers of $M_\pi L$ in the denominator and
depending on several low-energy constants (LEC)~\cite{Khan:2006de}. We
have symbolically represented these powers of $M_\pi L$ by $X(M_\pi
L)$. Since the variation of this factor is small
compared to the exponential over the range of $M_\pi L$ investigated,
we set $X(M_\pi L) = {\rm constant}$ and retain only the
appropriate overall factor $M_\pi^2 e^{-M_\pi L}$, common to all the
terms in the finite-volume expansion, in our fit ansatz. The, {\it a posteriori}, 
justification for this simplification is that no significant finite
volume dependence is observed in the data as shown in Fig.~\ref{fig:conUmD-extrap11}. 

We have carried out four fits with different selections of the
fourteen data points and for the two constructions of the renormalized
charges. Starting with the 14 calculations, we first construct a
weighted average of the pairs of points from the three $a09m130$,
$a06m310$ and $a06m220$ ensembles. For errors, we adopt the Schmelling
procedure~\cite{Schmelling:1994pz} assuming maximum correlation
between the two values from each ensemble.  This gives us eleven data points
to fit.

\begin{itemize}
\item
The fit with all the data points is called the 11-point fit. This is 
used to obtain the final results.
\item
Remove the coarsest $a15m310$ ensemble point from the analysis. This
is called the 10-point fit. 
\item
Remove the $a12m220S$ point as it has
the largest errors and the smallest volume.  This is called the
$10^\ast$-point fit.
\item
To compare results for $g_A^{u-d}$ with those from the CalLat collaboration~\cite{Chang:2018uxx}
(see Sec.~\ref{sec:comparison}), we perform an $8-$point fit that
neglects the data from the three $a\approx 0.06$~fm ensembles.
\end{itemize}
The results from these four fits and for the two ways of constructing
the renormalized isovector charges are given in
Table~\ref{tab:resultsrenormIV}. We find that the six estimates for
$g_A^{u-d}$ and $g_T^{u-d}$ from the 11-point, 10-point and
$10^\ast$-point fits with the two ways of renormalization overlap
within $1\sigma$. As discussed in Sec.~\ref{sec:comparison}, 
for $g_A^{u-d}$, the $a15m310$ point plays an
important role in the comparison with the CalLat results. 

For the final results, we use the 11-point fit to the isovector charges
renormalized using $g_\Gamma^{\rm bare}/g_V^{\rm bare} \times
Z_\Gamma/Z_V$ as some of the systematics cancel in the double
ratio. These fits are shown in Fig.~\ref{fig:conUmD-extrap11}.

The lattice artifact that has the most impact on the final values is
the dependence of $g_A^{u-d}$ and $g_S^{u-d}$ on the lattice spacing
$a$. As shown in Fig.~\ref{fig:conUmD-extrap11}, in these cases the
CCFV fit coincides with the fit versus just $a$ (pink and grey bands
overlap in such cases). On the other hand, one can see from the middle
panels, showing the variation versus $M_\pi^2$, that had we only
analyzed the data versus $M_\pi^2$ (grey band), we would have gotten a
higher value for $g_A^{u-d}$ and a lower one for $g_S^{u-d}$, and both
with smaller errors.  Our conclusion is that, even when the observed
variation is small, it is essential to perform a simultaneous CCFV fit
to remove the correlated contributions from the three lattice
artifacts.

The data for $g_T^{u-d}$ continues to show very little sensitivity to
the three variables and the extrapolated value is
stable~\cite{Bhattacharya:2016zcn}.  A large part of the error in the
individual data points, and thus in the extrapolated value, is now due
to the poorly behaved two-loop perturbation theory used to match the
RI-sMOM to the $\overline{\rm MS}$ scheme in the calculation of the
renormalization constant $Z_T$. Further precision in $g_T^{u-d}$,
therefore, requires developing more precise methods for calculating the
renormalization constants.

Overall, compared to the results presented in
Ref.~\cite{Bhattacharya:2016zcn}, our confidence in the CCFV fits for
all three charges has improved with the new higher precision data.
The final results for the isovector charges in the $\overline{\rm MS}$
scheme at 2~GeV from the 11-point fit to data given in
Table~\ref{tab:resultsrenormIV} and renormalized using $g_\Gamma^{\rm
  bare}/g_V^{\rm bare} \times Z_\Gamma/Z_V$ are:
\begin{align}
  g_A^{u-d}  &= 1.218(25) \,,   \nonumber \\
  g_S^{u-d}  &= 1.022(80) \,,   \nonumber \\
  g_T^{u-d}  &= 0.989(32) \,.
  \label{eq:gFinal}
\end{align}
These results for $g_S^{u-d}$ and $g_T^{u-d}$ meet the target 
ten percent uncertainty needed to leverage precision neutron decay
measurements of the helicity flip parameters $b$ and $b_\nu$ at the
$10^{-3}$ level to constrain novel scalar and tensor couplings,
$\epsilon_S$ and $\epsilon_T$, arising at the TeV
scale~\cite{Bhattacharya:2011qm,Bhattacharya:2016zcn}.

\begin{figure*}[tb]  
\subfigure{
    \includegraphics[width=0.32\linewidth]{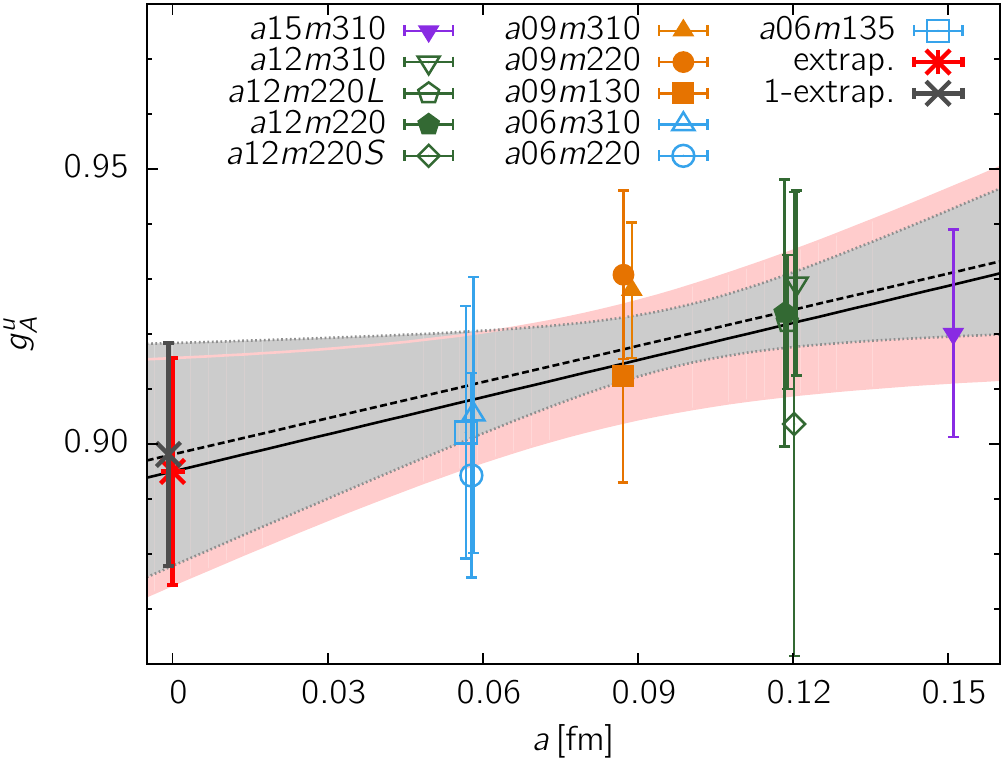}
    \includegraphics[width=0.32\linewidth]{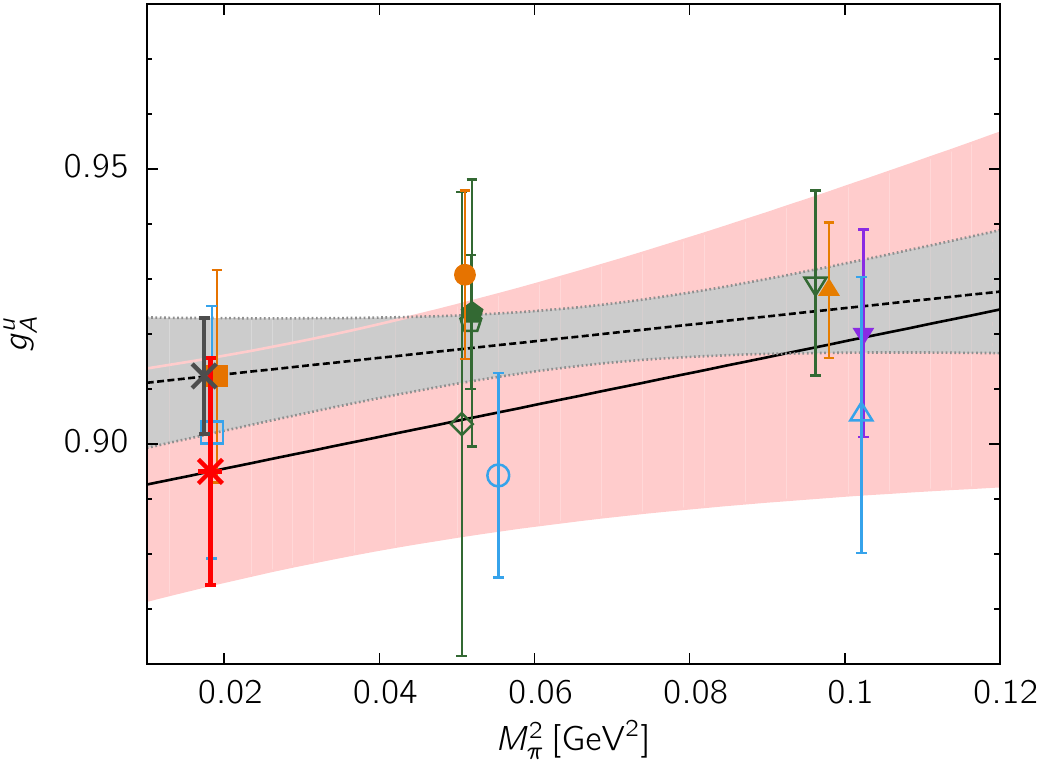}
    \includegraphics[width=0.32\linewidth]{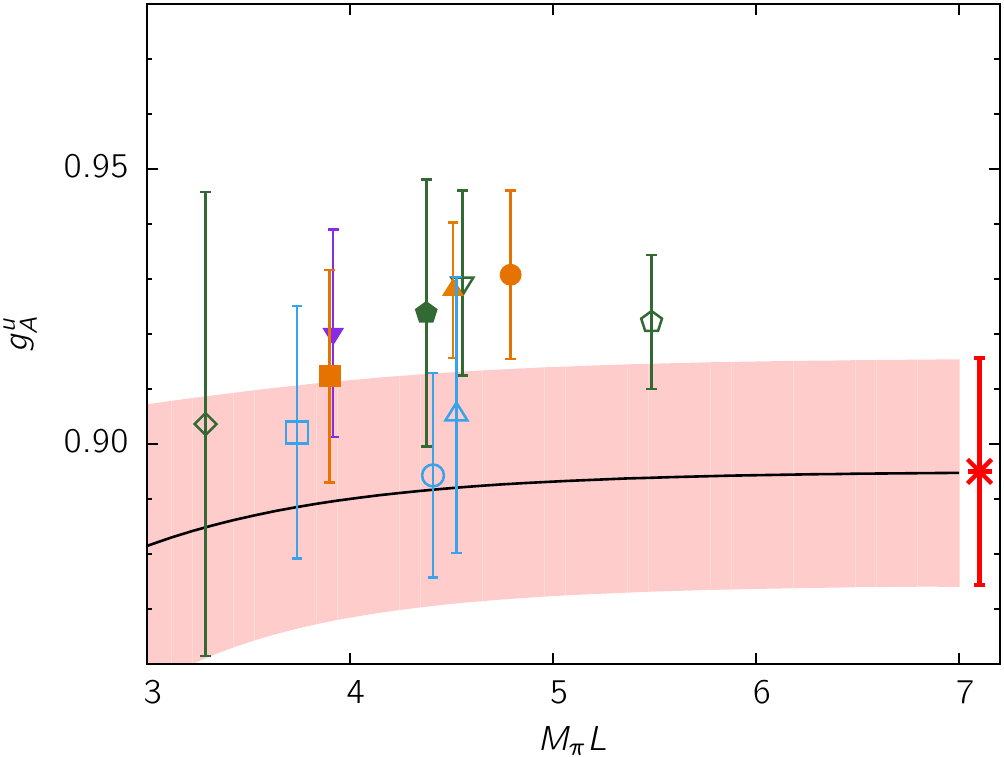}
}
\hspace{0.04\linewidth}
\subfigure{
    \includegraphics[width=0.32\linewidth]{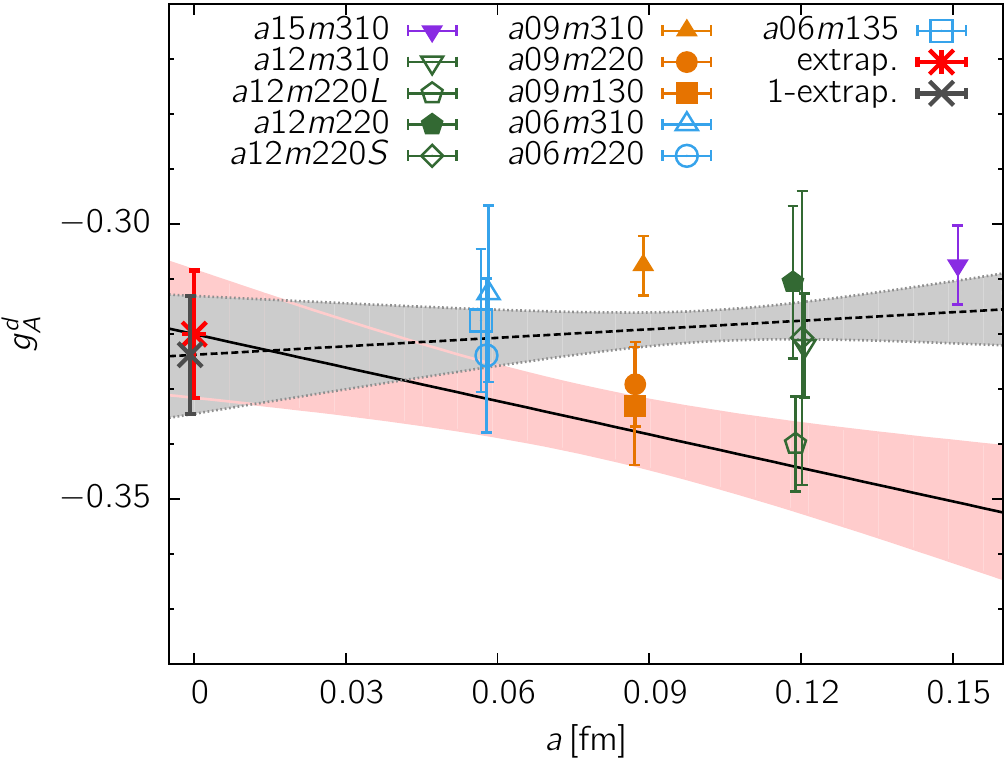}
    \includegraphics[width=0.32\linewidth]{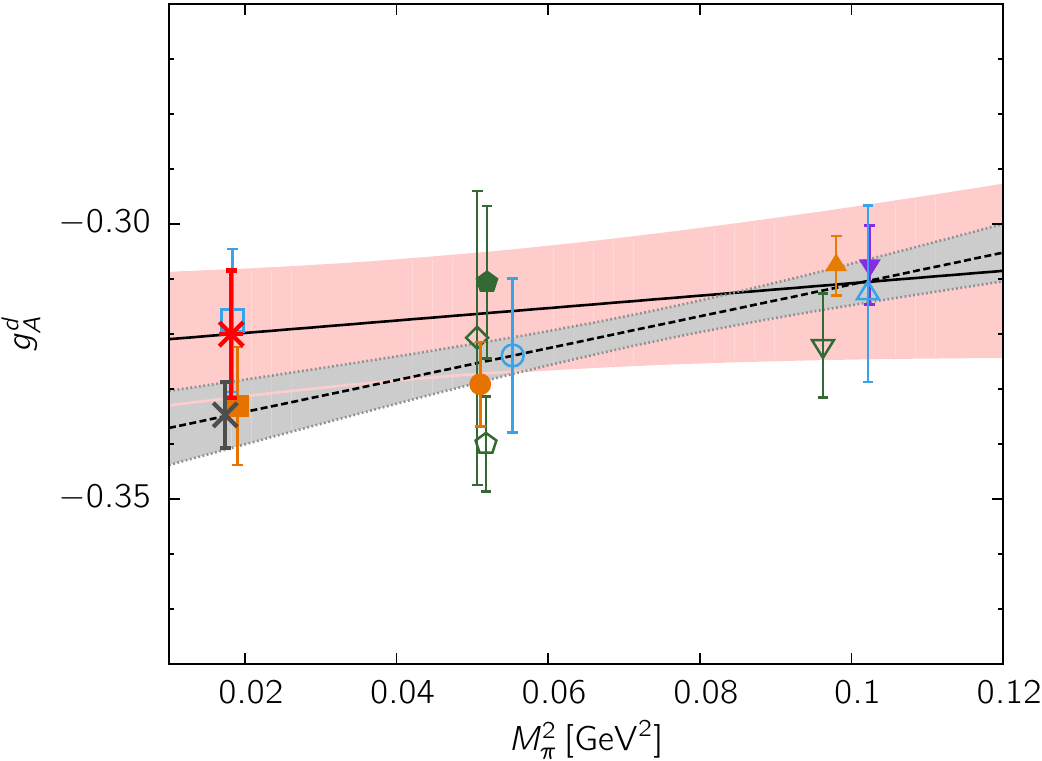}
    \includegraphics[width=0.32\linewidth]{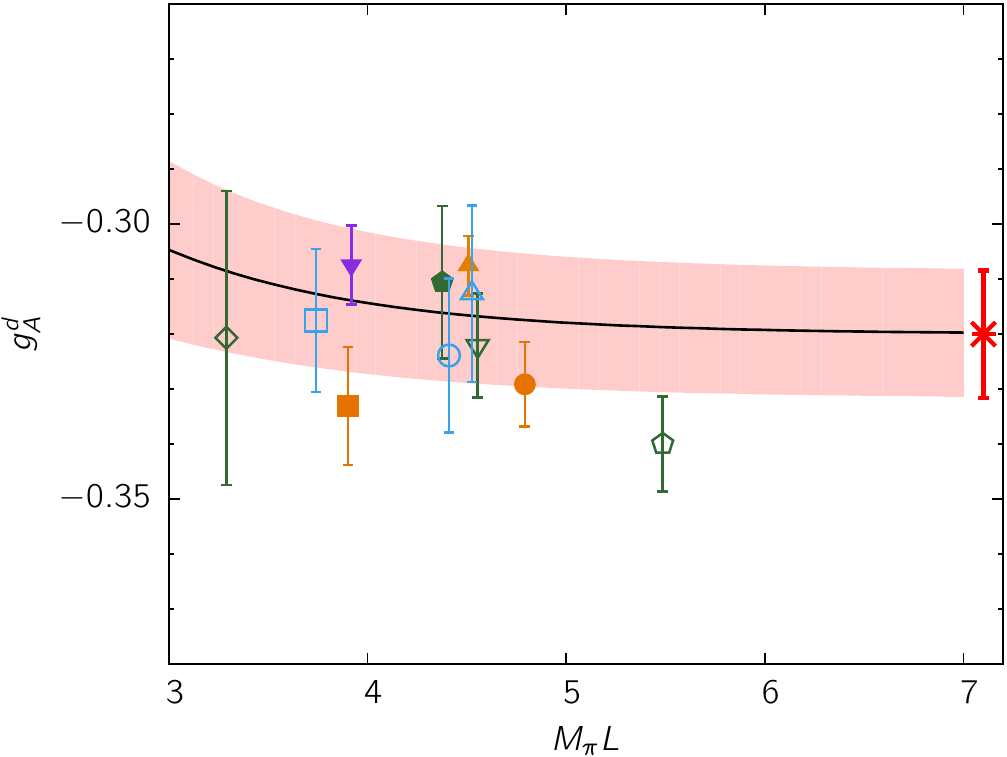}
}
\caption{The 11-point
  CCFV fit using Eq.~\protect\eqref{eq:extrapgAST} to the connected
  data for the flavor diagonal charges $g_A^{u}$ and $g_A^{d}$
  renormalized in the $\overline{{\rm MS}}$ scheme at 2~GeV.  Only the
  data for $g_A^u$ show a notable dependence on the lattice spacing
  $a$. The rest is the same as in
  Fig.~\protect\ref{fig:conUmD-extrap11}.\looseness-1
\label{fig:extrap-gA-diagonal}}
\end{figure*}


\begin{figure*}[tb] 
\subfigure{
    \includegraphics[width=0.32\linewidth]{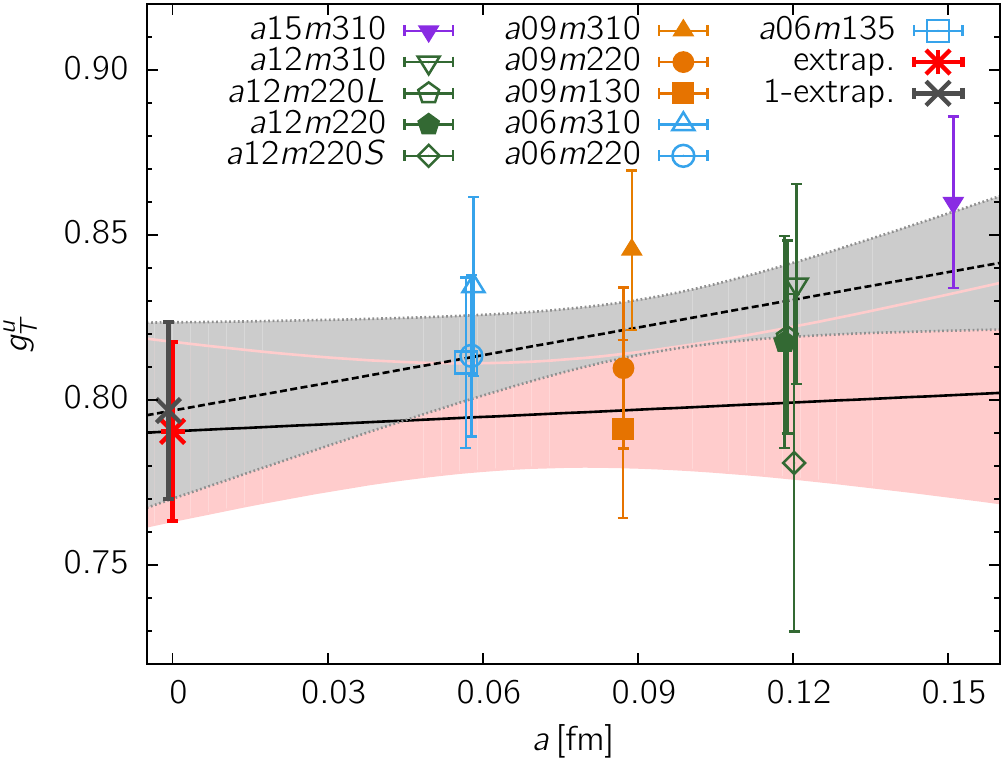}
    \includegraphics[width=0.32\linewidth]{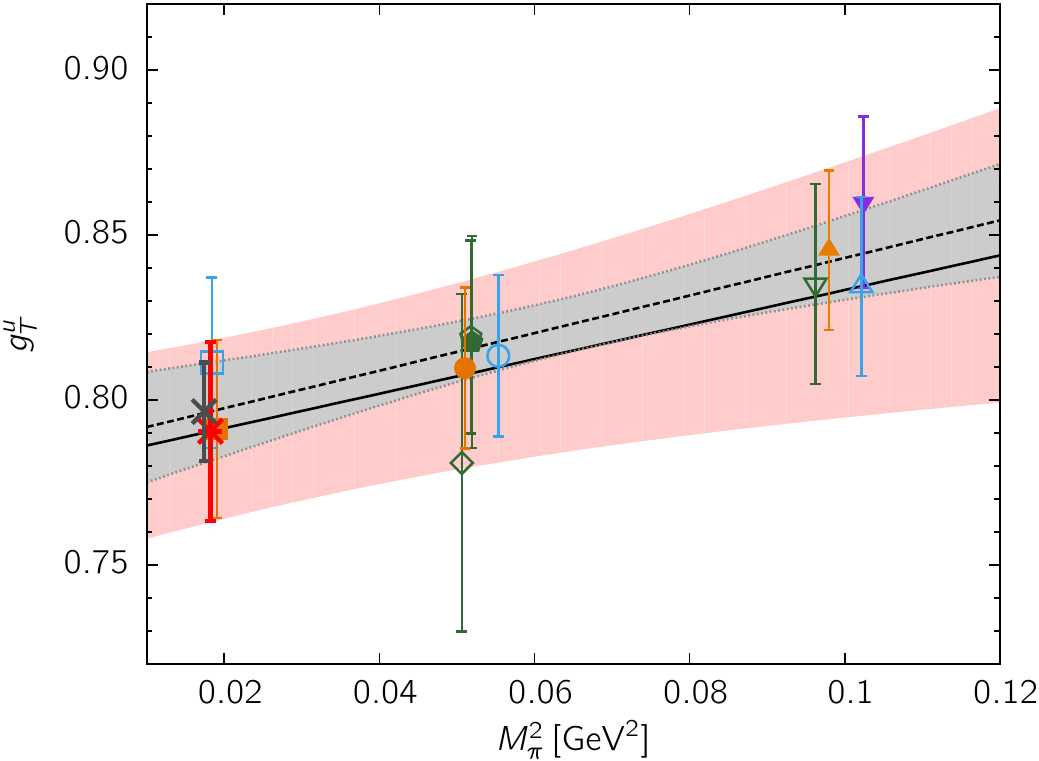}
    \includegraphics[width=0.32\linewidth]{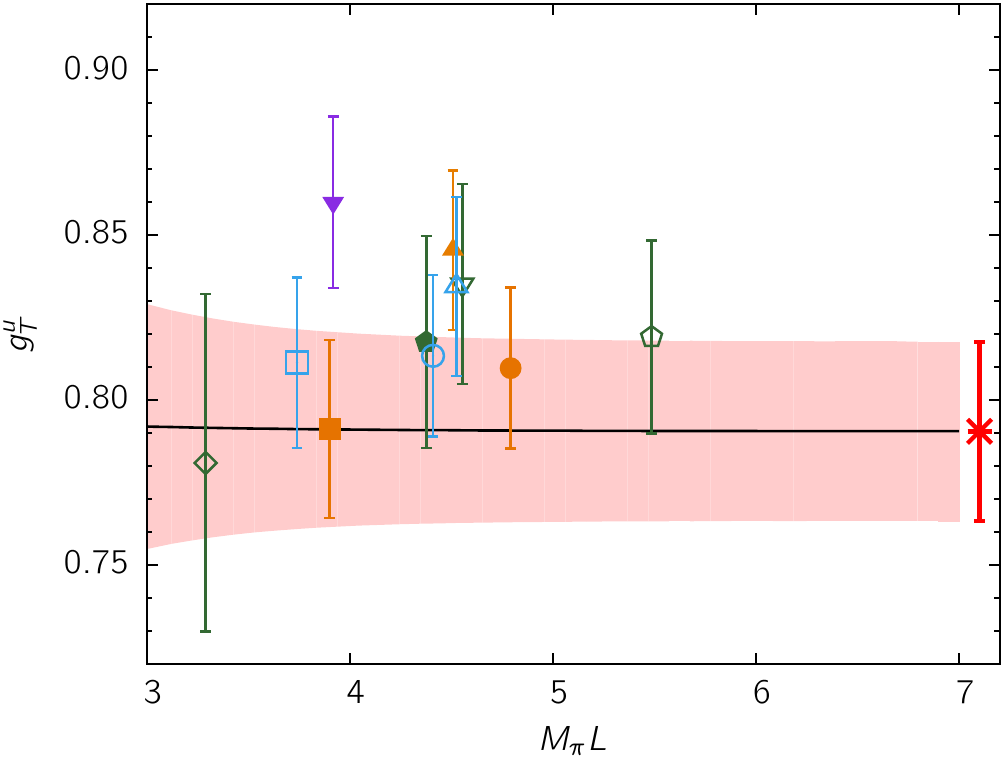}
}
\hspace{0.04\linewidth}
\subfigure{
    \includegraphics[width=0.32\linewidth]{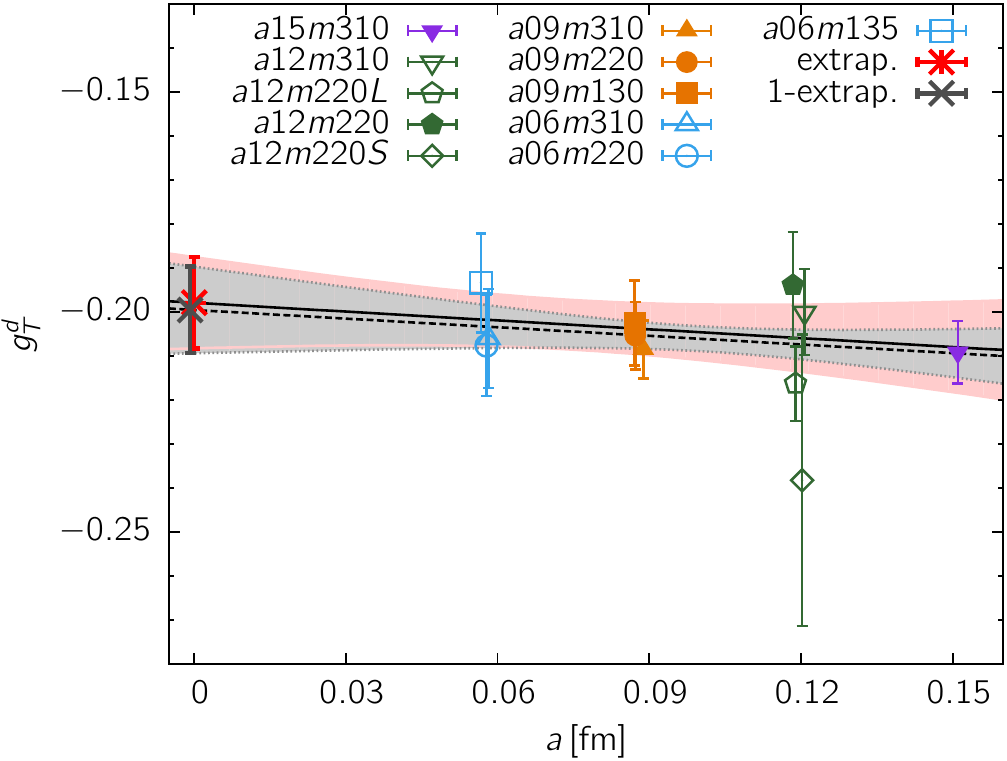}
    \includegraphics[width=0.32\linewidth]{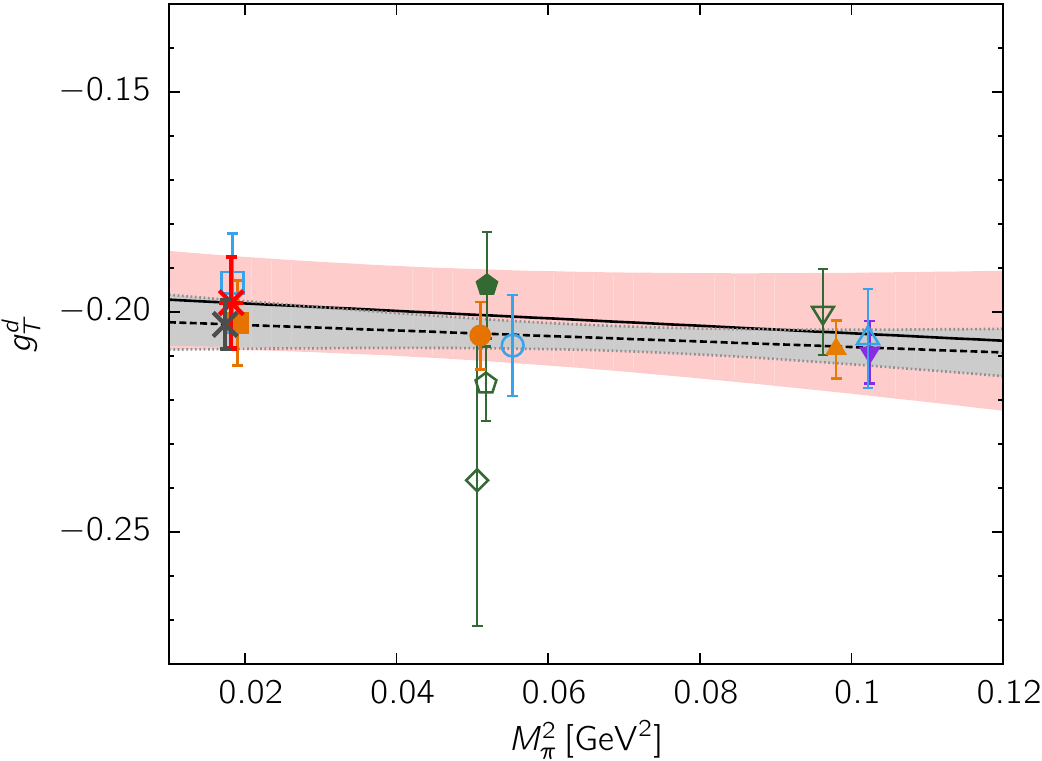}
    \includegraphics[width=0.32\linewidth]{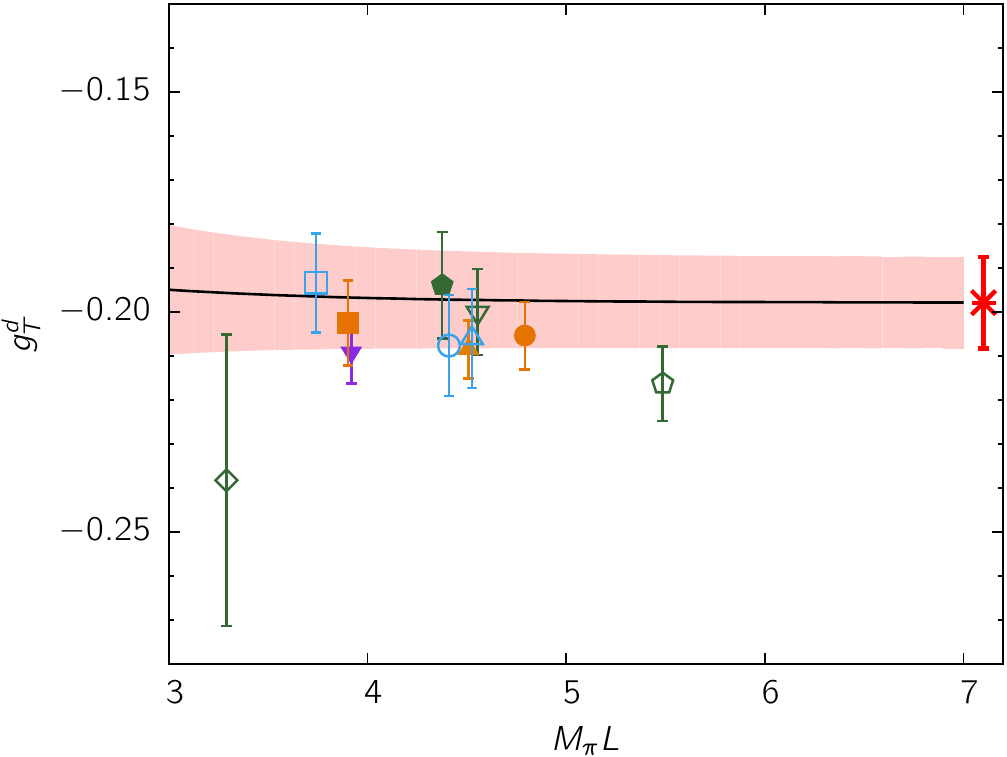}
}
\caption{The 11-point
  CCFV fit using Eq.~\protect\eqref{eq:extrapgAST} to the connected
  data for the flavor diagonal charges $g_T^{u}$ and $g_T^{d}$
  renormalized in the $\overline{{\rm MS}}$ scheme at 2~GeV.  Only the
  data for $g_T^u$ show a notable dependence on $M_\pi$. The rest is the same as in
  Fig.~\protect\ref{fig:conUmD-extrap11}.\looseness-1
  \label{fig:extrap-gT-diagonal}}
\end{figure*}

Results of the 11-point, 10-point, and $10^\ast$-point
fits to the connected contributions to the flavor-diagonal charges
$g_{A,T}^{u,d}$, using the isovector renormalization factor $Z_{A,T}^{\rm isovector}$, respectively, 
are given in Table~\ref{tab:resultsrenormFD}. 
Their behavior
versus the lattice spacing and the pion mass is shown in
Figs.~\ref{fig:extrap-gA-diagonal} 
and~\ref{fig:extrap-gT-diagonal} using the 11-point fits, again with
$c_3^{\rm log}=0$ in the ansatz given in Eq.~\eqref{eq:extrapgAST}.
The data exhibit the following features:
\begin{itemize}
\item
The noticeable variation in the axial charges is in $g_A^u$ versus $a$ 
which carries over to $g_A^{u-d}$. 
\item
The flavor diagonal charges $g_T^{u,d}$ show little variation except for the 
small dependence of $g_T^u$ on $M_\pi^2$ which carries over to $g_T^{u-d}$.
\end{itemize}
Our final results from the 11-point fits for the connected parts of
the flavor diagonal charges for the proton are \looseness-1
\begin{align}
  g_A^{u,{\rm conn}}  &= 0.895(21)  \qquad\      g_A^{d,{\rm conn}}  = -0.320(12)  \,,     \nonumber \\
  g_T^{u,{\rm conn}}  &= 0.790(27)  \qquad\      g_T^{d,{\rm conn}}  = -0.198(10)    \,.
  \label{eq:FDconnected}
\end{align}
Estimates for the neutron are given by the $u \leftrightarrow d$
interchange.  

We again remind the reader that the disconnected contributions for the
flavor diagonal axial charges are $O(15\%)$ and will be discussed elsewhere. The
disconnected contribution to $g_T^{u+d}$ is small (comparable to the
statistical errors) and $Z_T^{u-d} \approx Z_T^{u+d}$. Thus, the
results for $g_T^{u,d}$ and $g_T^{u+d}$ are a good approximation to 
the total contribution. The new estimates given here supersede
the values presented in
Refs.~\cite{Bhattacharya:2015wna,Bhattacharya:2015esa}.

\begin{table*}[tb]
\begin{center}
\renewcommand{\arraystretch}{1.2} 
\begin{ruledtabular}
\begin{tabular}{c|c|c|c|c|c}
             &   $c_1$      &  $c_2$       &   $c_3$        & $c_4$            & $g_\Gamma$   \\
             &              &  fm${}^{-1}$ & GeV${}^{-2}$   & GeV${}^{-2}$     &              \\
\hline                                                                                      
$g_A^{u-d}$  &   1.21(3)    &   0.41(26)   & 0.18(33)       &$-$32(19)         & 1.218(25)   \\ 
\hline                                                                      
$g_S^{u-d}$  &   1.02(1)    &$-$1.57(75)   & 0.22(1.12)     &  24(54)          & 1.022(80)   \\ 
\hline                                                                      
$g_T^{u-d}$  &   0.98(3)    &   0.11(38)   & 0.55(45)       &  5(29)           & 0.989(32)   \\ 
\end{tabular}
\end{ruledtabular}
\caption{Values of the fit parameters in the
  CCFV ansatz defined in Eq.~\eqref{eq:extrapgAST} with $c_3^{\rm
    log}=0$. The results are given for the  11-point fit used to 
    extract the three isovector charges. }
\label{tab:chiralfit}
\end{center}
\end{table*}

\section{Assessing additional error due to CCFV fit ansatz}
\label{sec:errors}

In this section we reassess the estimation of errors from various
sources and provide an additional systematic uncertainty in the
isovector charges due to using a CCFV ansatz with only the leading
order correction terms. We first briefly review the systematics that
are already addressed in our analysis leading to the results in Eq.~\eqref{eq:gFinal}:
\begin{itemize}
\item
Statistical and excited-state contamination (SESC): Errors from these
two sources are jointly estimated in the 2- and $3^\ast$ state
fits. The 2- and $3^\ast$ state fits for $g_A^{u-d}$ and $g_T^{u-d}$
give overlapping results and in most cases the error estimates from
the quoted $3^\ast$-state fits are larger. For $g_S^{u-d}$, we compare
the 2- and $2^\ast$-state fits. Based on these comparisons, an
estimate of the magnitude of possible residual ESC is given in the
first row of Table~\ref{tab:errors} for all three charges.
\item
Uncertainty in the determination of the renormalization constants
$Z_\Gamma$: The results for the $Z$'s and an estimate of the possible
uncertainty presented in Ref.~\cite{Bhattacharya:2016zcn} have not
changed. These are reproduced in Tables~\ref{tab:Zfinal}
and~\ref{tab:errors}, respectively. With the increase in statistical
precision of the bare charges, the uncertainty in the $Z_\Gamma$ is
now a significant fraction of the total uncertainty in
$g_{A,S,T}^{u-d}$.
\item
Residual uncertainties due to the three systematics, extrapolations to
$a\to 0$ and $M_\pi L \to \infty$ and the variation with $M_\pi$.
Estimates of errors in the simultaneous CCFV fit using the lowest
order corrections (see Eq.~\eqref{eq:extrapgAST}) are given in rows
3--5 in Table~\ref{tab:errors}.  These are, in most cases, judged to
be small because the variation with respect to the three variables,
displayed in Fig.~\ref{fig:conUmD-extrap11}, is small. With increased
statistics and the second physical mass ensemble, $a06m135$, our
confidence in the CCFV fits and the error estimates obtained with
keeping only the lowest-order corrections in each variable has
increased significantly. The exception is the dependence of
$g_S^{u-d}$ on $a$ as highlighted by the dependence of the
extrapolated value on whether the $a15m310$ point is included
(11-point fit) or excluded (10-point fit).
\end{itemize}
Adding the guesstimates for these five systematic uncertainties, given
in rows 1--5, in quadrature, leads to an error estimate given 
in the sixth row in Table~\ref{tab:errors}. This is 
consistent with the errors quoted in Eq.~\eqref{eq:gFinal} and
reproduced in the seventh row of Table~\ref{tab:errors}. We, therefore,
regard the fits and the error estimates given in Eq.~\eqref{eq:gFinal} as
adequately capturing the uncertainty due to the five systematics discussed above. 

The $\chi^2/{\rm d.o.f.}$ of all four fits for the axial and tensor
charges given in Table~\ref{tab:resultsrenormIV} are already very
small. Therefore, adding higher order terms to the ansatz is not
justified as per the Akaike Information
Criterion~\cite{Akaike:1100705}.  Nevertheless, to be conservative, we
quote an additional systematic uncertainty due to the truncation of
the CCFV fit ansatz at the leading order in each of the three
variables, by examining the variation in the data in
Fig.~\ref{fig:conUmD-extrap11}.

For $g^{u-d}_A$, the key reason for the difference between our extrapolated value and
the experimental results are the data on the $a\approx 0.06$~fm
lattices. As discussed in Sec.~\ref{sec:comparison}, an extrapolation
in $a$ with and without these ensembles gives $g^{u-d}_A=1.218(25)$
and $g^{u-d}_A=1.245(42)$, respectively. The difference, $0.03$, is
roughly half the total spread between the fourteen values of
$g^{u-d}_A$ given in Table~\ref{tab:resultsrenormIV}. We, therefore,
quote $0.03$ as the additional uncertainty due to the truncation of
the fit ansatz.

The dominant variation in $g^{u-d}_S$ is again versus $a$, and, as
stated above, the result depends on whether the $a15m310$ point is
included in the fit. We, therefore, take half the difference, $0.06$,
between the 11-point and 10-point fit values as the additional
systematic uncertainty. One gets a similar estimate by taking the
difference in the fit value at $a=0.06$~fm and $a=0$. For $g^{u-d}_T$,
the largest variation is versus $M_\pi^2$. Since we have data from two
ensembles at $M_\pi \approx 135$~MeV that anchor the chiral fit, we
take half the difference in the fit values at $M_\pi=135$ and $220$~MeV as
the estimate of the additional systematic uncertainty.\looseness-1

These error estimates, rounded up to two decimal places, are given in the last row of
Table~\ref{tab:errors}. Including them as a second systematic
error, our final results for the isovector charges in the
$\overline{\rm MS}$ scheme at 2~GeV are:
\begin{align}
  g_A^{u-d}  &= 1.218(25)(30) \,, \nonumber \\
  g_S^{u-d}  &= 1.022(80)(60) \,, \nonumber \\
  g_T^{u-d}  &= 0.989(32)(10) \,.
  \label{eq:gFinal2}
\end{align}
Similar estimates of possible extrapolation uncertainty apply also to
results for the connected contributions to the flavor diagonal charges
presented in Eq.~\eqref{eq:FDconnected}. Their final analysis, 
including disconnected contributions, 
will be presented in a separate publication. 

\begin{table}
\centering
\begin{ruledtabular}
\begin{tabular}{c|ccc}
Error From          &  $g_A^{u-d}$ & $g_S^{u-d}$ &   $g_T^{u-d}$     \\
\hline                                                      
SESC                &  $0.02$  $\Uparrow$    & $0.03$   $\Uparrow$     &  $0.01$   $\Downarrow$  \\
$Z$                 &  $0.01$  $\Downarrow$  & $0.04$   $\Uparrow$     &  $0.03$   $\Downarrow$  \\
$a$                 &  $0.02$  $\Downarrow$  & $0.04$   $\Uparrow$     &  $0.01$   $\Downarrow$  \\
Chiral              &  $0.01$  $\Uparrow$    & $0.01$  $\Downarrow$    &  $0.02$   $\Downarrow$  \\
Finite volume       &  $0.01$  $\Uparrow$    & $0.01$  $\Uparrow$      &  $0.01$   $\Uparrow$    \\
\hline                                                      
Guesstimate error   &  $0.033$               & $0.066$                 &  $0.04$                 \\
\hline                                                      
Error quoted        &  $0.025$               & $0.080$                 &  $0.032$                \\
\hline
Fit ansatz          &  $0.03$                & $0.06$                  &  $0.01$                 \\
\end{tabular}
\end{ruledtabular}
\caption{Estimates of the error budget for the three isovector charges
  due to each of the five systematic effects described in the text.
  The symbols $\Uparrow$ and $\Downarrow$ indicate the direction in
  which a given systematic is observed to drive the central value
  obtained from the 11-point fit. The sixth row gives a guesstimate of
  error obtained by combining these five systematics in quadrature.
  This guesstimate is consistent with the actual errors obtained from
  the 11-point fit and quoted in Eq.~\protect\ref{eq:gFinal} and
  reproduced in the seventh row.  The last row gives the additional
  systematic error assigned to account for possible uncertainty
  due to the using the CCFV fit ansatz with just the lowest order
  correction terms as described in the text.  }
\label{tab:errors}
\end{table}

Our new estimate $g_S^{u-d}= 1.022(80)(60)$ is in very good agreement
with $g_S^{u-d}= 1.02(8)(7)$ obtained by Gonzalez-Alonso and
Camalich~\cite{Gonzalez-Alonso:2013ura} using the conserved vector
current (CVC) relation $g_S/g_V = (M_N-M_P)^{\rm QCD}/ (m_d-m_u)^{\rm
  QCD}$ with the FLAG lattice-QCD estimates~\cite{FLAG:2016qm} for the
two quantities on the right hand side.  The superscript QCD denotes
that the results are in a theory with just QCD, i.e., neglecting
electromagnetic corrections. Using CVC in reverse, our predictions for
$(M_N-M_P)^{\rm QCD}$, using lattice QCD estimates for $m_u$ and
$m_d$, are given in Table~\ref{tab:Mn-Mp}. The uncertainty in these
estimates is dominated by that in $g_S^{u-d}$.\looseness-1

\begin{table}[ht]
\begin{center}
\renewcommand{\arraystretch}{1.2} 
\begin{ruledtabular}
\begin{tabular}{c|c|l}
$M_N-M_P$ &  $N_f$    & $\{m_d,m_u\}^{\rm QCD}$             \\
(MeV)     &  Flavors  & (MeV)                               \\
\hline
$2.58(32)$  & 2+1       & $m_d = 4.68(14)(7),m_u = 2.16(9)(7)$~\protect\cite{FLAG:2016qm}       \\
$2.73(44)$  & 2+1+1     & $m_d = 5.03(26),m_u = 2.36(24)$~\protect\cite{FLAG:2016qm}       \\
$2.41(27)$  & 2+1       & $m_d - m_u  = 2.41(6)(4)(9)$~\protect\cite{Fodor:2016bgu}           \\
$2.63(27)$  & 2+1+1     & $m_d = 4.690(54),m_u = 2.118(38)$~\protect\cite{Bazavov:2018omf}         
\end{tabular}
\end{ruledtabular}
\caption{Results for the mass difference $(M_N-M_P)^{\rm QCD}$ obtained using the CVC relation with 
  our estimate $g_S^{u-d}= 1.022(80)(60)$ and lattice results for the up and down quark masses 
  from the FLAG review~\cite{FLAG:2016qm} and recent results~\protect\cite{Fodor:2016bgu,Bazavov:2018omf}.  }
\label{tab:Mn-Mp}
\end{center}
\end{table}

\section{Comparison with Previous Work}
\label{sec:comparison}

\begin{figure}
\begin{center}
\includegraphics[width=.47\textwidth]{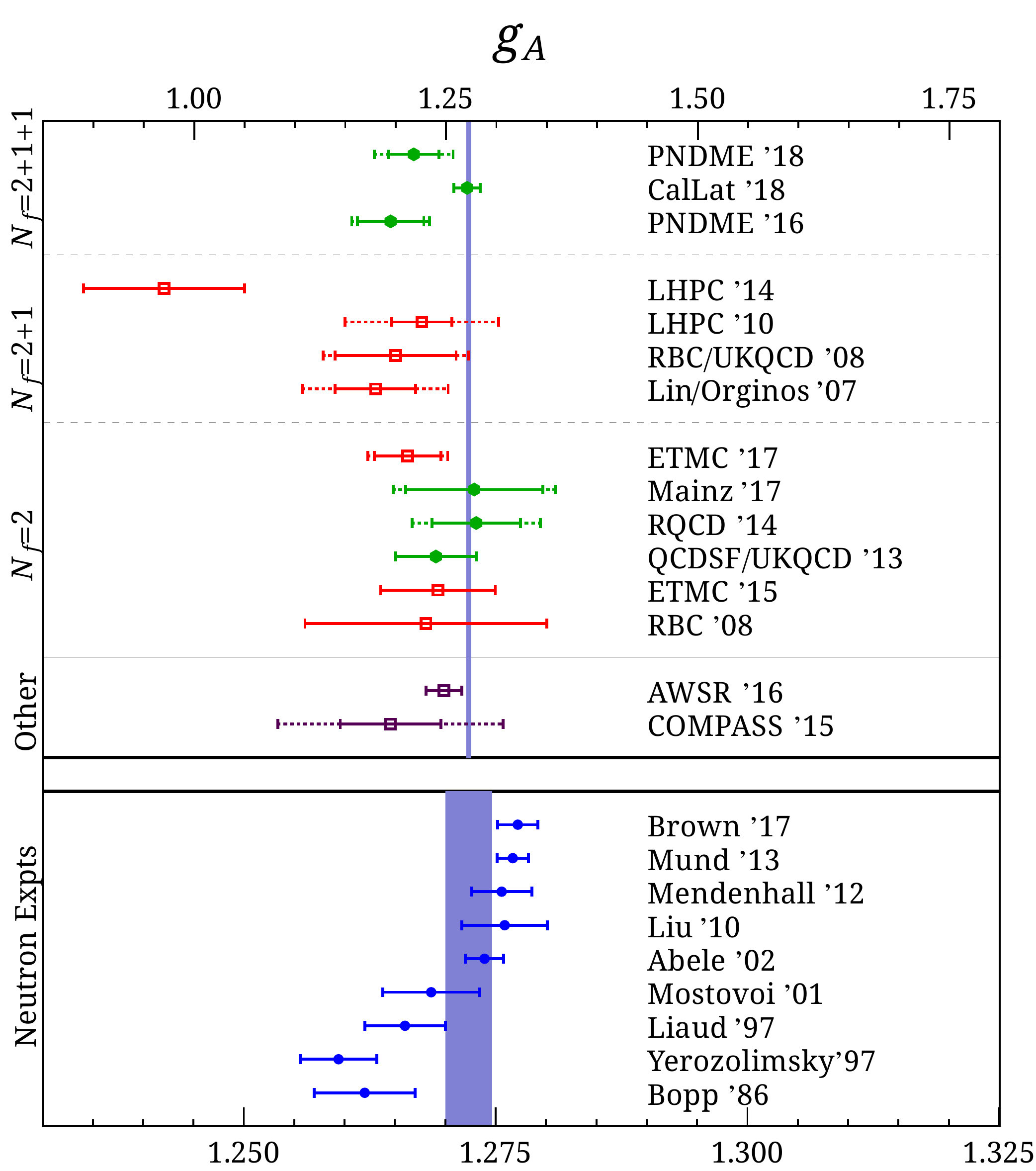}
\end{center}
\vspace{-0.5cm}
\caption{A summary of results for the axial isovector charge,
  $g_A^{u-d}$, for $N_f=2$- 2+1- and 2+1+1-flavors. Note the much
  finer x-axis scale for the plot showing experimental results for
  $g_A^{u-d}$. The lattice results (top panel) are from: PNDME'18
  (this work); PNDME'16~\protect\cite{Bhattacharya:2016zcn};
  CalLat'18~\protect\cite{Chang:2018uxx};
  LHPC'14~\protect\cite{Green:2012ud};
  LHPC'10~\protect\cite{Bratt:2010jn};
  RBC/UKQCD'08~\protect\cite{Lin:2008uz};
  Lin/Orginos'07~\protect\cite{Lin:2007ap};
  ETMC'17~\protect\cite{Alexandrou:2017oeh,Alexandrou:2017hac};
  Mainz'17~\protect\cite{Capitani:2017qpc}
  RQCD'14~\protect\cite{Bali:2014nma};
  QCDSF/UKQCD'13~\protect\cite{Horsley:2013ayv};
  ETMC'15~\protect\cite{Abdel-Rehim:2015owa} and
  RBC'08~\protect\cite{Yamazaki:2008py}.  Phenomenological and other
  experimental results (middle panel) are from:
  AWSR'16~\protect\cite{Beane:2016lcm} and
  COMPASS'15~\protect\cite{Adolph:2015saz}.  The results from neutron
  decay experiments (bottom panel) have been taken from:
  Brown'17~\protect\cite{Brown:2017mhw};
  Mund'13~\protect\cite{Mund:2012fq};
  Mendenhall'12~\protect\cite{Mendenhall:2012tz};
  Liu'10~\protect\cite{Liu:2010ms};
  Abele'02~\protect\cite{Abele:2002wc};
  Mostovoi'01~\protect\cite{Mostovoi:2001ye};
  Liaud'97~\protect\cite{Liaud:1997vu};
  Yerozolimsky'97~\protect\cite{Erozolimsky:1997wi} and
  Bopp'86~\protect\cite{Bopp:1986rt}.  The lattice-QCD estimates in
  red indicate that estimates of excited-state contamination, or
  discretization errors, or chiral extrapolation were not
  presented. When available, systematic errors have been added to
  statistical ones as outer error bars marked with dashed lines.  }
\label{fig:PASTgA}
\end{figure}

\begin{figure}
\begin{center}
\includegraphics[width=.47\textwidth,clip]{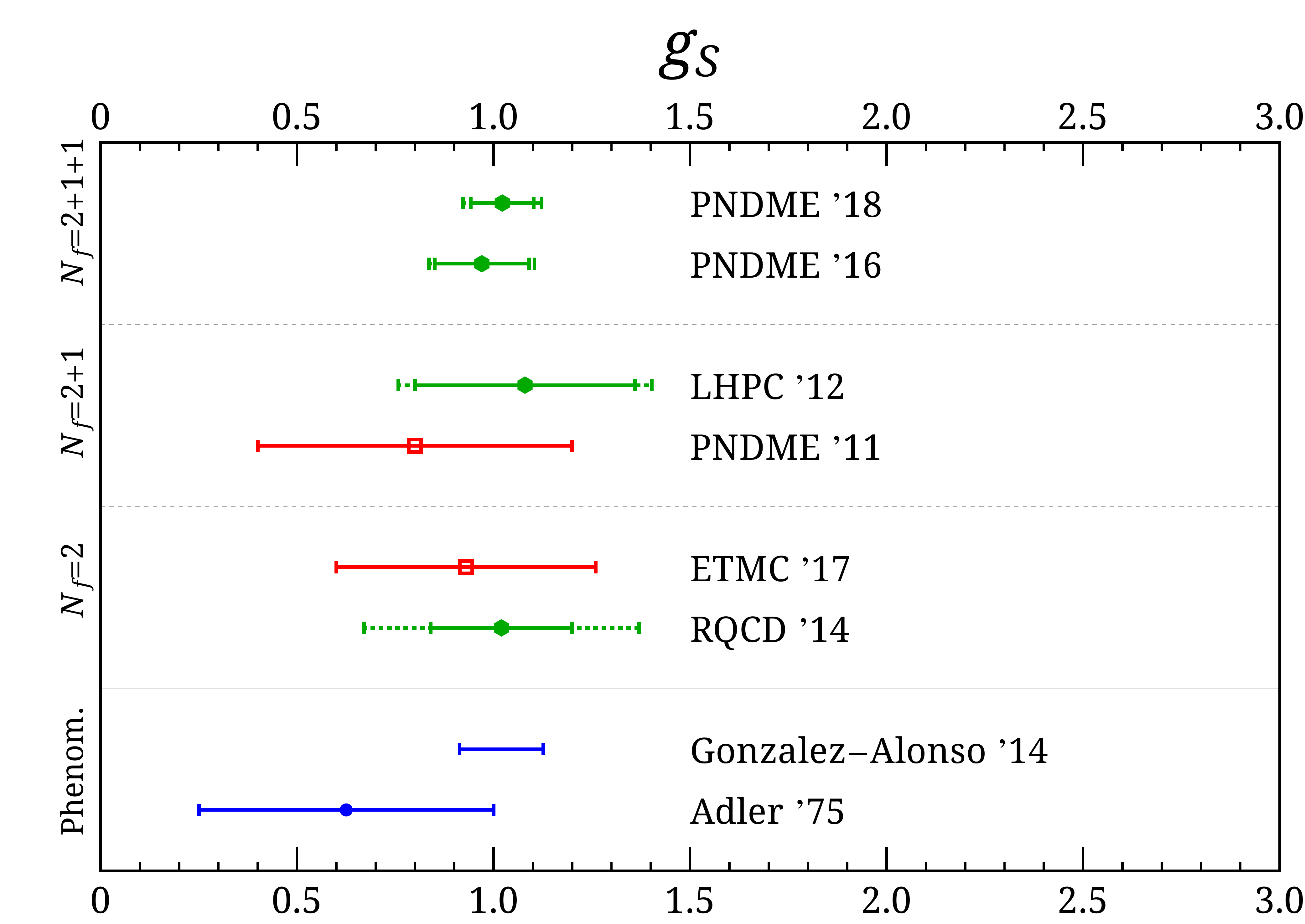}
\end{center}
\vspace{-0.5cm}
\caption{A summary of results for the isovector scalar charge, $g_S^{u-d}$, for
  $N_f=2$- 2+1- and 2+1+1-flavors. The lattice results are from: 
  PNDME'18 (this work);
  PNDME'16~\protect\cite{Bhattacharya:2016zcn};
  LHPC'12~\protect\cite{Green:2012ej};
  PNDME'11~\protect\cite{Bhattacharya:2011qm};  
  ETMC'17~\protect\cite{Alexandrou:2017qyt} and 
  RQCD'14~\protect\cite{Bali:2014nma}.  The estimates based on the
  conserved vector current and phenomenology are taken from 
  Gonzalez-Alonso'14~\protect\cite{Gonzalez-Alonso:2013ura} and
  Adler'75~\protect\cite{Adler:1975he}. The rest is the same as in Fig.~\protect\ref{fig:PASTgA}. 
}
\label{fig:PASTgS}
\end{figure}

\begin{figure}
\begin{center}
\includegraphics[width=.47\textwidth,clip]{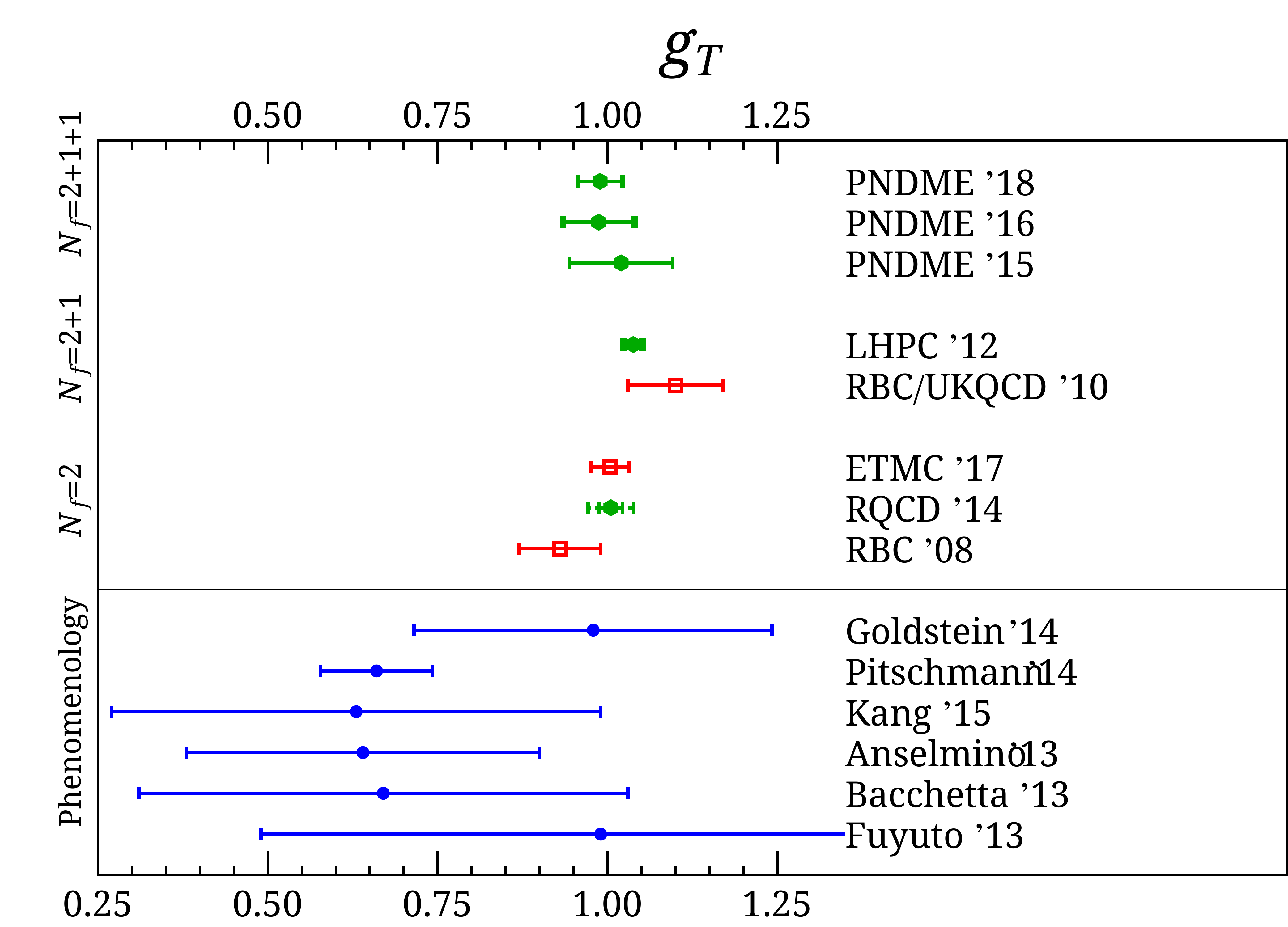}
\end{center}
\vspace{-0.5cm}
\caption{A summary of results for the isovector tensor charge, $g_T^{u-d}$, for
  $N_f=2$- 2+1- and 2+1+1-flavors. The lattice and phenomenology 
  results quoted from: 
  PNDME'18 (this work); 
  PNDME'16~\protect\cite{Bhattacharya:2016zcn}; 
  PNDME'15~\protect\cite{Bhattacharya:2015wna}
  LHPC'12~\protect\cite{Green:2012ej};
  RBC/UKQCD'10~\protect\cite{Aoki:2010xg}; 
  ETMC'17~\protect\cite{Alexandrou:2017qyt};
  RQCD'14~\protect\cite{Bali:2014nma} and 
  RBC'08~\protect\cite{Yamazaki:2008py}. 
  The phenomenological estimates are taken from the following sources: 
  Kang'15~\protect\cite{Kang:2015msa}; 
  Goldstein'14~\protect\cite{Goldstein:2014aja}; 
  Pitschmann'14~\protect\cite{Pitschmann:2014jxa}; 
  Anselmino'13~\protect\cite{Anselmino:2013vqa}; 
  Bacchetta'13~\protect\cite{Bacchetta:2012ty} and 
  Fuyuto'13~\protect\cite{Fuyuto:2013gla}.  
  Rest same as in Fig.~\protect\ref{fig:PASTgA}. 
}
\label{fig:PASTgT}
\end{figure}

A summary of lattice results for the three isovector charges for
$N_f=2$-, 2+1- and 2+1+1-flavors is shown in Figs.~\ref{fig:PASTgA},~\ref{fig:PASTgS}  and~\ref{fig:PASTgT}.
They show the steady improvement in results from lattice QCD.  In this
section we compare our results with two calculations published after
the analysis and the comparison presented in
Ref.~\cite{Bhattacharya:2016zcn}, and that include data from physical
pion mass ensembles. These are the
ETMC~\cite{Alexandrou:2017oeh,Alexandrou:2017qyt,Alexandrou:2017hac}
and CalLat results~\cite{Chang:2018uxx}.

The ETMC results $g_A^{u-d}=1.212(40)$, $g_S^{u-d}=0.93(33)$ and
$g_T^{u-d}=1.004(28)$~\cite{Alexandrou:2017oeh,Alexandrou:2017qyt,Alexandrou:2017hac}
were obtained from a single physical mass ensemble generated with
2-flavors of maximally twisted mass fermions with a clover term at
$a=0.0938(4)$~fm, $M_\pi=130.5(4)$~MeV and $M_\pi L = 2.98$. Assuming
that the number of quark flavors and finite volume corrections do not
make a significant difference, one could compare them against our results
from the $a09m130W$ ensemble with similar lattice parameters:
$g_A^{u-d}=1.249(21)$, $g_S^{u-d}=0.952(74)$ and
$g_T^{u-d}=1.011(30)$. We remind the
reader that this comparison is at best qualitative since estimates
from different lattice actions are only expected to agree in the
continuum limit.\looseness-1

Based on the trends observed in our CCFV fits shown in
Figs.~\ref{fig:conUmD-extrap11}--\ref{fig:extrap-gT-diagonal}, we
speculate where one may expect to see a difference due to the lack of
a continuum extrapolation in the ETMC results. The quantities that
exhibit a significant slope versus $a$ are $g_A^{u-d}$ and
$g_S^{u-d}$. Again, under the assumptions stated above, we would
expect ETMC values $g_A^{u-d}=1.212(40)$ to be larger and
$g_S^{u-d}=0.93(33)$ to be smaller than our extrapolated values given
in Eq.~\eqref{eq:gFinal}. We find that the scalar charge (ignoring the large error) fits the
expected pattern, but the axial charge does not.

We also point out that the ETMC error estimates are taken from a
single ensemble and a single value of the source-sink separation using
the plateau method. Our results from the comparable calculation on the
$a09m130W$ ensemble with $\tau=14$ (see Figs.~\ref{fig:gA2v3a09}
and~\ref{fig:gT2v3a09} and results in
Table~\ref{tab:results3bareu-d}), have much smaller errors.

The more detailed comparison we make is against the CalLat result
$g_A^{u-d} = 1.271(13)$~\cite{Chang:2018uxx} that agrees with the
latest experimental average, $g_A^{u-d} = 1.2766(20)$. The important
question is, since the CalLat calculations were also done using the
same 2+1+1-flavor HISQ ensembles, why are the two results, after CCFV
fits, different?  

To understand why the results can be different, we first review the
notable differences between the two calculations. CalLat uses (i)
M\"obius domain wall versus clover for the valence quark action. This
means that their discretization errors start at $a^2$ versus $a$ for
PNDME. They also have no uncertainty due to the renormalization factor
since $Z_A/Z_V=1$ for the M\"obius domain wall on HISQ formalism. (ii)
They use gradient flow smearing with $t_{gf}/a=1$ versus one HYP
smearing to smooth high frequency fluctuations in the gauge
configurations. This can impact the size of statistical errors. (iii)
Different construction of the sequential propagator. CalLat inserts a
zero-momentum projected axial current simultaneously at all time slices
on the lattice to construct the sequential propagator. Their data are,
therefore, for the sum of contributions from insertions on {\it all}
time slices on the lattice, i.e., including contact terms and insertion
on time slices outside the interval between the source and the sink.
CalLat fits this summed three-point function versus only 
the source-sink separation $\tau$ using the 2-state fit
ansatz. (iv) The ranges of $\tau$ for which the data have the maximum
weight in the respective n-state fits are very different in the two
calculations. The CalLat results are obtained from data at much
smaller values of $\tau$, which accounts for the smaller error
estimates in the data for $g_A^{u-d}$.  (v) CalLat analyze the coarser
$a\approx 0.15$, $0.12$ and $0.09$~fm ensembles. At $a \approx 0.15$~fm,
we can only analyze the $a15m310$ ensemble due to the presence of
exceptional configurations in the clover-on-HISQ formulation at
lighter pion masses. On the other hand, computing resources have so far
limited CalLat from analyzing the three fine $a\approx 0.06$~fm and
the physical mass $a09m130$ ensembles.

A combination of these factors could easily explain the $\approx 5\%$
difference in the final values.  The surprising result, shown in
Table~\ref{tab:CalLat}, is that estimates on the seven ensembles
analyzed by both collaborations are consistent and do not show a
systematic difference. (Note again that results from two different lattice
formulations are not, {\it a priori}, expected to agree at finite
$a$.) These data suggest that differences at the $1\sigma$ level (see
also our analysis in Table~\ref{tab:errors}) are conspiring to produce
a 5\% difference in the extrapolated value. Thus, one should look for
differences in the details of the CCFV fit.

We first examine the extrapolation in $a$. A CCFV fit keeping our data
from only the eight $a\approx 0.15$, $0.12$ and $0.09$~fm ensembles
gives a larger value, $g_A^{u-d} = 1.245(42)$, since the sign of the
slope versus $a$ changes sign as is apparent from the data shown in
the top three panels of Fig.~\ref{fig:conUmD-extrap11}. Thus the three
$a\approx 0.06$~fm ensembles play an important role in our continuum
extrapolation.

Our initial concern was possible underestimation of statistical errors
in results from the $a \approx 0.06$~fm lattices.  This prompted us to
analyze three crucial ensembles, $a09m130$, $a06m310$ and $a06m220$, a
second time with different smearing sizes and different random
selection of source points.  The consistency between the pairs of data
points on these ensembles suggests that statistical fluctuations are
not a likely explanation for the size of the undershoot in
$g_A^{u-d}$. The possibility that these ensembles are not large enough
to have adequately explored the phase space of the functional
integral, and the results are possibly biased, can only be checked
with the generation and analysis of additional lattices.

The chiral fits are also different in detail. In our data, the errors
in the points at $M_\pi \approx 310$, 220 and 130 MeV are similar,
consequently all points contribute with similar weight in the fits. The
errors in the CalLat data from the two physical mass ensembles
$a15m130$ and $a12m130$ are much larger and the fits are predominately
weighted by the data at the heavier masses $M_\pi \approx 400$, 350
310 and 220 MeV.  Also, CalLat finds a significant change in the value
between the $M_\pi \approx \{400,\ 350,\ 310\}$~ MeV and $M_\pi
\approx 220$~MeV points, and this concerted change, well within
$1\sigma$ errors in individual points, produces a larger dependence on
$M_\pi$. In other words, it is the uniformly smaller values on the
$M_\pi \approx \{400,\ 350,\ 310\}$~MeV ensembles compared to the data
at $M_\pi\approx 220$~MeV that makes the CalLat chiral fits different and 
the final value of $g_A^{u-d}$ larger. 

\begin{figure*}
\begin{center}
\includegraphics[width=.49\textwidth,clip]{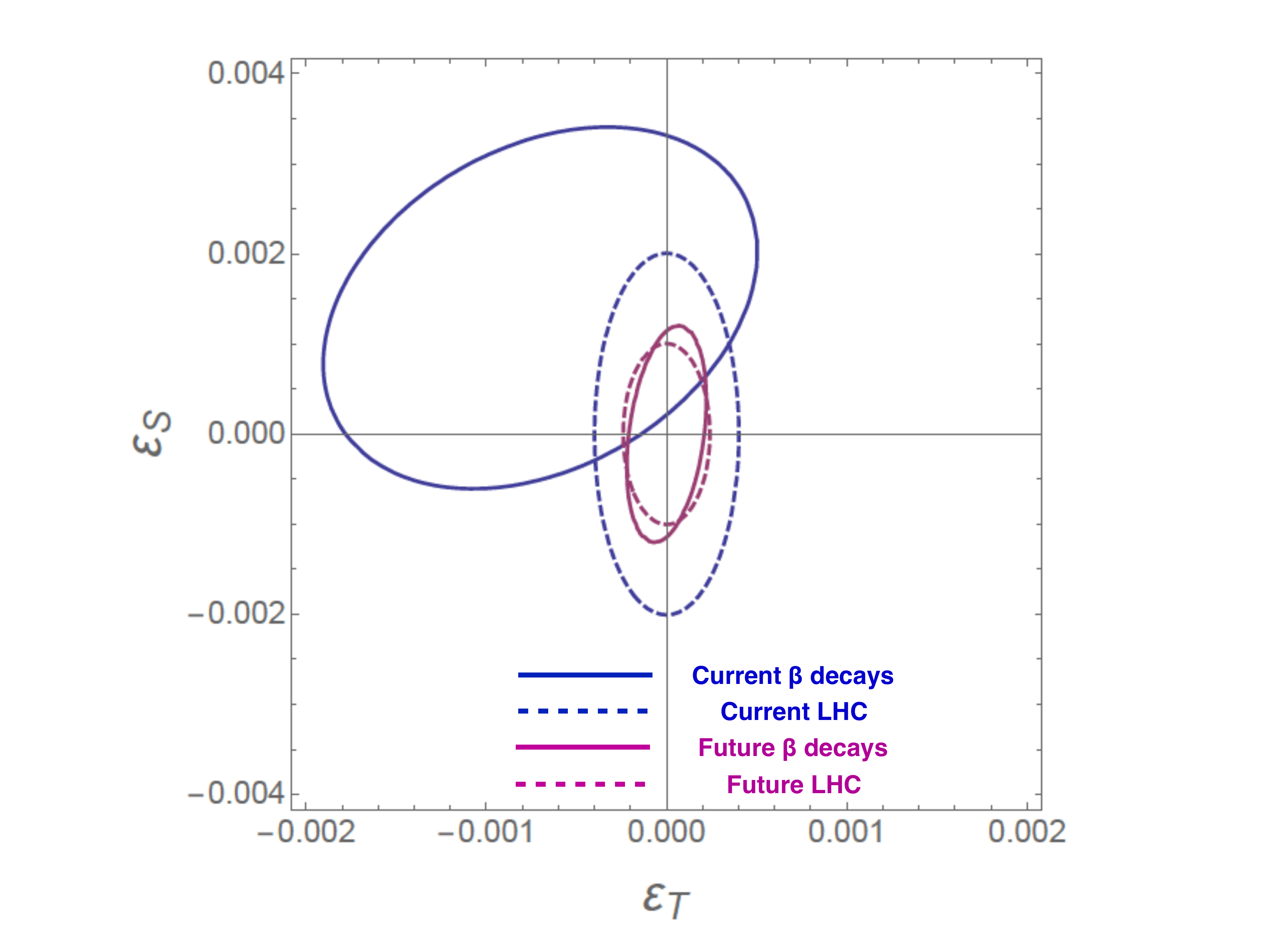}
\includegraphics[width=.49\textwidth,clip]{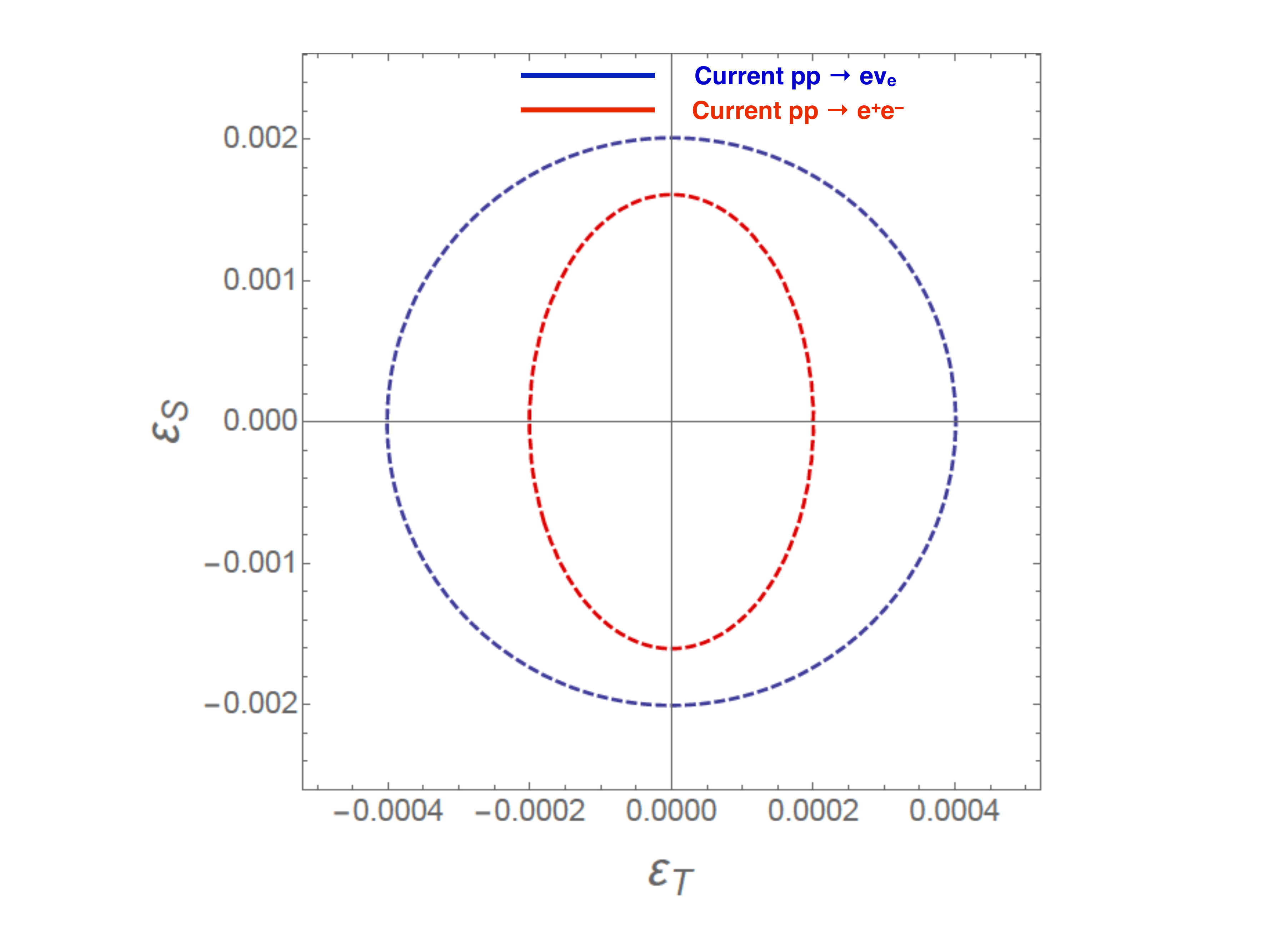}
\end{center}
\vspace{-0.5cm}
\caption{Current and projected $90 \%$ C.L. constraints on
  $\epsilon_S$ and $\epsilon_T$ defined at 2~GeV in the
  $\overline{MS}$ scheme.  (Left) The beta-decay constraints are
  obtained from the recent review article
  Ref.~\protect\cite{Gonzalez-Alonso:2018omy}. The current and future
  LHC bounds are obtained from the analysis of the $pp \to e + MET +
  X$.  We have used
  the ATLAS results~\protect\cite{Aaboud:2017efa}, at $\sqrt{s} =
  13$~TeV and integrated luminosity of 36 fb$^{-1}$.  We find that the
  strongest bound comes from the cumulative distribution with a cut on
  the transverse mass at 2 TeV.  The projected future LHC bounds are
  obtained by assuming that no events are observed at transverse mass
  greater than 3 TeV with an integrated luminosity of 300
  fb$^{-1}$. (Right) Comparison of current LHC bounds from $pp \to e +
  MET + X$ versus $pp \to e^+ e^- + X$.  }
\label{fig:eSeT}
\end{figure*}

To summarize, the difference between our and CalLat results comes from the
chiral fit and the continuum extrapolation.  The difference in the
chiral fit is a consequence of the ``jump'' in the CalLat data between
$M_\pi = \{400,\ 350,\ 310\}$ and the $220$~MeV data. The CalLat data
at $M_\pi \approx 130$~MeV do not contribute much to the fit because
of the much larger errors.  We do not see a similar jump between our
$M_\pi \approx 310$ and $220$~MeV or between the 220 and the 130~MeV
data as is evident from Fig.~\ref{fig:conUmD-extrap11}.  Also, our
four data points at $M_\pi \approx 310$~MeV show a larger spread.  The
difference in the continuum extrapolation is driven by the smaller
estimates on all three fine $a \approx 0.06$~fm ensembles that we have
analyzed.  Unfortunately, neither of these two differences in the fits
can be resolved with the current data, especially since the data on 7
ensembles, shown in Table~\ref{tab:CalLat}, agree within
$1\sigma$. Our two conclusions are: (i) figuring out why the $a\approx
0.06$~fm ensembles give smaller estimates is crucial to understanding
the difference, and (ii) with present data, a total error estimate of
$\approx 5\%$ in $g_A^{u-d}$ is realistic.

\begin{table}[ht]
\begin{center}
\renewcommand{\arraystretch}{1.2} 
\begin{ruledtabular}
\begin{tabular}{l|c|c}
           & This Work  &   CalLat           \\
\hline
$a15m310$  & 1.228(25)  & 1.215(12)          \\
$a12m310$  & 1.251(19)  & 1.214(13)          \\
$a12m220S$ & 1.224(44)  & 1.272(28)          \\
$a12m220$  & 1.234(25)  & 1.259(15)          \\
$a12m220L$ & 1.262(17)  & 1.252(21)          \\
$a09m310$  & 1.235(15)  & 1.236(11)          \\
$a09m220$  & 1.260(19)  & 1.253(09)          \\
\end{tabular}
\end{ruledtabular}
\caption{The data for the renormalized axial charge $g_A^{u-d}$ for
  the proton on the seven 2+1+1-flavor HISQ ensembles that have been
  analyzed by us and the CalLat collaboration~\protect\cite{Chang:2018uxx}. The 
  results are consistent within $1\sigma$ in most cases.  }
\label{tab:CalLat}
\end{center}
\end{table}

Even with the high statistics calculation presented here, the
statistical and ESC errors in the calculation of the scalar charge are between
5\%--15\% on individual ensembles. As a result, the error after the
continuum extrapolation is about $10\%$.  Over time, results for
$g_S^{u-d}$, presented in Fig.~\ref{fig:PASTgS}, do show significant
reduction in the error with improved higher-statistics calculations.

The variation of the tensor charge $g_T^{u-d}$ with $a$ or $M_\pi $ or
$M_\pi L$ is small.  As a result, the lattice estimates have been
stable over time as shown in Fig.~\ref{fig:PASTgT}. The first error
estimate in our result, $g_T^{u-d} = 0.989(32)(10) $, is now dominated
by the error in $Z_T$.


\section{Constraining new physics using precision beta decay measurements}
\label{sec:est}

Nonstandard scalar and tensor
charged-current interactions are parametrized by the dimensionless
couplings $\epsilon_{S,T}$~\cite{Bhattacharya:2011qm,Cirigliano:2012ab}:
\begin{eqnarray}
{\cal L}_{\rm CC}  &=&
- \frac{G_F^{(0)} V_{ud}}{\sqrt{2}} \  \Big[ \
 \epsilon_S  \  \bar{e}  (1 - \gamma_5) \nu_{\ell}  \cdot  \bar{u} d 
 \nonumber \\
&+  & 
\epsilon_T \   \bar{e}   \sigma_{\mu \nu} (1 - \gamma_5) \nu_{\ell}    \cdot  \bar{u}   \sigma^{\mu \nu} (1 - \gamma_5) d
\Big] ~. 
\end{eqnarray}
These couplings can be constrained by a
combination of low energy precision beta-decay measurements (of the
pion, neutron, and nuclei) combined with our results for the isovector
charges $g_{S}^{\rm u-d}$ and $g_T^{\rm u-d}$, as well at the
Large Hadron Collider (LHC) through the reaction $pp \to e \nu +
X$ and $pp \to e^+ e^- + X$. The LHC constraint is valid provided the mediator of the new
interaction is heavier than a few TeV.

In Fig.~\ref{fig:eSeT} (left) we show current and projected bounds on
$\{\epsilon_S, \epsilon_T\}$ defined at 2~GeV in the $\overline{MS}$
scheme.  The beta decays constraints are obtained from the recent
review article Ref.~\cite{Gonzalez-Alonso:2018omy}.  The current
analysis includes all existing neutron and nuclear decay measurements,
while the future projection assumes measurements of the various decay
correlations with fractional uncertainty of $0.1\%$, the Fierz
interference term at the $10^{-3}$ level, and neutron lifetime with
uncertainty $\delta \tau_n = 0.1 s$.  The current LHC bounds are
obtained from the analysis of the $pp \to e + MET + X$, where $MET$
stands for missing transverse energy.  We have used the ATLAS
results~\cite{Aaboud:2017efa}, at $\sqrt{s} = 13$~TeV and integrated
luminosity of 36 fb$^{-1}$.  We find that the strongest bound comes by
the cumulative distribution with a cut on the transverse mass at 2
TeV.  The projected future LHC bounds are obtained by assuming that no
events are observed at transverse mass greater than 3~TeV with an
integrated luminosity of 300 fb$^{-1}$.

The LHC bounds become tighter on the inclusion of $Z$-like mediated
process $pp \to e^+ e^- + X$. As shown in Fig.~\ref{fig:eSeT} (right),
including both $W$-like and $Z$-like mediated processes, the current
LHC bounds are comparable to future low energy ones, motivating 
more precise low energy experiments.  In this analysis
we have neglected the NLO QCD corrections~\cite{Alioli:2018ljm}, which
would further strengthen the LHC bounds by $O(10\%)$. Similar bounds are
obtained using the CMS data~\cite{Sirunyan:2018mpc,Sirunyan:2018exx}.

\section{Conclusions}
\label{sec:conclusions}

We have presented a high-statistics study of the isovector and
flavor-diagonal charges of the nucleon using clover-on-HISQ lattice
QCD formulation. By using the truncated solver with bias correction
error-reduction technique with the multigrid solver, we have significantly improved the
statistical precision of the data.  Also, we show stability in the
isolation and mitigation of excited-state contamination by keeping up
to three states in the analysis of data at multiple values of
source-sink separation $\tsep$. Together, these two improvements allow
us to demonstrate that the excited-state contamination in the axial
and the tensor channels has been reduced to the 1\%--2\% level.  The
high-statistics analysis of eleven ensembles covering the range
0.15--0.06~fm in the lattice spacing, $M_\pi =$ 135--320~MeV in the
pion mass, and $M_\pi L =$ 3.3--5.5 in the lattice size allowed us to
analyze the three systematic uncertainties due to lattice
discretization, dependence on the quark mass and finite lattice
size, by making a simultaneous fit in the three variables $a$,
$M_\pi^2$ and $M_\pi L$.  Data from the two physical mass ensembles,
$a09m130$ and $a06m135$, anchor the improved chiral fit.  Our final
estimates for the isovector charges are given in
Eq.~\eqref{eq:gFinal2}.

One of the largest sources of uncertainty now is from the calculation
of the renormalization constants for the quark bilinear operators.
These are calculated nonperturbatively in the RI-sMOM scheme over a
range of values of the scale $Q^2$. As discussed in
Ref.~\cite{Bhattacharya:2016zcn}, the dominant systematics in the
calculation of the $Z$'s comes from the breaking of the rotational
symmetry on the lattice and the 2-loop perturbative matching between
the RI-sMOM and the $\overline{\text{MS}}$ schemes.

Our estimate $g_A^{u-d}=1.218(25)(30)$ is about $1.5 \sigma$ (about
$5\%$) below the experimental value $g_A/g_V = 1.2766(20)$.  Such low
values are typical of most lattice QCD calculations. The recent
calculation by the CalLat collaboration, also using the 2+1+1-flavor
HISQ ensembles, gives $g_A^{u-d}=1.271(13)$~\cite{Chang:2018uxx}. A
detailed comparison between the two calculations is presented in
Sec~\ref{sec:comparison}. We show in Table~\ref{tab:CalLat} that
results from seven ensembles, which have been analyzed by both
collaborations, agree within $1\sigma$ uncertainty. Our analysis
indicates that the majority of the difference comes from the chiral
and continuum extrapolations, with $1\sigma$ differences in individual
points getting amplified.  Given that CalLat have not analyzed the
fine $0.06$~fm ensembles and their data on the two physical pion mass
ensembles, $a15m130$ and $a12m130$ have much larger errors and do not
contribute significantly to their chiral fit, we conclude that our error
estimate is more realistic. Further work is, therefore, required to
resolve the difference between the two results.

Our results for the isovector scalar and tensor charges,
$g_S^{u-d}=1.022(80)(60)$ and $g_T^{u-d}=0.989(32)(10)$, have achieved
the target accuracy of 10\% needed to put bounds on scalar and tensor
interactions, $\epsilon_S$ and $\epsilon_T$, arising at the TeV scale
when combined with experimental measurements of $b$ and $b_\nu$
parameters in neutron decay experiments with $10^{-3}$
sensitivity~\cite{Bhattacharya:2011qm}. In Sec.~\ref{sec:est}, we
update the constraints on $\epsilon_S$ and $\epsilon_T$ from both low
energy experiments combined with our new lattice results on
$g_S^{u-d}$ and $g_T^{u-d}$, and from the ATLAS and the CMS
experiments at the LHC.  We find that the constraints from low energy
experiments combined with matrix elements from lattice QCD are
comparable to those from the LHC.

For the tensor charges, we find that the dependence on the lattice
size, the lattice spacing and the light-quark mass is small, and the 
simultaneous fit in these three variables, keeping just the lowest-order
corrections, has improved over that presented in 
Ref.~\cite{Bhattacharya:2015wna}.

We have also updated our estimates for the connected parts of the
flavor-diagonal charges. For the tensor charges, the contribution of
the disconnected diagram is consistent with
zero~\cite{Bhattacharya:2015wna,Bhattacharya:2015esa}, so the
connected contribution, $g_T^{u} = 0.790(27)$ and $g_T^{d} = -
0.198(10)$ for the proton, is a good approximation to the full result that 
will be discussed elsewhere.

The extraction of the scalar charge of the proton has larger
uncertainty.  The statistical errors in the lattice data for
$g_S^{u-d}(a, M_\pi, M_\pi L)$ are 3--5 times larger than those in
$g_T^{u-d}(a,M_\pi,M_\pi L)$, and the data show significant dependence
on the lattice spacing $a$ and a weaker dependence on the pion mass
$M_\pi$.  Our estimate, $g_S^{u-d}=1.022(80)(60)$, is in very good
agreement with the estimate $g_S^{u-d}=1.02(8)(7)$ obtained using the
CVC relation $g_S/g_V = (M_N-M_P)^{\rm QCD}/ (m_d-m_u)^{\rm QCD}$ in
Ref.~\cite{Gonzalez-Alonso:2013ura}.  In Table~\ref{tab:Mn-Mp}, we
used our new estimate to update the results for the mass difference
$(M_N-M_P)^{\rm QCD}$ obtained by using the 
CVC relation. Taking the recent 2+1 flavor value $m_d - m_u =
2.41(6)(4)(9)$~MeV from the BMW collaboration~\cite{Fodor:2016bgu} gives
$(M_N-M_P)^{\rm 2+1QCD} = 2.41(27)$~MeV, while the 2+1+1-flavor
estimates $m_u=2.118(38)$~MeV and $m_d=4.690(54)$~MeV from the
MILC/Fermilab/TUMQCD collaboration~\cite{Bazavov:2018omf} give 
$(M_N-M_P)^{\rm 2+1+1QCD} = 2.63(27)$~MeV.

\appendix
\section{ESC in the extraction of the isovector charges}
\label{appendix:axial}

\begin{figure*}
\centering
  \subfigure{
    \includegraphics[height=1.5in,trim={0.0cm 0.00cm 0 0},clip]{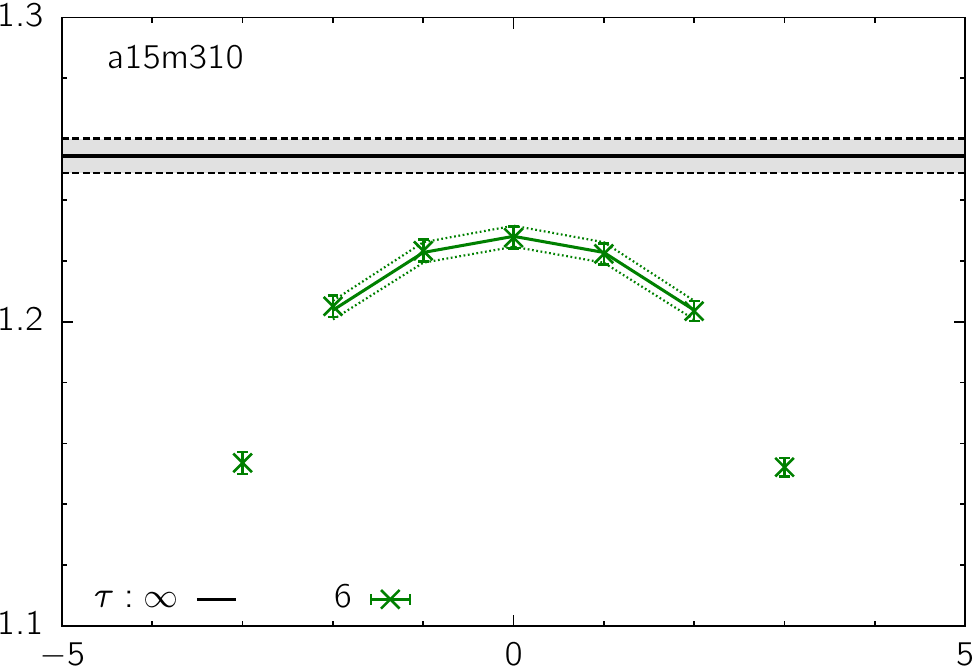}
    \includegraphics[height=1.5in,trim={0.0cm 0.00cm 0 0},clip]{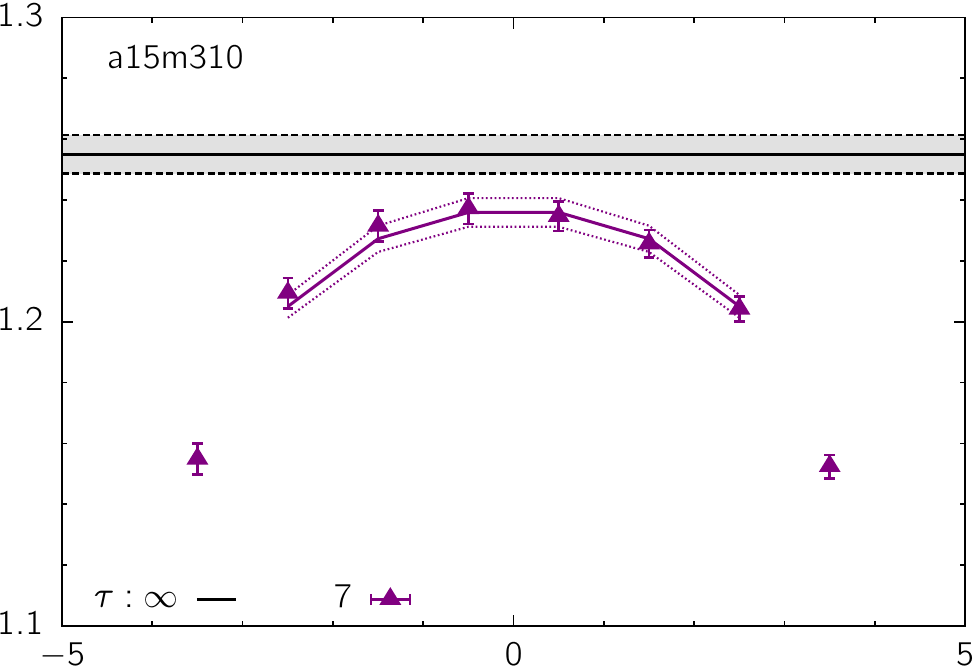}
    \includegraphics[height=1.5in,trim={0.0cm 0.00cm 0 0},clip]{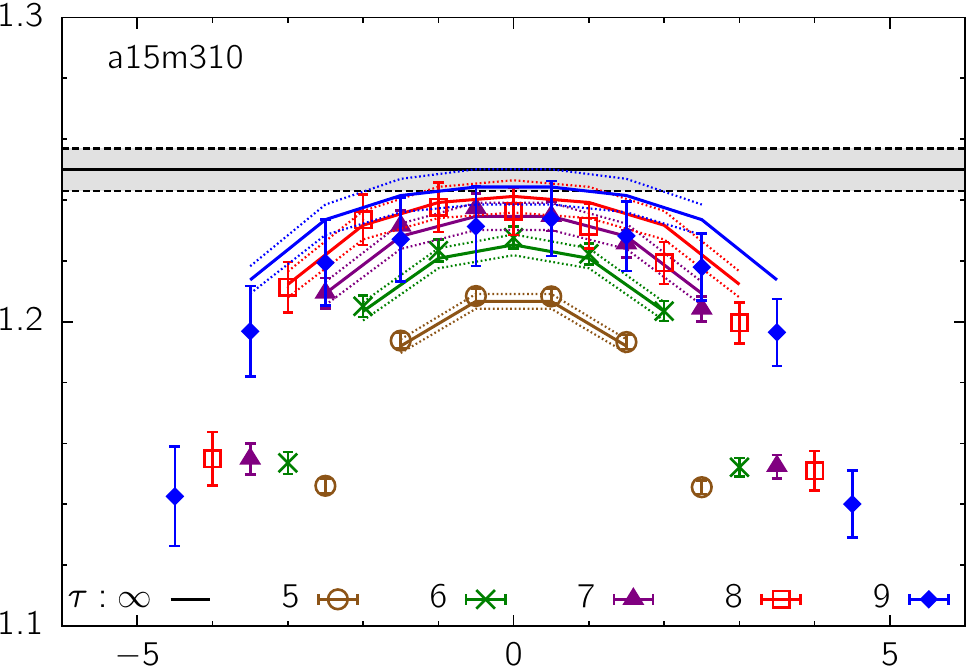}
  }
  \hspace{0.04\linewidth}
  \subfigure{
    \includegraphics[height=1.5in,trim={0.0cm 0.00cm 0 0},clip]{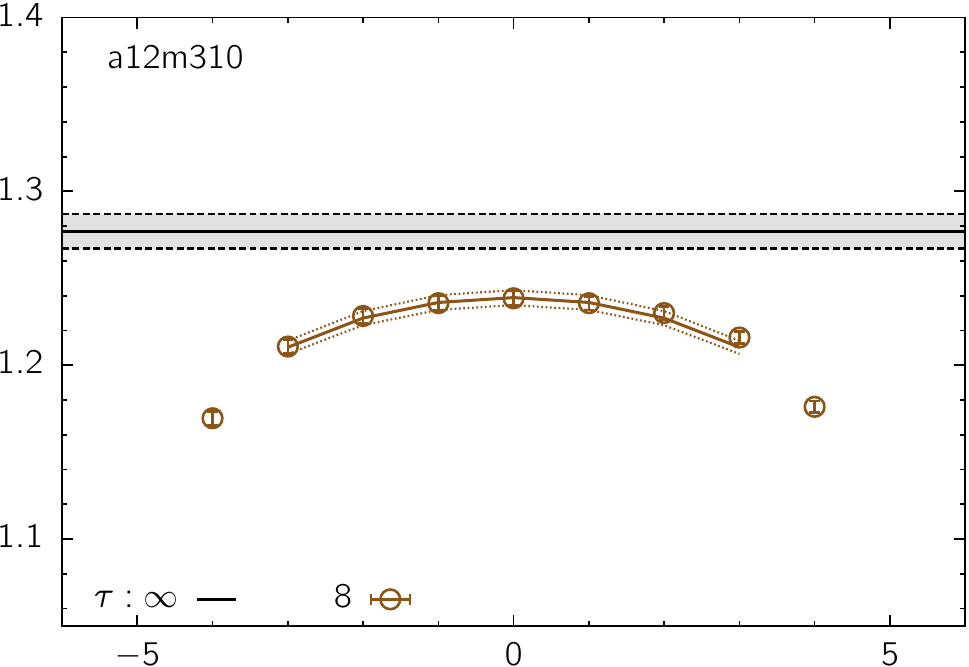}
    \includegraphics[height=1.5in,trim={0.0cm 0.00cm 0 0},clip]{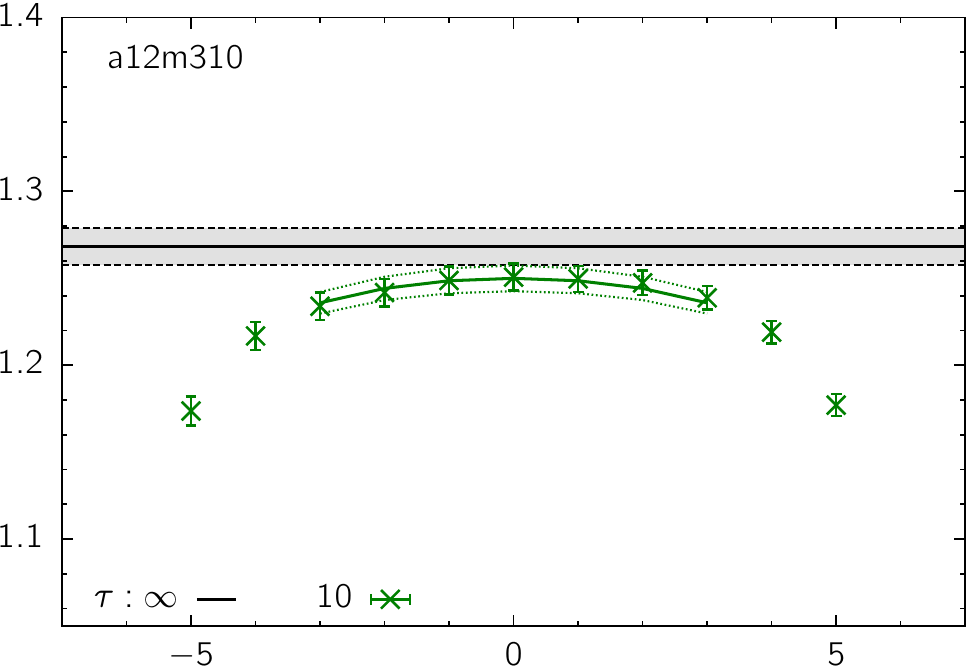}
    \includegraphics[height=1.5in,trim={0.0cm 0.00cm 0 0},clip]{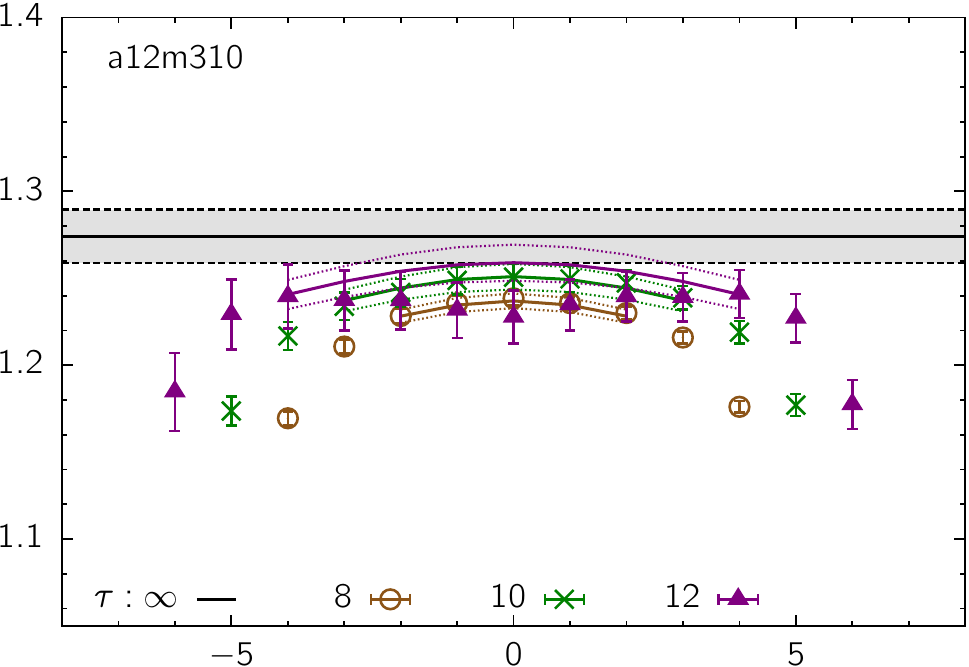}
  }
  \hspace{0.04\linewidth}
  \subfigure{
    \includegraphics[height=1.5in,trim={0.0cm 0.00cm 0 0},clip]{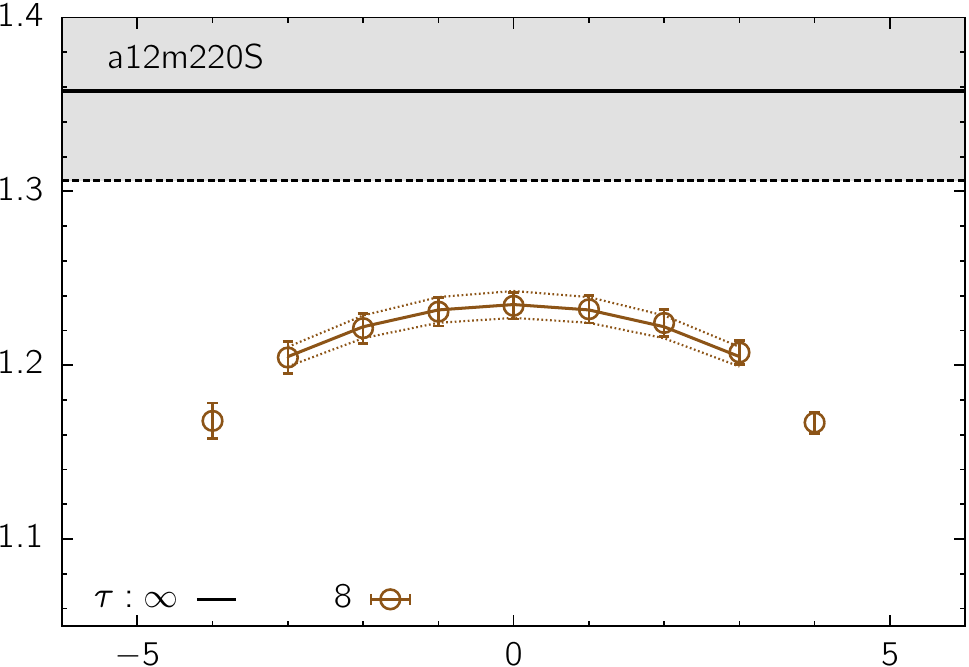}
    \includegraphics[height=1.5in,trim={0.0cm 0.00cm 0 0},clip]{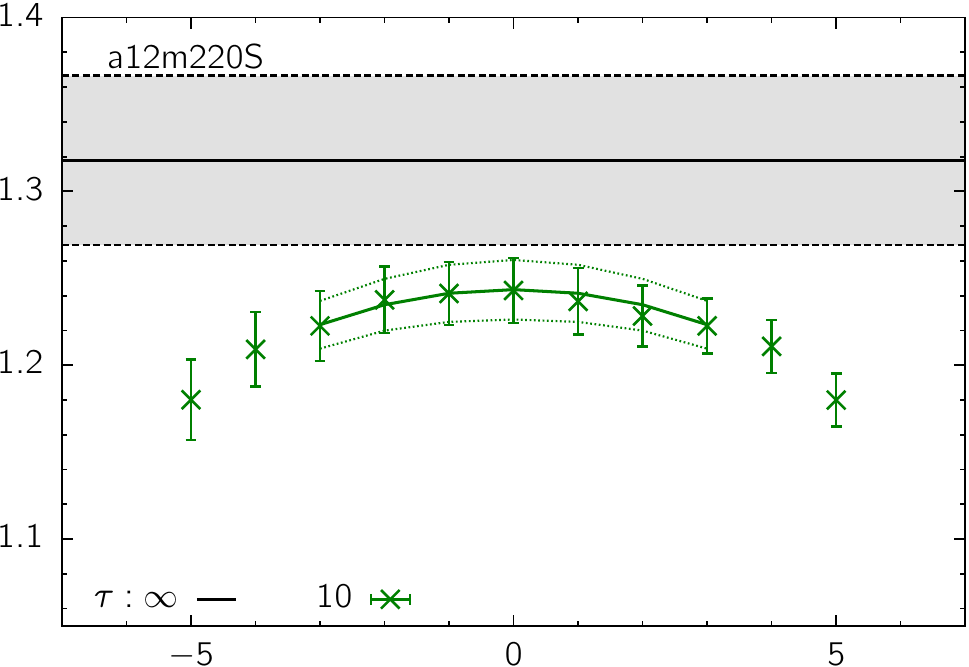}
    \includegraphics[height=1.5in,trim={0.0cm 0.00cm 0 0},clip]{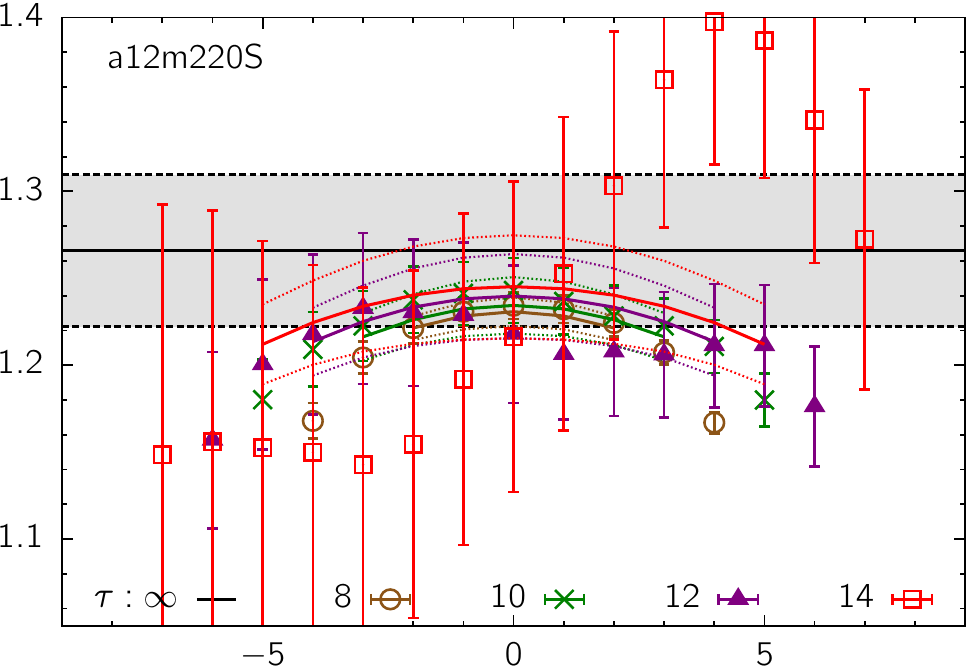}
  }
  \hspace{0.04\linewidth}
  \subfigure{
    \includegraphics[height=1.5in,trim={0.0cm 0.00cm 0 0},clip]{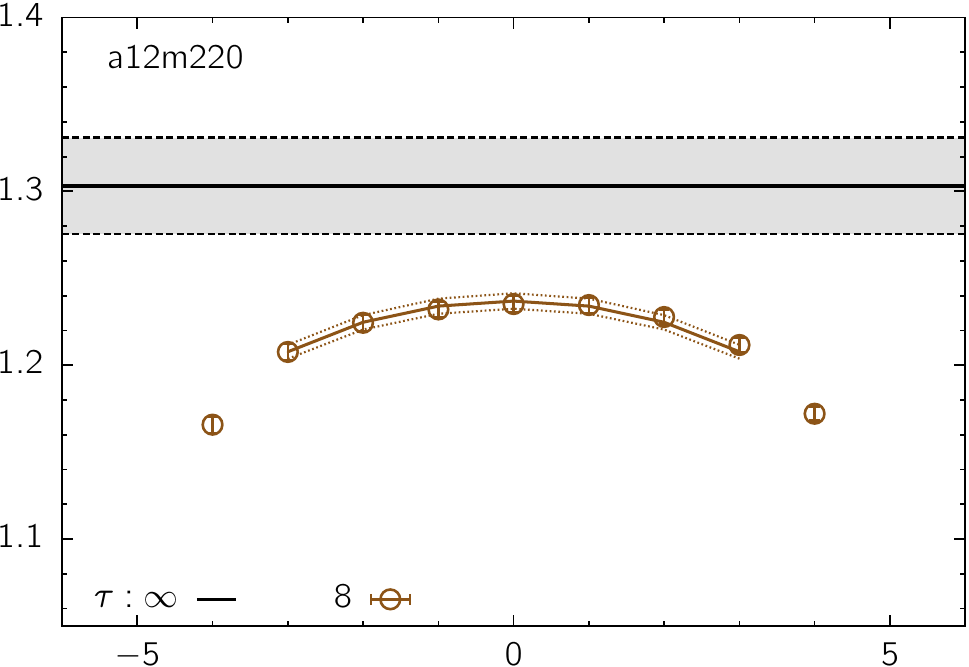}
    \includegraphics[height=1.5in,trim={0.0cm 0.00cm 0 0},clip]{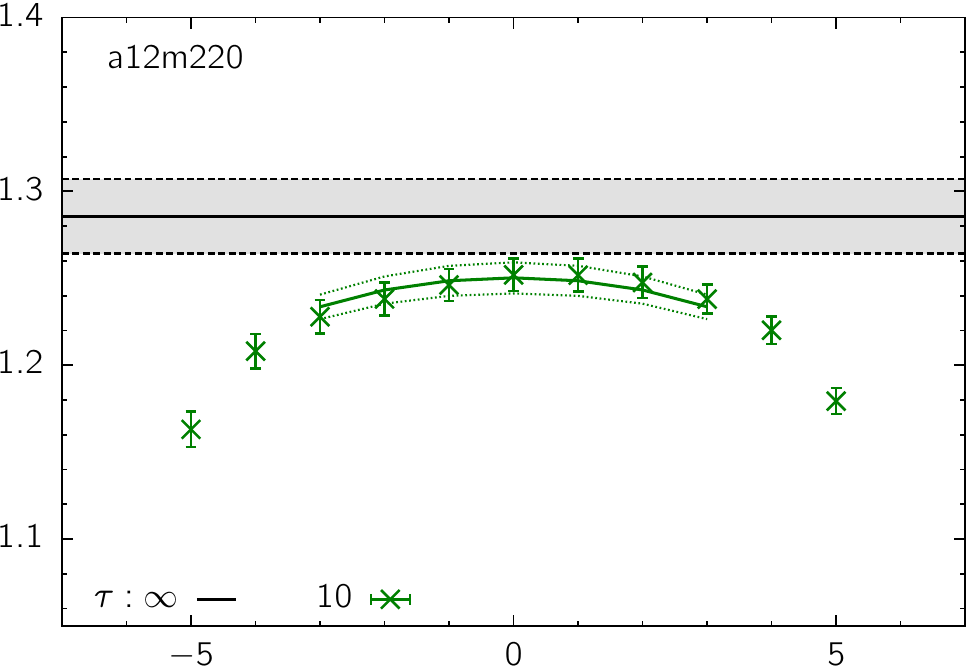}
    \includegraphics[height=1.5in,trim={0.0cm 0.00cm 0 0},clip]{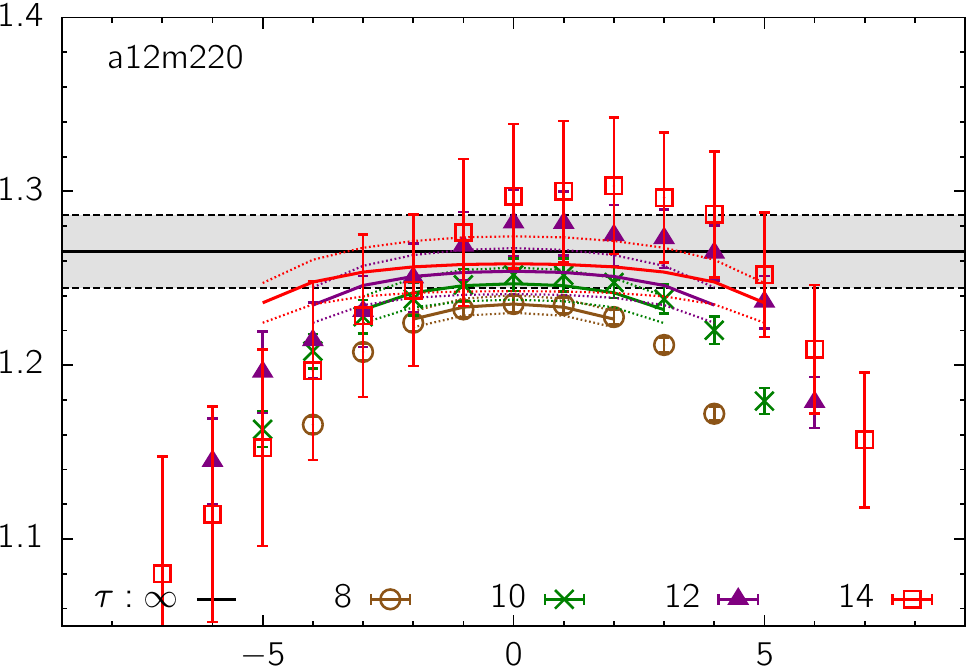}
  }
  \hspace{0.04\linewidth}
  \subfigure{
    \includegraphics[height=1.5in,trim={0.0cm 0.00cm 0 0},clip]{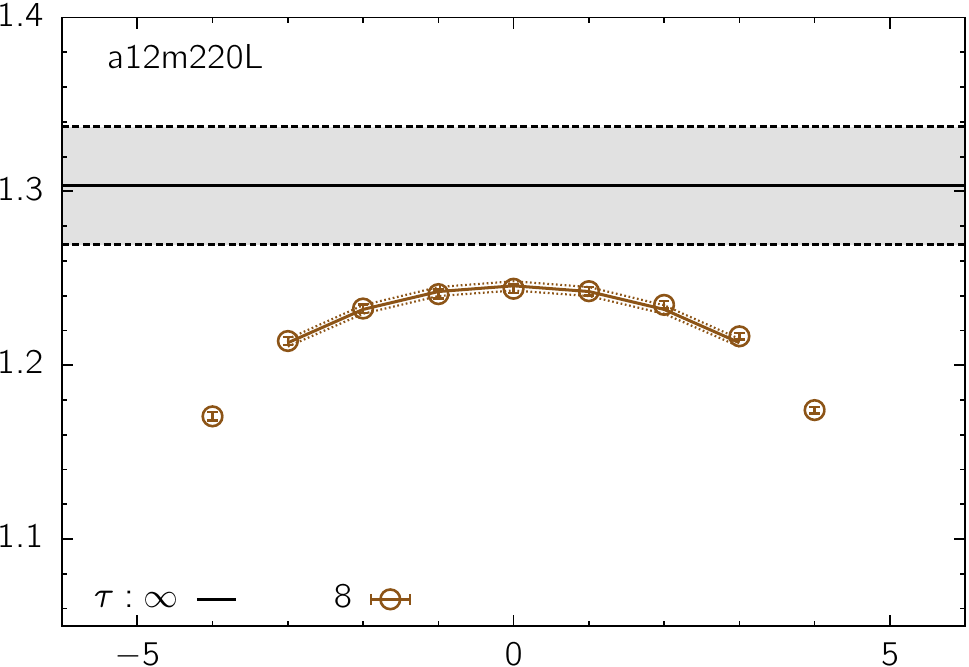}
    \includegraphics[height=1.5in,trim={0.0cm 0.00cm 0 0},clip]{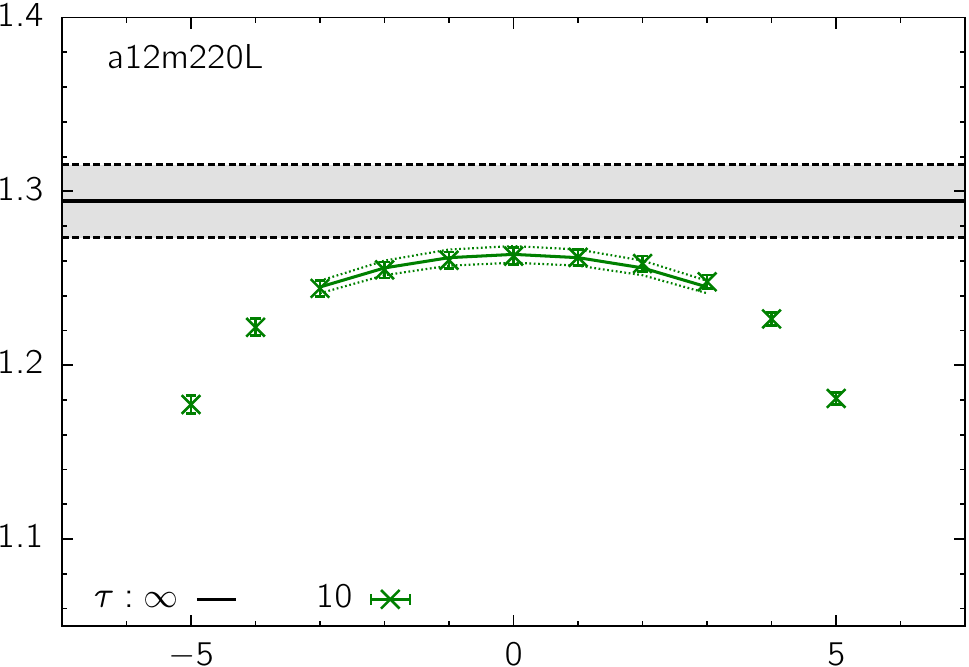}
    \includegraphics[height=1.5in,trim={0.0cm 0.00cm 0 0},clip]{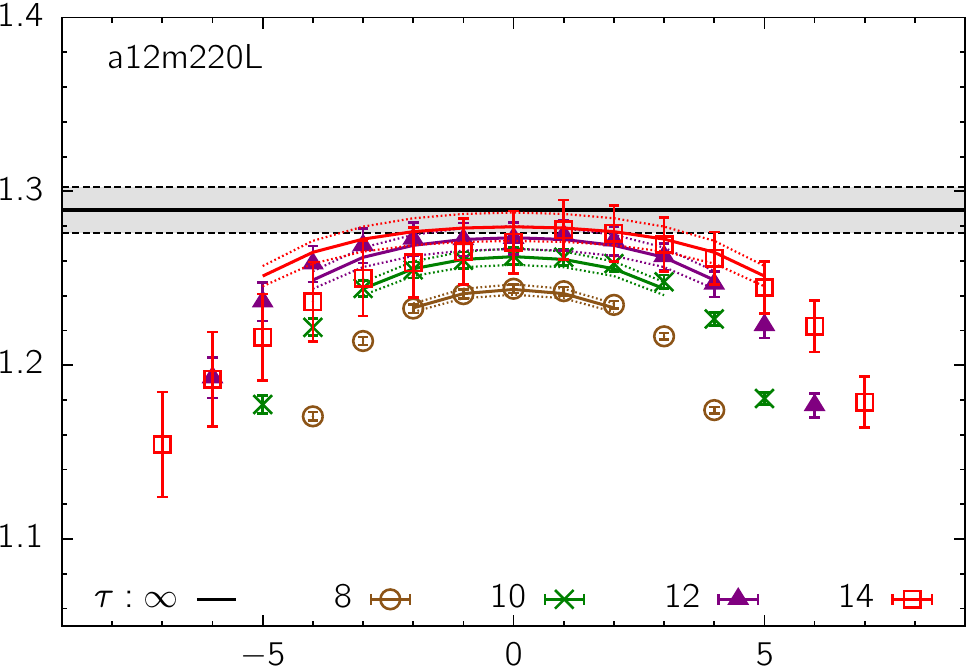}
  }
\caption{Comparison between the $2^\ast$ and $3^\ast$ fits to the
  axial charge $g_A^{u-d}$ data from the $a \approx 0.15$~fm (top row)
  and $a \approx 0.12$~fm (bottom 4 rows) ensembles.  The results of
  the fits are summarized in Table~\protect\ref{tab:results3bareu-d}
  along with the number of points $\tskip$ skipped. The first two
  columns show $2^\ast$ fits to data versus $t$ at a single value of
  $\tau$, while the third panel shows the $3^\ast$ fit using data at
  multiple values of $\tau$. The labels give the ensemble ID, and the
  values of $\tau$ used in the fits.  The $\tsepi$ value is given by
  the grey band in each panel.
  \label{fig:gA2v3a12}}
\end{figure*}

\begin{figure*}
\centering
  \subfigure{
    \includegraphics[height=1.5in,trim={0.0cm 0.00cm 0 0},clip]{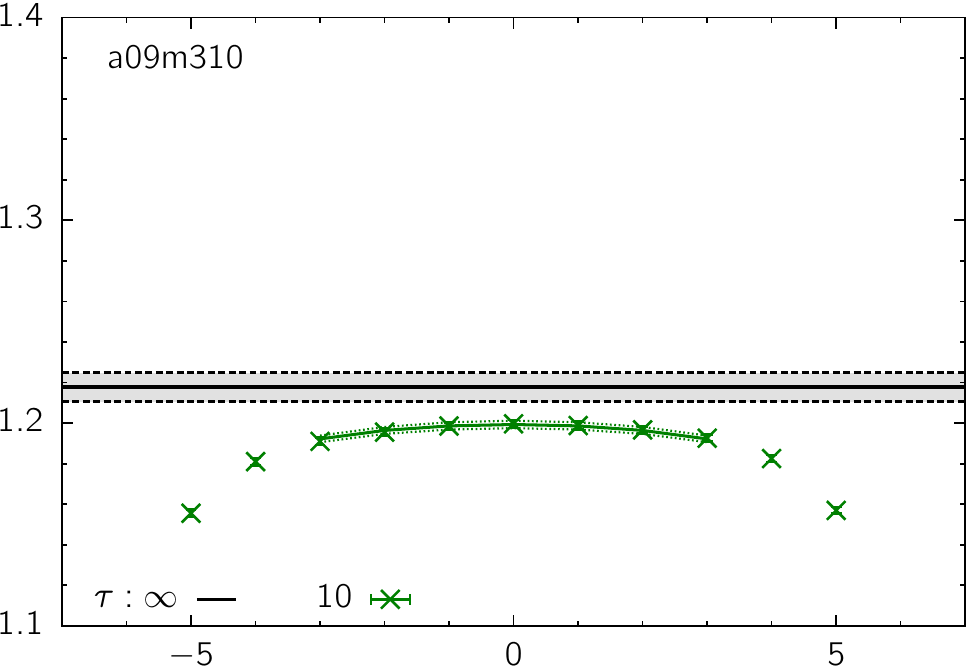}
    \includegraphics[height=1.5in,trim={0.0cm 0.00cm 0 0},clip]{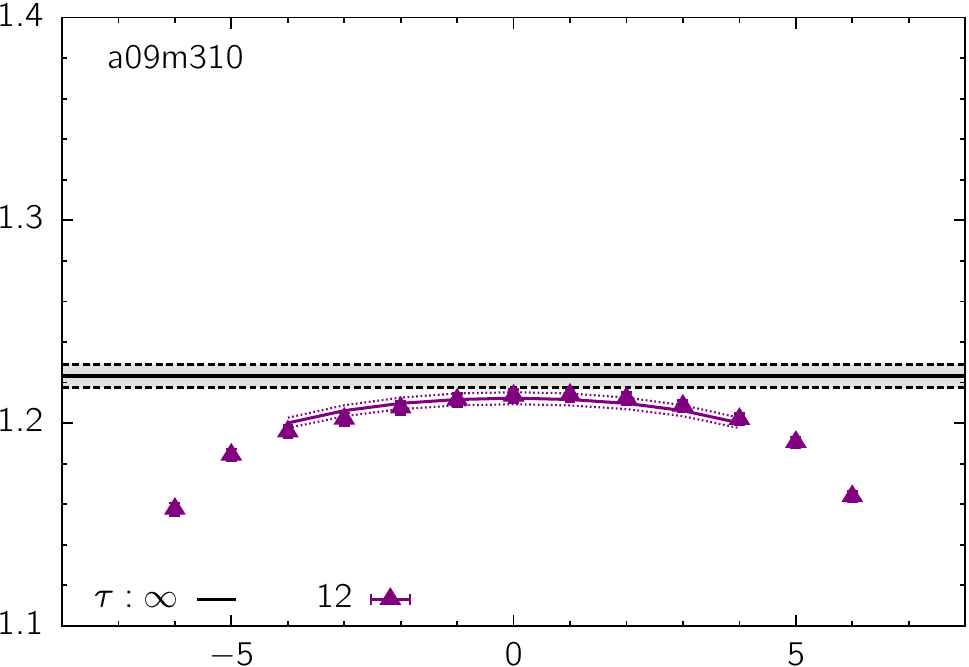}
    \includegraphics[height=1.5in,trim={0.0cm 0.00cm 0 0},clip]{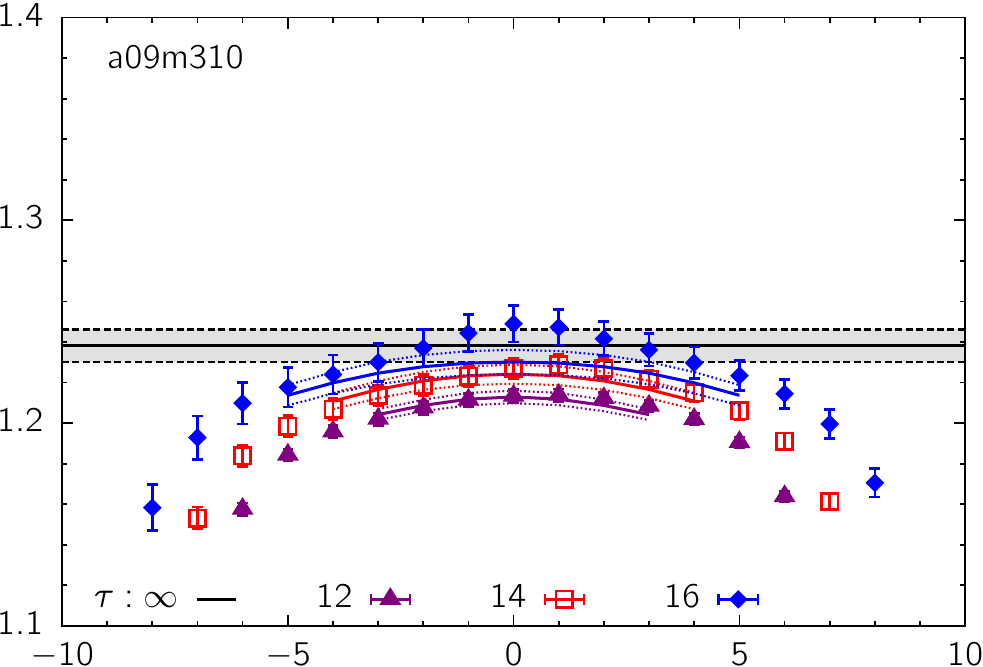}
  }
  \hspace{0.04\linewidth}
  \subfigure{
    \includegraphics[height=1.5in,trim={0.0cm 0.00cm 0 0},clip]{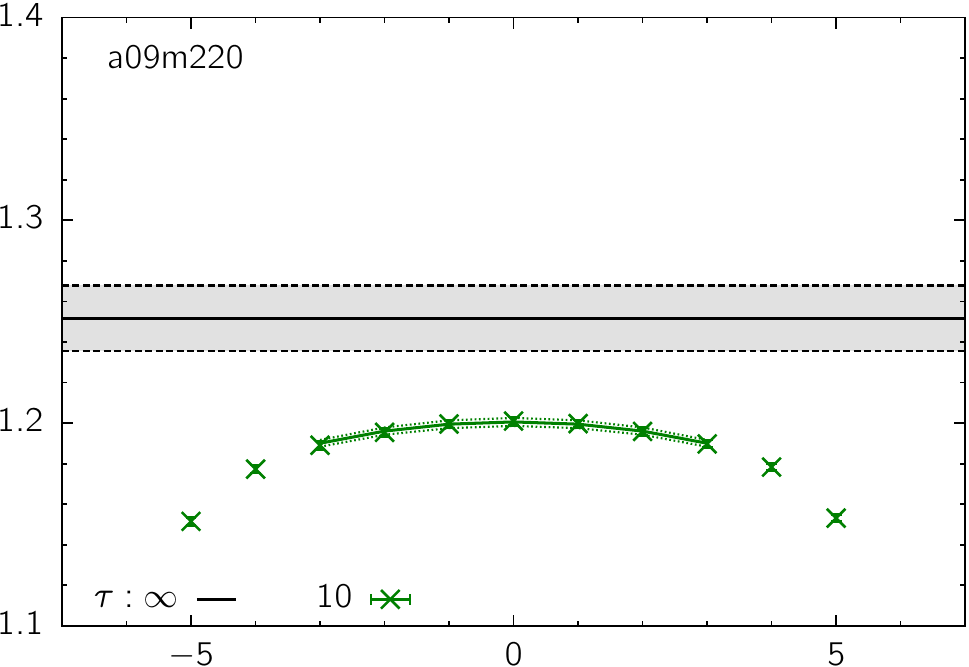}
    \includegraphics[height=1.5in,trim={0.0cm 0.00cm 0 0},clip]{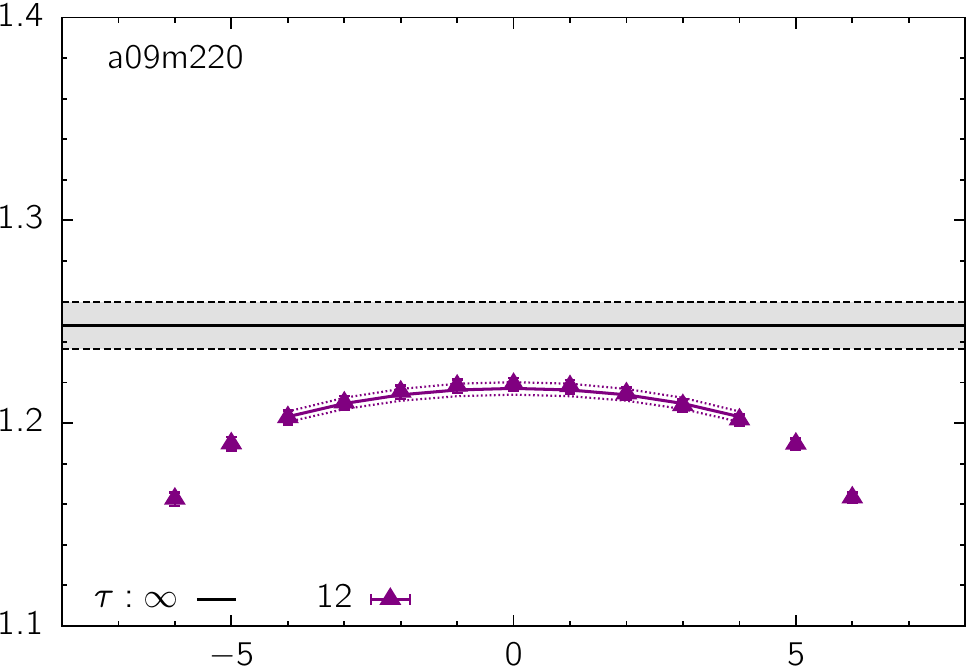}
    \includegraphics[height=1.5in,trim={0.0cm 0.00cm 0 0},clip]{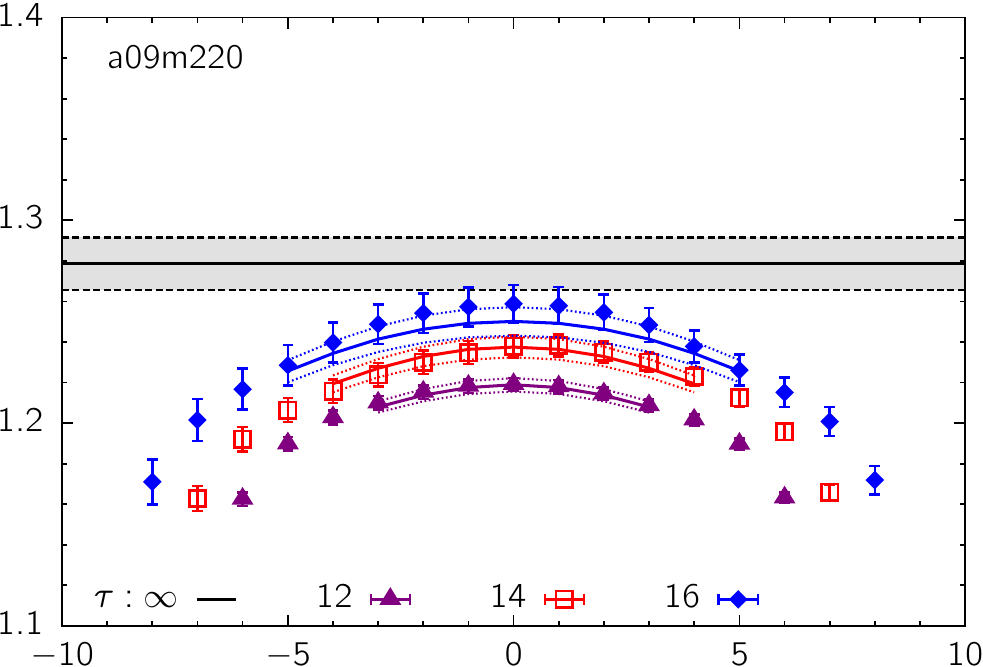}
  }
  \hspace{0.04\linewidth}
  \subfigure{
    \includegraphics[height=1.5in,trim={0.0cm 0.00cm 0 0},clip]{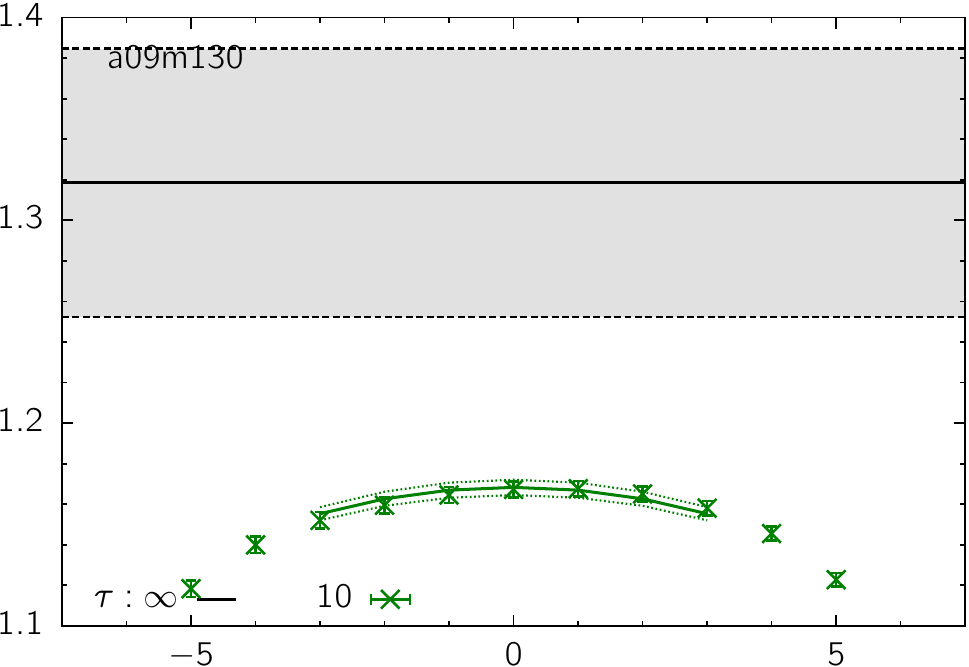}
    \includegraphics[height=1.5in,trim={0.0cm 0.00cm 0 0},clip]{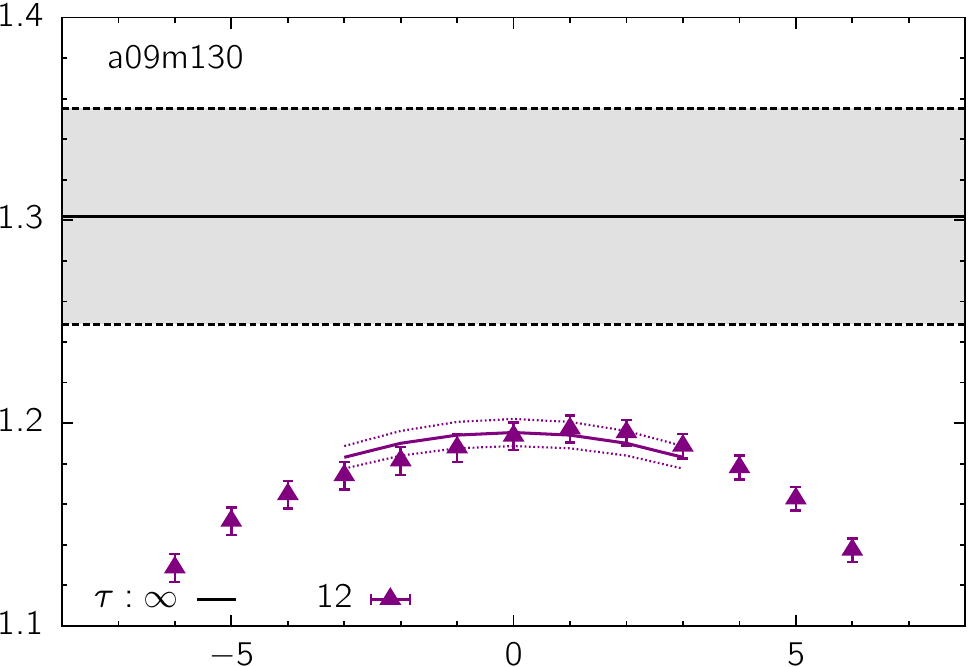}
    \includegraphics[height=1.5in,trim={0.0cm 0.00cm 0 0},clip]{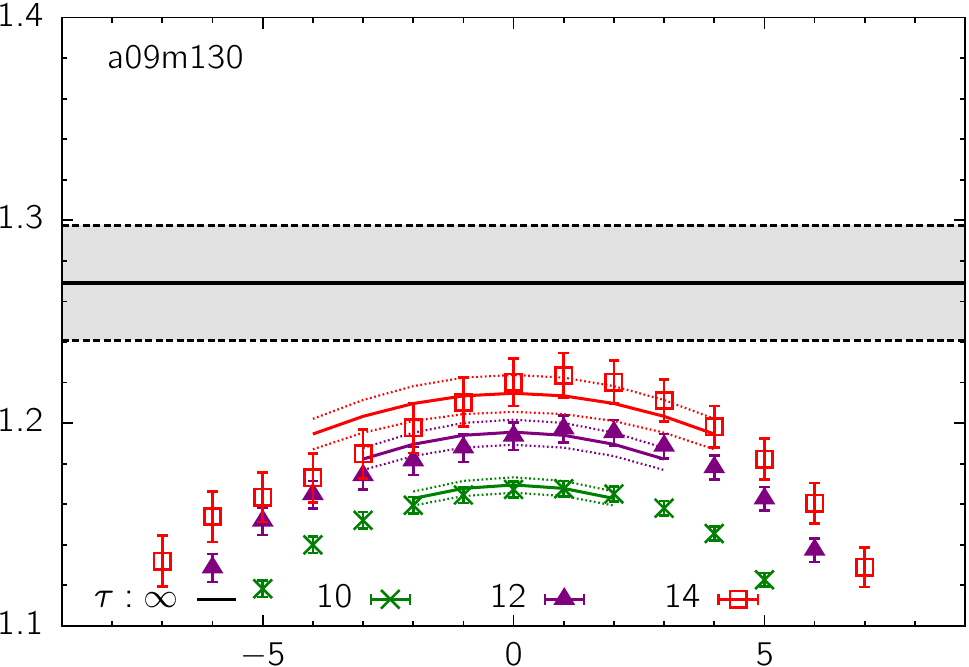}
  }
  \hspace{0.04\linewidth}
  \subfigure{
    \includegraphics[height=1.5in,trim={0.0cm 0.00cm 0 0},clip]{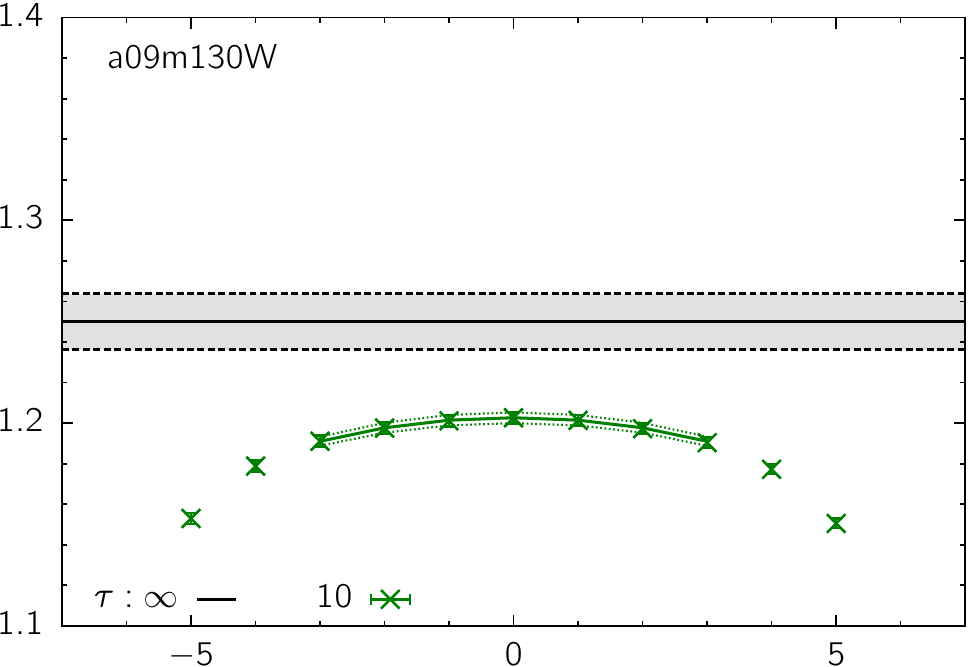}
    \includegraphics[height=1.5in,trim={0.0cm 0.00cm 0 0},clip]{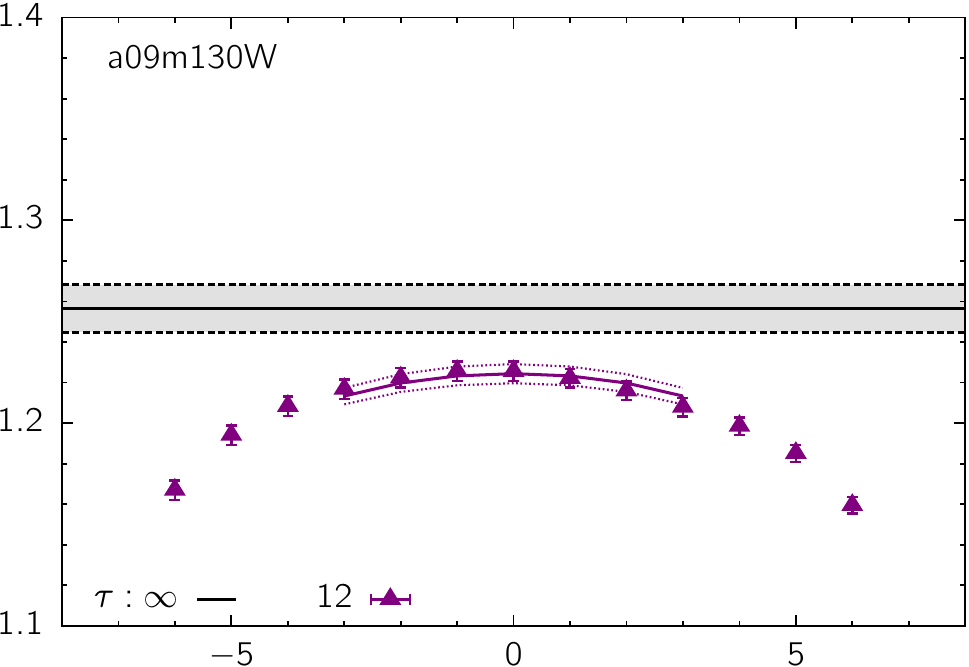}
    \includegraphics[height=1.5in,trim={0.0cm 0.00cm 0 0},clip]{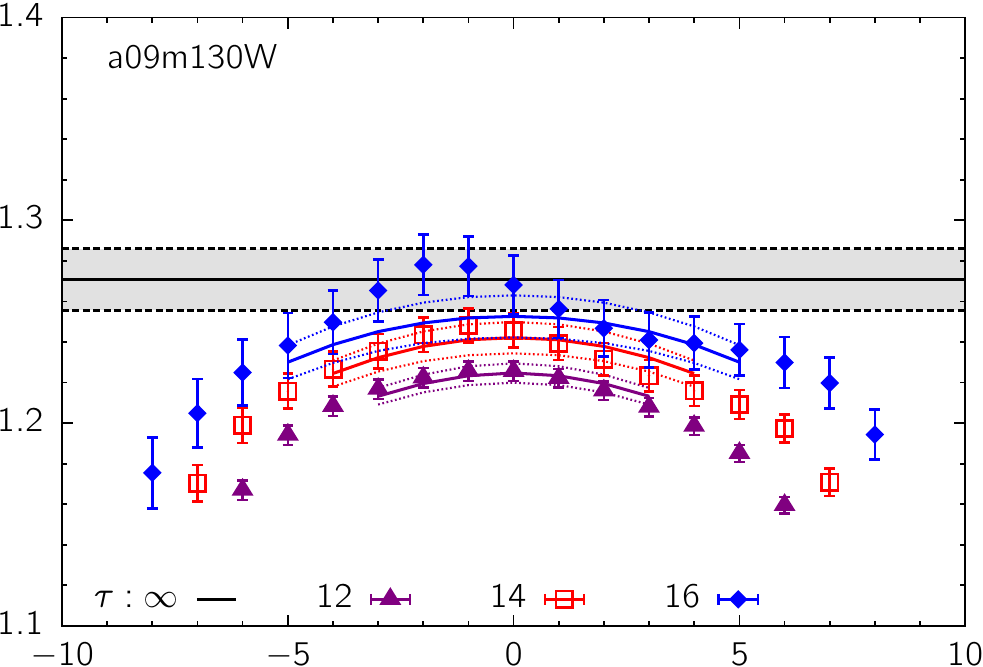}
  }
\caption{Comparison between the $2^\ast$ and
  $3^\ast$ fits to the axial charge $g_A^{u-d}$ data from the 
  $a \approx 0.09$~fm ensembles. The rest is the same as in Fig.~\ref{fig:gA2v3a12}.
  \label{fig:gA2v3a09}}
\end{figure*}

\begin{figure*}
\centering
  \subfigure{
    \includegraphics[height=1.5in,trim={0.0cm 0.00cm 0 0},clip]{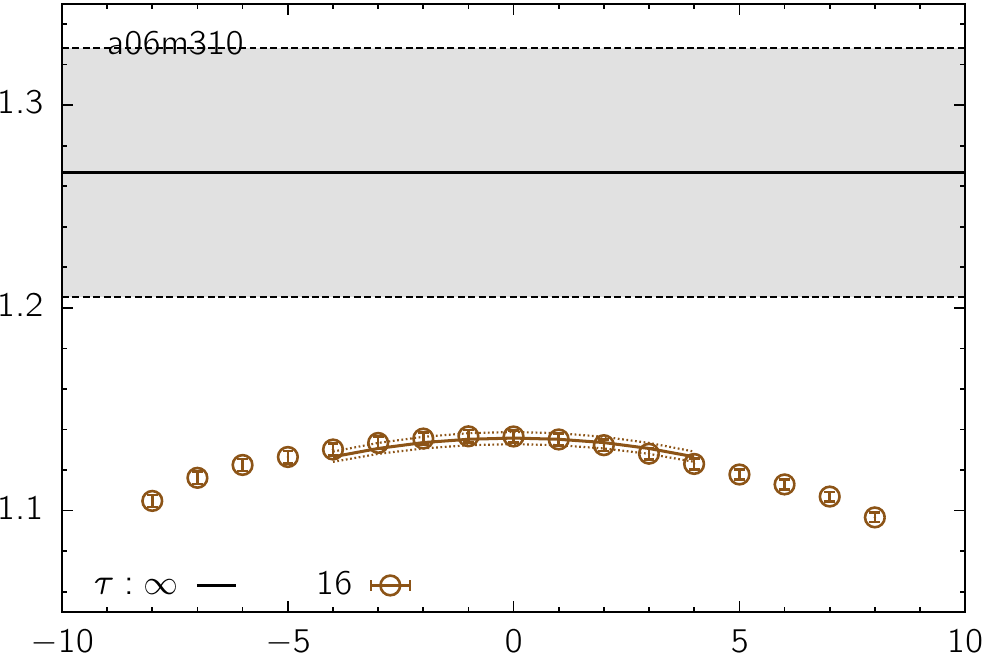}
    \includegraphics[height=1.5in,trim={0.0cm 0.00cm 0 0},clip]{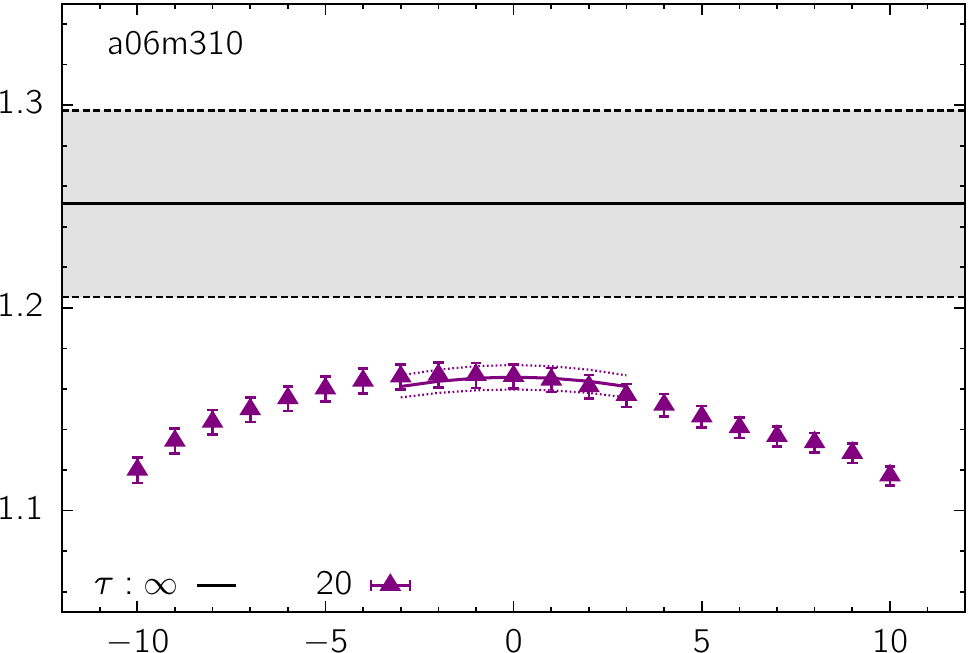}
    \includegraphics[height=1.5in,trim={0.0cm 0.00cm 0 0},clip]{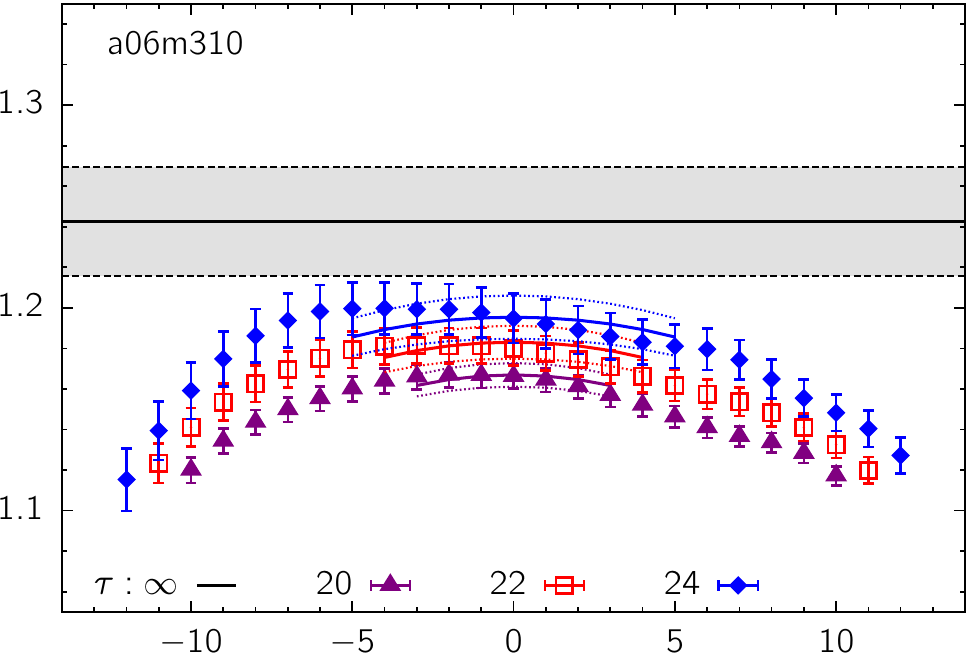}
  }
  \subfigure{
    \includegraphics[height=1.5in,trim={0.0cm 0.00cm 0 0},clip]{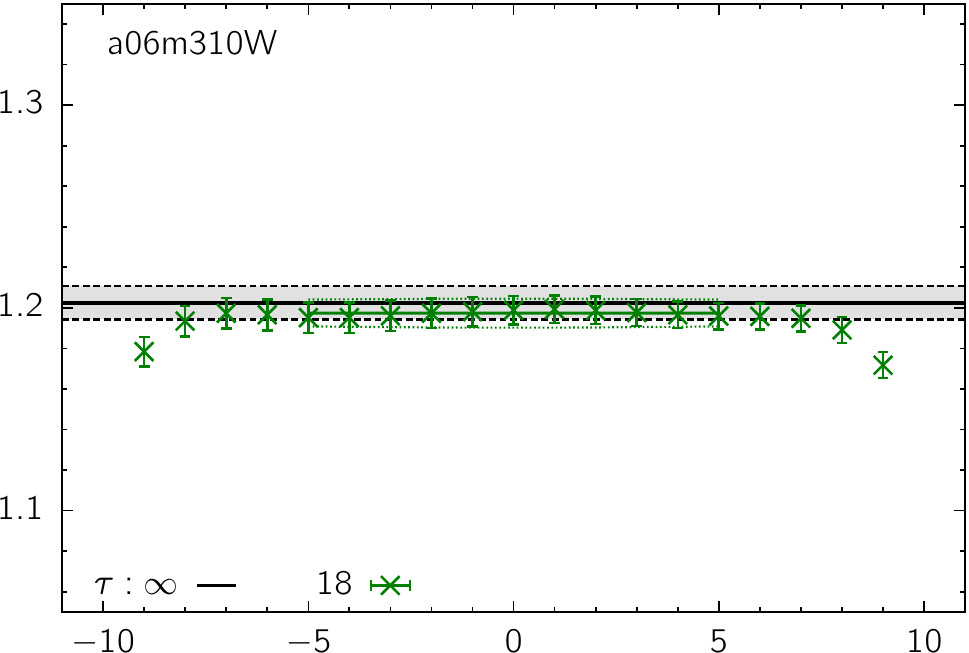}
    \includegraphics[height=1.5in,trim={0.0cm 0.00cm 0 0},clip]{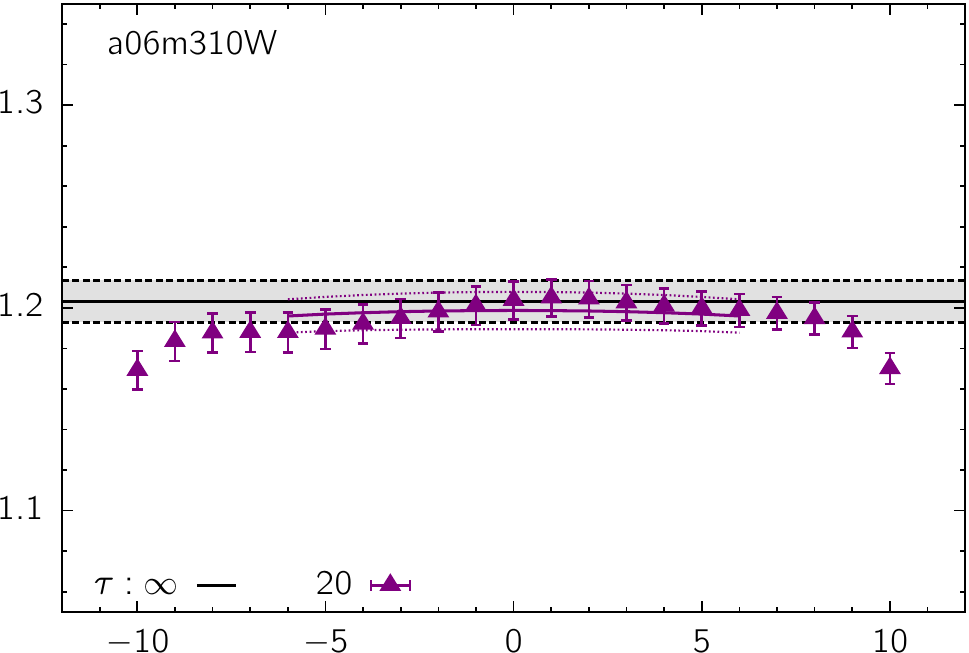}
    \includegraphics[height=1.5in,trim={0.0cm 0.00cm 0 0},clip]{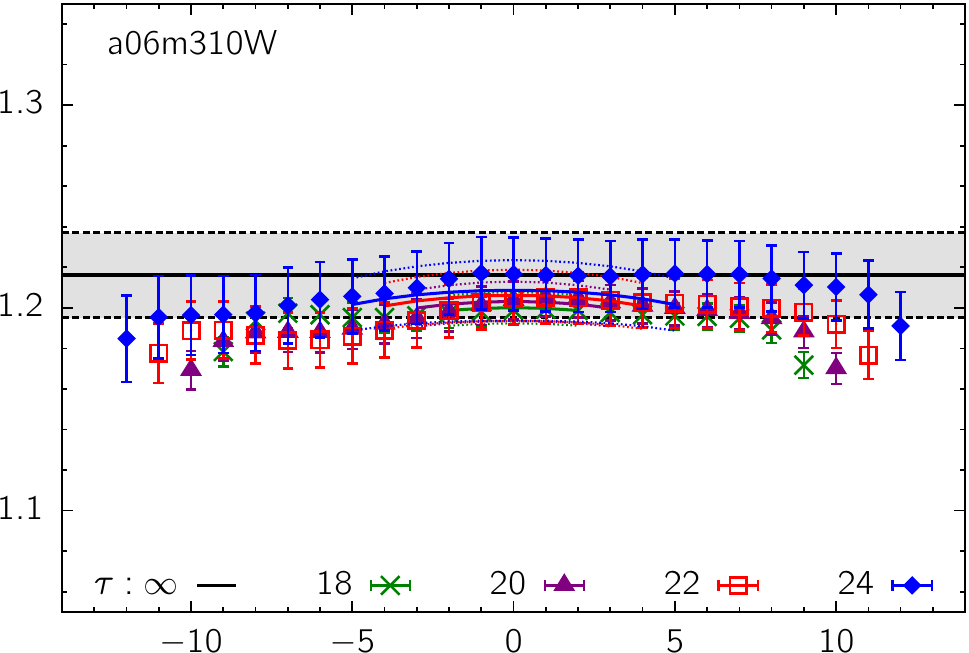}
  }
  \hspace{0.04\linewidth}
  \subfigure{
    \includegraphics[height=1.5in,trim={0.0cm 0.00cm 0 0},clip]{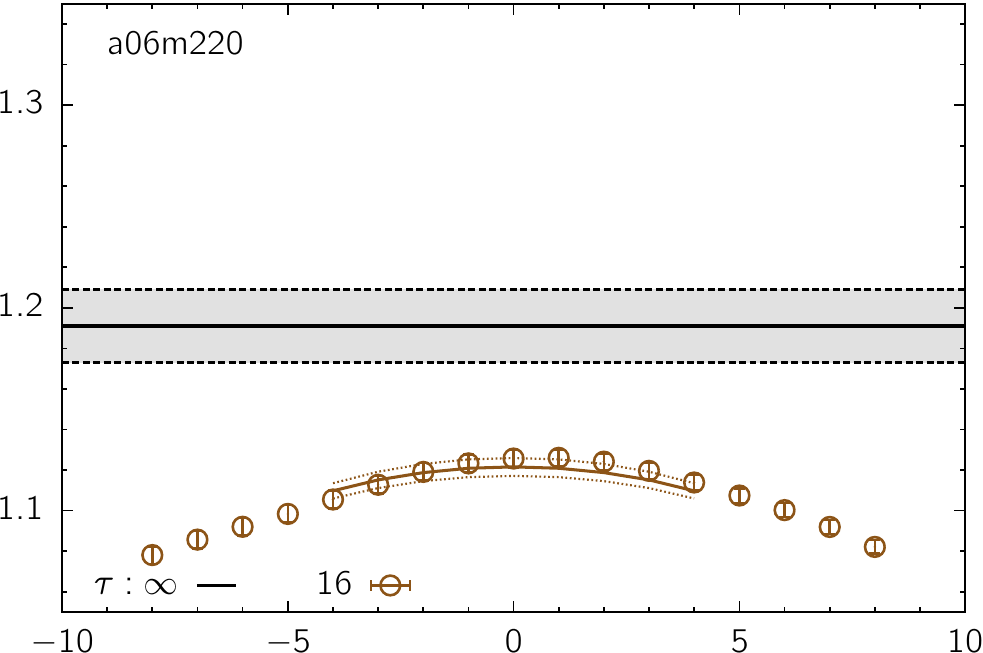}
    \includegraphics[height=1.5in,trim={0.0cm 0.00cm 0 0},clip]{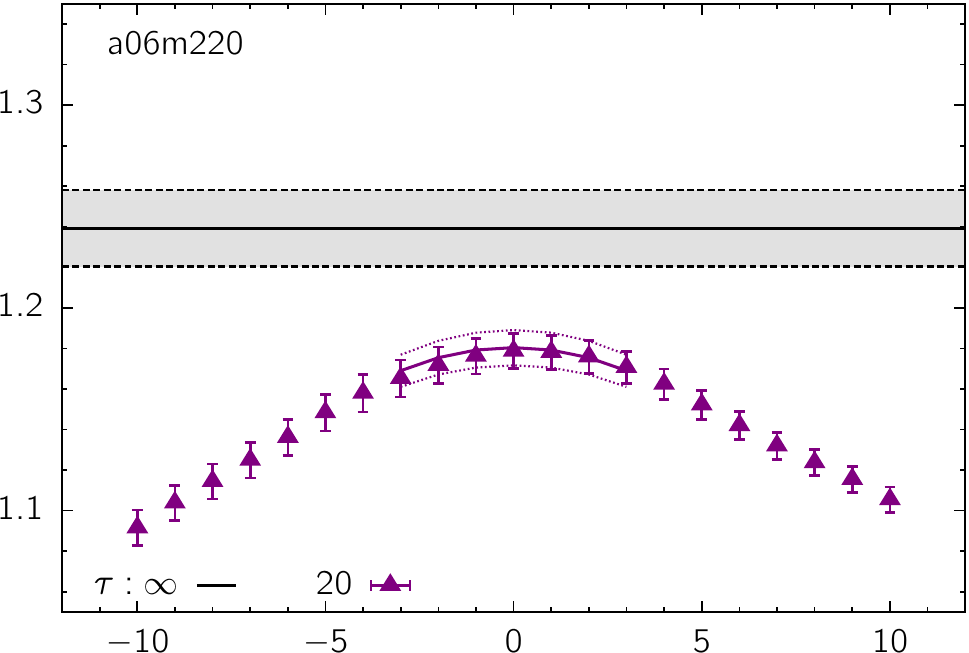}
    \includegraphics[height=1.5in,trim={0.0cm 0.00cm 0 0},clip]{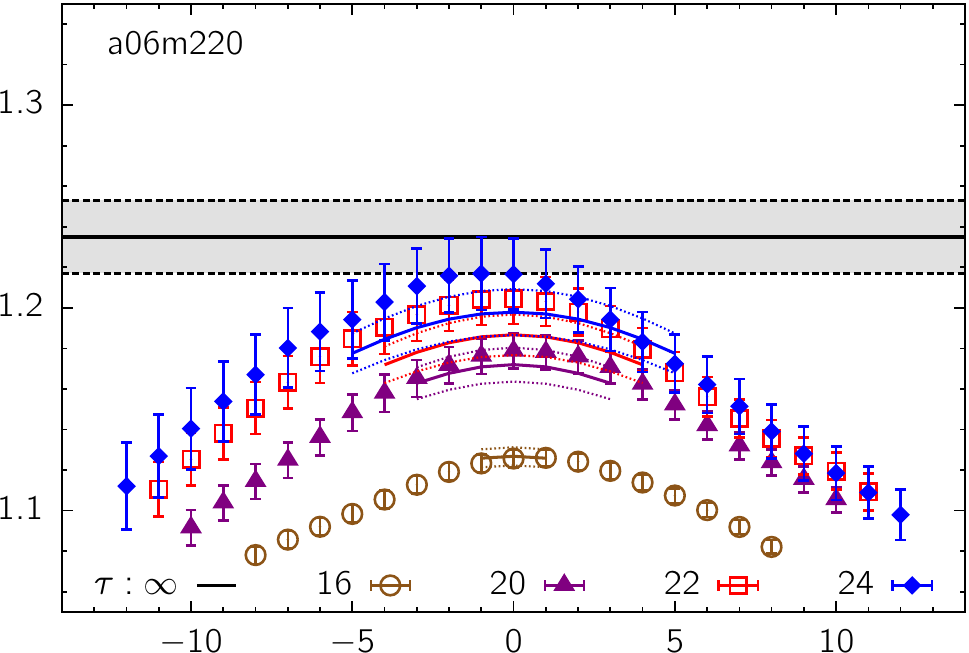}
  }
  \hspace{0.04\linewidth}
  \subfigure{
    \includegraphics[height=1.5in,trim={0.0cm 0.00cm 0 0},clip]{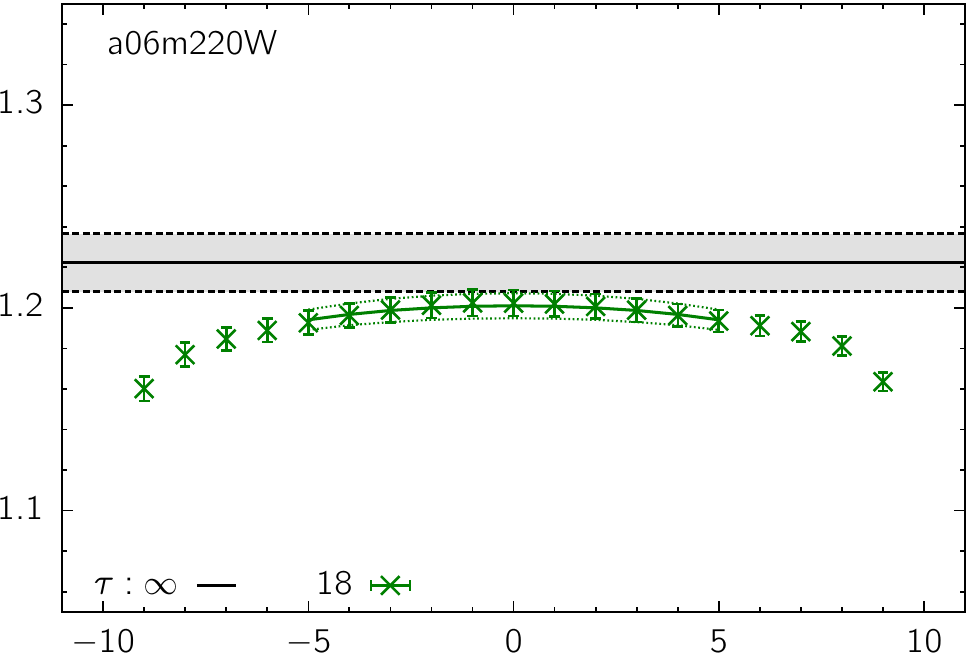}
    \includegraphics[height=1.5in,trim={0.0cm 0.00cm 0 0},clip]{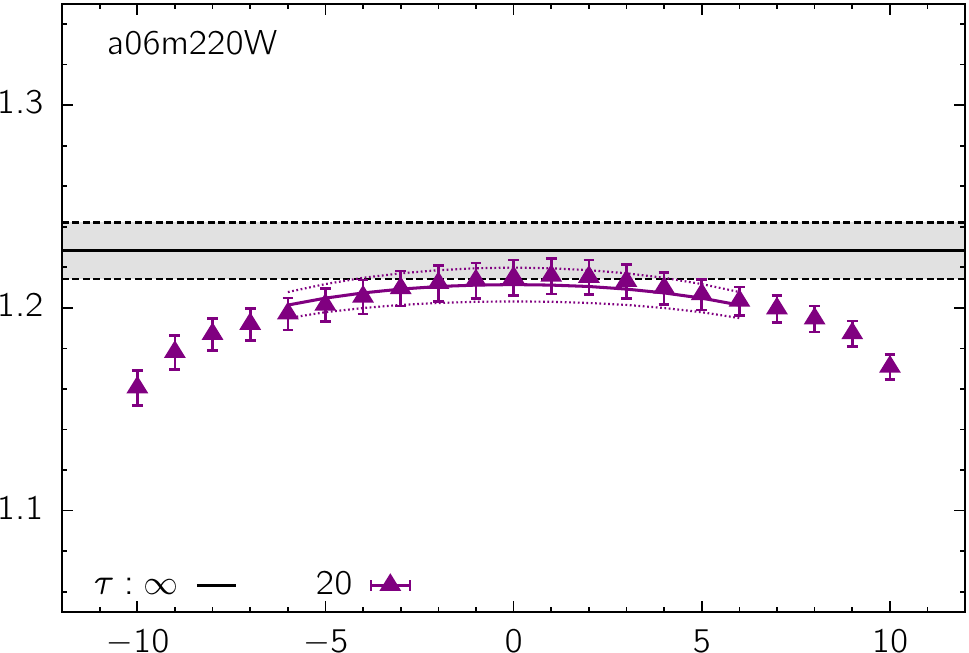}
    \includegraphics[height=1.5in,trim={0.0cm 0.00cm 0 0},clip]{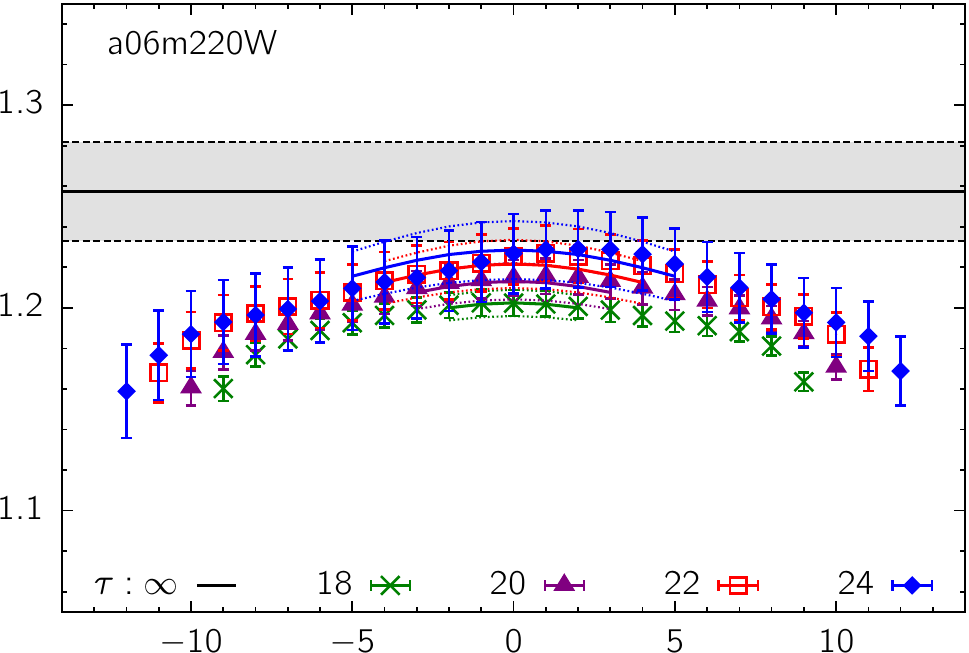}
  }
  \hspace{0.04\linewidth}
  \subfigure{
    \includegraphics[height=1.5in,trim={0.0cm 0.00cm 0 0},clip]{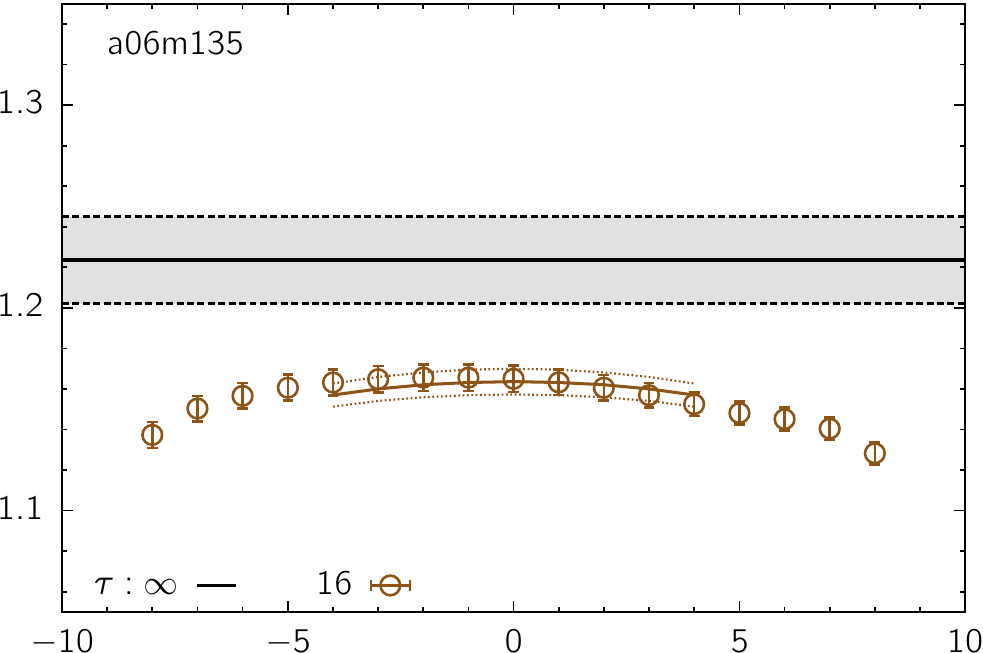}
    \includegraphics[height=1.5in,trim={0.0cm 0.00cm 0 0},clip]{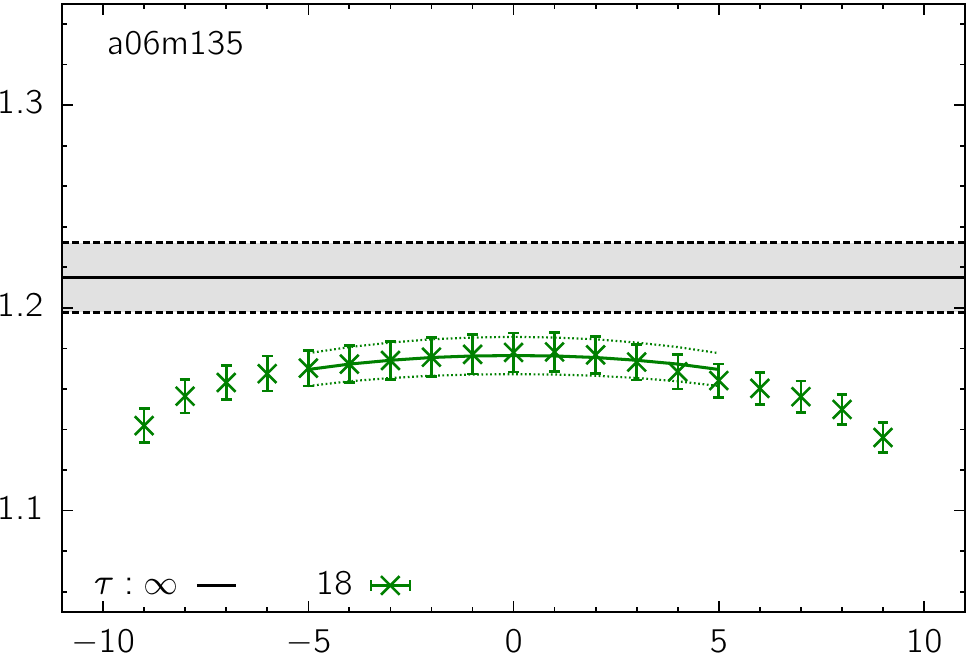}
    \includegraphics[height=1.5in,trim={0.0cm 0.00cm 0 0},clip]{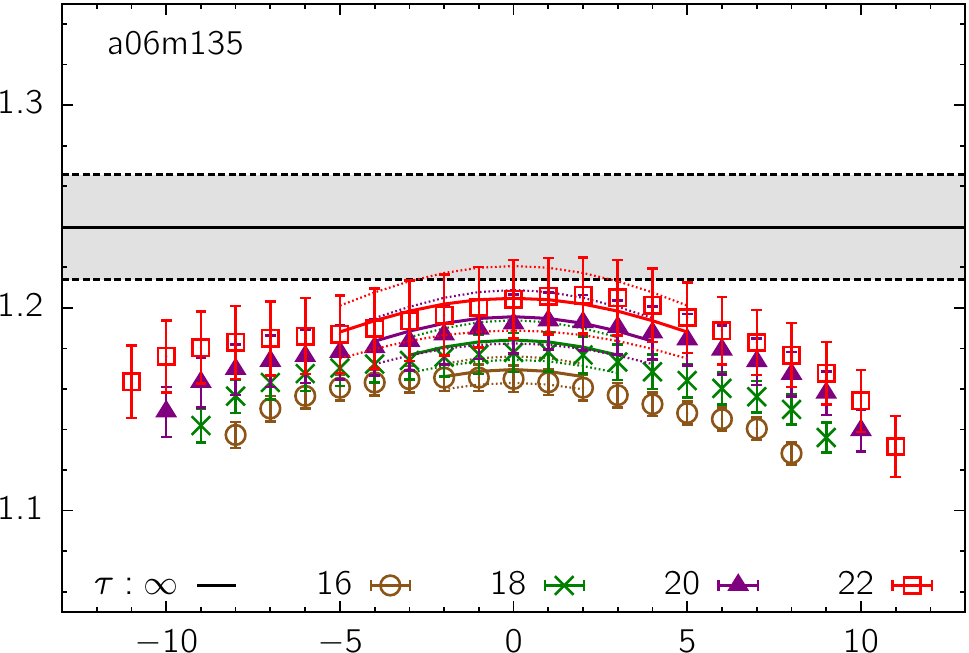}
  }
\caption{Comparison between the $2^\ast$ and
  $3^\ast$ fits to the axial charge $g_A^{u-d}$ data from the 
  $a \approx 0.06$~fm ensembles. The rest is the same as in Fig.~\ref{fig:gA2v3a12}. 
  \label{fig:gA2v3a06}}
\end{figure*}

\begin{figure*}
\centering
  \subfigure{
    \includegraphics[height=1.5in,trim={0.0cm 0.00cm 0 0},clip]{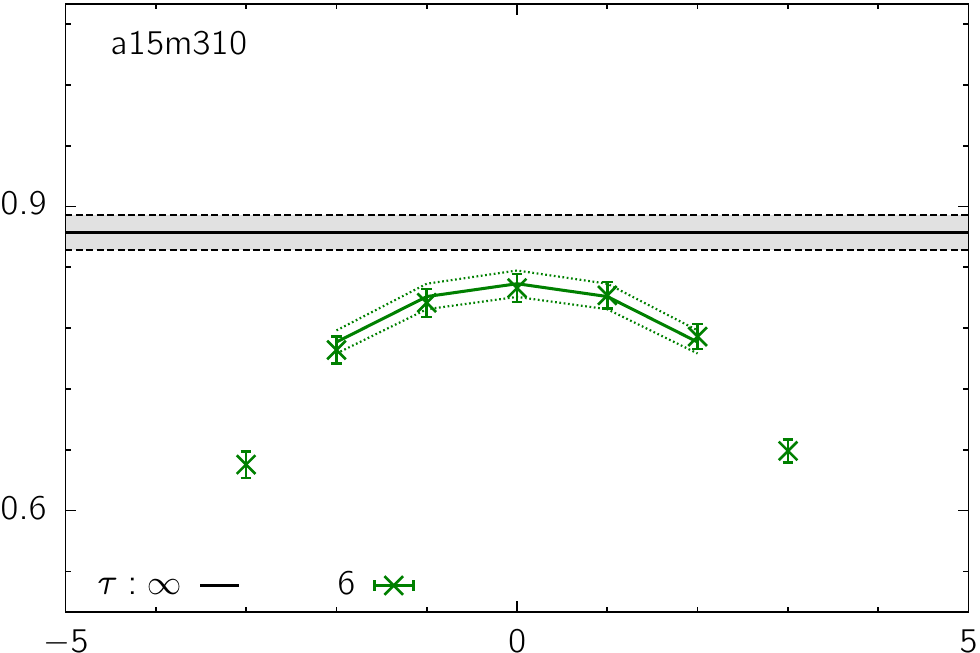}
    \includegraphics[height=1.5in,trim={0.1cm 0.00cm 0 0},clip]{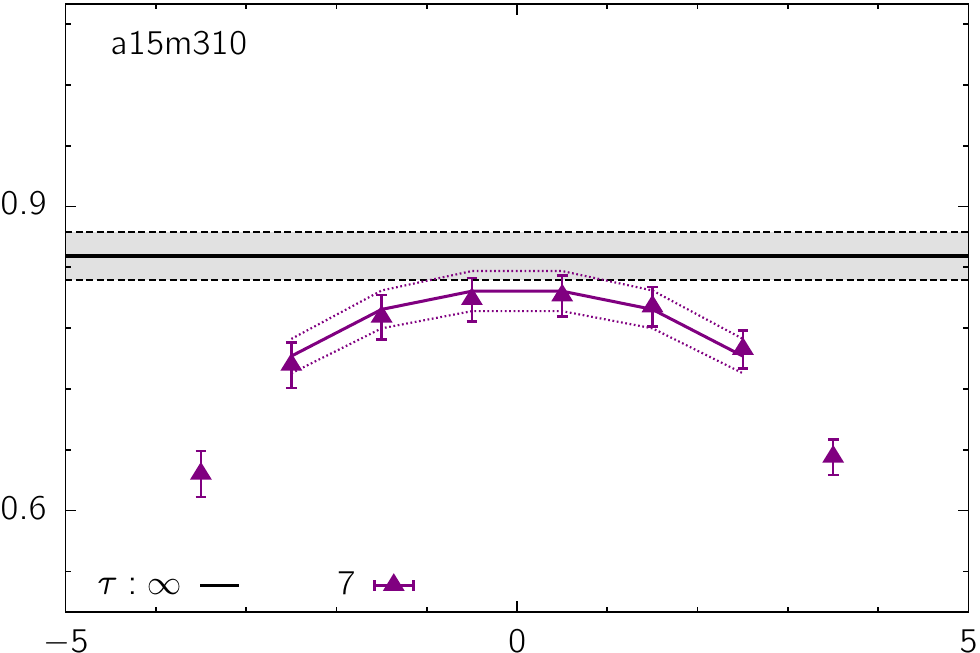}
    \includegraphics[height=1.5in,trim={0.1cm 0.00cm 0 0},clip]{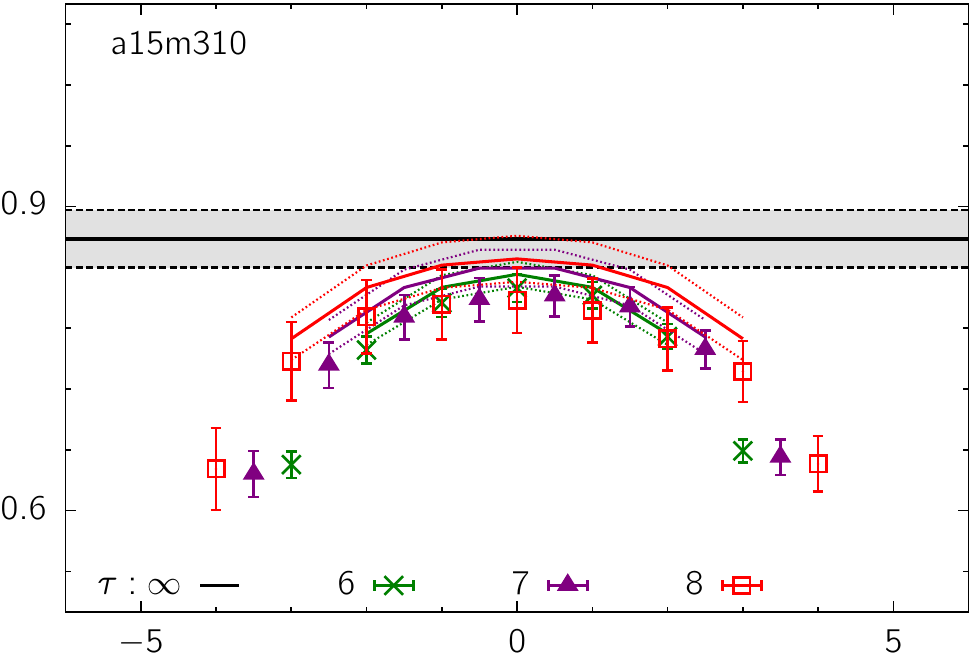}
  }
  \hspace{0.04\linewidth}
  \subfigure{
    \includegraphics[height=1.5in,trim={0.0cm 0.00cm 0 0},clip]{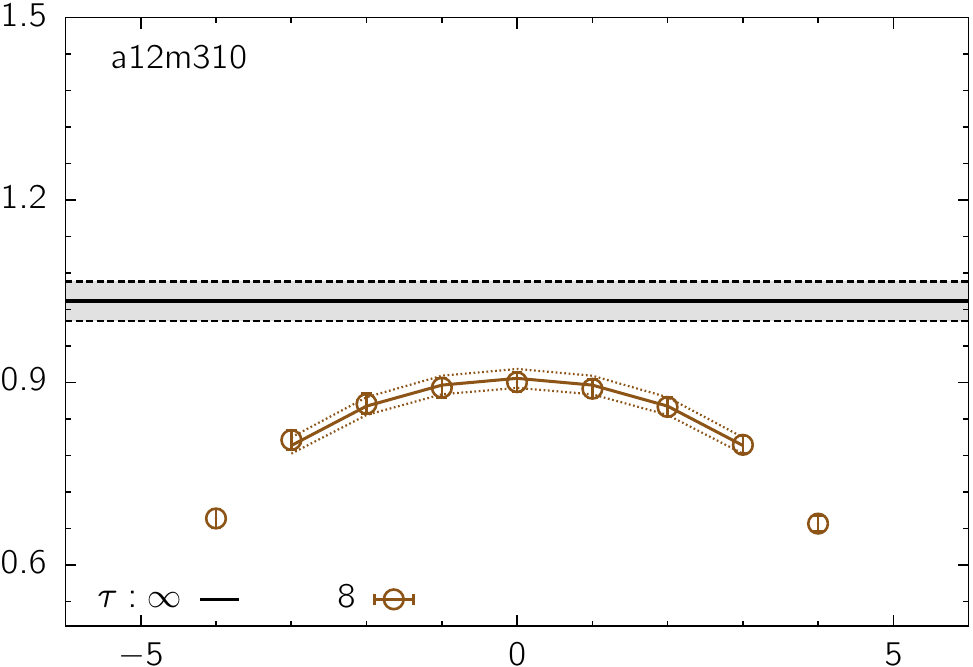}
    \includegraphics[height=1.5in,trim={0.1cm 0.00cm 0 0},clip]{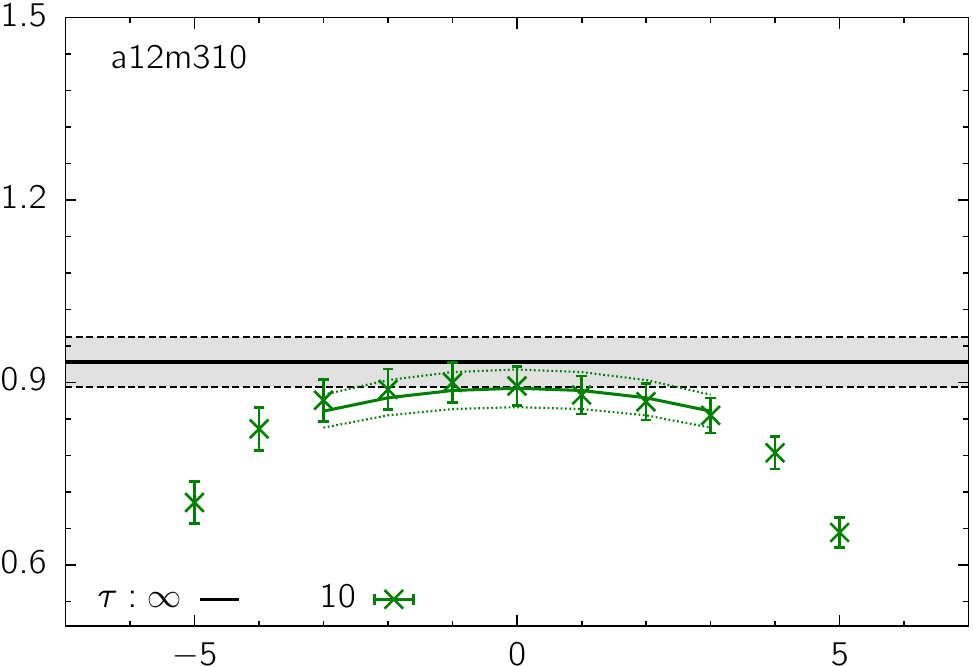}
    \includegraphics[height=1.5in,trim={0.1cm 0.00cm 0 0},clip]{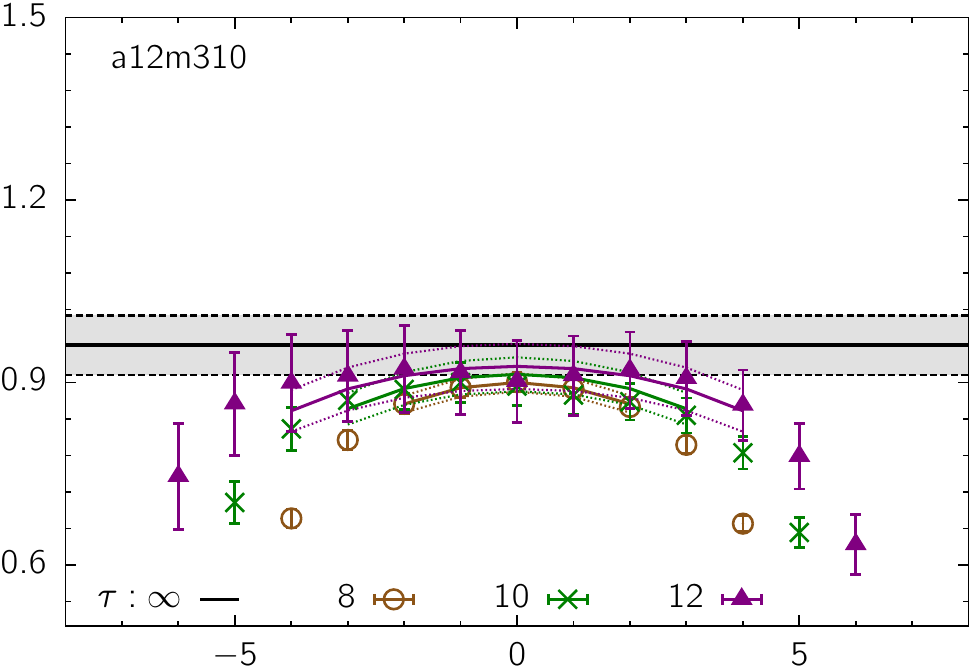}
  }
  \hspace{0.04\linewidth}
  \subfigure{
    \includegraphics[height=1.5in,trim={0.0cm 0.00cm 0 0},clip]{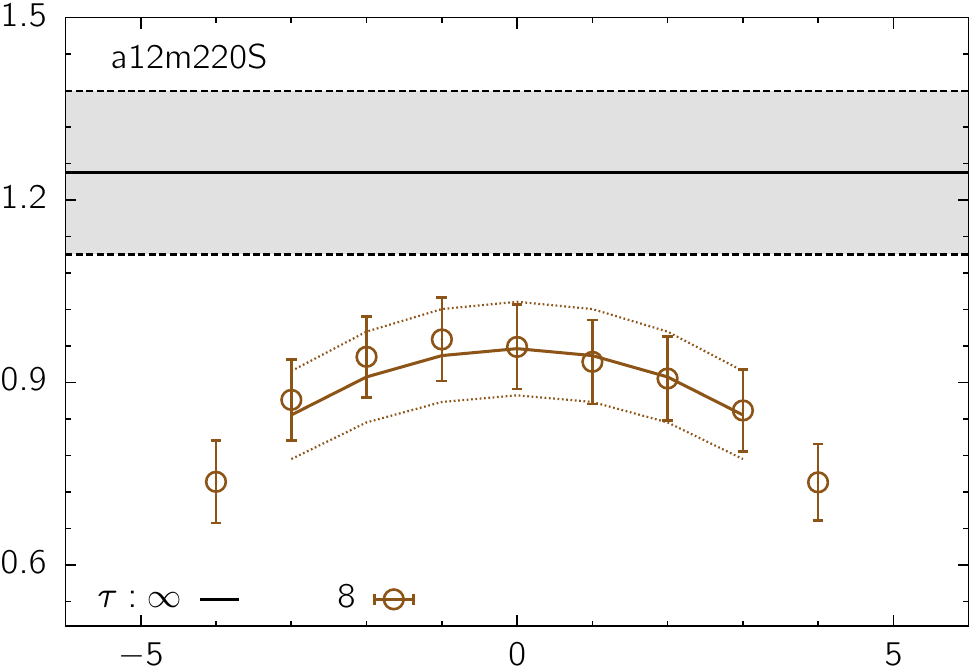}
    \includegraphics[height=1.5in,trim={0.1cm 0.00cm 0 0},clip]{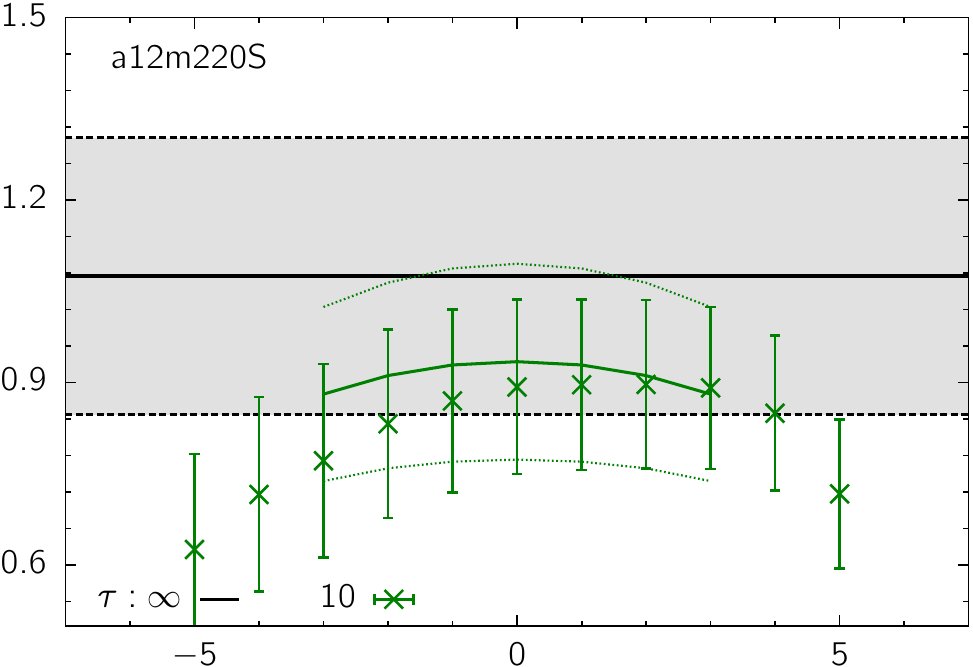}
    \includegraphics[height=1.5in,trim={0.1cm 0.00cm 0 0},clip]{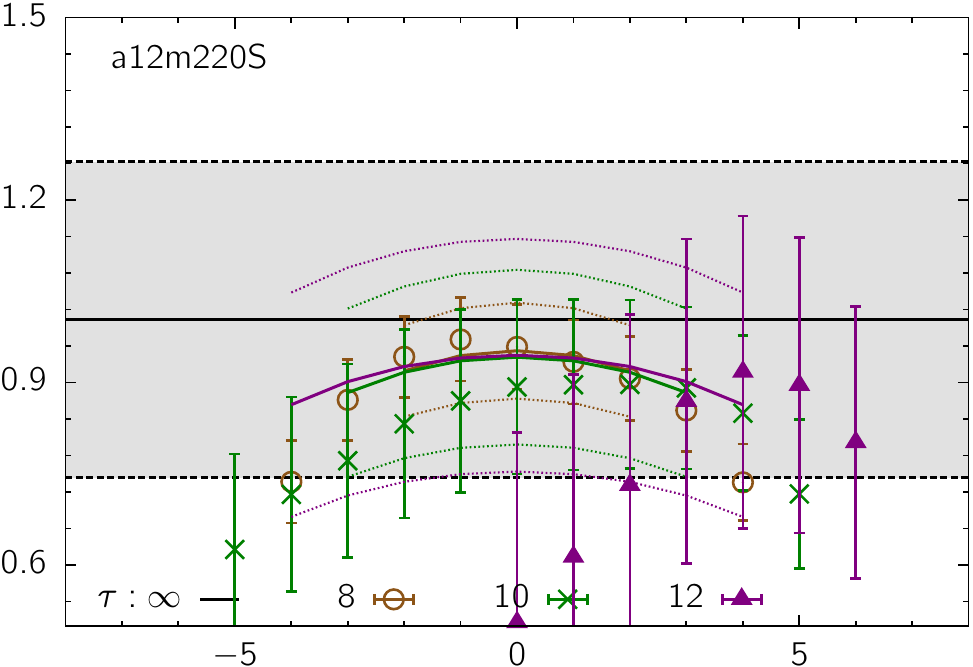}
  }
  \hspace{0.04\linewidth}
  \subfigure{
    \includegraphics[height=1.5in,trim={0.0cm 0.00cm 0 0},clip]{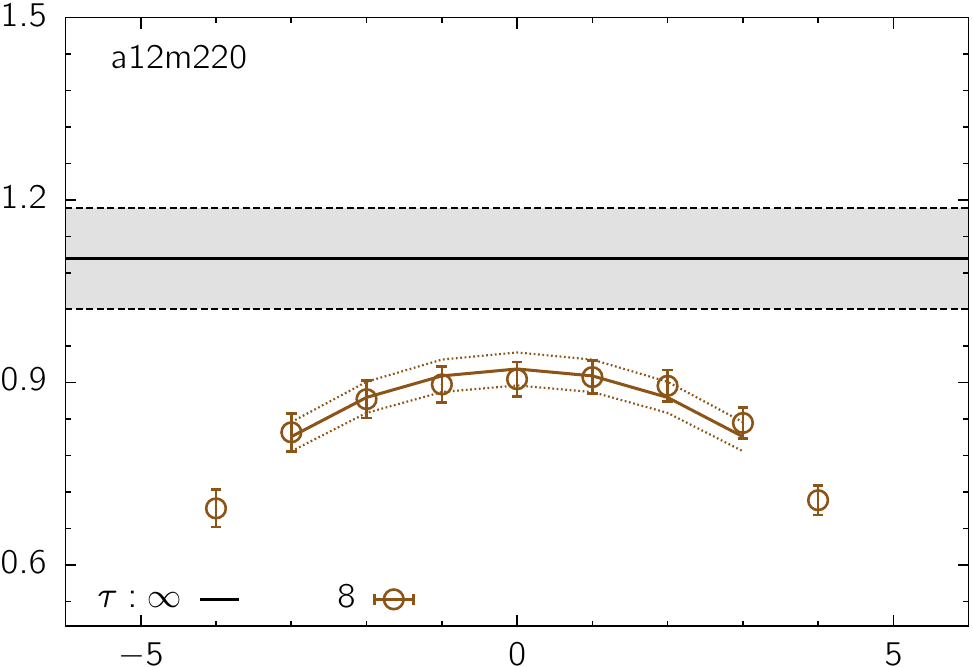}
    \includegraphics[height=1.5in,trim={0.0cm 0.00cm 0 0},clip]{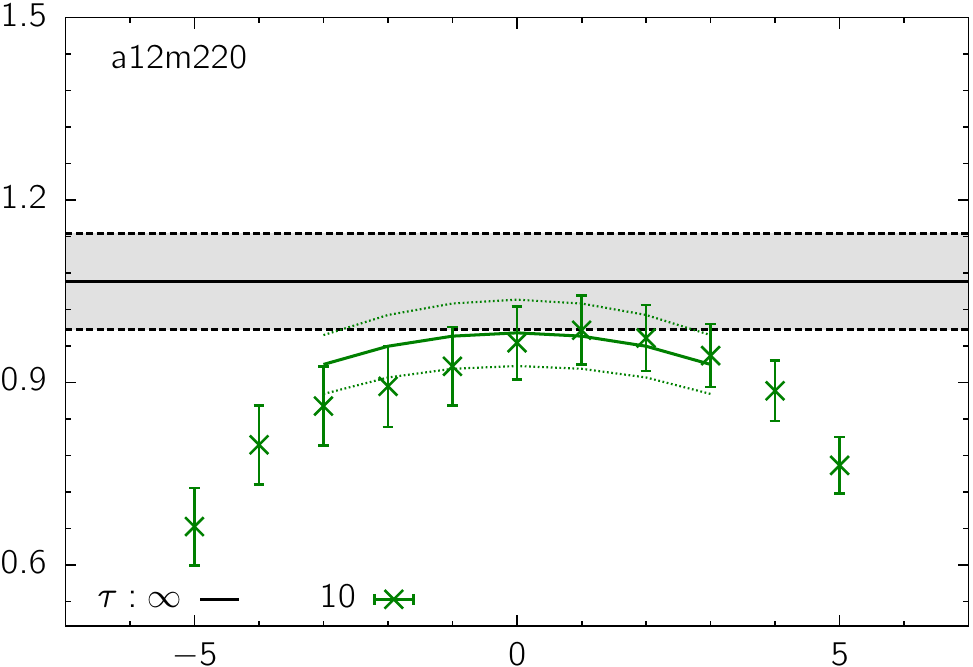}
    \includegraphics[height=1.5in,trim={0.0cm 0.00cm 0 0},clip]{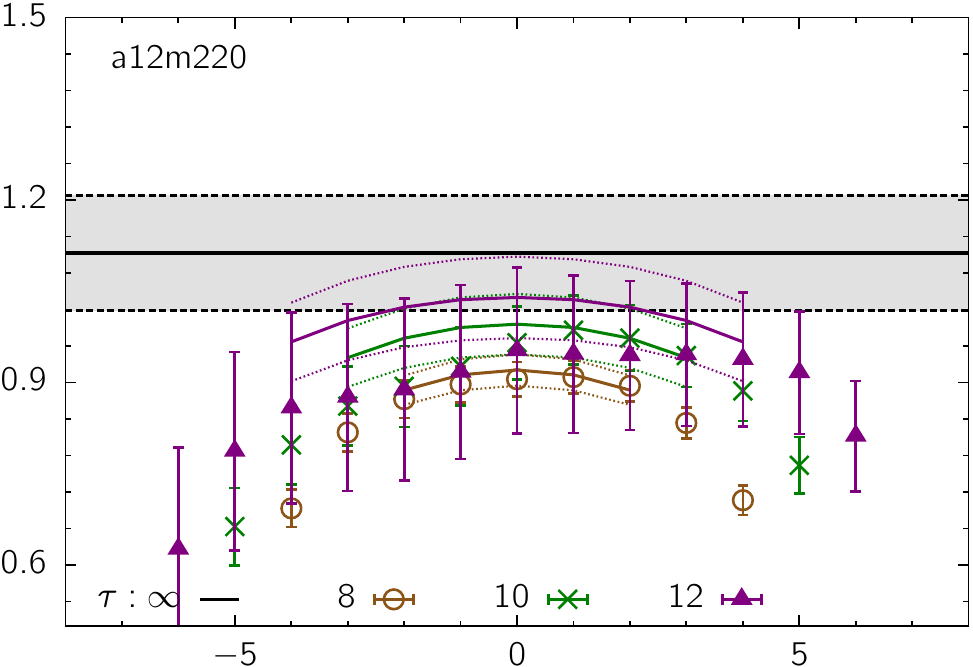}
  }
  \hspace{0.04\linewidth}
  \subfigure{
    \includegraphics[height=1.5in,trim={0.0cm 0.00cm 0 0},clip]{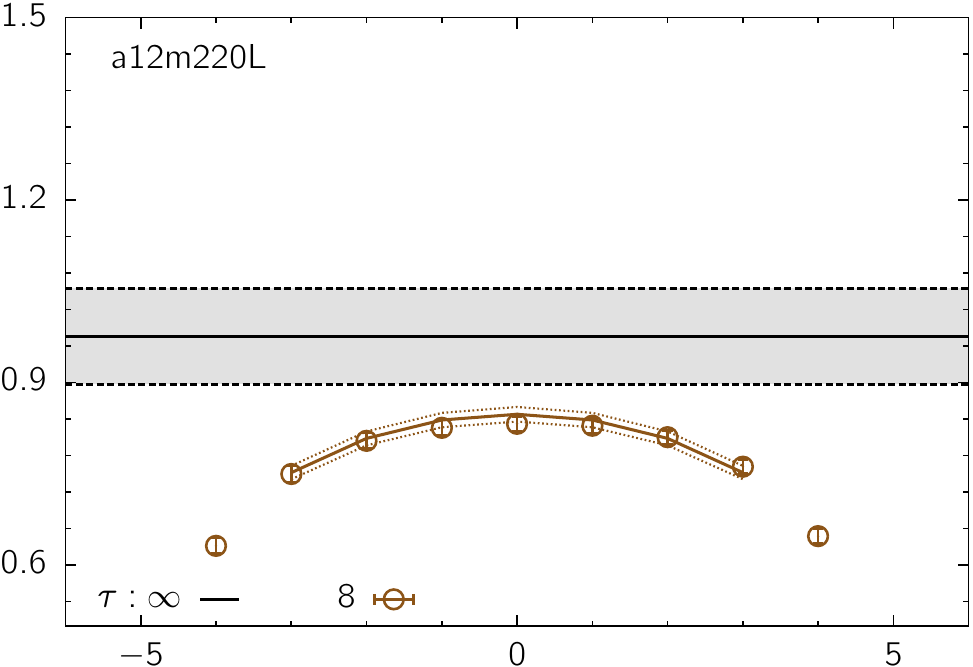}
    \includegraphics[height=1.5in,trim={0.0cm 0.00cm 0 0},clip]{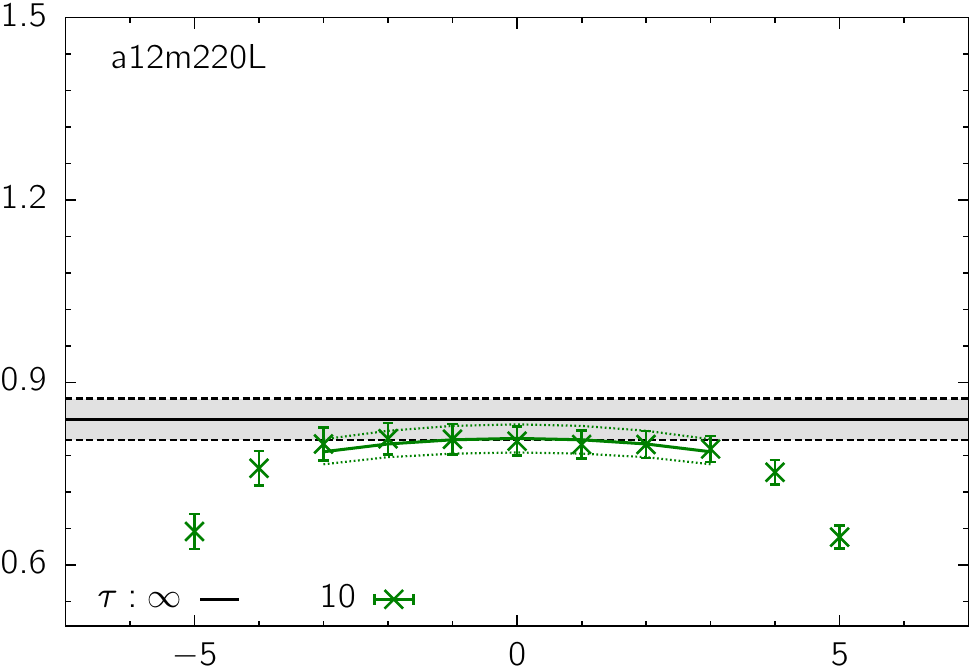}
    \includegraphics[height=1.5in,trim={0.0cm 0.00cm 0 0},clip]{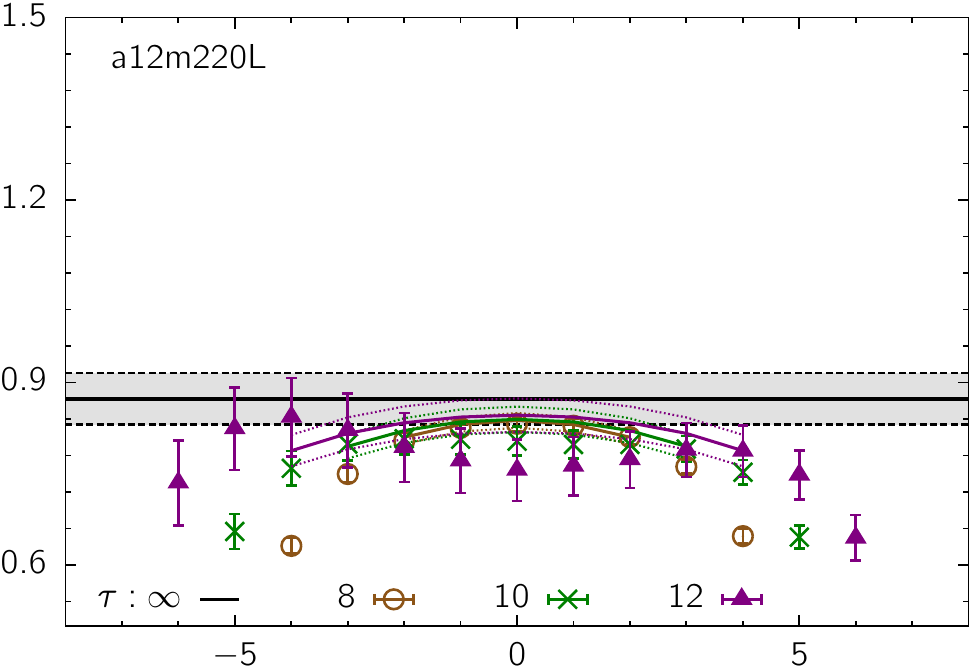}
  }
\caption{Comparison between the $2^\ast$ and
  $2-$state fits to the scalar charge $g_S^{u-d}$ data from the 
  $a \approx 0.15$~fm (top row) and  $a \approx 0.12$~fm (bottom 4 rows) ensembles. 
  The rest is the same as in Fig.~\ref{fig:gA2v3a12}. 
  \label{fig:gS2v3a12}}
\end{figure*}

\begin{figure*}
\centering
  \subfigure{
    \includegraphics[height=1.5in,trim={0.0cm 0.00cm 0 0},clip]{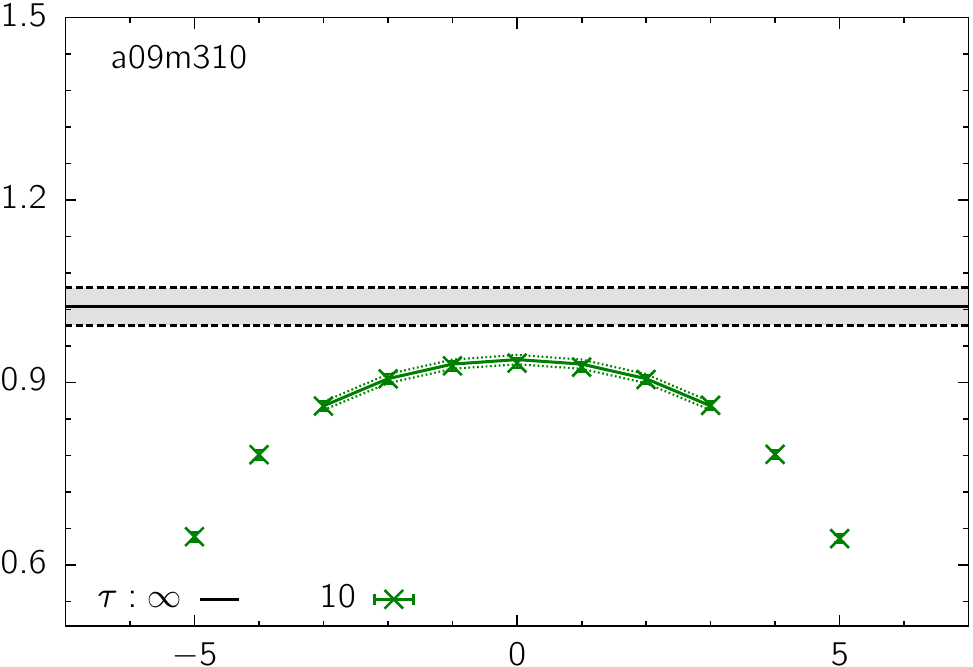}
    \includegraphics[height=1.5in,trim={0.0cm 0.00cm 0 0},clip]{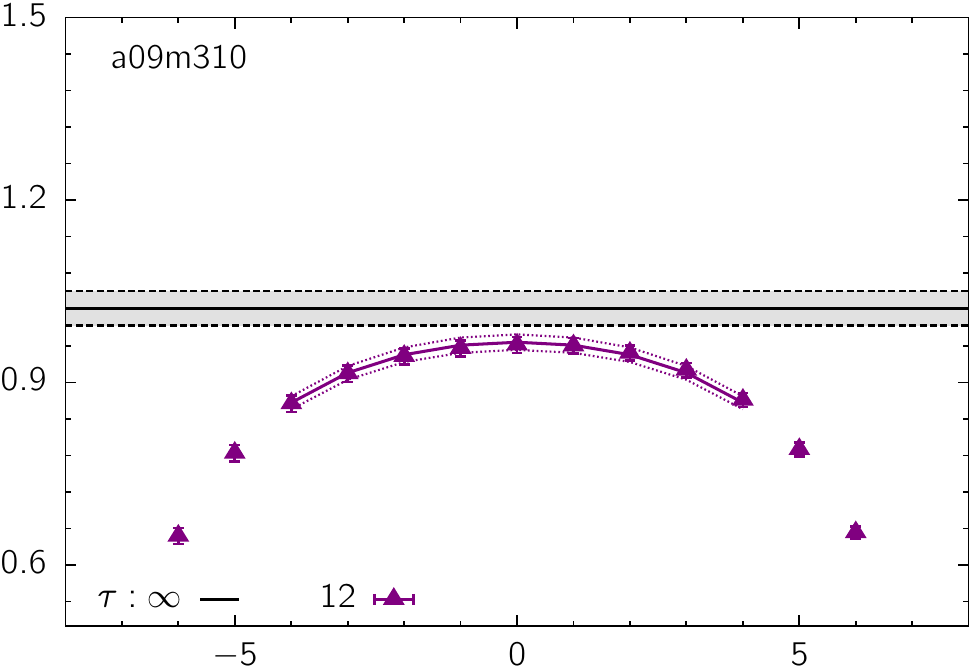}
    \includegraphics[height=1.5in,trim={0.0cm 0.00cm 0 0},clip]{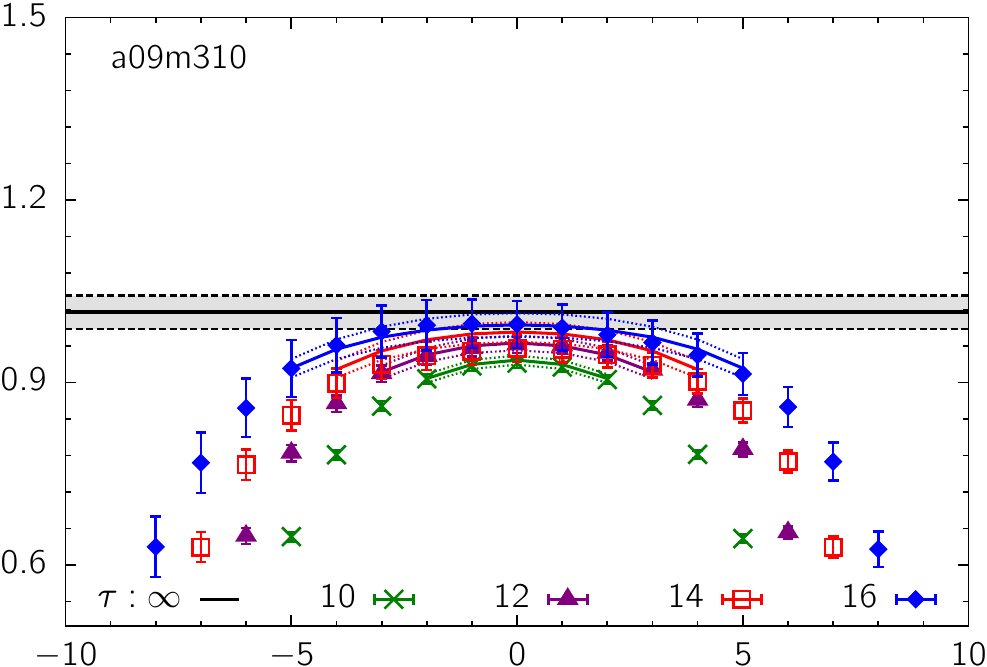}
  }
  \hspace{0.04\linewidth}
  \subfigure{
    \includegraphics[height=1.5in,trim={0.0cm 0.00cm 0 0},clip]{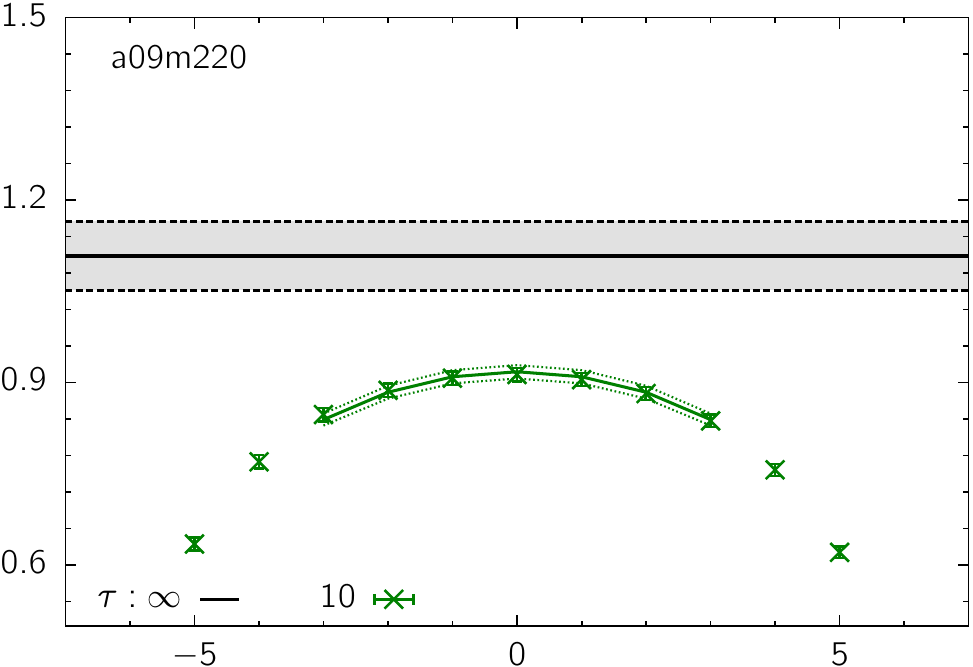}
    \includegraphics[height=1.5in,trim={0.0cm 0.00cm 0 0},clip]{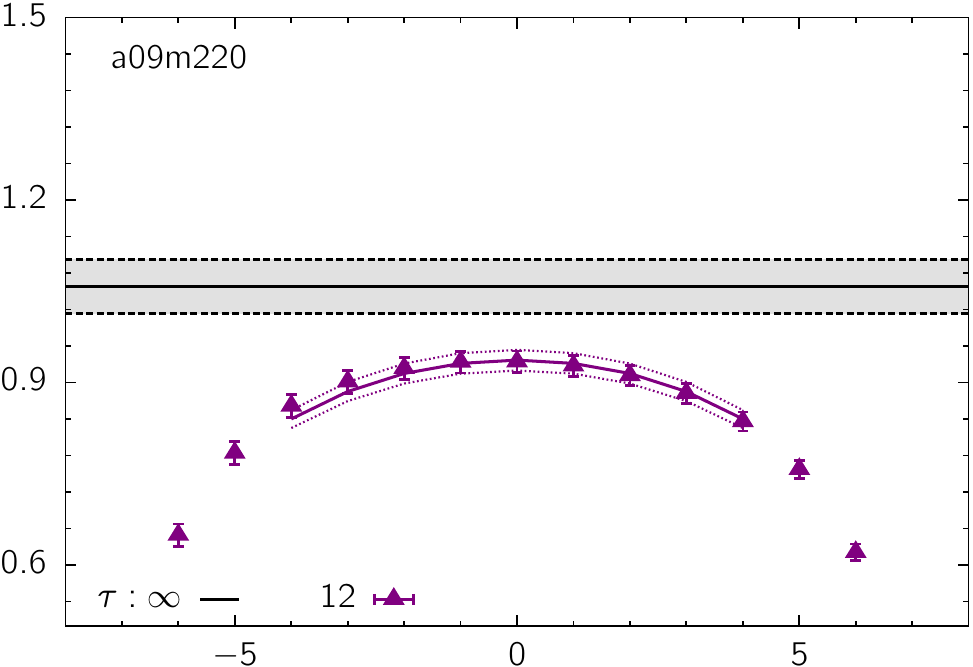}
    \includegraphics[height=1.5in,trim={0.0cm 0.00cm 0 0},clip]{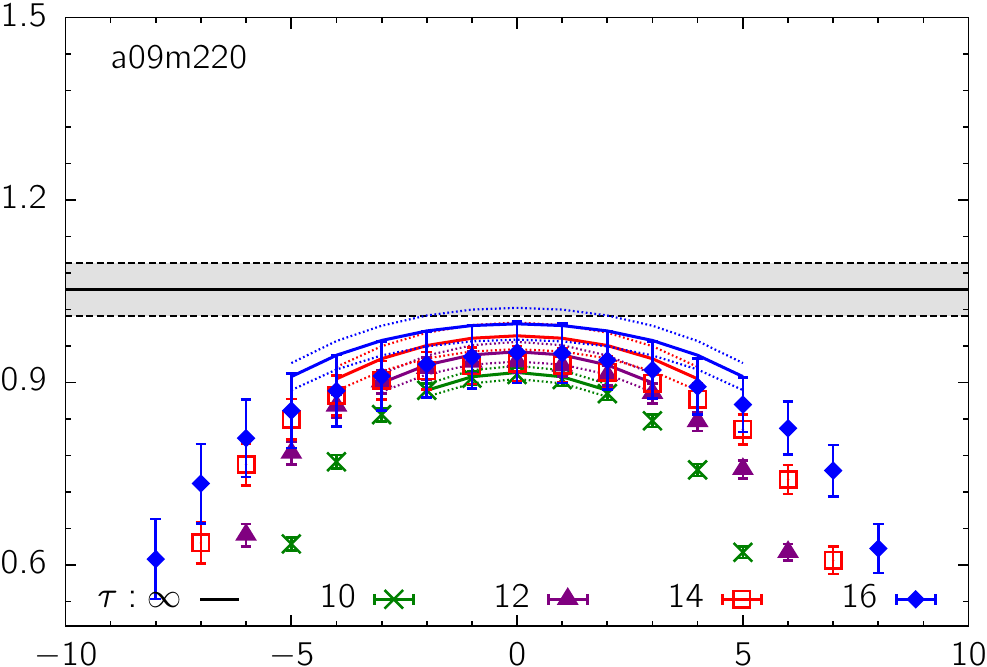}
  }
  \hspace{0.04\linewidth}
  \subfigure{
    \includegraphics[height=1.5in,trim={0.0cm 0.00cm 0 0},clip]{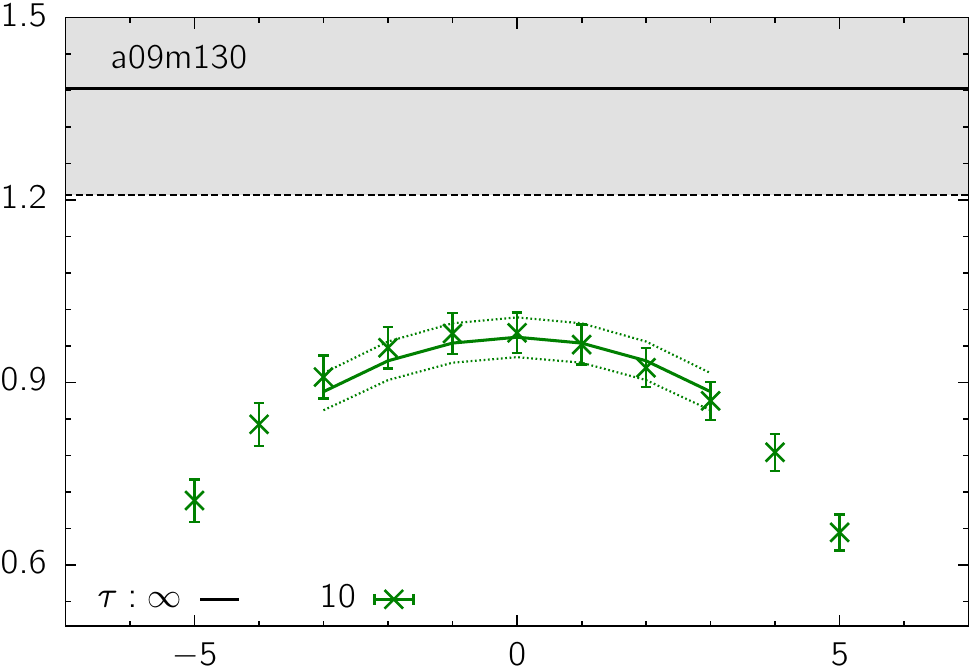}
    \includegraphics[height=1.5in,trim={0.0cm 0.00cm 0 0},clip]{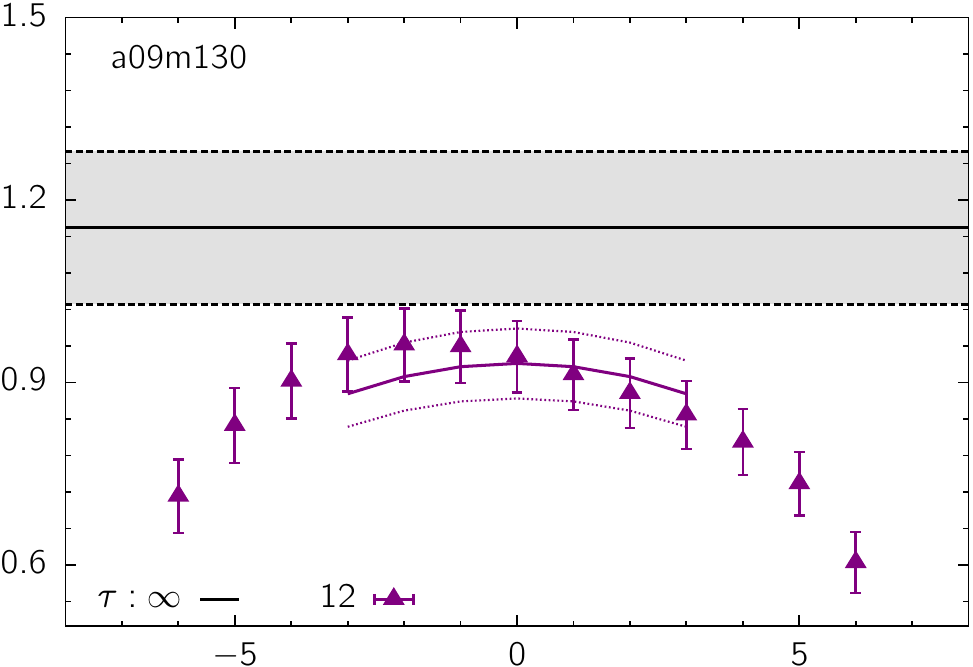}
    \includegraphics[height=1.5in,trim={0.0cm 0.00cm 0 0},clip]{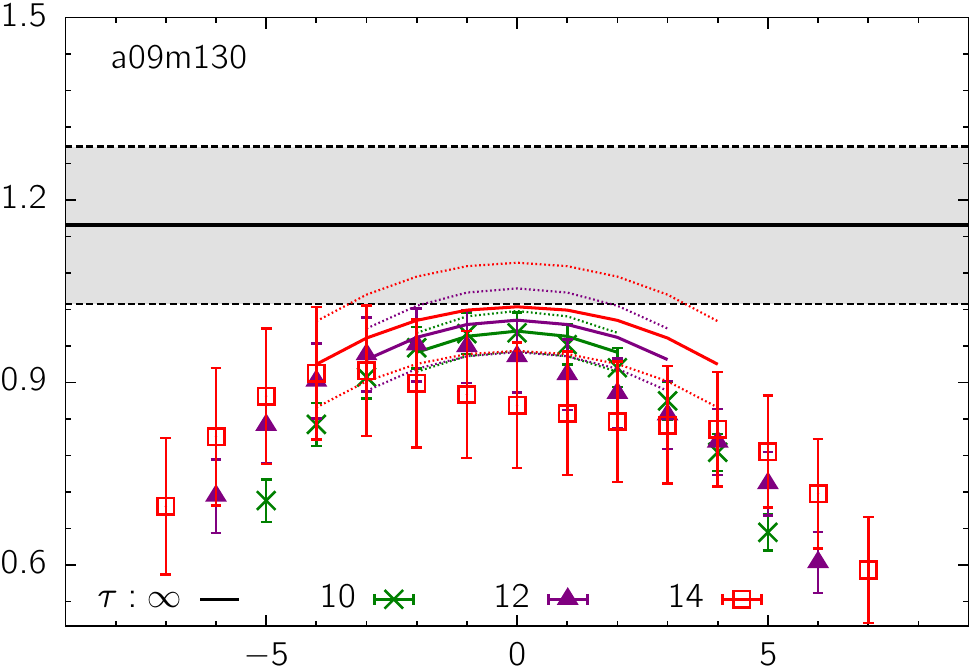}
  }
  \hspace{0.04\linewidth}
  \subfigure{
    \includegraphics[height=1.5in,trim={0.0cm 0.00cm 0 0},clip]{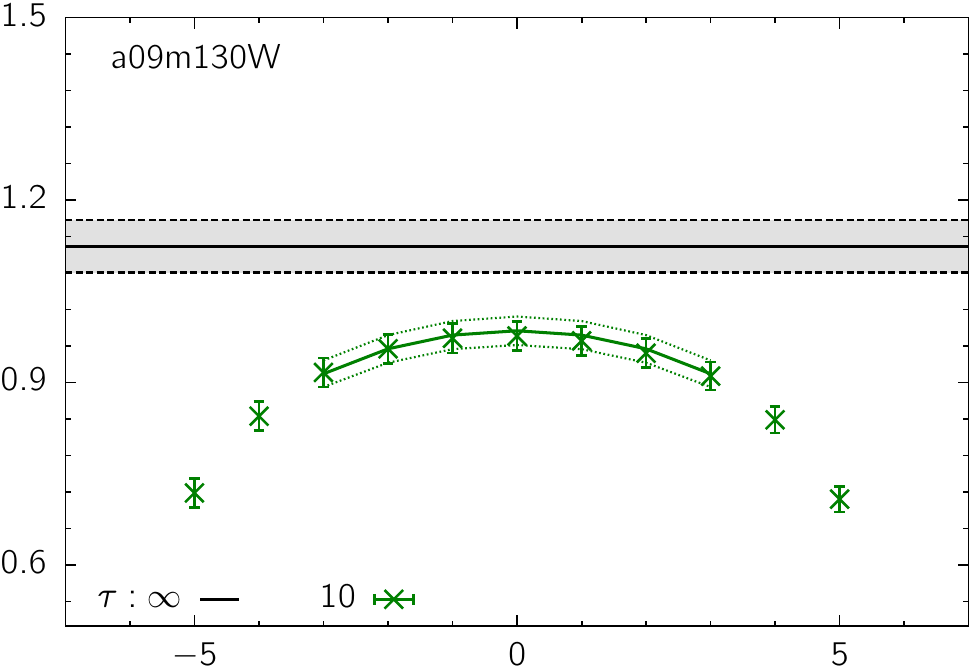}
    \includegraphics[height=1.5in,trim={0.0cm 0.00cm 0 0},clip]{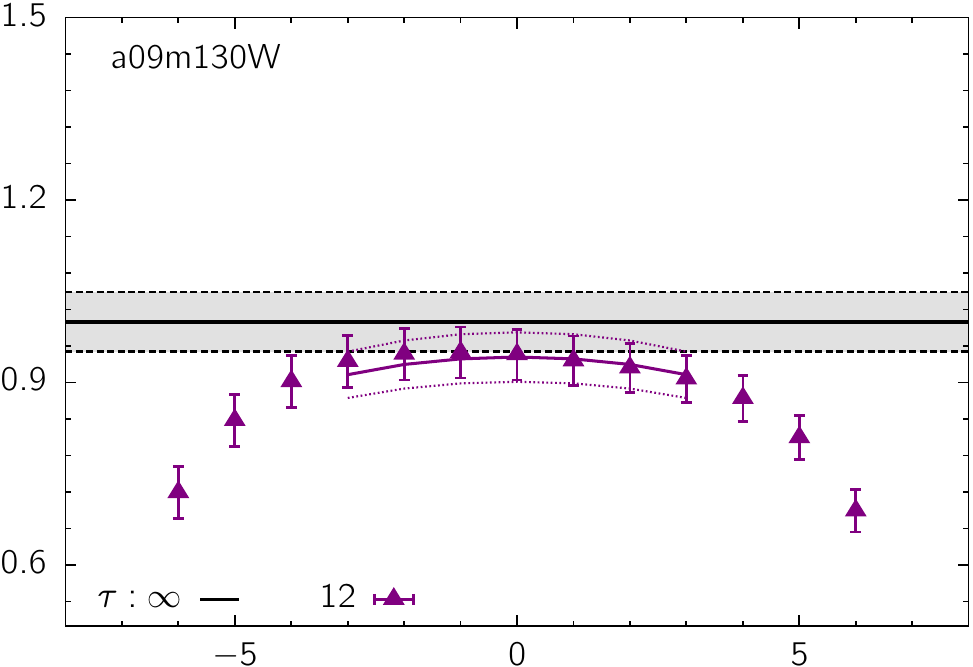}
    \includegraphics[height=1.5in,trim={0.0cm 0.00cm 0 0},clip]{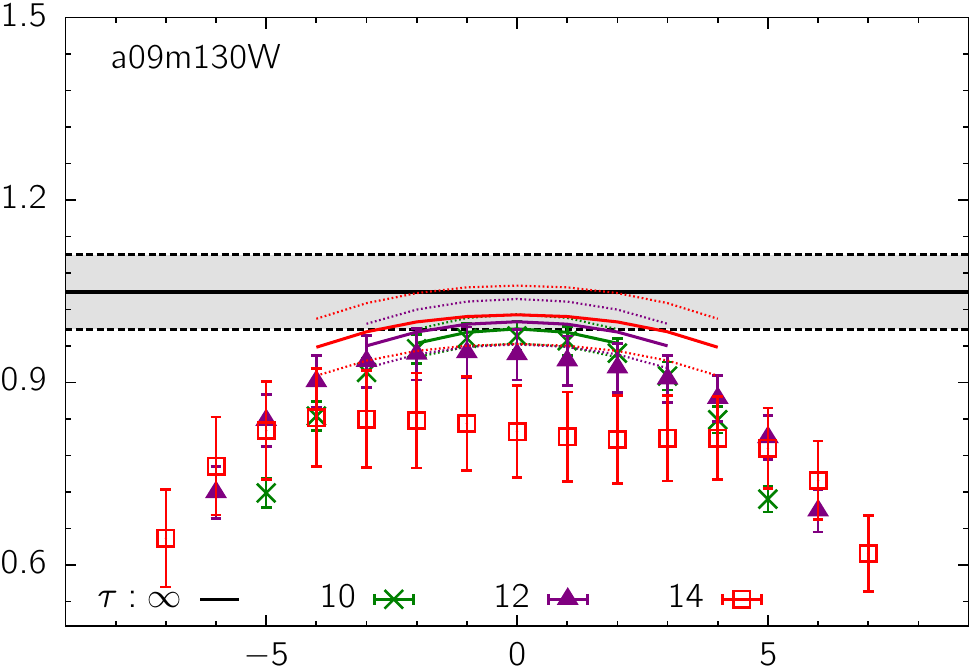}
  }
\caption{Comparison between the $2^\ast$ and
  $2-$state fits to the scalar charge $g_S^{u-d}$ data from the 
  $a \approx 0.09$~fm ensembles. The rest is the same as in Fig.~\ref{fig:gA2v3a12}. 
  \label{fig:gS2v3a09}}
\end{figure*}

\begin{figure*}
\centering
  \subfigure{
    \includegraphics[height=1.5in,trim={0.0cm 0.00cm 0 0},clip]{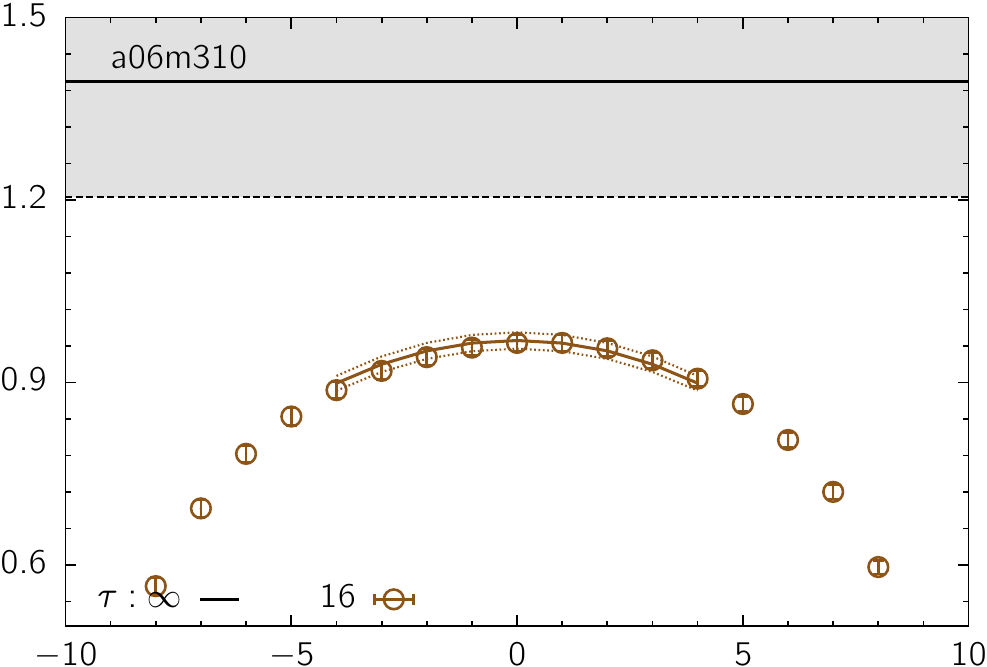}
    \includegraphics[height=1.5in,trim={0.0cm 0.00cm 0 0},clip]{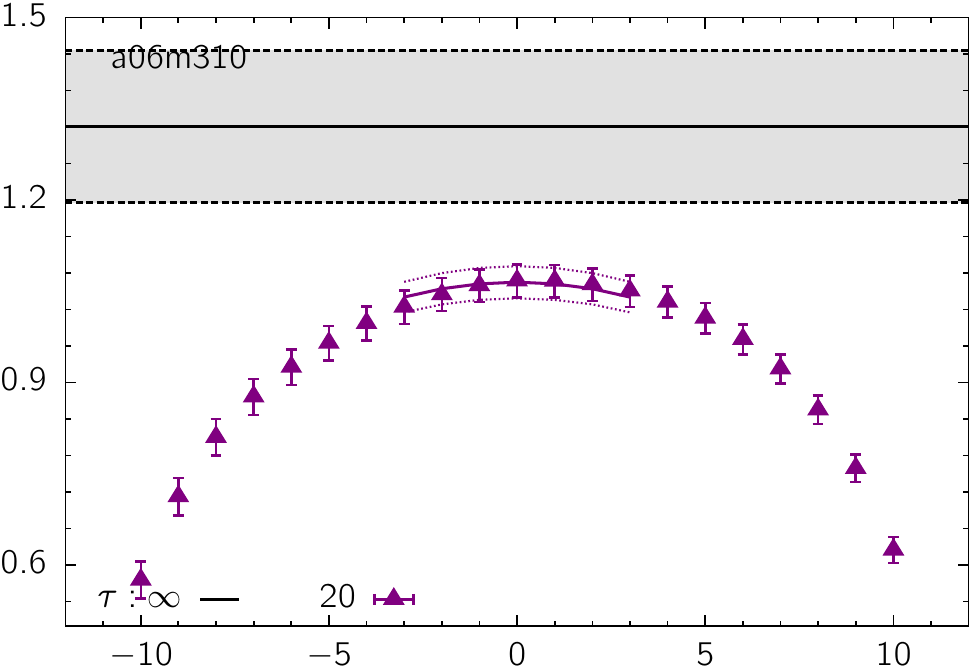}
    \includegraphics[height=1.5in,trim={0.0cm 0.00cm 0 0},clip]{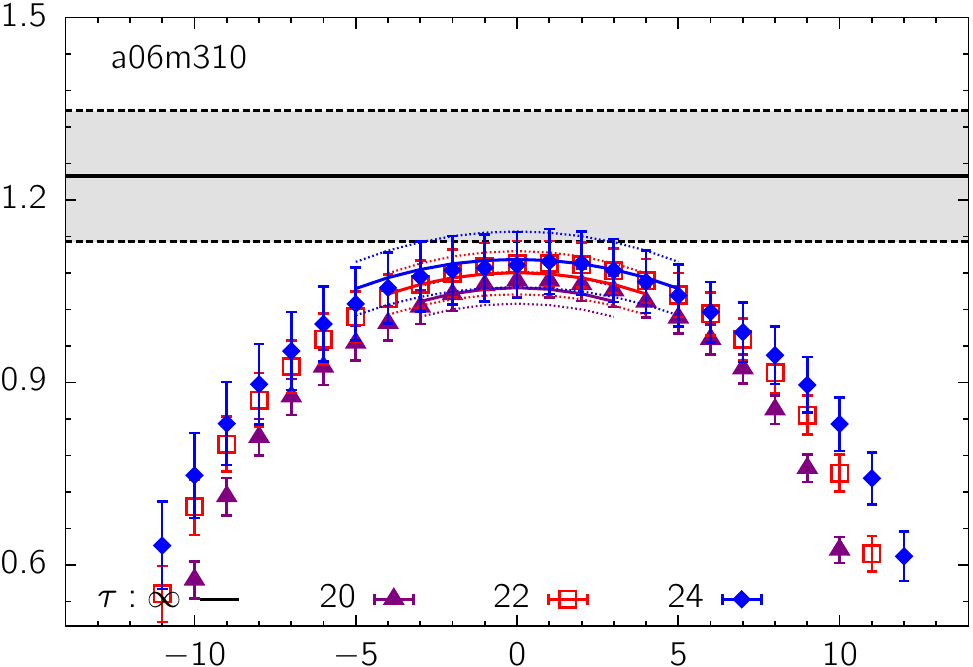}
  }
  \subfigure{
    \includegraphics[height=1.5in,trim={0.0cm 0.00cm 0 0},clip]{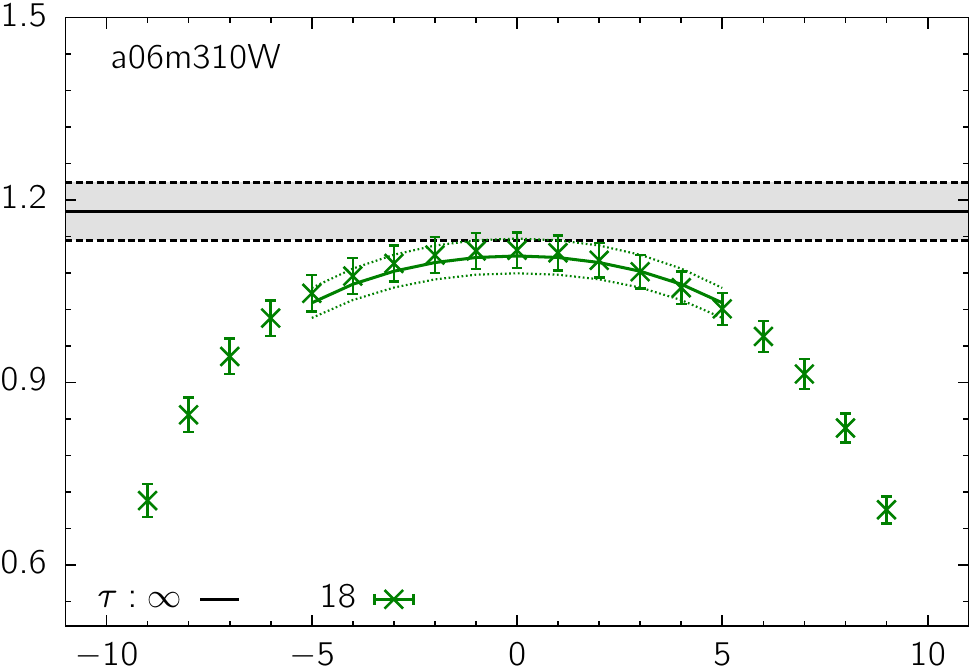}
    \includegraphics[height=1.5in,trim={0.0cm 0.00cm 0 0},clip]{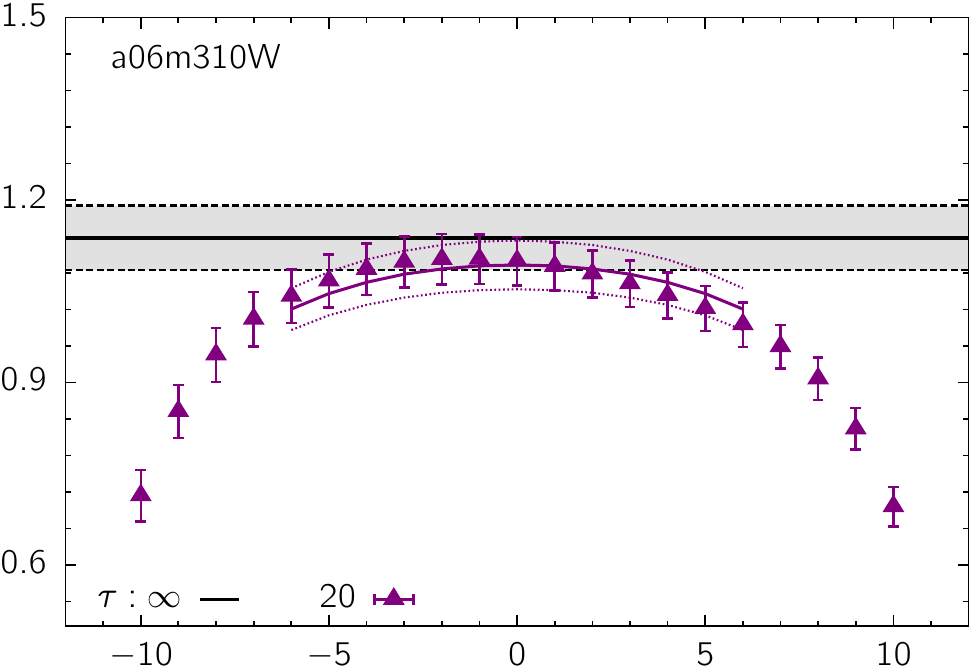}
    \includegraphics[height=1.5in,trim={0.0cm 0.00cm 0 0},clip]{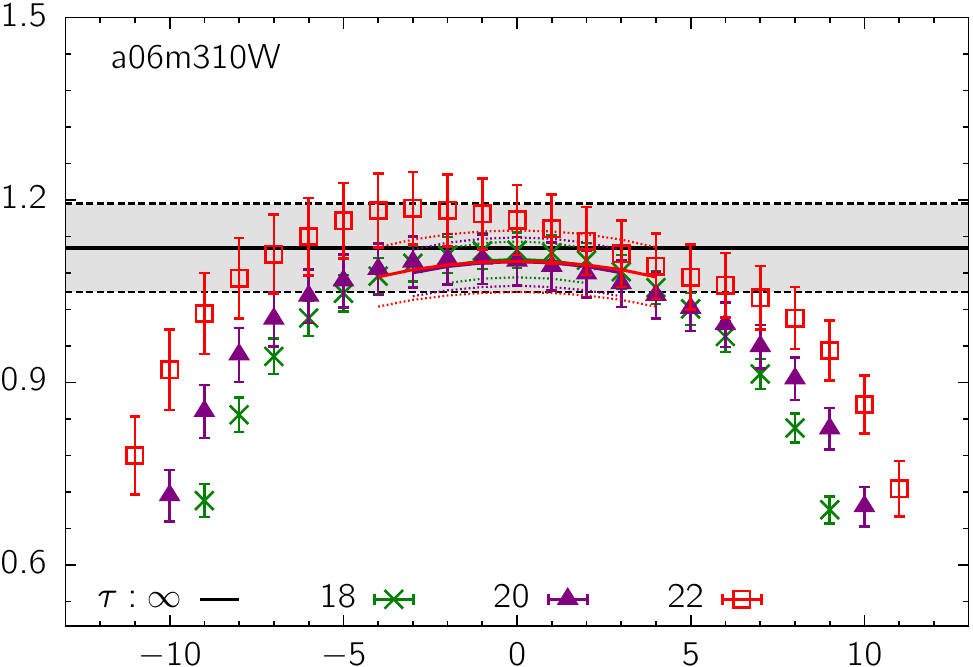}
  }
  \hspace{0.04\linewidth}
  \subfigure{
    \includegraphics[height=1.5in,trim={0.0cm 0.00cm 0 0},clip]{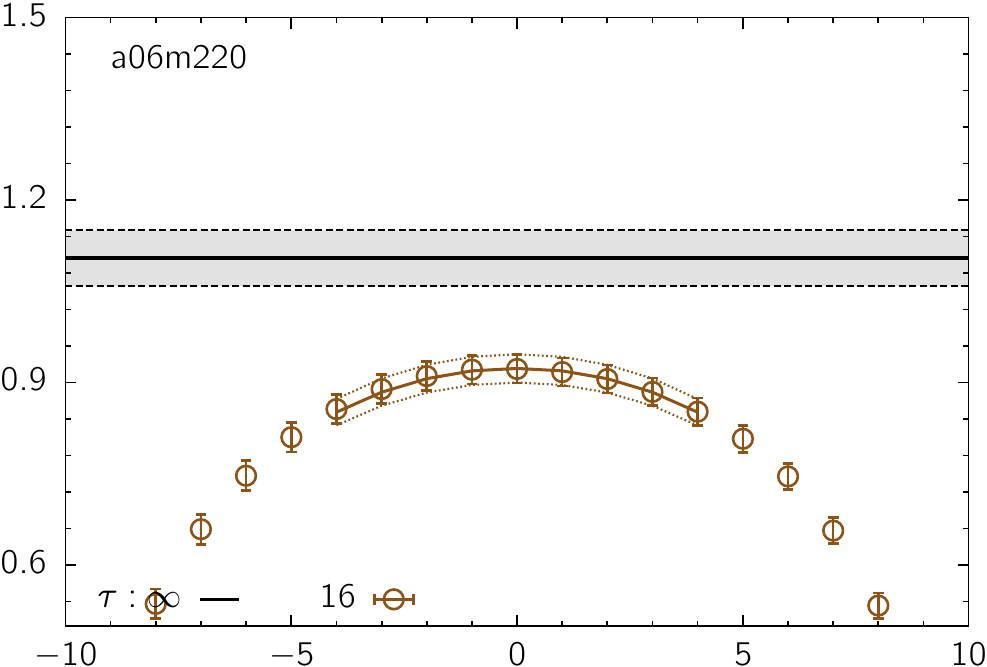}
    \includegraphics[height=1.5in,trim={0.0cm 0.00cm 0 0},clip]{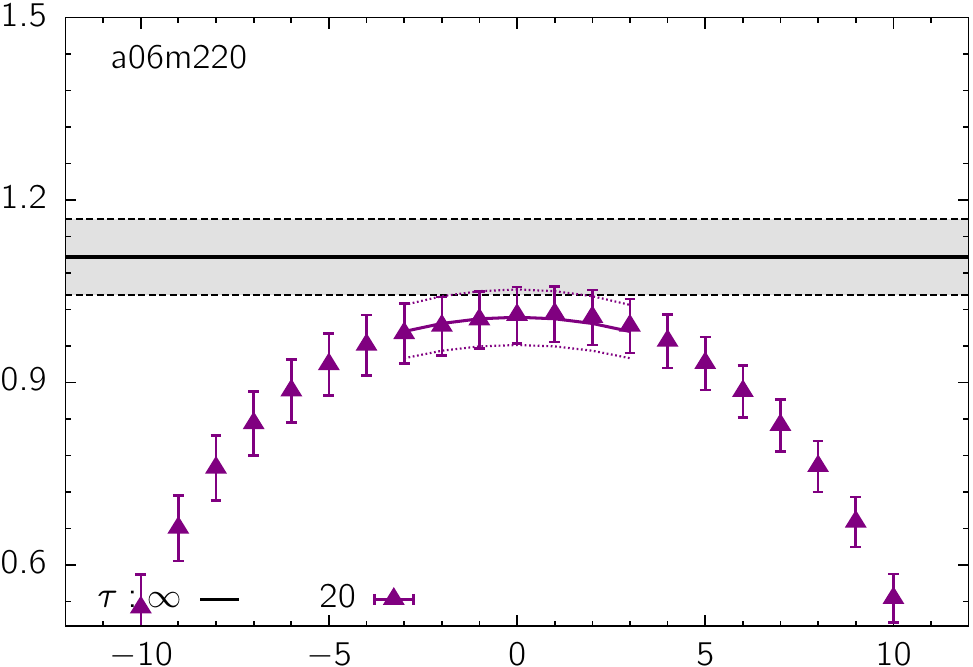}
    \includegraphics[height=1.5in,trim={0.0cm 0.00cm 0 0},clip]{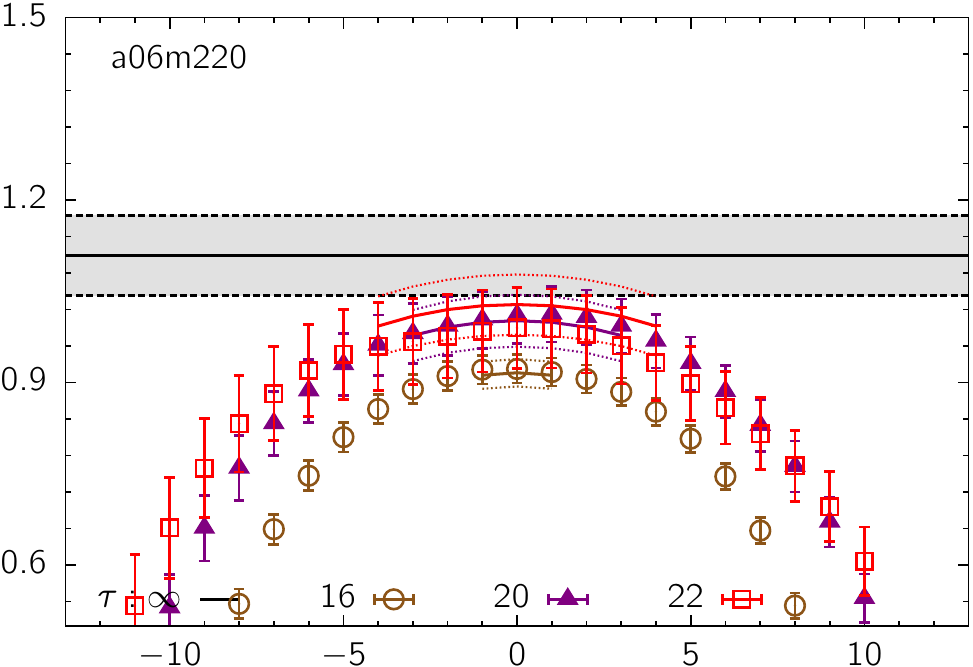}
  }
  \hspace{0.04\linewidth}
  \subfigure{
    \includegraphics[height=1.5in,trim={0.0cm 0.00cm 0 0},clip]{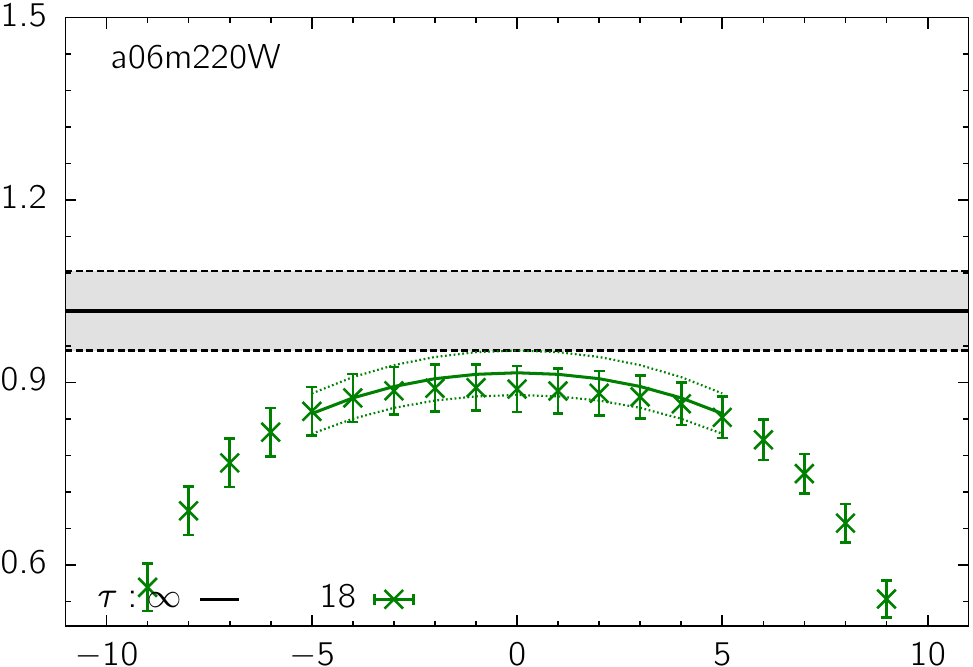}
    \includegraphics[height=1.5in,trim={0.0cm 0.00cm 0 0},clip]{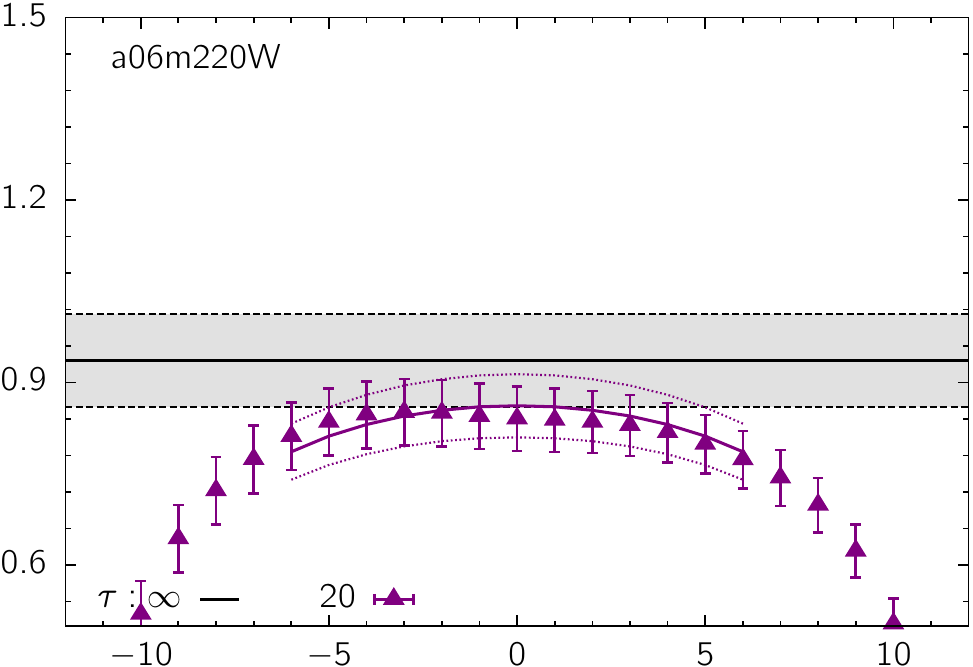}
    \includegraphics[height=1.5in,trim={0.0cm 0.00cm 0 0},clip]{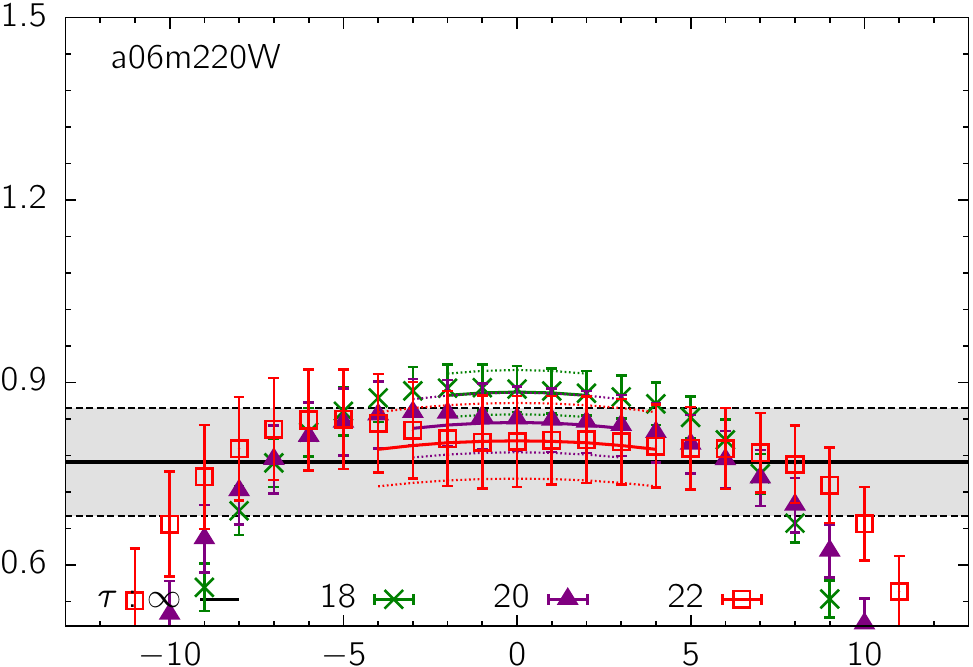}
  }
  \hspace{0.04\linewidth}
  \subfigure{
    \includegraphics[height=1.5in,trim={0.0cm 0.00cm 0 0},clip]{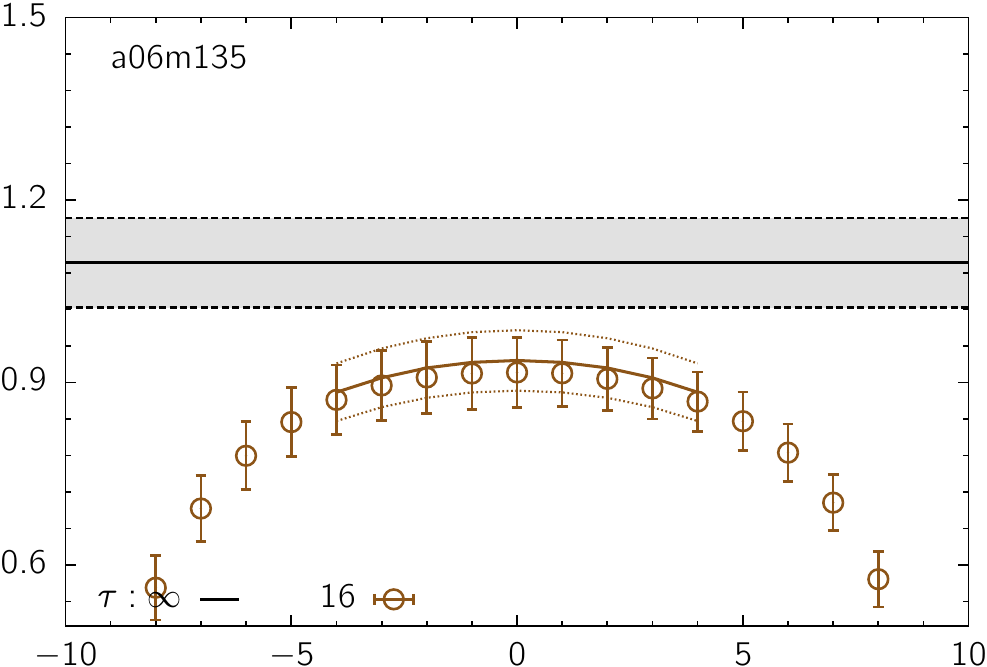}
    \includegraphics[height=1.5in,trim={0.0cm 0.00cm 0 0},clip]{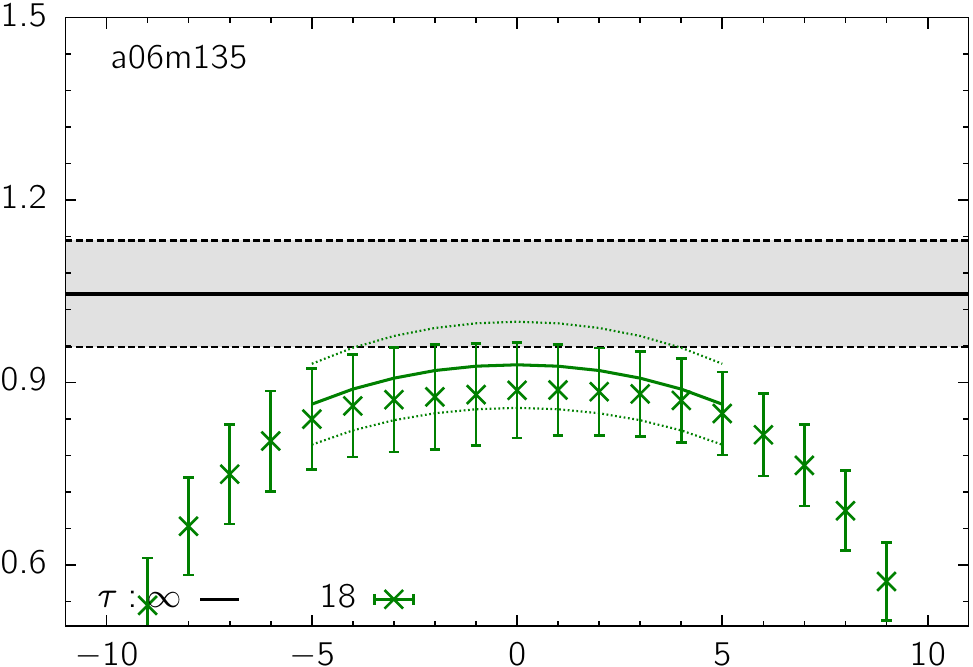}
    \includegraphics[height=1.5in,trim={0.0cm 0.00cm 0 0},clip]{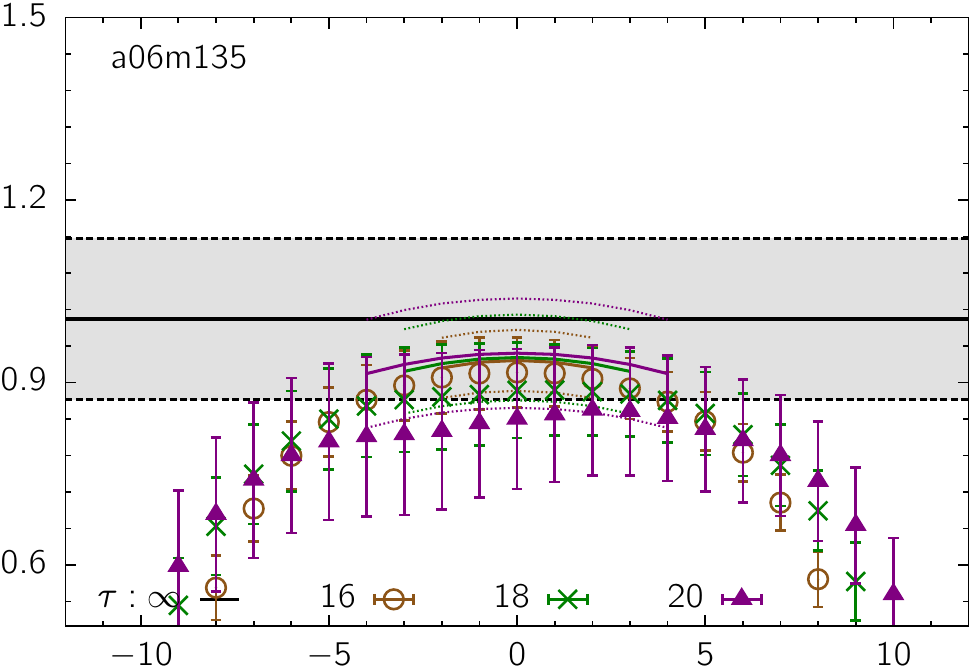}
  }
\caption{Comparison between the $2^\ast$ and
  $2-$state fits to the scalar charge $g_S^{u-d}$ data from the 
  $a \approx 0.06$~fm ensembles. The rest is the same as in Fig.~\ref{fig:gA2v3a12}. 
  \label{fig:gS2v3a06}}
\end{figure*}

\begin{figure*}
\centering
  \subfigure{
    \includegraphics[height=1.5in,trim={0.0cm 0.00cm 0 0},clip]{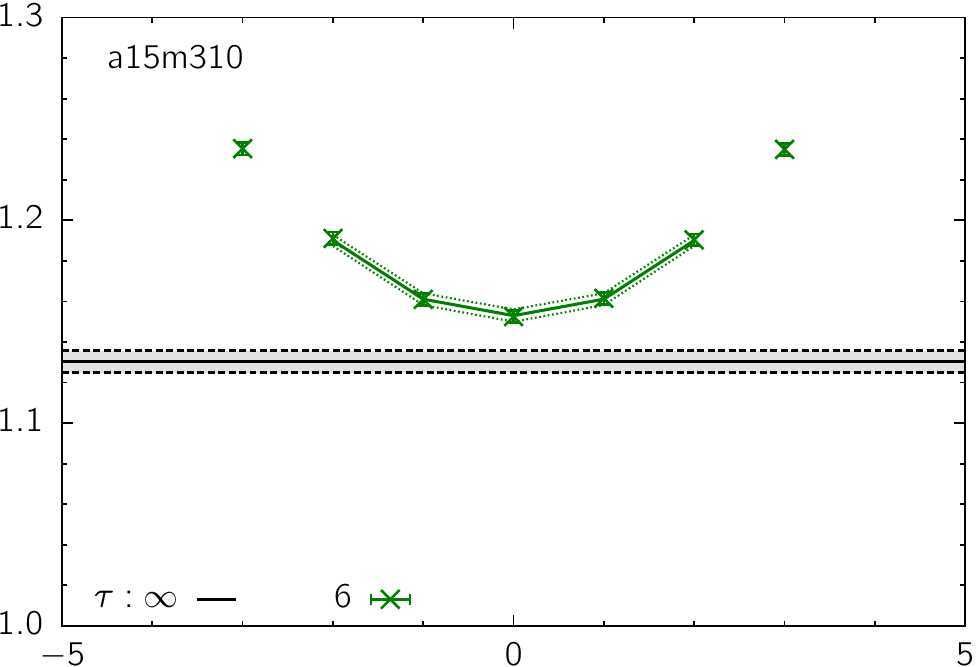}
    \includegraphics[height=1.5in,trim={0.1cm 0.00cm 0 0},clip]{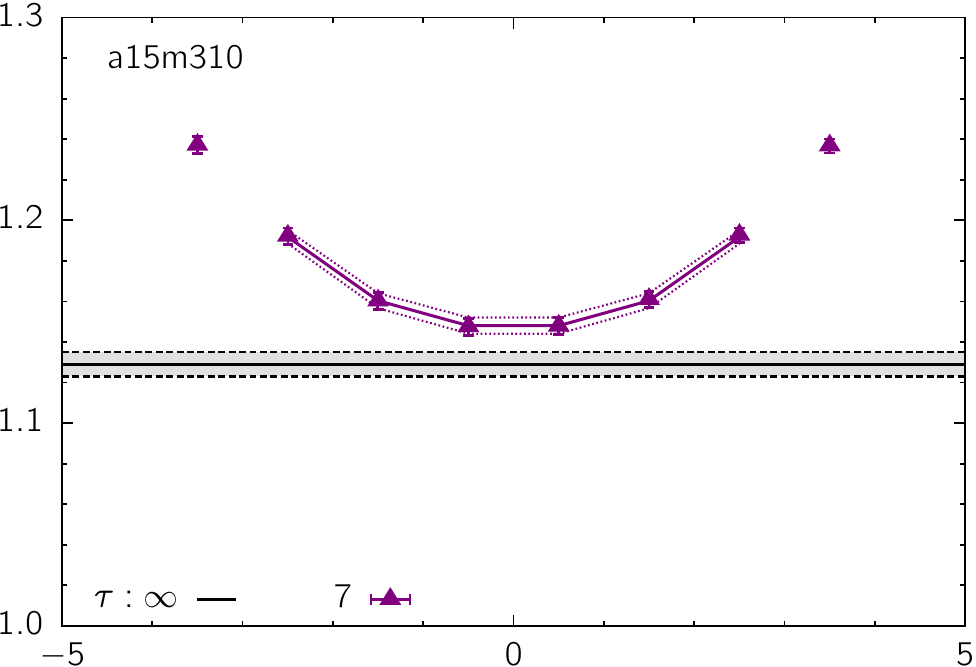}
    \includegraphics[height=1.5in,trim={0.1cm 0.00cm 0 0},clip]{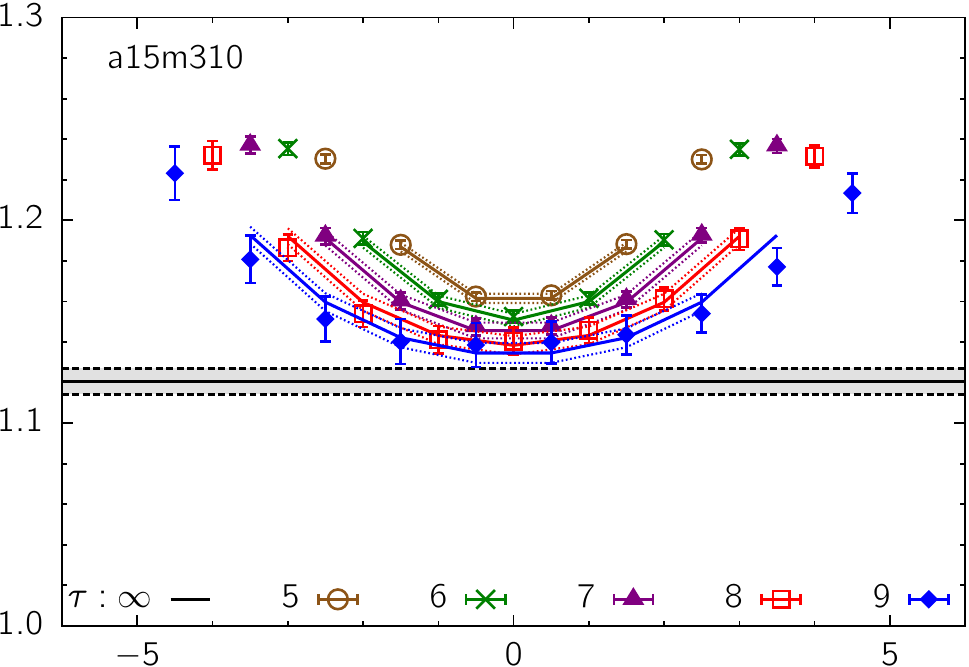}
  }
  \hspace{0.04\linewidth}
  \subfigure{
    \includegraphics[height=1.5in,trim={0.0cm 0.00cm 0 0},clip]{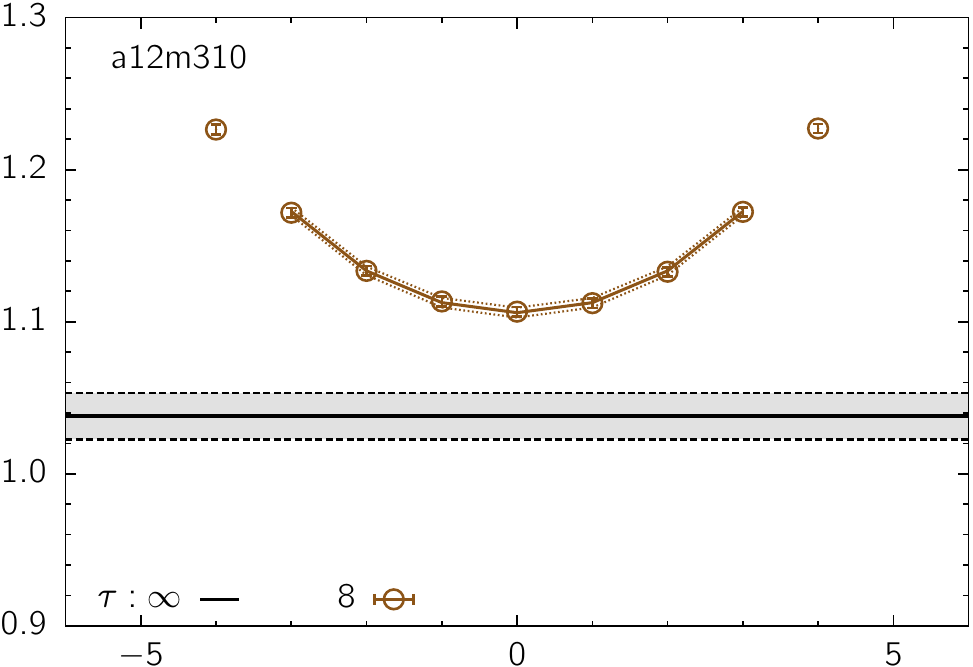}
    \includegraphics[height=1.5in,trim={0.1cm 0.00cm 0 0},clip]{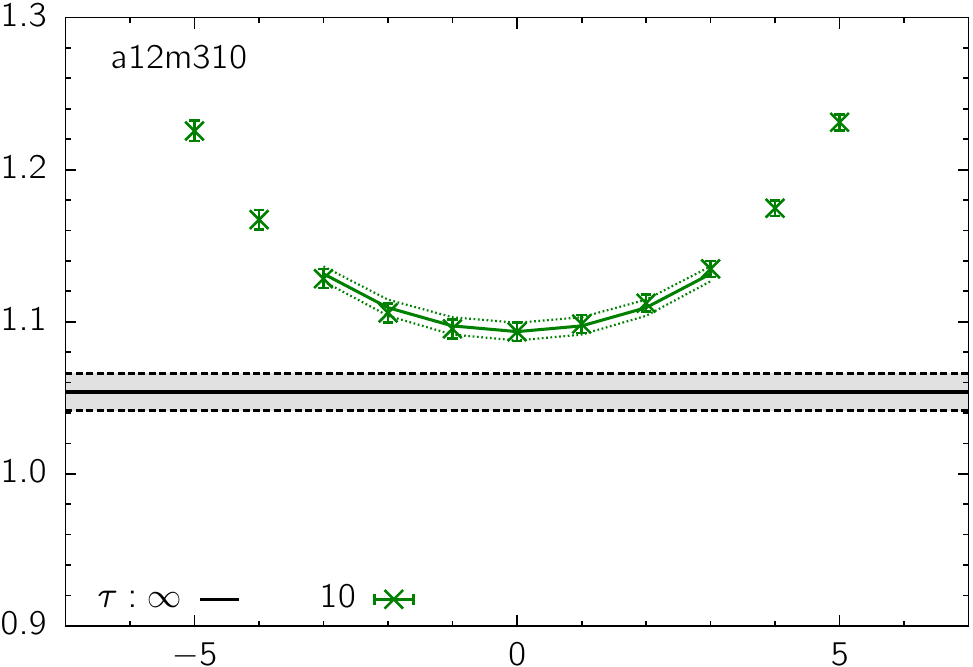}
    \includegraphics[height=1.5in,trim={0.1cm 0.00cm 0 0},clip]{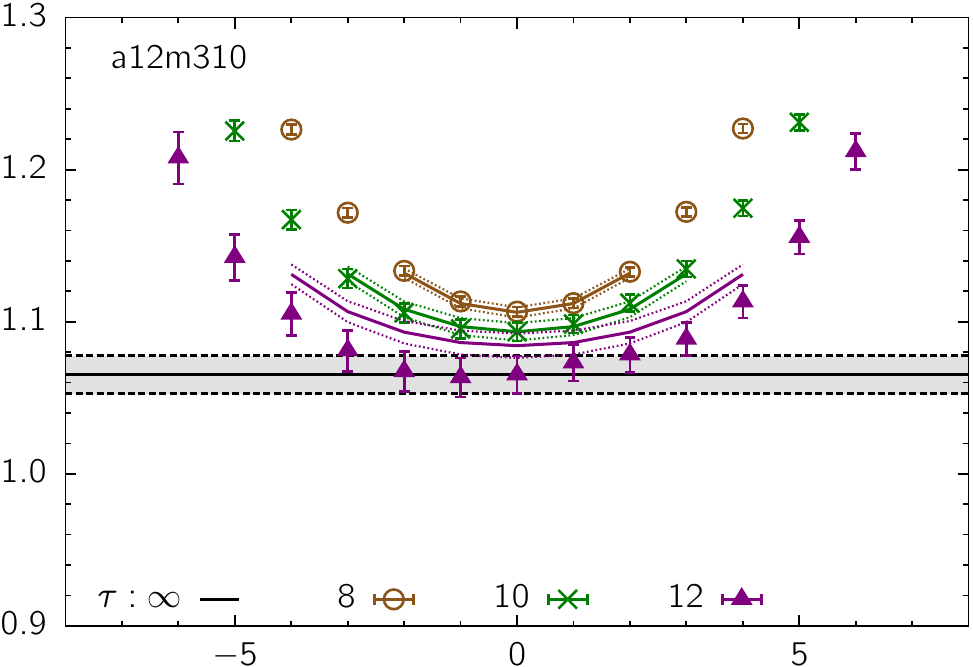}
  }
  \hspace{0.04\linewidth}
  \subfigure{
    \includegraphics[height=1.5in,trim={0.0cm 0.00cm 0 0},clip]{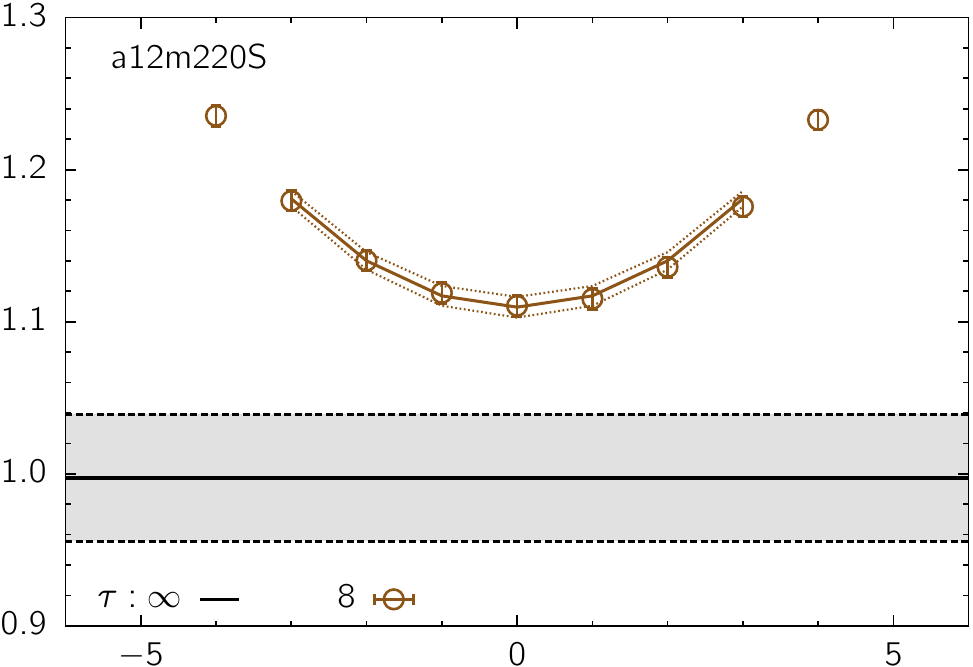}
    \includegraphics[height=1.5in,trim={0.1cm 0.00cm 0 0},clip]{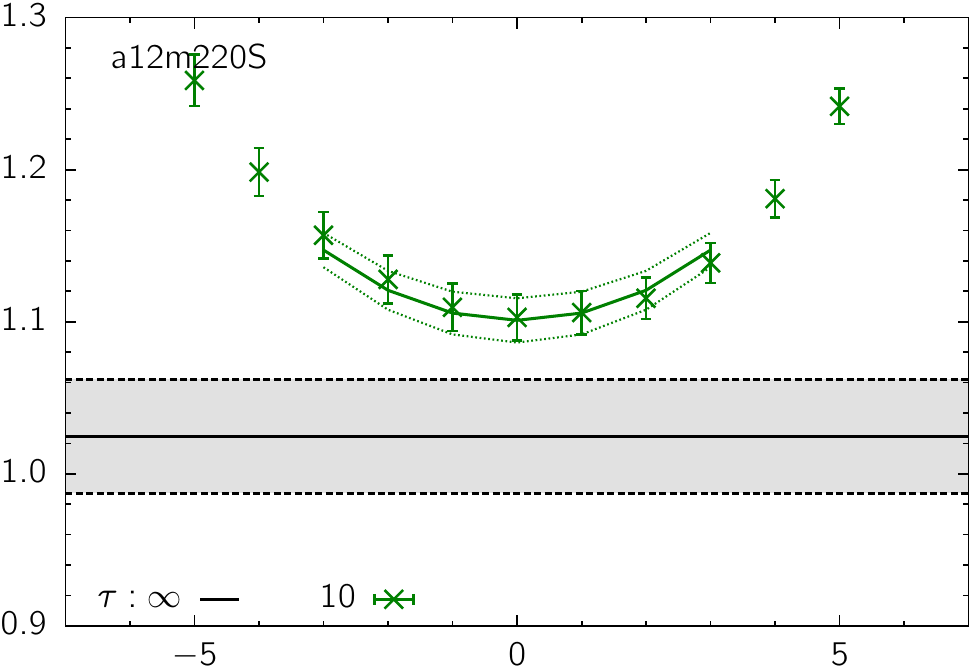}
    \includegraphics[height=1.5in,trim={0.1cm 0.00cm 0 0},clip]{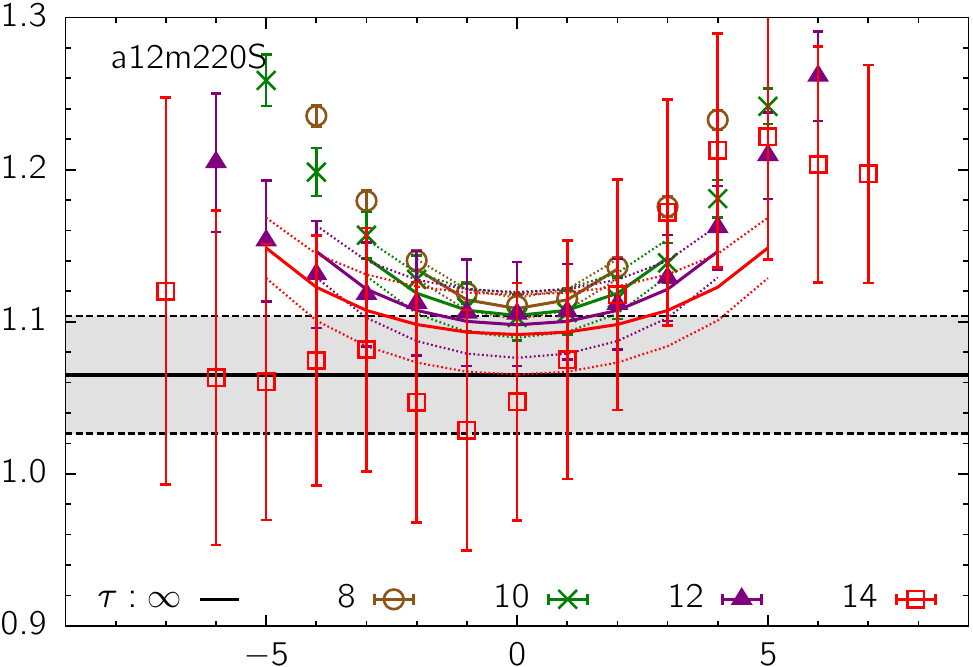}
  }
  \hspace{0.04\linewidth}
  \subfigure{
    \includegraphics[height=1.5in,trim={0.0cm 0.00cm 0 0},clip]{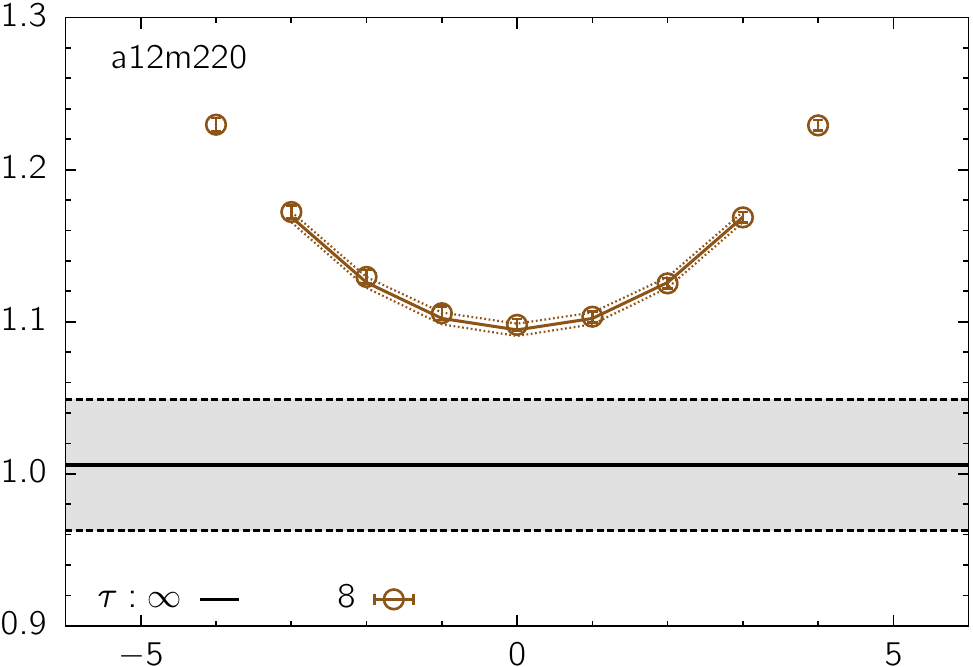}
    \includegraphics[height=1.5in,trim={0.0cm 0.00cm 0 0},clip]{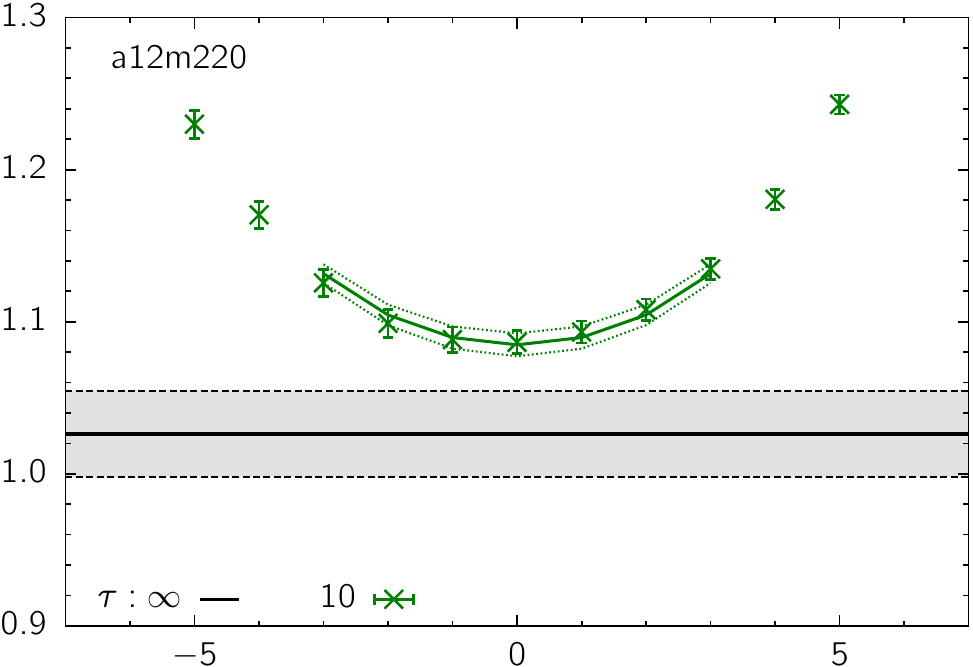}
    \includegraphics[height=1.5in,trim={0.0cm 0.00cm 0 0},clip]{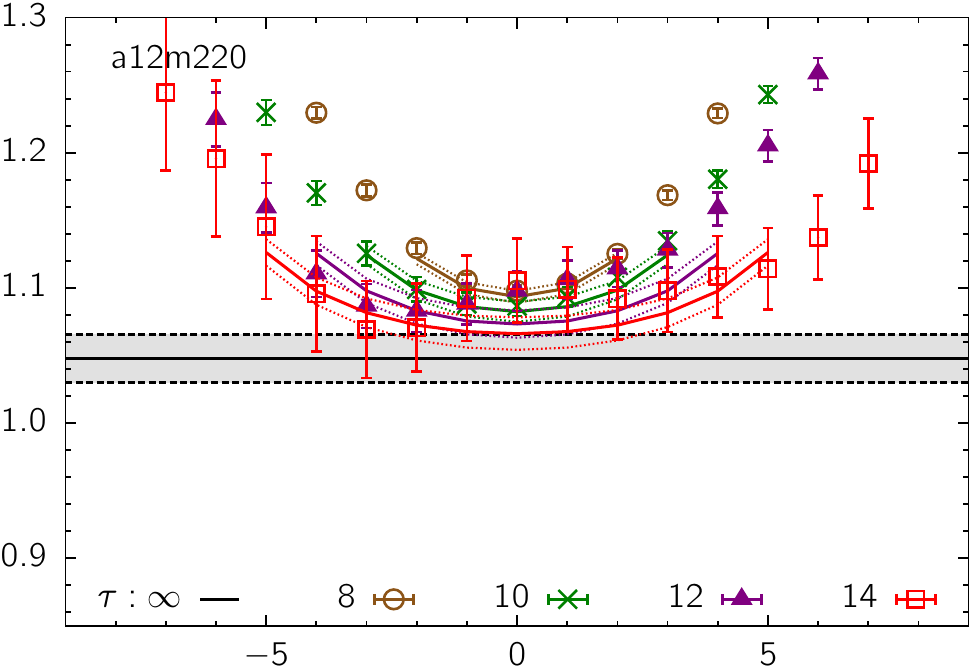}
  }
  \hspace{0.04\linewidth}
  \subfigure{
    \includegraphics[height=1.5in,trim={0.0cm 0.00cm 0 0},clip]{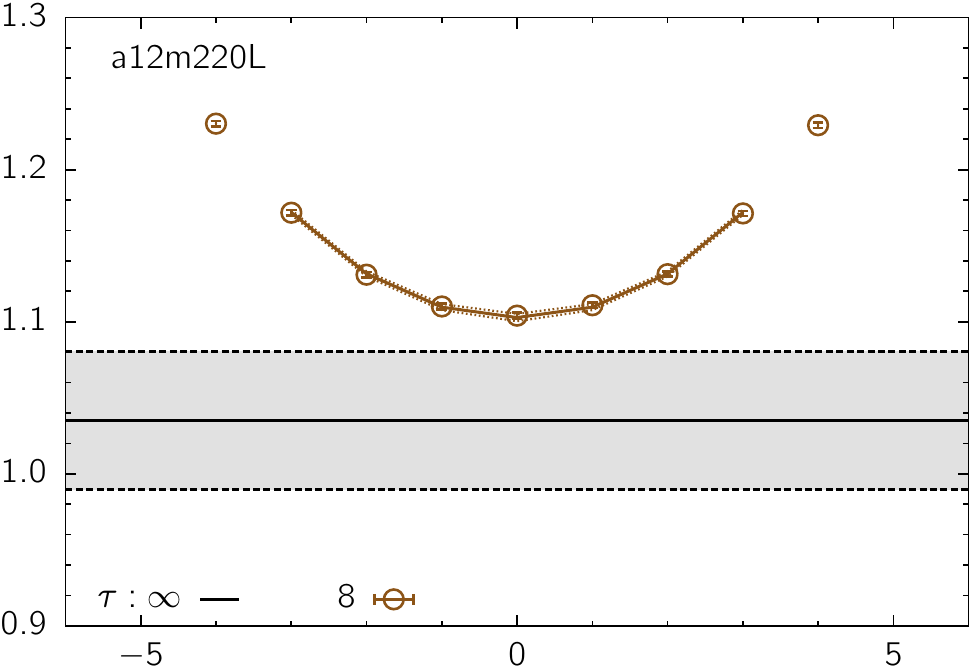}
    \includegraphics[height=1.5in,trim={0.0cm 0.00cm 0 0},clip]{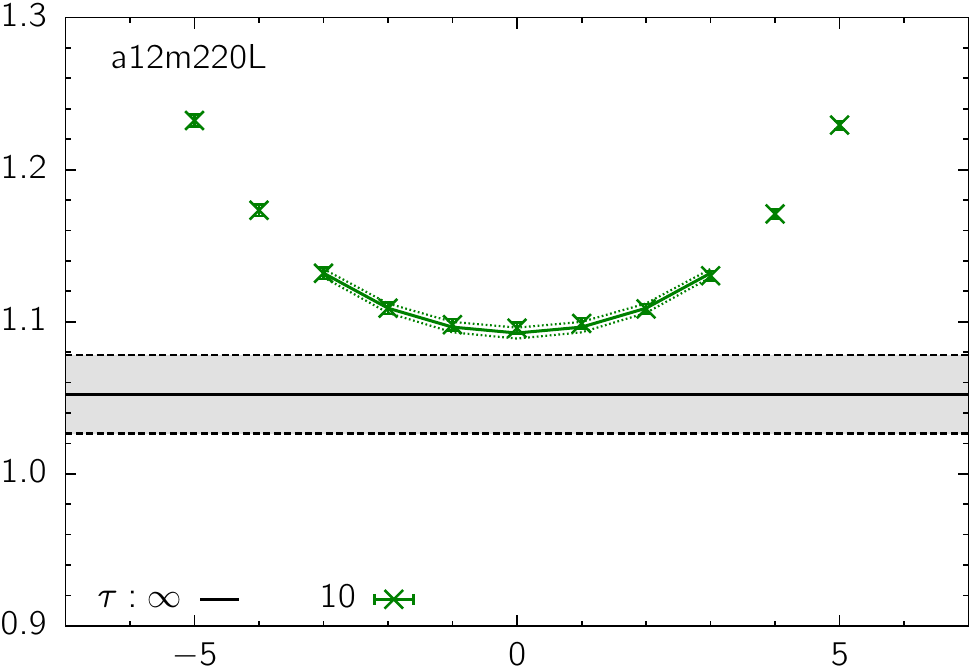}
    \includegraphics[height=1.5in,trim={0.0cm 0.00cm 0 0},clip]{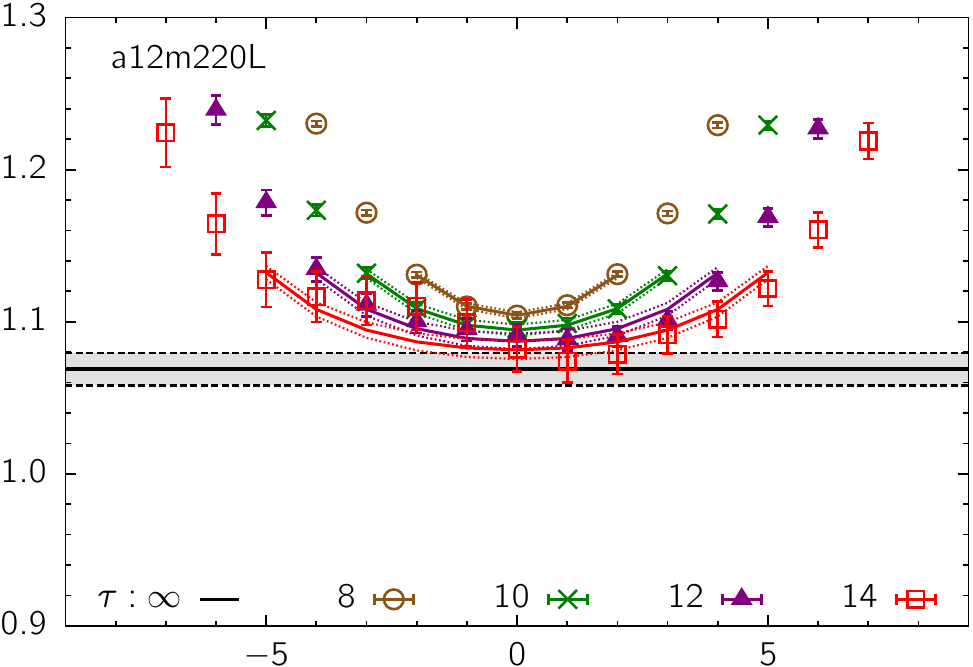}
  }
\caption{Comparison between the $2^\ast$ and
  $3^\ast$ fits to the tensor charge $g_T^{u-d}$ data from the 
  $a \approx 0.15$~fm (top row) and  $a \approx 0.12$~fm (bottom 4 rows) ensembles. 
  The rest is the same as in Fig.~\ref{fig:gA2v3a12}. 
  \label{fig:gT2v3a12}}
\end{figure*}

\begin{figure*}
\centering
  \subfigure{
    \includegraphics[height=1.5in,trim={0.0cm 0.00cm 0 0},clip]{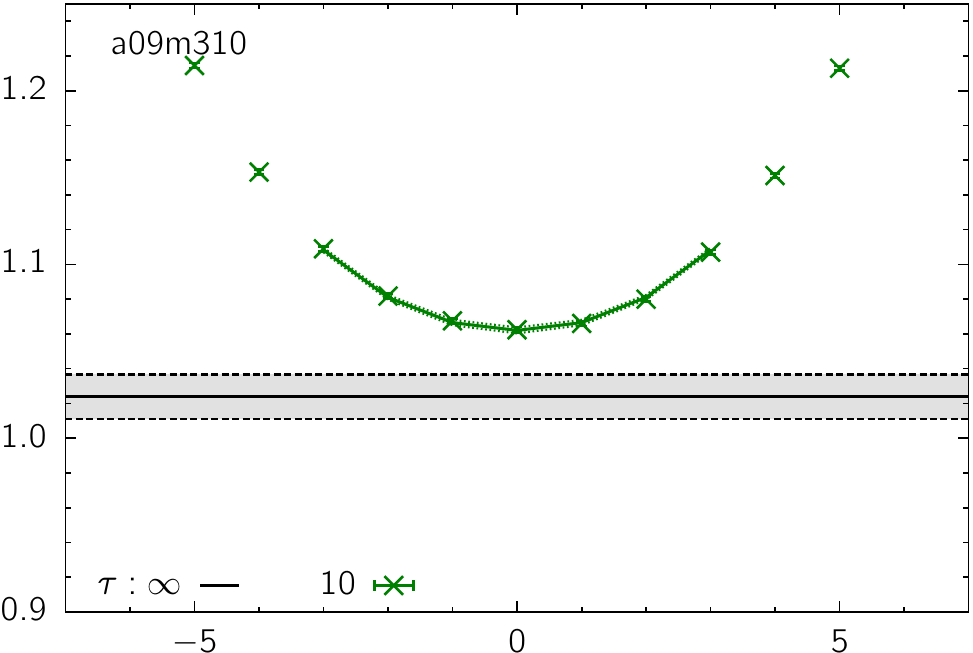}
    \includegraphics[height=1.5in,trim={0.0cm 0.00cm 0 0},clip]{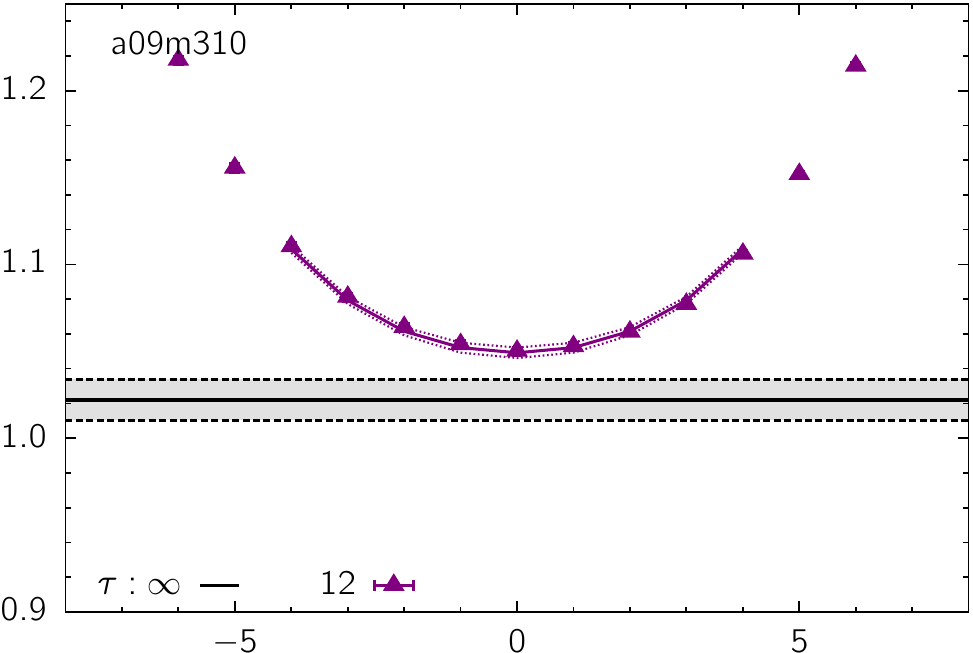}
    \includegraphics[height=1.5in,trim={0.0cm 0.00cm 0 0},clip]{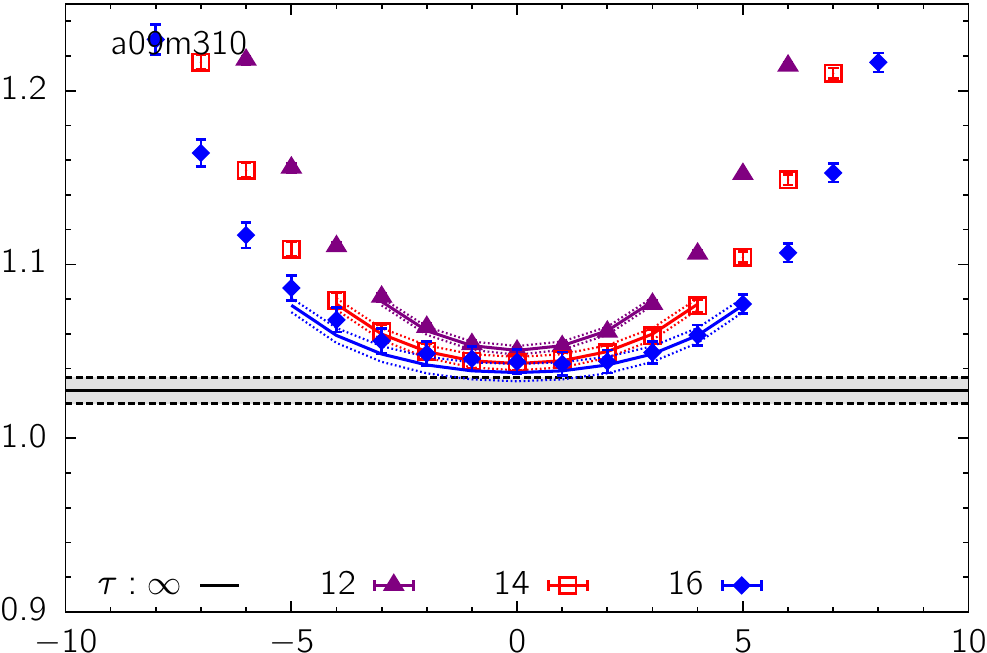}
  }
  \hspace{0.04\linewidth}
  \subfigure{
    \includegraphics[height=1.5in,trim={0.0cm 0.00cm 0 0},clip]{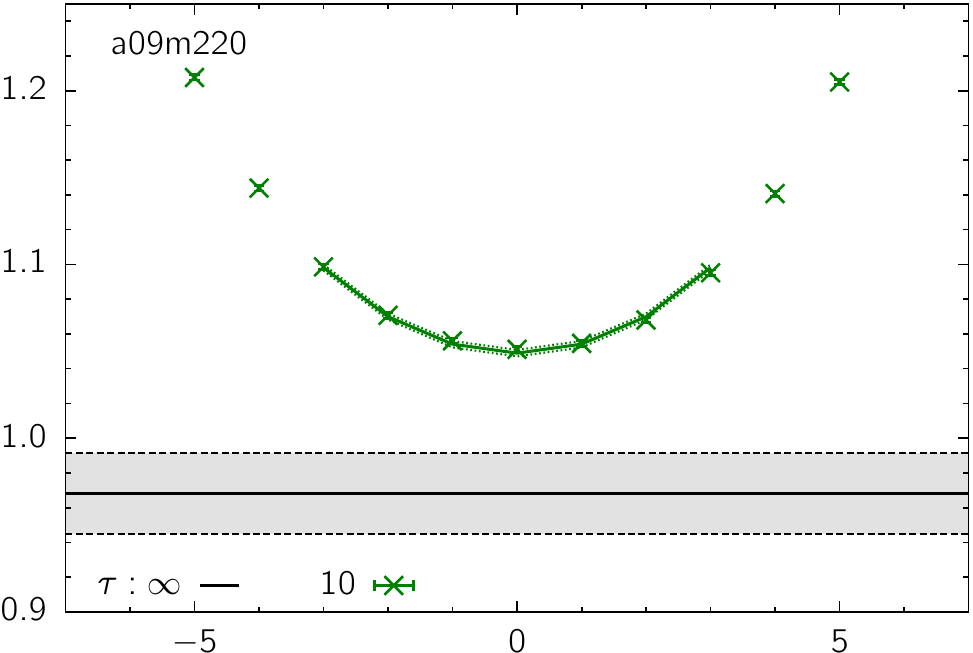}
    \includegraphics[height=1.5in,trim={0.0cm 0.00cm 0 0},clip]{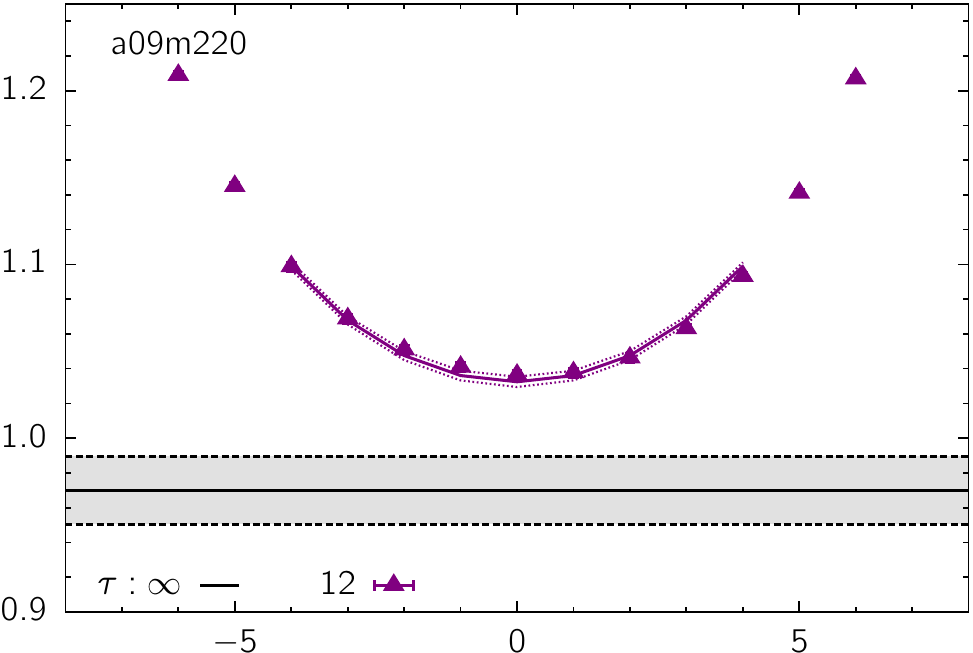}
    \includegraphics[height=1.5in,trim={0.0cm 0.00cm 0 0},clip]{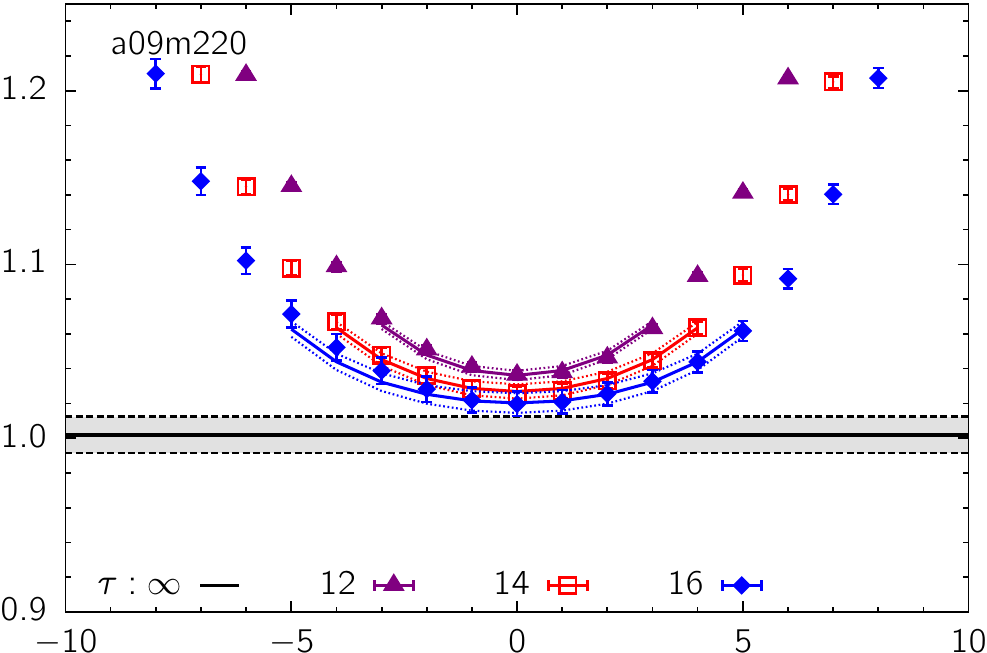}
  }
  \hspace{0.04\linewidth}
  \subfigure{
    \includegraphics[height=1.5in,trim={0.0cm 0.00cm 0 0},clip]{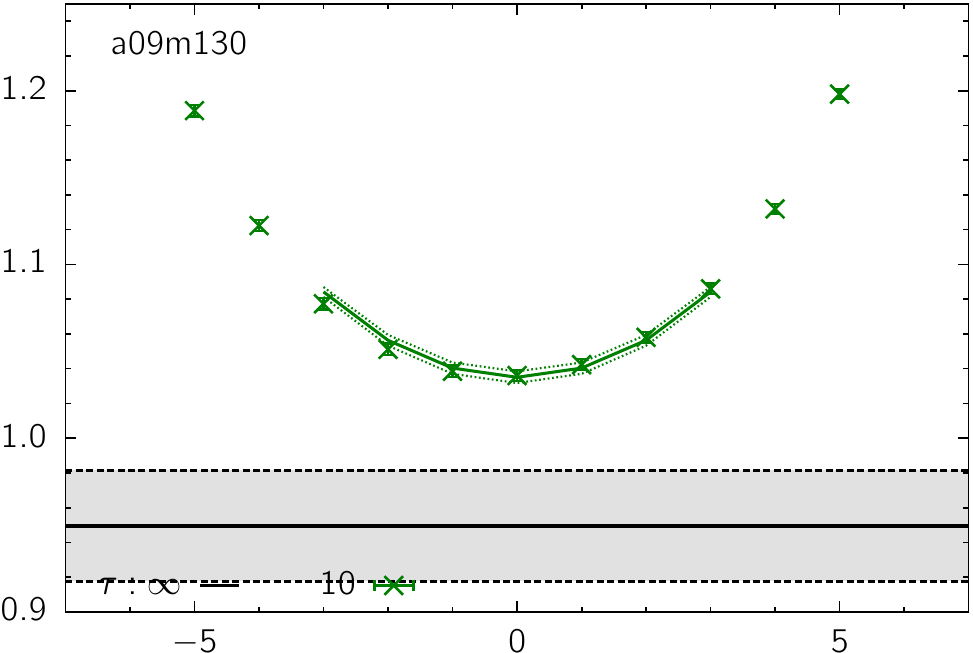}
    \includegraphics[height=1.5in,trim={0.0cm 0.00cm 0 0},clip]{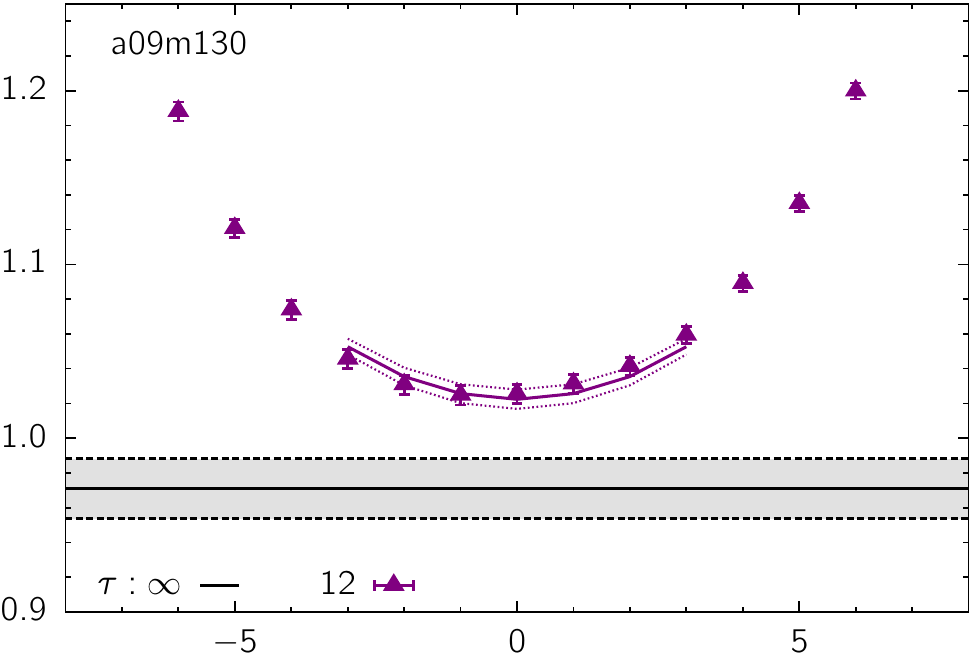}
    \includegraphics[height=1.5in,trim={0.0cm 0.00cm 0 0},clip]{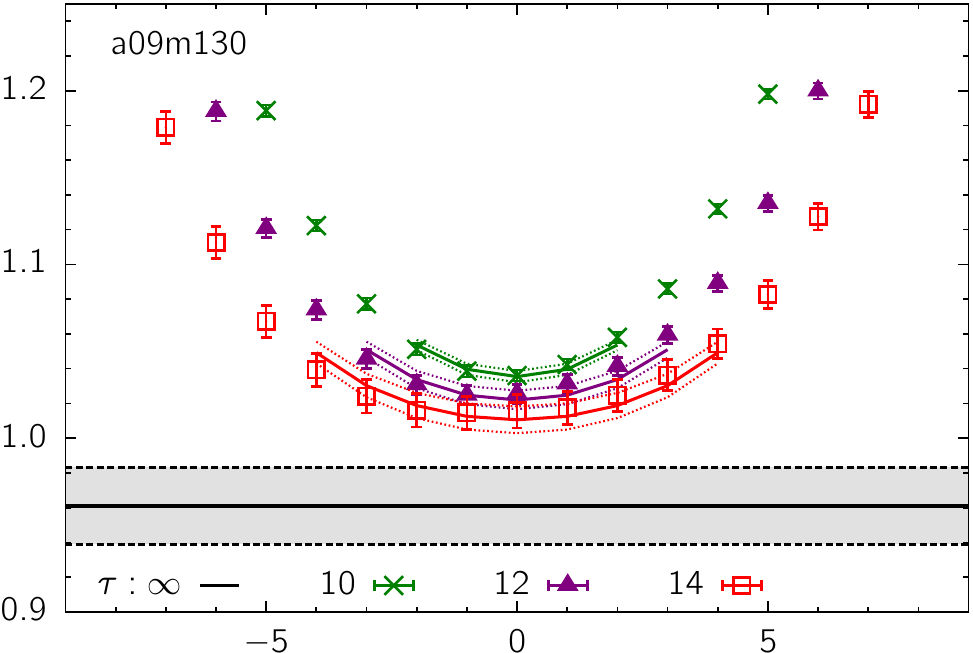}
  }
  \hspace{0.04\linewidth}
  \subfigure{
    \includegraphics[height=1.5in,trim={0.0cm 0.00cm 0 0},clip]{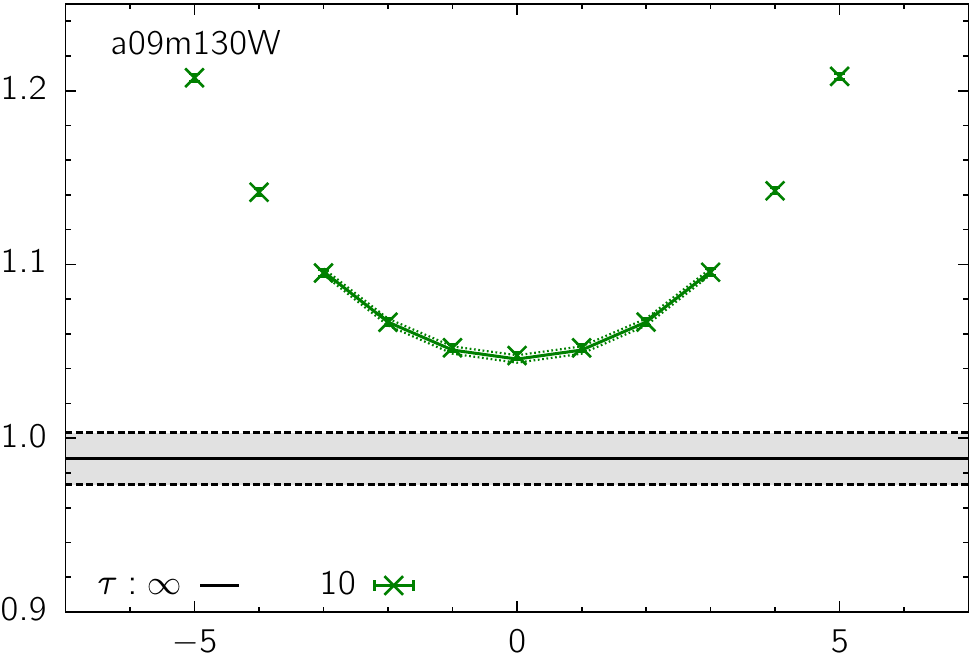}
    \includegraphics[height=1.5in,trim={0.0cm 0.00cm 0 0},clip]{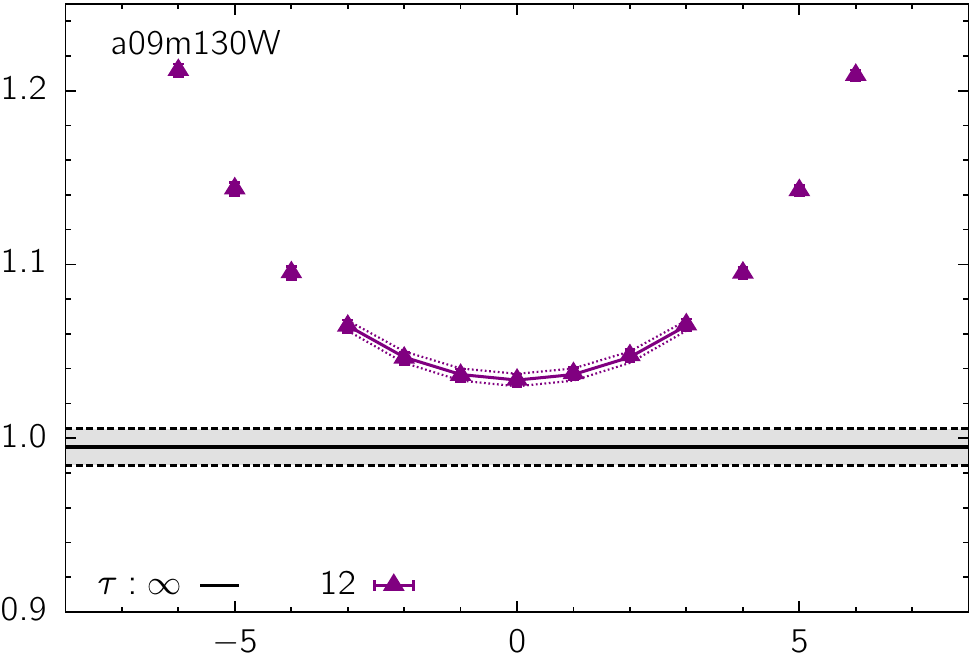}
    \includegraphics[height=1.5in,trim={0.0cm 0.00cm 0 0},clip]{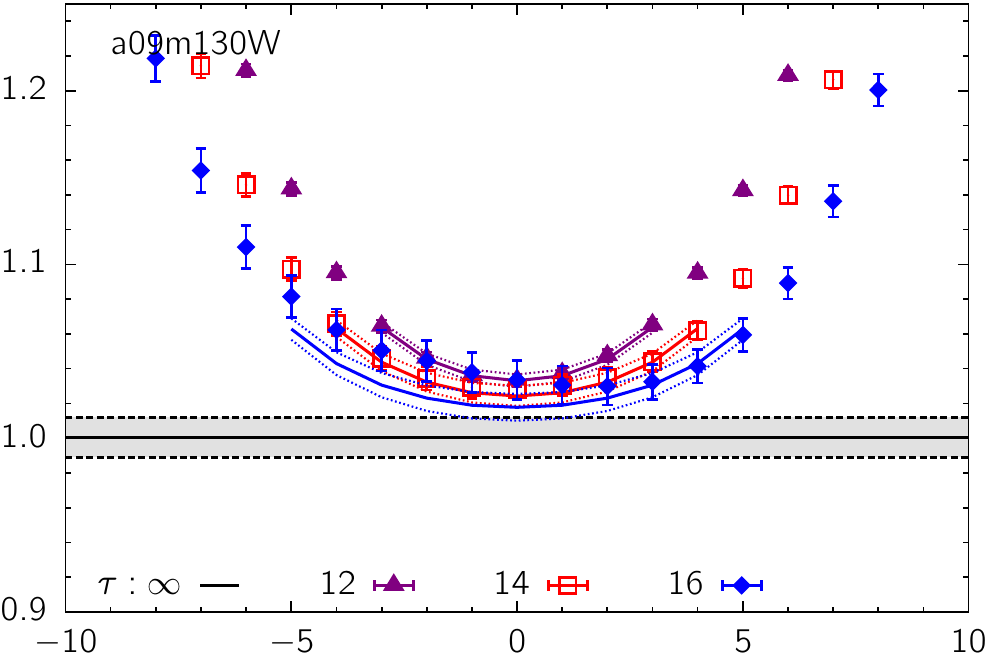}
  }
\caption{Comparison between the $2^\ast$ and
  $3^\ast$ fits to the tensor charge $g_T^{u-d}$ data from the 
  $a \approx 0.09$~fm ensembles. The rest is the same as in Fig.~\ref{fig:gA2v3a12}. 
  \label{fig:gT2v3a09}}
\end{figure*}

\begin{figure*}
\centering
  \subfigure{
    \includegraphics[height=1.5in,trim={0.0cm 0.00cm 0 0},clip]{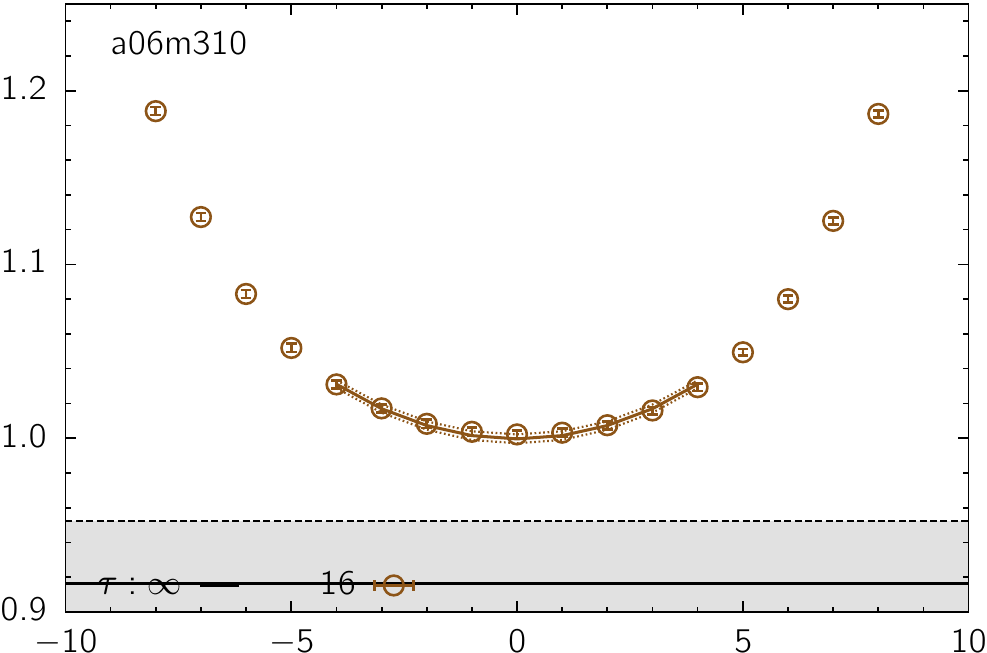}
    \includegraphics[height=1.5in,trim={0.0cm 0.00cm 0 0},clip]{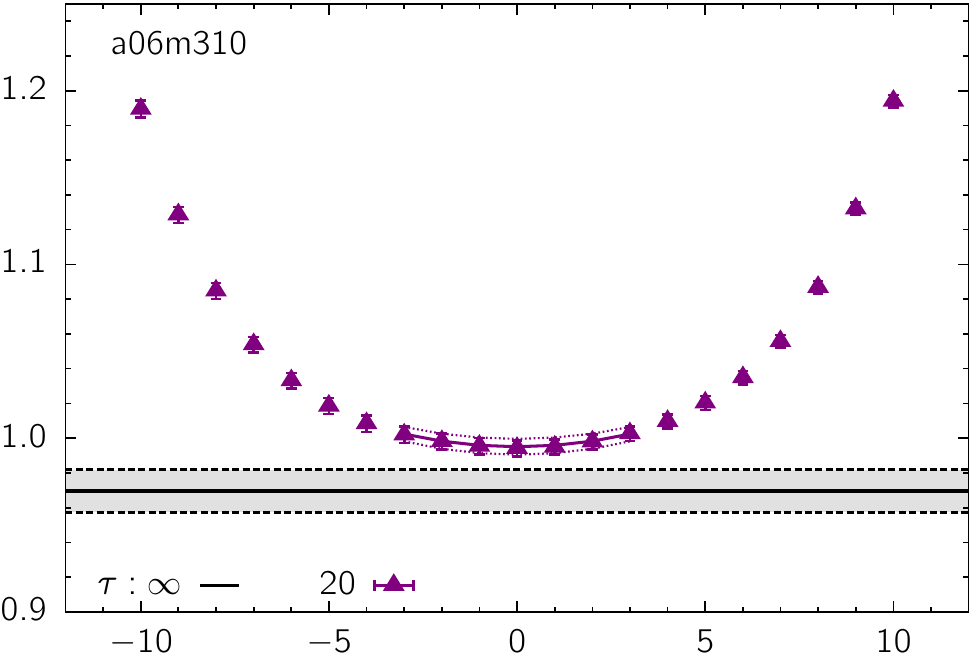}
    \includegraphics[height=1.5in,trim={0.0cm 0.00cm 0 0},clip]{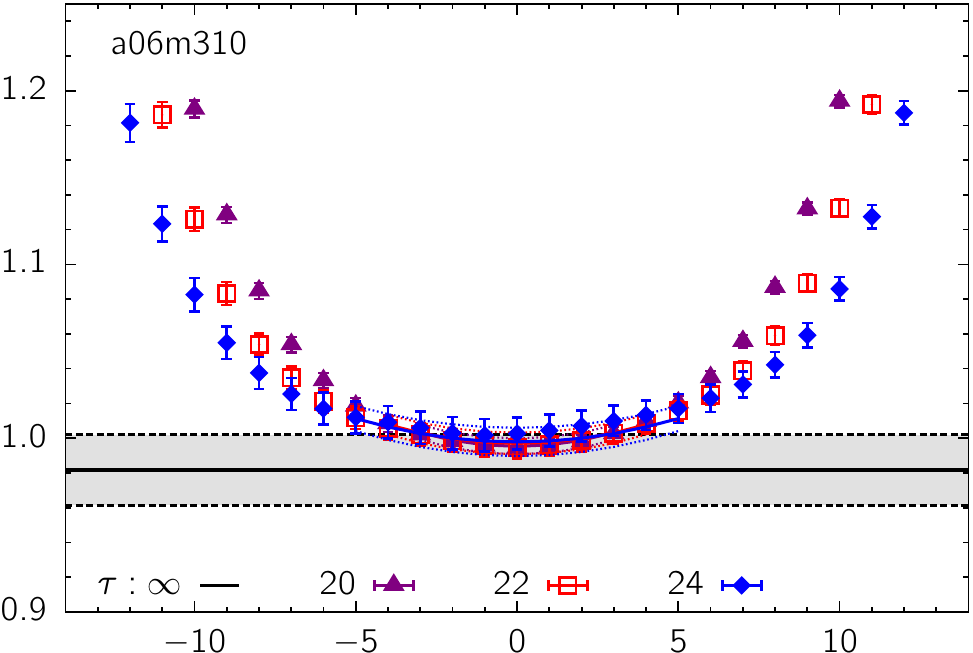}
  }
  \subfigure{
    \includegraphics[height=1.5in,trim={0.0cm 0.00cm 0 0},clip]{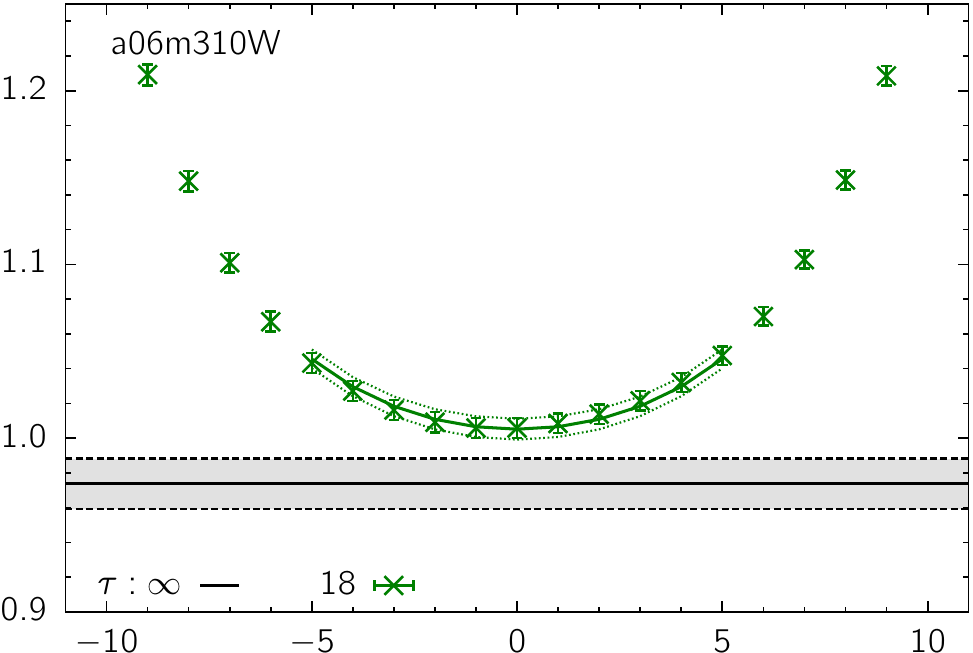}
    \includegraphics[height=1.5in,trim={0.0cm 0.00cm 0 0},clip]{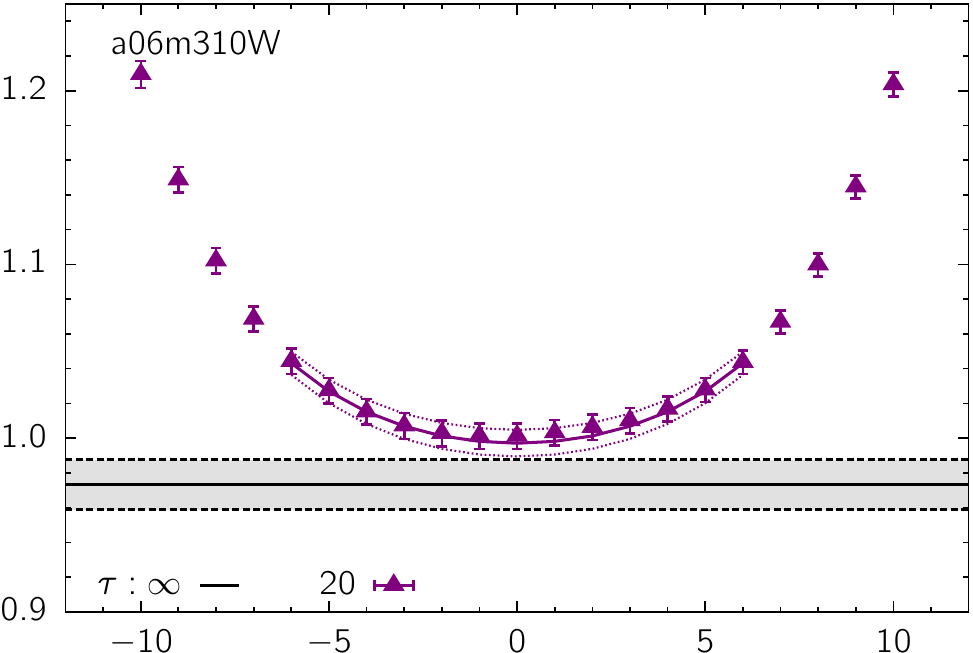}
    \includegraphics[height=1.5in,trim={0.0cm 0.00cm 0 0},clip]{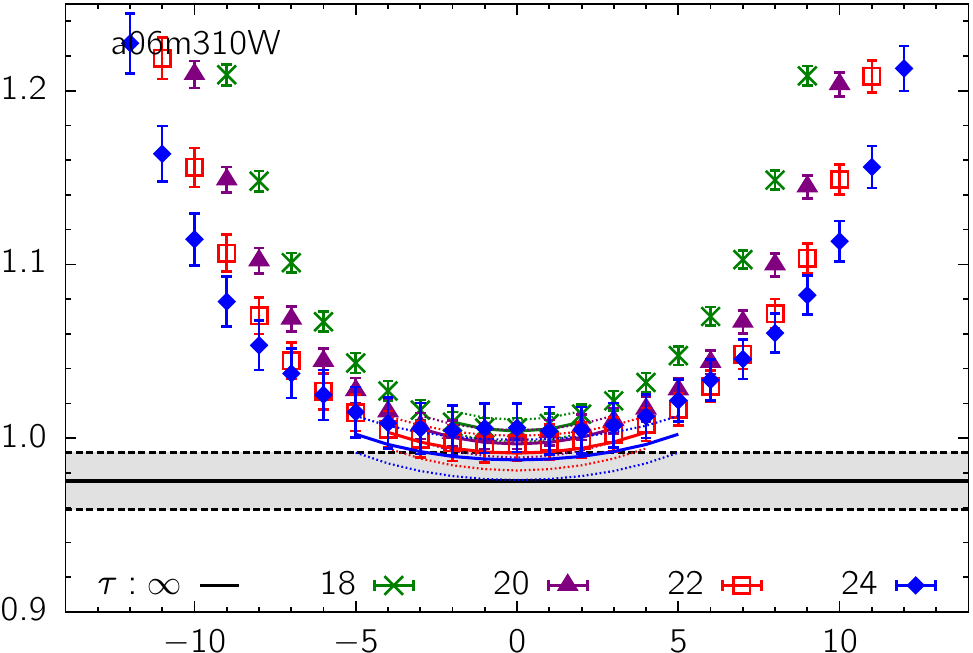}
  }
  \hspace{0.04\linewidth}
  \subfigure{
    \includegraphics[height=1.5in,trim={0.0cm 0.00cm 0 0},clip]{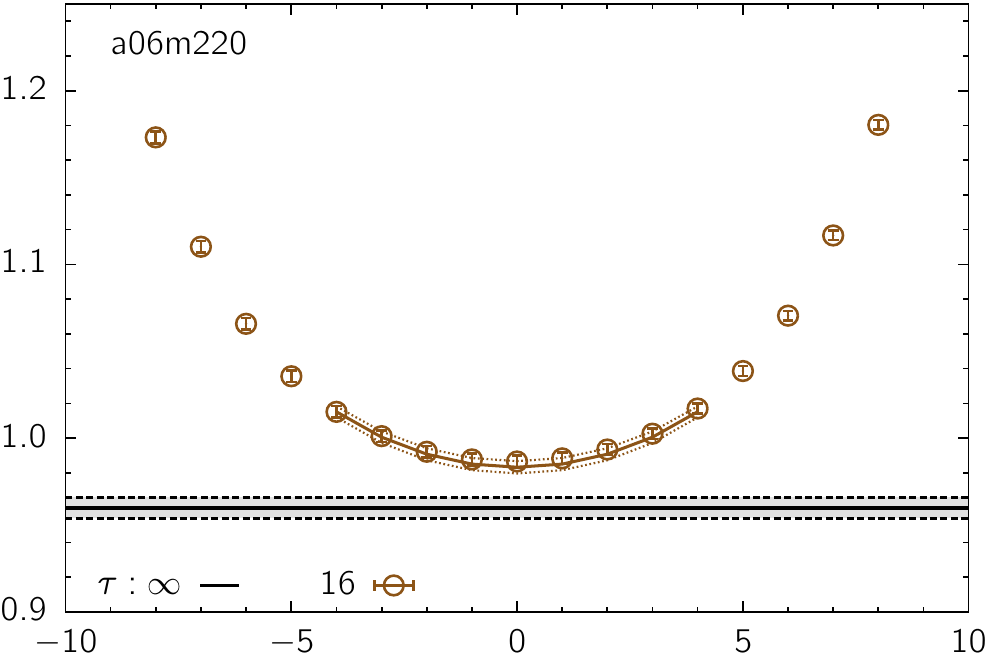}
    \includegraphics[height=1.5in,trim={0.0cm 0.00cm 0 0},clip]{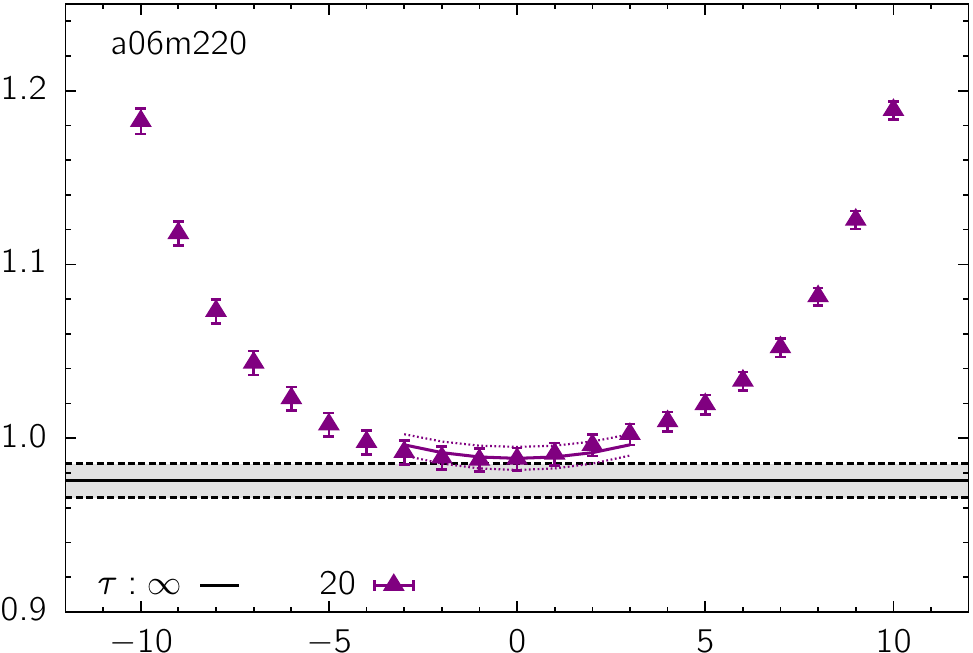}
    \includegraphics[height=1.5in,trim={0.0cm 0.00cm 0 0},clip]{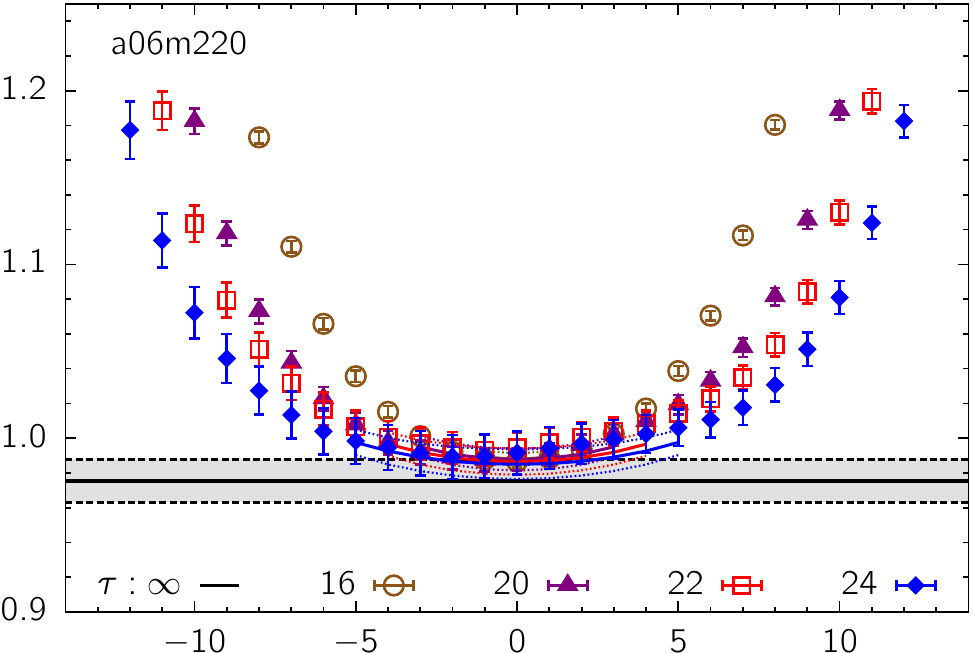}
  }
  \hspace{0.04\linewidth}
  \subfigure{
    \includegraphics[height=1.5in,trim={0.0cm 0.00cm 0 0},clip]{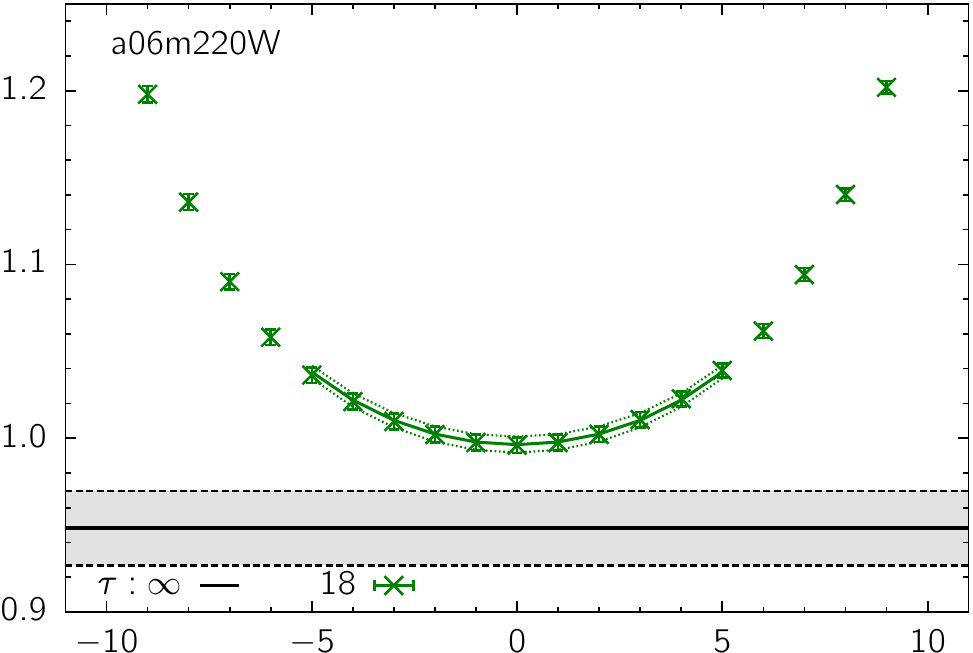}
    \includegraphics[height=1.5in,trim={0.0cm 0.00cm 0 0},clip]{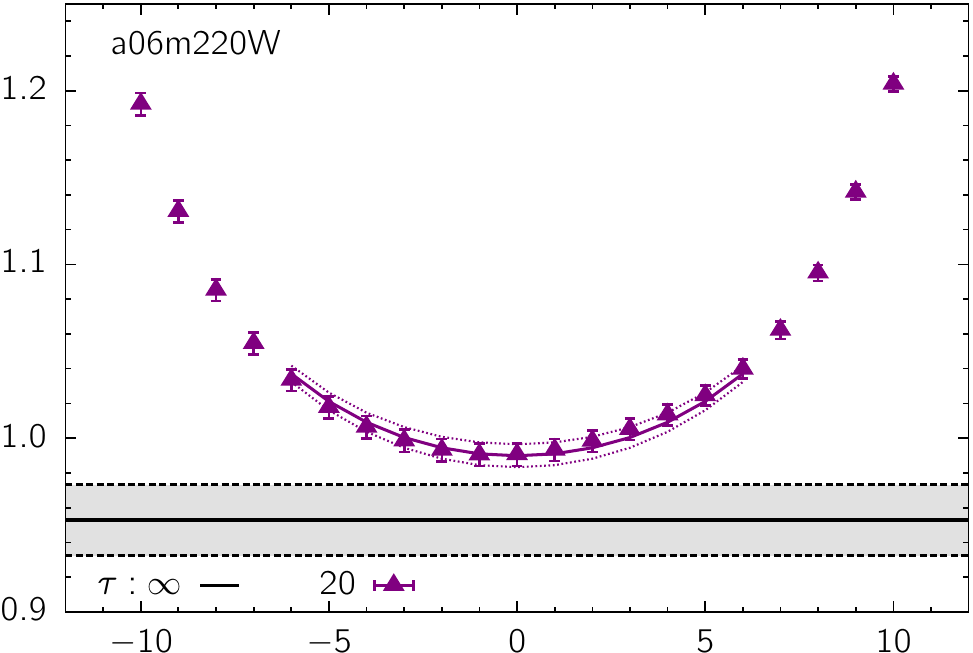}
    \includegraphics[height=1.5in,trim={0.0cm 0.00cm 0 0},clip]{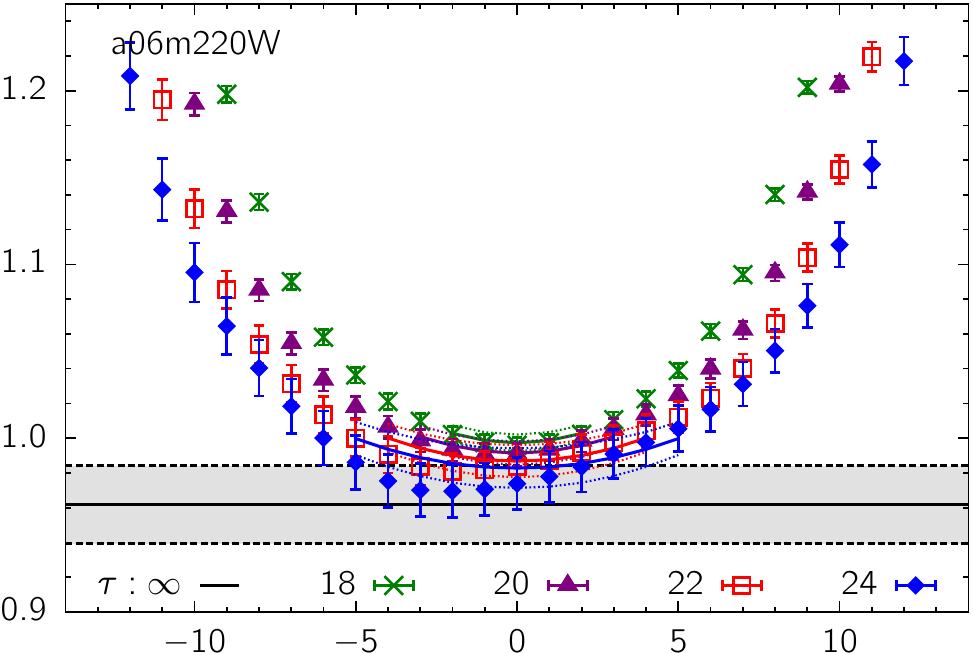}
  }
  \hspace{0.04\linewidth}
  \subfigure{
    \includegraphics[height=1.5in,trim={0.0cm 0.00cm 0 0},clip]{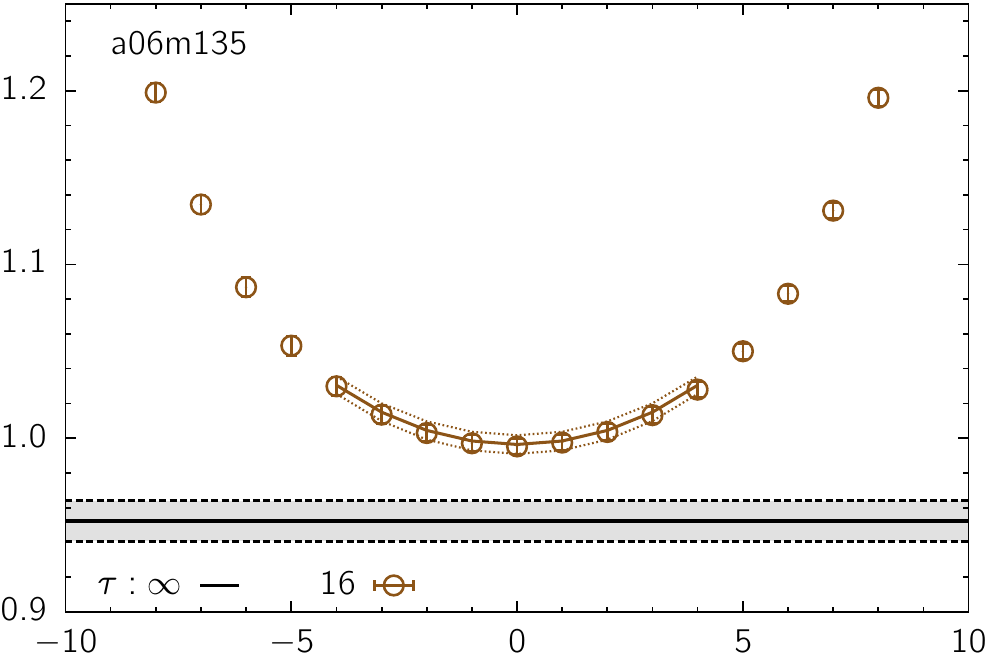}
    \includegraphics[height=1.5in,trim={0.0cm 0.00cm 0 0},clip]{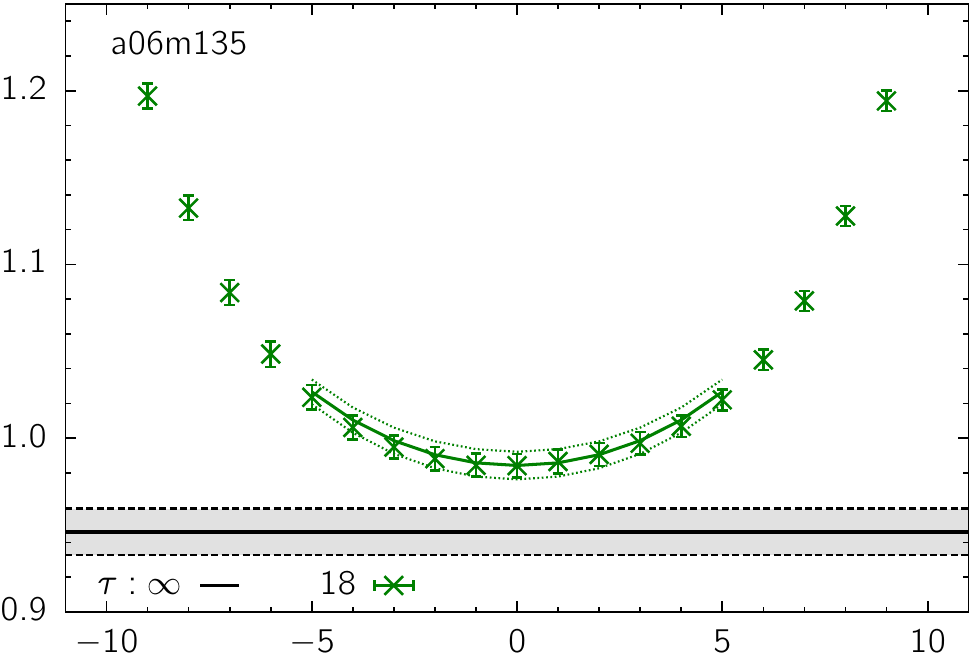}
    \includegraphics[height=1.5in,trim={0.0cm 0.00cm 0 0},clip]{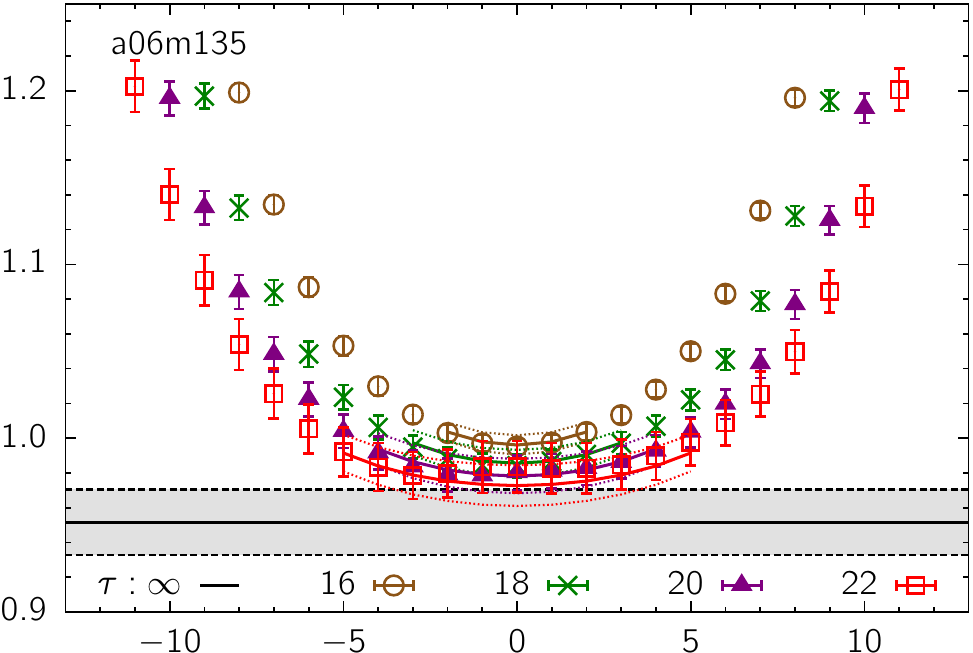}
  }
\caption{Comparison between the $2^\ast$ and
  $3^\ast$ fits to the tensor charge $g_T^{u-d}$ data from the 
  $a \approx 0.06$~fm ensembles. The rest is the same as in Fig.~\ref{fig:gA2v3a12}. 
  \label{fig:gT2v3a06}}
\end{figure*}

In this Appendix, we first present the masses and amplitudes obtained from fits
to the 2-point function using the spectral decomposition, 
Eq.~\eqref{eq:2pt}, in Table~\ref{tab:2ptmulti}. These are used as inputs
in the fits to the 3-point functions using Eq.~\eqref{eq:3pt}.  We then give in
Tables~\ref{tab:results3bareu-d} and~\ref{tab:results3bareu+d} the
results of $2^\ast$-, 2- and $3^\ast$-state fits used to control the
ESC in the extraction of the isovector and the connected contribution
to the isoscalar axial, scalar and tensor charges for the fourteen
calculations. The data and the $2^\ast$-, 2- and $3^\ast$-state fits
are shown in Figs.~\ref{fig:gA2v3a12}--\ref{fig:gT2v3a06}.  In each
case, we compare the $2^\ast$ fit on data from two source-sink
separations with $\tau \approx 1$~fm with the $2$- or $3^\ast$-state
fit using data from multiple values of $\tau$.


\begin{longtable*}{c|cc|cc|cc|cc|c}      
\hline
\hline
   &
  \multicolumn{1}{c}{$\mathcal{A}_0^2$} &
  \multicolumn{1}{c|}{$aM_0$} &
  \multicolumn{1}{c}{$r_1$} &
  \multicolumn{1}{c|}{$a\Delta M_1$} &
  \multicolumn{1}{c}{$r_2$} &
  \multicolumn{1}{c|}{$a\Delta M_2$} &
  \multicolumn{1}{c}{$r_3$} &
  \multicolumn{1}{c|}{$a\Delta M_3$} &
  \multicolumn{1}{c}{$\chi^2/\text{d.o.f}$}
  \\\hline
  & \multicolumn{9}{c} {\underline{$a15m310$ Smearing $\sigma=4.2$}}     \\  
Priors      & & & 0.5(3) & 0.7(4) & 0.3(2) & 0.4(2) & 0.3(2) & 0.4(2) &       \\
\{2,3--10\} & $8.76(17) \times 10^{-9}$  & 0.833(003) & 0.750(279) & 0.926(194) &  &  &  &  & 1.304 \\                  
\{3,1--10\} & $8.61(09) \times 10^{-9}$  & 0.831(002) & 0.479(013) & 0.769(026) & 0.251(013) & 0.316(047) &  &  & 0.892 \\  
\{4,1--10\} & $8.58(10) \times 10^{-9}$  & 0.830(002) & 0.420(042) & 0.729(048) & 0.241(011) & 0.281(034) & 0.084(061) & 0.366(016) & 1.146 \\
\hline
  & \multicolumn{9}{c} {\underline{$a12m310$ Smearing $\sigma=5.5$}}     \\  
Priors      & & & 0.15(10) & 0.4(2) & 0.8(6) & 0.6(3) & 0.6(4) & 0.4(2) &       \\
\{2,3--15\} & $6.86(11) \times 10^{-11}$ & 0.671(2) & 1.011(186) & 0.837(098) &  &  &  &  & 0.916 \\
\{3,2--15\} & $6.78(10) \times 10^{-11}$ & 0.670(2) & 0.143(028) & 0.450(038) & 1.137(063) & 0.563(075) &  &  & 0.747 \\
\{4,2--15\} & $6.75(10) \times 10^{-11}$ & 0.669(2) & 0.137(030) & 0.420(037) & 0.732(038) & 0.500(066) & 0.518(066) & 0.396(023) & 0.738 \\
\hline
  & \multicolumn{9}{c} {\underline{$a12m220S$ Smearing $\sigma=5.5$}}     \\  
Priors      & & & 0.4(3) & 0.3(2) & 1.0(8) & 0.8(4) & 0.8(6) & 0.4(2) &         \\
\{2,4--15\} & $5.69(51) \times 10^{-11}$ & 0.607(8) & 0.681(086) & 0.419(132) &  &  &  &  & 0.124 \\
\{3,2--15\} & $5.53(26) \times 10^{-11}$ & 0.605(4) & 0.488(079) & 0.310(036) & 1.591(226) & 0.968(110) &  &  & 0.181 \\
\{4,2--15\} & $5.46(32) \times 10^{-11}$ & 0.604(5) & 0.525(095) & 0.309(047) & 0.994(167) & 0.913(126) & 0.853(130) & 0.405(006) & 0.136 \\
\hline
  & \multicolumn{9}{c} {\underline{$a12m220$ Smearing $\sigma=5.5$}}     \\  
Priors      & & & 0.4(3) & 0.3(2) & 1.0(8) & 0.8(4) & 0.8(6) & 0.4(2) &        \\
\{2,4--15\} & $6.09(18) \times 10^{-11}$ & 0.612(3) & 0.832(303) & 0.637(157) &  &  &  &  & 0.285 \\
\{3,2--15\} & $5.86(19) \times 10^{-11}$ & 0.608(3) & 0.376(071) & 0.386(056) & 1.304(164) & 0.770(128) &  &  & 0.234 \\
\{4,2--15\} & $5.83(18) \times 10^{-11}$ & 0.608(3) & 0.365(131) & 0.372(070) & 0.801(071) & 0.670(145) & 0.631(124) & 0.404(011) & 0.254 \\
\hline
  & \multicolumn{9}{c} {\underline{$a12m220L_O$ Smearing $\sigma=5.5$}}     \\  
Priors      & & & 0.4(3) & 0.3(2) & 1.0(8) & 0.8(4) & 0.8(6) & 0.4(2) &       \\
\{2,4--15\} & $5.97(18) \times 10^{-11}$ & 0.612(3) & 0.669(118) & 0.529(100) &  &  &  &  & 1.363 \\
\{3,2--15\} & $5.75(22) \times 10^{-11}$ & 0.609(3) & 0.400(067) & 0.350(071) & 1.461(171) & 0.878(102) &  &  & 0.885 \\
\{4,2--15\} & $5.74(23) \times 10^{-11}$ & 0.609(3) & 0.400(091) & 0.349(085) & 0.873(099) & 0.775(117) & 0.725(107) & 0.405(010) & 0.881 \\
\hline
  & \multicolumn{9}{c} {\underline{$a12m220L$ Smearing $\sigma=5.5$}}     \\  
Priors      & & & 0.4(3) & 0.3(2) & 1.0(8) & 0.8(4) & 0.8(6) & 0.4(2) &       \\
\{2,4--15\} & $6.14(11) \times 10^{-11}$ & 0.615(2) & 0.825(165) & 0.642(088) &  &  &  &  & 0.216 \\                   
\{3,2--15\} & $5.96(13) \times 10^{-11}$ & 0.613(2) & 0.391(114) & 0.420(082) & 1.258(114) & 0.759(105) &  &  & 0.223 \\ 
\{4,2--15\} & $5.94(14) \times 10^{-11}$ & 0.612(2) & 0.371(152) & 0.406(106) & 0.763(064) & 0.645(112) & 0.611(083) & 0.411(011) & 0.233 \\
\hline
\pagebreak
\hline
   &
  \multicolumn{1}{c}{$\mathcal{A}_0^2$} &
  \multicolumn{1}{c|}{$aM_0$} &
  \multicolumn{1}{c}{$r_1$} &
  \multicolumn{1}{c|}{$a\Delta M_1$} &
  \multicolumn{1}{c}{$r_2$} &
  \multicolumn{1}{c|}{$a\Delta M_2$} &
  \multicolumn{1}{c}{$r_3$} &
  \multicolumn{1}{c|}{$a\Delta M_3$} &
  \multicolumn{1}{c}{$\chi^2/\text{d.o.f}$}
  \\
\hline
& \multicolumn{9}{c} {\underline{$a09m310$ Smearing $\sigma=7$}}     \\  
Priors      & & & 0.7(4) & 0.40(25) & 1.0(5) & 0.70(35) & 1.0(6) & 0.5(3) &        \\
\{2,4--18\} & $3.64(04) \times 10^{-13}$ & 0.496(1) & 0.924(052) & 0.500(029) &  &  &  &  & 1.438 \\
\{3,2--18\} & $3.60(06) \times 10^{-13}$ & 0.495(1) & 0.697(092) & 0.432(044) & 1.425(111) & 0.810(086) &  &  & 1.191 \\
\{4,2--18\} & $3.60(06) \times 10^{-13}$ & 0.495(1) & 0.702(140) & 0.434(058) & 0.854(051) & 0.696(133) & 0.807(129) & 0.526(024) & 1.146 \\
\hline
  & \multicolumn{9}{c} {\underline{$a09m220$ Smearing $\sigma=7$}}     \\  
Priors      &                            &            & 0.6(3)     & 0.30(15)   & 0.8(5)     & 0.4(2)     & 0.7(4)     & 0.4(2)     &          \\
\{2,5--20\} & $3.02(06) \times 10^{-13}$ & 0.451(2) & 0.937(067) & 0.407(034) &  &  &  &  & 0.466 \\
\{3,3--20\} & $2.99(07) \times 10^{-13}$ & 0.450(2) & 0.566(061) & 0.329(036) & 1.097(139) & 0.453(061) &  &  & 0.509 \\
\{4,3--20\} & $2.97(07) \times 10^{-13}$ & 0.450(2) & 0.529(076) & 0.314(040) & 0.723(074) & 0.370(056) & 0.591(098) & 0.386(031) & 0.502 \\
\hline
  & \multicolumn{9}{c} {\underline{$a09m130$ Smearing $\sigma=5.5$}}     \\  
Priors      &                    &            & 1.0(5)     & 0.20(15)   & 2.0(1.5)   & 0.6(3)     & 1.7(1.2)   & 0.4(2)     &       \\
\{2,6--20\} & $8.92(51) \times 10^{-11}$ & 0.417(4) & 1.322(083) & 0.329(041) &            &            &  &  & 0.727 \\
\{3,4--20\} & $8.15(72) \times 10^{-11}$ & 0.412(5) & 1.067(100) & 0.244(043) & 2.572(522) & 0.666(079) &  &  & 0.627 \\
\{4,4--20\} & $8.21(71) \times 10^{-11}$ & 0.412(5) & 1.104(089) & 0.253(043) & 1.924(389) & 0.661(082) & 1.771(242) & 0.402(020) & 0.597 \\
\hline
  & \multicolumn{9}{c} {\underline{$a09m130W$ Smearing $\sigma=7.0$}}     \\  
Priors      &                            &            & 0.7(4) & 0.35(20) & 0.7(5) & 0.5(3) & 1.0(6) & 0.35(20) &       \\
\{2,6--20\} & $2.74(07) \times 10^{-13}$ & 0.422(2) & 1.071(165) & 0.415(052) &  &  &  &  & 0.670 \\
\{3,4--20\} & $2.70(06) \times 10^{-13}$ & 0.421(2) & 0.794(076) & 0.359(031) & 0.857(214) & 0.482(051) &  &  & 0.533 \\
\{4,4--20\} & $2.69(06) \times 10^{-13}$ & 0.421(2) & 0.833(108) & 0.356(039) & 0.623(106) & 0.538(051) & 0.942(124) & 0.367(021) & 0.524 \\
\hline
& \multicolumn{9}{c} {\underline{$a06m310$ Smearing $\sigma=6.5$}}     \\  
Priors      &                            &            & 1.0(5)     & 0.16(10)   & 2.4(1.5)   & 0.3(2)     & 2.2(1.5)   & 0.3(2)     &       \\
\{2,0--30\} & $5.56(35) \times 10^{-12}$ & 0.326(3) & 1.362(097) & 0.199(026) &  &  &  &  & 1.371 \\
\{3,7--30\} & $5.46(39) \times 10^{-12}$ & 0.325(3) & 0.936(109) & 0.163(028) & 3.368(597) & 0.356(035) &  &  & 1.268 \\
\{4,7--30\} & $5.40(43) \times 10^{-12}$ & 0.325(3) & 0.964(116) & 0.161(031) & 2.554(366) & 0.338(037) & 2.323(334) & 0.276(042) & 1.238 \\
\hline
& \multicolumn{9}{c} {\underline{$a06m310W$ Smearing $\sigma=12.0$}}     \\  
Priors      &                            &            & 0.7(4)     & 0.3(2)     &   0.7(4)   &     0.3(2) &     0.7(4) &     0.3(2) &            \\
\{2,6--25\} & $1.34(04) \times 10^{-22}$ & 0.329(2) & 1.229(252) & 0.377(059) &  &  &  &  & 1.488 \\
\{3,4--25\} & $1.33(03) \times 10^{-22}$ & 0.328(1) & 0.769(042) & 0.317(031) & 0.860(135) & 0.293(061) &  &  & 1.162 \\
\{4,4--25\} & $1.31(04) \times 10^{-22}$ & 0.328(2) & 0.646(084) & 0.278(043) & 0.685(102) & 0.278(057) & 0.655(136) & 0.293(033) & 1.150 \\
\hline
& \multicolumn{9}{c} {\underline{$a06m220$ Smearing $\sigma=5.5$}}     \\  
Priors      &                            &            & 2.0(1.0)   & 0.25(20)   & 3.0(1.5)   & 0.3(2)     & 2.8(1.8)   & 0.3(2)     &     \\
\{2,0--30\} & $1.08(04) \times 10^{-10}$ & 0.305(2) & 2.900(348) & 0.286(025) &  &  &  &  & 1.774 \\
\{3,7--30\} & $1.06(04) \times 10^{-10}$ & 0.304(2) & 2.035(225) & 0.249(019) & 3.919(681) & 0.342(021) &  &  & 1.591 \\
\{4,7--30\} & $1.05(04) \times 10^{-10}$ & 0.304(2) & 2.066(240) & 0.245(021) & 3.185(345) & 0.344(022) & 3.078(406) & 0.267(048) & 1.548 \\
\hline
& \multicolumn{9}{c} {\underline{$a06m220W$ Smearing $\sigma=11.0$}}     \\  
Priors      &                            &            & 0.70(35)   & 0.25(15)   & 1.0(5)     & 0.3(2)     & 1.0(5)     & 0.3(2)     &       \\
\{2,7--20\} & $2.69(08) \times 10^{-20}$ & 0.305(2) & 1.174(176) & 0.321(043) &  &  &  &  & 0.289 \\
\{3,4--20\} & $2.66(08) \times 10^{-20}$ & 0.304(2) & 0.698(053) & 0.262(028) & 1.228(185) & 0.310(060) &  &  & 0.197 \\
\{4,4--20\} & $2.60(11) \times 10^{-20}$ & 0.303(2) & 0.592(090) & 0.229(042) & 0.862(094) & 0.260(056) & 0.779(157) & 0.290(056) & 0.223 \\
\hline
& \multicolumn{9}{c} {\underline{$a06m135$ Smearing $\sigma=9$}}     \\  
Priors      &      & & 1.3(7) & 0.20(15) & 1.3(1.0) & 0.3(2) & 1.1(9) & 0.3(2) &       \\
\{2,8--30\} & $2.89(17) \times 10^{-16}$ & 0.274(3) & 1.685(098) & 0.249(026) &  &  &  &  & 1.047 \\
\{3,6--30\} & $2.89(16) \times 10^{-16}$ & 0.274(3) & 1.371(121) & 0.229(024) & 1.683(448) & 0.373(031) &  &  & 1.010 \\
\{4,6--30\} & $2.86(16) \times 10^{-16}$ & 0.273(3) & 1.380(131) & 0.225(025) & 1.328(314) & 0.365(028) & 1.247(183) & 0.290(030) & 0.983 \\
\hline
  \caption{Results of 2-, 3- and 4-state fits to the two-point
    correlation function data for the fourteen calculations. The first column
    specifies the parameters, \{$N_\text{2pt}$,
    $t_\text{min}$--\,$t_\text{max}$\}, where $N_\text{2pt}$ is number of
    states used in the fits to the two-point correlators, and
    $[t_\text{min},t_\text{max}]$ is the fit interval in lattice units. The following
    columns give the nucleon ground state amplitude $\mathcal{A}_0^2$
    and mass $aM_0$, followed by the ratio of the excited state
    amplitudes $r_i = (\mathcal{A}_i/\mathcal{A}_0)^2$, and the mass gaps
    $a\Delta M_i = a(M_i-M_{i-1})$. For each ensemble, the first row
    gives the values of the priors used in the final 3- and 4-state fits.}
\vspace{2.0cm}
  \label{tab:2ptmulti}
\end{longtable*}

\begin{table*}   
\centering
\begin{ruledtabular}
\begin{tabular}{c|ccc|ccc}
ID              &  Fit Type  & $\tau$ values       & $\tskip$ & $g_A^{u-d}$ &      $g_S^{u-d}$       & $g_T^{u-d}$    \\ 
\hline
$a15m310 $      &  $3^\ast$  & \{5,6,7,8,9\}       &   1      &  1.250(07)  &    $ 0.80( 3)     $    &     1.121(06)   \\   
$a15m310 $      &  $2$       & \{6,7,8\}           &   1      &  1.250(07)  &    $ 0.87( 3)^\dag$    &     1.130(07)   \\   
$a15m310 $      &  $2^\ast$  & \{7\}               &   1      &  1.255(06)  &    $ 0.85( 2)     $    &     1.129(06)   \\   
$a15m310 $      &  $2^\ast$  & \{6\}               &   1      &  1.255(06)  &    $ 0.87( 2)     $    &     1.130(05)   \\   
\hline
$a12m310 $      &  $3^\ast$  & \{8,10,12\}         &   2      &  1.274(15)  &    $ 0.91( 6)     $    &     1.065(13)   \\   
$a12m310 $      &  $2$       & \{8,10,12\}         &   2      &  1.270(12)  &    $ 0.96( 5)^\dag$    &     1.051(13)   \\   
$a12m310 $      &  $2^\ast$  & \{10\}              &   2      &  1.268(11)  &    $ 0.93( 4)     $    &     1.054(12)   \\   
$a12m310 $      &  $2^\ast$  & \{8\}               &   1      &  1.277(10)  &    $ 1.03( 3)     $    &     1.038(15)   \\   
$a12m220S$      &  $3^\ast$  & \{8,10,12,14\}      &   2      &  1.266(44)  &    $ 1.04(29)     $    &     1.065(39)   \\  
$a12m220S$      &  $2$       & \{8,10,12\}         &   2      &  1.266(33)  &    $ 1.00(26)^\dag$    &     1.025(37)   \\  
$a12m220S$      &  $2^\ast$  & \{10\}              &   2      &  1.318(49)  &    $ 1.07(23)     $    &     1.025(37)   \\  
$a12m220S$      &  $2^\ast$  & \{8\}               &   1      &  1.358(52)  &    $ 1.25(13)     $    &     0.997(42)   \\  
$a12m220 $      &  $3^\ast$  & \{8,10,12,14\}      &   2      &  1.265(21)  &    $ 1.00(11)     $    &     1.048(18)   \\ 
$a12m220 $      &  $2$       & \{8,10,12\}         &   2      &  1.275(18)  &    $ 1.11( 9)^\dag$    &     1.030(28)   \\ 
$a12m220 $      &  $2^\ast$  & \{10\}              &   2      &  1.286(21)  &    $ 1.07( 8)     $    &     1.026(28)   \\ 
$a12m220 $      &  $2^\ast$  & \{8\}               &   1      &  1.303(28)  &    $ 1.10( 8)     $    &     1.006(43)   \\ 
$a12m220L_O$    &  $3^\ast$  & \{8,10,12,14\}      &   2      &  1.303(19)  &    $ 0.82( 6)     $    &     1.043(20)   \\   
$a12m220L_O$    &  $2$       & \{8,10,12\}         &   2      &  1.305(20)  &    $ 0.85( 6)^\dag$    &     1.017(38)   \\   
$a12m220L_O$    &  $2^\ast$  & \{10\}              &   2      &  1.315(30)  &    $ 0.94( 7)     $    &     1.023(34)   \\   
$a12m220L_O$    &  $2^\ast$  & \{8\}               &   1      &  1.337(50)  &    $ 1.04(11)     $    &     0.997(60)   \\   \
$a12m220L$      &  $3^\ast$  & \{8,10,12,14\}      &   2      &  1.289(13)  &    $ 0.75( 5)     $    &     1.069(11)   \\   
$a12m220L$      &  $2$       & \{8,10,12\}         &   2      &  1.291(17)  &    $ 0.87( 4)^\dag$    &     1.047(29)   \\   
$a12m220L$      &  $2^\ast$  & \{10\}              &   2      &  1.294(21)  &    $ 0.84( 3)     $    &     1.052(26)   \\   
$a12m220L$      &  $2^\ast$  & \{8\}               &   1      &  1.303(34)  &    $ 0.98( 8)     $    &     1.035(46)   \\   
\hline                                                                                                               
$a09m310 $      &  $3^\ast$  & \{12,14,16\}        &   3      &  1.238( 8)  &    $ 0.96( 3)     $    &     1.027( 7)   \\   
$a09m310 $      &  $2$       & \{10,12,14,16\}     &   3      &  1.221( 8)  &    $ 1.02( 3)^\dag$    &     1.022(10)   \\   
$a09m310 $      &  $2^\ast$  & \{12\}              &   2      &  1.223( 6)  &    $ 1.02( 3)     $    &     1.022(12)   \\   
$a09m310 $      &  $2^\ast$  & \{10\}              &   2      &  1.218( 7)  &    $ 1.03( 3)     $    &     1.024(13)   \\
$a09m220 $      &  $3^\ast$  & \{12,14,16\}        &   3      &  1.279(13)  &    $ 0.97( 6)     $    &     1.002(10)   \\   
$a09m220 $      &  $2$       & \{10,12,14,16\}     &   3      &  1.247( 9)  &    $ 1.05( 4)^\dag$    &     0.976(19)   \\   
$a09m220 $      &  $2^\ast$  & \{12\}              &   2      &  1.248(12)  &    $ 1.06( 4)     $    &     0.970(20)   \\   
$a09m220 $      &  $2^\ast$  & \{10\}              &   2      &  1.252(16)  &    $ 1.11( 6)     $    &     0.968(23)   \\   
$a09m130 $      &  $3^\ast$  & \{10,12,14\}        &   3      &  1.269(28)  &    $ 1.02(13)     $    &     0.961(22)   \\
$a09m130 $      &  $2$       & \{10,12,14\}        &   3      &  1.259(24)  &    $ 1.16(13)^\dag$    &     0.917(42)   \\
$a09m130 $      &  $2^\ast$  & \{12\}              &   3      &  1.302(53)  &    $ 1.15(13)     $    &     0.971(17)   \\
$a09m130 $      &  $2^\ast$  & \{10\}              &   2      &  1.319(66)  &    $ 1.38(18)     $    &     0.950(32)   \\
$a09m130W$      &  $3^\ast$  & \{12,14,16\}        &   3      &  1.271(15)  &    $ 0.72(12)     $    &     1.000(11)   \\
$a09m130W$      &  $2$       & \{10,12,14\}        &   3      &  1.247(12)  &    $ 1.05( 6)^\dag$    &     0.984(14)   \\
$a09m130W$      &  $2^\ast$  & \{12\}              &   3      &  1.257(12)  &    $ 1.00( 5)     $    &     0.995(11)   \\
$a09m130W$      &  $2^\ast$  & \{10\}              &   2      &  1.250(14)  &    $ 1.12( 4)     $    &     0.988(15)   \\
\hline                                                                                                               
$a06m310 $      &  $3^\ast$  & \{20,22,24\}        &   7      &  1.243(27)  &    $ 1.27(13)     $    &     0.982(20)   \\   
$a06m310 $      &  $2$       & \{20,22,24\}        &   7      &  1.250(26)  &    $ 1.24(11)^\dag$    &     0.965(21)   \\   
$a06m310 $      &  $2^\ast$  & \{20\}              &   4      &  1.239(41)  &    $ 1.34(14)     $    &     0.935(22)   \\   
$a06m310 $      &  $2^\ast$  & \{16\}              &   4      &  1.267(61)  &    $ 1.40(19)     $    &     0.916(36)   \\   
$a06m310W$      &  $3^\ast$  & \{18,20,22,24\}     &   7      &  1.216(21)  &    $ 1.10( 8)     $    &     0.975(16)   \\   
$a06m310W$      &  $2$       & \{18,20,22\}        &   7      &  1.208(15)  &    $ 1.12( 7)^\dag$    &     0.972(15)   \\   
$a06m310W$      &  $2^\ast$  & \{20\}              &   4      &  1.203(10)  &    $ 1.14( 5)     $    &     0.973(14)   \\   
$a06m310W$      &  $2^\ast$  & \{18\}              &   4      &  1.203( 8)  &    $ 1.18( 5)     $    &     0.974(15)   \\   
$a06m220 $      &  $3^\ast$  & \{16,20,22,24\}     &   7      &  1.235(18)  &    $ 1.18( 8)     $    &     0.975(12)   \\  
$a06m220 $      &  $2$       & \{16,20,22\}        &   7      &  1.208(14)  &    $ 1.11( 7)^\dag$    &     0.966(10)   \\  
$a06m220 $      &  $2^\ast$  & \{20\}              &   4      &  1.213(15)  &    $ 1.12( 6)     $    &     0.969( 8)   \\  
$a06m220 $      &  $2^\ast$  & \{16\}              &   4      &  1.191(18)  &    $ 1.10( 5)     $    &     0.960( 6)   \\  
$a06m220W$      &  $3^\ast$  & \{18,20,22,24\}     &   7      &  1.257(24)  &    $ 0.78(12)     $    &     0.962(22)   \\  
$a06m220W$      &  $2$       & \{18,20,22\}        &   7      &  1.239(17)  &    $ 0.77( 9)^\dag$    &     0.959(20)   \\  
$a06m220W$      &  $2^\ast$  & \{20\}              &   4      &  1.228(14)  &    $ 0.94( 8)     $    &     0.953(20)   \\  
$a06m220W$      &  $2^\ast$  & \{18\}              &   4      &  1.222(14)  &    $ 1.02( 7)     $    &     0.948(21)   \\  
$a06m135 $      &  $3^\ast$  & \{16,18,20,22\}     &   6      &  1.240(26)  &    $ 0.92(15)     $    &     0.952(19)   \\  
$a06m135 $      &  $2$       & \{16,18,20\}        &   6      &  1.218(17)  &    $ 1.00(13)^\dag$    &     0.925(21)   \\  
$a06m135 $      &  $2^\ast$  & \{18\}              &   4      &  1.215(17)  &    $ 1.05( 9)     $    &     0.946(13)   \\  
$a06m135 $      &  $2^\ast$  & \{16\}              &   4      &  1.224(21)  &    $ 1.10( 7)     $    &     0.952(12)   \\  
\end{tabular}
\end{ruledtabular}
\caption{ Estimates of the unrenormalized connected contribution to
  the isovector charges $g_{A,S,T}^{u-d}$.  Results from four
  different fits to control ESC are shown: the $3^\ast$-state fits to
  multiple values of $\tau$ listed in the third column from which the
  axial and tensor charges are extracted, the $2$-state fits from
  which the scalar charge is determined, and the two $2^\ast$-state
  fits to data with the smallest values of $\tau$. The $2$-state fit
  values of $g_S^{u-d}$ used in our final analysis are marked with a
  ${}^\dag$.}
\label{tab:results3bareu-d}
\end{table*}

\begin{table*}
\centering
\begin{ruledtabular}
\begin{tabular}{c|ccc|ccc}
ID           &  Fit Type   & $\tau$ values     & $\tskip$ &  $g_A^{u+d}$ & $g_S^{u+d}$           & $g_T^{u+d}$   \\ 
\hline                                                   
$a15m310 $   &  $3^\ast$   &  \{5,6,7,8,9\}    &    1  &      0.624( 8)  &   $  5.42( 7)      $  &  0.682( 8)    \\
$a15m310 $   &  $2     $   &  \{8,10,12\}      &    1  &      0.621( 7)  &   $  5.34(14)^\dag $  &  0.688( 6)    \\
$a15m310 $   &  $2^\ast$   &  \{10\}           &    1  &      0.629( 7)  &   $  5.42(10)      $  &  0.688( 6)    \\
$a15m310 $   &  $2^\ast$   &  \{8\}            &    1  &      0.624( 5)  &   $  5.38(10)      $  &  0.687( 5)    \\
\hline                                                   
$a12m310 $   &  $3^\ast$   &  \{8,10,12\}      &    2  &      0.618(19)  &   $  6.21(15)      $  &  0.654(16)    \\
$a12m310 $   &  $2     $   &  \{8,10,12\}      &    2  &      0.615(11)  &   $  6.34(22)^\dag $  &  0.641(10)    \\
$a12m310 $   &  $2^\ast$   &  \{10\}           &    2  &      0.606(13)  &   $  6.30(18)      $  &  0.647(11)    \\
$a12m310 $   &  $2^\ast$   &  \{8\}            &    1  &      0.614( 7)  &   $  6.44(23)      $  &  0.634( 9)    \\
$a12m220S$   &  $3^\ast$   &  \{8,10,12,14\}   &    2  &      0.603(57)  &   $  9.42(71)      $  &  0.567(68)    \\
$a12m220S$   &  $2     $   &  \{8,10,12\}      &    2  &      0.601(38)  &   $  9.46(85)^\dag $  &  0.539(39)    \\
$a12m220S$   &  $2^\ast$   &  \{10\}           &    2  &      0.577(46)  &   $  9.91(82)      $  &  0.569(50)    \\
$a12m220S$   &  $2^\ast$   &  \{8\}            &    1  &      0.611(19)  &   $ 10.45(111)     $  &  0.576(35)    \\
$a12m220 $   &  $3^\ast$   &  \{8,10,12,14\}   &    2  &      0.629(30)  &   $  8.23(41)      $  &  0.646(23)    \\
$a12m220 $   &  $2     $   &  \{8,10,12\}      &    2  &      0.621(16)  &   $  8.55(63)^\dag $  &  0.623(20)    \\
$a12m220 $   &  $2^\ast$   &  \{10\}           &    2  &      0.607(19)  &   $  8.33(50)      $  &  0.632(20)    \\
$a12m220 $   &  $2^\ast$   &  \{8\}            &    1  &      0.625(10)  &   $  8.52(78)      $  &  0.619(24)    \\
$a12m220L_O$ &  $3^\ast$   &  \{8,10,12,14\}   &    2  &      0.613(16)  &   $  7.62(38)      $  &  0.625(17)    \\
$a12m220L_O$ &  $2     $   &  \{8,10,12\}      &    2  &      0.604(13)  &   $  7.79(63)^\dag $  &  0.611(25)    \\
$a12m220L_O$ &  $2^\ast$   &  \{10\}           &    2  &      0.606(13)  &   $  7.97(72)      $  &  0.614(22)    \\
$a12m220L_O$ &  $2^\ast$   &  \{8\}            &    1  &      0.629( 9)  &   $  8.48(121)     $  &  0.611(30)    \\
$a12m220L$   &  $3^\ast$   &  \{8,10,12,14\}   &    2  &      0.595(11)  &   $  7.34(27)      $  &  0.622(13)    \\
$a12m220L$   &  $2     $   &  \{8,10,12\}      &    2  &      0.601( 8)  &   $  7.55(55)^\dag $  &  0.619(20)    \\
$a12m220L$   &  $2^\ast$   &  \{10\}           &    2  &      0.595( 9)  &   $  7.54(51)      $  &  0.610(22)    \\
$a12m220L$   &  $2^\ast$   &  \{8\}            &    1  &      0.606( 7)  &   $  7.84(89)      $  &  0.617(27)    \\
\hline                                                                                           
$a09m310 $   &  $3^\ast$   &  \{12,14,16\}     &    3  &      0.622( 9)  &   $  6.29(14)      $  &  0.621( 8)    \\
$a09m310 $   &  $2     $   &  \{10,12,14,16\}  &    3  &      0.609( 4)  &   $  6.17(22)^\dag $  &  0.619( 7)    \\
$a09m310 $   &  $2^\ast$   &  \{12\}           &    2  &      0.611( 4)  &   $  6.23(19)      $  &  0.621( 7)    \\
$a09m310 $   &  $2^\ast$   &  \{10\}           &    2  &      0.609( 3)  &   $  6.16(23)      $  &  0.621( 7)    \\
$a09m220 $   &  $3^\ast$   &  \{12,14,16\}     &    3  &      0.611(14)  &   $  7.75(19)      $  &  0.597(11)    \\
$a09m220 $   &  $2     $   &  \{10,12,14,16\}  &    3  &      0.592( 6)  &   $  7.87(35)^\dag $  &  0.585(12)    \\
$a09m220 $   &  $2^\ast$   &  \{12\}           &    2  &      0.587( 6)  &   $  7.89(36)      $  &  0.583(12)    \\
$a09m220 $   &  $2^\ast$   &  \{10\}           &    2  &      0.591( 4)  &   $  7.94(48)      $  &  0.588(12)    \\
$a09m130 $   &  $3^\ast$   &  \{10,12,14\}     &    3  &      0.569(24)  &   $ 10.37(57)      $  &  0.569(23)    \\
$a09m130 $   &  $2     $   &  \{10,12,14\}     &    3  &      0.565(15)  &   $ 10.58(89)^\dag $  &  0.548(27)    \\
$a09m130 $   &  $2^\ast$   &  \{12\}           &    3  &      0.594(23)  &   $ 11.06(109      $  &  0.600(17)    \\
$a09m130 $   &  $2^\ast$   &  \{10\}           &    2  &      0.616(22)  &   $ 11.54(148)     $  &  0.581(19)    \\
$a09m130W$   &  $3^\ast$   &  \{12,14,16\}     &    3  &      0.599(18)  &   $  9.58(30)      $  &  0.594(15)    \\
$a09m130W$   &  $2     $   &  \{10,12,14\}     &    3  &      0.586( 8)  &   $  9.52(30)^\dag $  &  0.589(10)    \\
$a09m130W$   &  $2^\ast$   &  \{12\}           &    3  &      0.595( 8)  &   $  9.56(29)      $  &  0.600( 9)    \\
$a09m130W$   &  $2^\ast$   &  \{10\}           &    2  &      0.587( 5)  &   $  9.61(37)      $  &  0.586(10)    \\
 \hline                                                                                           
$a06m310 $   &  $3^\ast$   &  \{20,22,24\}     &    7  &      0.603(32)  &   $  7.59(39)      $  &  0.588(25)    \\
$a06m310 $   &  $2     $   &  \{20,22,24\}     &    7  &      0.593(20)  &   $  7.71(57)^\dag $  &  0.586(20)    \\
$a06m310 $   &  $2^\ast$   &  \{20\}           &    4  &      0.601(17)  &   $  8.21(88)      $  &  0.571(17)    \\
$a06m310 $   &  $2^\ast$   &  \{16\}           &    4  &      0.621(23)  &   $  8.44(120)     $  &  0.561(21)    \\
$a06m310W$   &  $3^\ast$   &  \{18,20,22,24\}  &    7  &      0.596(33)  &   $  7.14(21)      $  &  0.592(18)    \\
$a06m310W$   &  $2     $   &  \{18,20,22\}     &    7  &      0.607(16)  &   $  7.00(26)^\dag $  &  0.595(13)    \\
$a06m310W$   &  $2^\ast$   &  \{20\}           &    4  &      0.596(12)  &   $  6.86(24)      $  &  0.592(12)    \\
$a06m310W$   &  $2^\ast$   &  \{18\}           &    4  &      0.596( 9)  &   $  6.86(26)      $  &  0.587(10)    \\
$a06m220 $   &  $3^\ast$   &  \{16,20,22,24\}  &    7  &      0.588(19)  &   $  7.92(20)      $  &  0.582(15)    \\
$a06m220 $   &  $2     $   &  \{16,20,22\}     &    7  &      0.592(13)  &   $  7.69(21)^\dag $  &  0.585(10)    \\
$a06m220 $   &  $2^\ast$   &  \{20\}           &    4  &      0.586(12)  &   $  7.71(22)      $  &  0.598( 9)    \\
$a06m220 $   &  $2^\ast$   &  \{16\}           &    4  &      0.593(10)  &   $  7.49(27)      $  &  0.595( 6)    \\
$a06m220W$   &  $3^\ast$   &  \{18,20,22,24\}  &    7  &      0.576(32)  &   $  7.83(28)      $  &  0.567(25)    \\
$a06m220W$   &  $2     $   &  \{18,20,22\}     &    7  &      0.608(19)  &   $  7.87(38)      $  &  0.579(16)    \\
$a06m220W$   &  $2^\ast$   &  \{20\}           &    4  &      0.596(13)  &   $  7.95(45)      $  &  0.576(14)    \\
$a06m220W$   &  $2^\ast$   &  \{18\}           &    4  &      0.596(11)  &   $  7.98(48) ^\dag$  &  0.577(13)    \\
$a06m135 $   &  $3^\ast$   &  \{16,18,20,22\}  &    6  &      0.594(26)  &   $  9.42(31)      $  &  0.585(21)    \\
$a06m135 $   &  $2     $   &  \{16,18,20\}     &    6  &      0.590(17)  &   $  9.52(35)      $  &  0.563(16)    \\
$a06m135 $   &  $2^\ast$   &  \{18\}           &    4  &      0.574(15)  &   $  9.64(42)      $  &  0.568(13)    \\
$a06m135 $   &  $2^\ast$   &  \{16\}           &    4  &      0.575(13)  &   $  9.67(50) ^\dag$  &  0.572(11)    \\
\end{tabular}
\end{ruledtabular}
\caption{
  Estimates of the unrenormalized connected contribution to the
  isoscalar charges $g_{A,S,T}^{u+d}$.  Results from four different
  fits to control ESC are shown: the $3^\ast$-state fits to multiple
  values of $\tau$ listed in the third column from which the axial
  and tensor charges are extracted, the $2$-state fits from which
  the scalar charge is determined, and the two $2^\ast$-state fits to data with the
  smallest values of $\tau$. The $2$-state fit values of $g_S^{u+d}$
  used in our final analysis are marked with a ${}^\dag$.}
\label{tab:results3bareu+d}
\end{table*}


\begin{acknowledgments}
We thank the MILC collaboration for providing the 2+1+1-flavor HISQ
lattices used in our calculations. The calculations used the CHROMA
software suite~\cite{Edwards:2004sx}. Simulations were carried out on
computer facilities at (i) the National Energy Research Scientific
Computing Center, a DOE Office of Science User Facility supported by
the Office of Science of the U.S. Department of Energy under Contract
No. DE-AC02-05CH11231; and, (ii) the Oak Ridge Leadership Computing
Facility at the Oak Ridge National Laboratory, which is supported by
the Office of Science of the U.S. Department of Energy under Contract
No. DE-AC05-00OR22725; (iii) the USQCD collaboration, which are funded
by the Office of Science of the U.S. Department of Energy, and (iv)
Institutional Computing at Los Alamos National Laboratory.
T. Bhattacharya and R. Gupta were partly supported by the
U.S. Department of Energy, Office of Science, Office of High Energy
Physics under Contract No.~DE-AC52-06NA25396.  T. Bhattacharya,
V. Cirigliano, R. Gupta, Y.-C. Jang and B.Yoon were partly supported by
the LANL LDRD program.  The work of H.-W. Lin is supported by the US National
Science Foundation under grant PHY 1653405 ``CAREER: Constraining 
Parton Distribution Functions for New-Physics Searches''.
\end{acknowledgments}

\clearpage
%
\bibliography{ref} 

\end{document}